%% file: main.tex
\documentclass[12pt,titlepage,makeidx]{book}
\usepackage{slivro}
\usepackage{emlines}
\usepackage{emlines2}
\usepackage{latexsym}
\usepackage{amsmath}
\usepackage{graphics}
\usepackage{epsfig}
\usepackage{ifpdf} 

\catcode`@=11
\def\cases#1{\left\{\,\vcenter{\normalbaselines\m@th
    \ialign{$##\hfil$&\quad##\hfil\crcr#1\crcr}}\right.}
\catcode`@=12

\renewcommand{\baselinestretch}{1.1}
\def\argmin \mathop{\rm argmin}
\def\Re{I\kern -0.37 em R}
\def\Na{I\kern -0.37 em N}

\def\fto{\colon\!} 
\def\g{\,\vert \,}
\def\ate{\,\mbox{:}\,}

\def\Vec{\,\mbox{Vec}} 
 
\def\kron{\otimes} 
 
\def\arg{\mbox{arg}} 
\def\diag{\mbox{diag}}

\def\uno{\mbox{\bf 1}}

\def\g{\,\vert\,}

\def\arg{\mbox{arg}}

\def\uno{\mbox{\bf 1}}

\def\Re{{\mathcal{R}}} 

\def\Then{\mbox{ ent\~{a}o }}

\def\Or{\mbox{ ou }}

\def\ate{\colon\!} 
\def\para{\ // \ } 
\def\lambdab{\overline{\lambda}}

\makeindex

\begin{document}


\begin{titlepage}
\thispagestyle{empty}
\begin{center}
\Huge\bf
\null\vfill
\vspace{1.0cm}
Otimiza\c{c}\~ao e Processos Estoc\'{a}sticos
Aplicados \`{a} Economia e Finan\c{c}as \\
\vspace{1.0cm}
\large 
Compilado em 07 de Setembro de 2007. \\ 
\vspace{1.5cm}
\large

Julio Michael Stern  \footnote{ Julio M.  Stern (jstern@ime.usp.br) \'e
Ph.D.  em Pesquisa Operacional e Engenharia Industrial pela Universidade
de Cornell (Ithaca, NY, USA).
 Atualmente \'{e} Professor Livre Docente do Departamento de Ci\^{e}ncia da
Computa\c{c}\~{a}o do Instituto de Matem\'{a}tica e Estat\'{\i}stica da
Universidade de S\~{a}o Paulo, MAC-IME-USP, e consultor na \'{a}rea de
Pesquisa Operacional.
 }
\\
Carlos Alberto de Bragan\c{c}a Pereira
\footnote{
Carlos Alberto de Bragan\c{c}a Pereira (cpereira@ime.usp.br) \'e Ph.D. em
Estat\'{\i}stica Pela Universidade da Florida, (Talahasse, FL, EUA),
Professor Titular do Departamento de Estat\'{\i}stica do IME-USP,
e consultor na \'{a}rea de Pesquisa Operacional..
 }
\\
Celma de Oliveira Ribeiro
\footnote{
Celma de O. Ribeiro (celma@ime.usp.br) \'e Doutora em
Engenharia de Produ\c{c}\~ao pela Escola Polit\'ecnica
da USP (EPUSP). \'{E} docente do Departamento de Engenharia
de Produ\c{c}\~ao da EPUSP, atuando em \'areas ligadas a
otimiza\c{c}\~ao e finan\c{c}as.
 }
\\
Cibele Dunder
\footnote{
Cibele Dunder (dunder@ime.usp.br) \'e Mestre em Matem\'{a}tica
Aplicada pelo Instituto de Matem\'atica da USP, e consultora
em \'areas ligadas a otimiza\c{c}\~ao e finan\c{c}as.
 }
\\
Fabio Nakano
\footnote{
Fabio Nakano (nakano@ime.usp.br) \'e Mestre em Computa\c{c}\~{a}o
pelo Instituto de Matem\'atica da USP, e s\'{o}cio da Supremum
Assessoria e Consultoria, que atua em \'areas ligadas a
Engenharia de Sistemas e Pesquisa Operacional.
 }
\\
Marcelo Lauretto
\footnote{
Marcelo Lauretto (lauretto@ime.usp.br) \'e Mestre em Computa\c{c}\~{a}o
pelo Instituto de Matem\'atica da USP, e s\'{o}cio da Supremum
Assessoria e Consultoria, que atua em \'areas ligadas a
Engenharia de Sistemas e Pesquisa Operacional.
 }

\vspace{2.0cm}

\vfill
\end{center}
\end{titlepage}


\addtocounter{page}{1}

\tableofcontents
\markboth{}{}

\include{prefacio}

\setcounter{chapter}{0}
\input{transfig}

\include{cap1}

\include{cap2}

\include{cap3}

\include{cap4}

\include{cap5}

\include{cap6}

\include{cap7}

\include{cap8}

\include{cap9}

\include{cap12}

\include{apen1}

\include{apen2}

\include{apen3}

\include{apen4}

\include{apen5}

\include{bibbook}

\end{document}

%% file: prefacio.tex
\chapter*{Pref\'{a}cio} 
\addcontentsline{toc}{chapter}{Pref\'{a}cio}  
\pagestyle{myheadings} 
\markboth{Pref\'{a}cio}{Pref\'{a}cio}


 Este \'{e} o livro texto do curso Otimiza\c{c}\~{a}o e Processos
Estoc\'{a}sticos Aplicados \`{a} Economia e Finan\c{c}as. 
 Parte deste material \'{e} usado no curso M\'etodos de Otimiza\c{c}\~ao
em Finan\c{c}as, ministrado regularmente no Instituto de Matem\'atica e
Estat\'{\i}stica da Universidade de S\~ao Paulo, desde 1993.  
 Este curso foi tamb\'{e}m oferecido no 
 XIX Congresso  Nacional de Matem\'atica Aplicada
e Computacional, realizado  pela SBMAC em 1996,  em Goi\^{a}nia. 
 Parte deste material foi  tamb\'{e}m utilizado em cursos oferecidos na
 FIPE - Funda\c{c}\~ao Instituto  de Pesquisas Econ\^omicas da FEA-USP -
Faculdade de  Economia, Administra\c{c}\~{a}o e Contabilidade, e no 
 ILA - Instituto de Log\'{\i}stica  da Aeron\'autica. 
 
 Parte deste texto e o software que o acompanha, 
 {\em Critical-Point\copyright }, foi desenvolvido no NOPEF-USP - 
 N\'{u}cleo de Apoio \`{a} Pesquisa em Otimiza\c{c}\~{a}o e Processos
Estoc\'{a}sticos Aplicados \`{a} Economia e Finan\c{c}as, com
patroc\'{\i}nio da {\em BM\&F} - Bolsa de Mercadorias e de Futuros de
S\~{a}o Paulo.  
 Os direitos autorais, do texto e do software, pertencem aos respectivos
autores. \'{E} proibida a reprodu\c{c}\~{a}o do texto e do software que
o acompanha, sem a permiss\~{a}o do primeiro autor. \'{E} proibido o uso
de trechos do c\'{o}digo objeto, a desassemblagem, decompila\c{c}\~{a}o,
ou engenharia reversa do software. A vers\~{a}o estudantil do software, 
{\it Student Critical Point} destina-se exclusivamente para fins 
educacionais, sendo vedado seu uso em apica\c{c}\~{o}es com fins
lucrativos.  

 Os cap\'{\i}tulos 1 e 2 prov\^{e}m uma introdu\c{c}\~{a}o b\'{a}sica  
a programa\c{c}\~{a}o Linear e N\~{a}o Linear. 
 Os cap\'{i}tulos 3 e 4 estudam em detalhe o problema de 
programa\c{c}\~{a}o quadr\'{a}tica e sua aplica\c{c}\~{a}o 
\`a teoria matem\'{a}tica da forma\c{c}\~{a}o de portf\'{o}lios.  
 Os modelos abordados baseiam-se na caracteriza\c{c}\~{a}o da fronteira
eficiente, do modelo de Markowitz e de suas propriedades.  N\~{a}o
h\'{a} maneira de computar a fronteira eficiente e deduzir suas
caracter\'{\i}sticas sem compreender um dos v\'{a}rios algoritmos de
programa\c{c}\~{a}o quadr\'{a}tica param\'{e}trica.  
 No cap\'{\i}tulo 3 apresentamos um algoritmo did\'{a}tico, fundamentado
por alguns fatos da teoria de programa\c{c}\~{a}o linear, apresentados
no cap\'{\i}tulo 1.  Este algoritmo, uma vers\~{a}o modificada de
algoritmos de Cottle, Dantzig, Lemke e Wolfe, \'{e} muito mais
sint\'{e}tico do que o algoritmo originalmente apresentado por
Markowitz, e a demonstra\c{c}\~{a}o de sua corretude muito mais
f\'{a}cil e curta. Todos os fatos deduzidos nestes cap\'{\i}tulos
s\~{a}o reapresentados (sem demonstra\c{c}\~{a}o) quando necess\'{a}rios
aos modelos financeiros dos cap\'{\i}tulos subseq\"{u}entes; um leitor 
menos rigoroso pode omitir estas demonstra\c{c}\~{o}es numa primeira
leitura.  
 O cap\'{\i}tulo 4 apresenta a pedra fundamental de toda a moderna
teoria de finan\c{c}as, o modelo de Markowitz, e uma s\'{e}rie de
modelos derivados, como os modelos de Tobin, Brennan e modelos de
\'{\i}ndices, bem como modelos de equil\'{\i}brio, como CAPM e APT.  

 Os cap\'{\i}tulos 5 e 6 exp\~oem os princ\'{\i}pios b\'asicos de 
Progama\c{c}\~ao Din\^amica. O cap\'{\i}tulo 5, inclui Cadeias de Markov,   
nos d\'a uma alternativa simples para modelagem de problemas 
multiper\'{\i}odo, exemplificadas pela precifica\c{c}\~ao de contratos de 
derivativos e pol\'{\i}ticas de Scarf. 
 O cap\'{\i}tulo 6 expande estes conceitos introduzindo no\c{c}\~oes da
teoria de  Estima\c{c}\~ao e Controle. O cap\'{\i}tulo 7 discute alguns
m\'etodos de Intelig\^encia  Artificial, destacando \'{a}rvores de 
decis\~{a}o. 
 O cap\'{\i}tulo 8 aborda fundos de pens\~{a}o, an\'{a}lise atuarial, 
programa\c{c}\~{a}o estoc\'{a}stica e outros modelos para gest\~{a}o 
dos ativos de fundos com objetivos de longo prazo.

 Os ap\^endices A e B s\~ao manuais resumidos da linguagem Matlab e do
software Critical-Point, que s\~ao amplamente  utilizados no curso. 
 O ap\^endice C apresenta resumidamente alguns conceitos de \'Algebra
Linear Computacional, o ap\^endice D  conceitos de probabilidade,
estat\'{\i}stica e teoria da utilidade, e o ap\^{e}ndice E alguns
c\'{o}digos fonte de programas.  

 Material suplementar est\'{a} \`{a} disposi\c{c}\~{a}o no site 
 \ {\em www.ime.usp.br/$\sim$ jstern}

\cleardoublepage 
\pagestyle{headings}

%% file: transfig.tex
\typeout{TransFig: figures in PiCTeX.}
\ifx\fivrm\undefined
  \font\fivrm=cmr5\relax
\fi
\input{prepictex}
\input{pictex}
\input{postpictex}

\begingroup\makeatletter
\def\x#1#2#3#4#5#6#7\relax{\def\x{#1#2#3#4#5#6}}
\expandafter\x\fmtname xxxxxx\relax \def\y{splain}
\ifx\x\y   
\gdef\SetFigFont#1#2#3{%
  \ifnum #1<17 \tiny\else \ifnum #1<20 \small\else
  \ifnum #1<24 \normalsize\else \ifnum #1<29 \large\else
  \ifnum #1<34 \Large\else \ifnum #1<41 \LARGE\else
     \huge\fi\fi\fi\fi\fi\fi
  \csname #3\endcsname}
\else
\gdef\SetFigFont#1#2#3{\begingroup
  \count@#1\relax \ifnum 25<\count@ \count@25 \fi
  \def\x{\endgroup\@setsize\SetFigFont{#2pt}}%
  \expandafter\x
    \csname \romannumeral\the\count@ pt\expandafter\endcsname
    \csname @\romannumeral\the\count@ pt\endcsname
  \csname #3\endcsname}
\fi
\endgroup

%% file: cap1.tex
\chapter{Programa\c{c}\~{a}o Linear}  

No processo de modelagem de sistemas, invariavelmente deparamos com 
duas quest\~oes: se o modelo se ajusta adequadamente ao problema 
sendo analisado e se o modelo proposto \'e implement\'avel sob o ponto 
de vista computacional. Algumas classes de modelos conseguem atingir um 
relativo equil\'{\i}brio entre estes dois aspectos, sendo aplic\'aveis 
a uma extensa categoria de problemas reais, facilmente implement\'aveis 
e fornecendo solu\c{c}\~oes robustas. Programa\c{c}\~ao linear insere-se 
nesta categoria. 

Em um problema de programa\c{c}\~ao linear procura-se encontrar uma 
solu\c{c}\~ao que maximize ou minimize um funcional linear, dentro de 
um conjunto descrito a partir de restri\c{c}\~oes lineares. In\'umeros 
problemas podem ser modelados desta forma, tanto em finan\c{c}as quanto em 
engenharia, mas essencialmente a principal vantagem destes modelos 
reside nas propriedades decorrentes da sua estrutura, que possibilitam a 
constru\c{c}\~ao de algoritmos bastante simples e eficientes. 

Iremos explorar  algumas caracter\'{\i}sticas dos problemas de 
programa\c{c}\~ao linear, apresentando um algoritmo para sua 
resolu\c{c}\~ao denominado simplex. Este algoritmo destaca-se 
tanto por sua simplicidade quanto ampla utiliza\c{c}\~ao.
Estudaremos ainda o  conceito de 
dualidade e procuraremos apresentar um exemplo de aplica\c{c}\~ao em 
finan\c{c}as.

 \section{Nota\c{c}\~{a}o Matricial} 

 Inicialmente definimos algumas nota\c{c}\~{o}es matriciais. 
 O operador  $r\fto s\fto t$, l\^{e}-se - 
 {\it de} $r$ {\it at\'{e}} $t$ com {\it passo} $s$,
 indica o vetor $[r,r+s,r+2s,\ldots t]$  
 no correspondente dom\'{\i}nio de \'{\i}ncices. 
 $r\fto t$ \'{e} uma abrevia\c{c}\~{a}o de $r\fto 1\fto t$.  
 Usualmente escrevemos uma matriz, $A$, como o \'{\i}ndice 
 de linha subscrito, e o \'{\i}ndice de coluna superscrito. 
 Assim, $A_i^j$ \'{e} o elemento na $i$-\'{e}sima linha e 
 $j$-\'{e}sima coluna da matriz $A$.  
 Vetores de \'{\i}ndices podem ser usados para montar uma matriz 
 extraindo de uma matriz maior um determinado sub-conjunto de 
 linhas e colunas.  
 Por exemplo $A_{1:m/2}^{n/2:n}$ \'{e} o bloco nordeste, 
 i.e. o bloco com as primeiras linhas e ultimas colunas, 
 de $A$. 
 Alternativamente, podemos escrever uma matriz com \'{\i}ndices 
 de linha e coluna entre parenteses, i.e. podemos escrever o 
 bloco nordeste como $A({1\ate m/2} , {n/2\ate n})$.

 Exemplo:
 Dadas as matrizes  
 \[ 
 A=\left[ \begin{array}{ccc} 11 & 12 & 13 \\ 21 & 22 & 23 \\
 31 & 32 & 33 \end{array} \right] \ , \  \ 
 r=\left[ \begin{array}{cc} 1 & 3 \end{array} \right] \ , \  \ 
 s=\left[ \begin{array}{ccc} 3 & 1 & 2 \end{array} \right] \ ,
 \]  
 \[  
 A_r^s=\left[ \begin{array}{ccc} 13 & 11 & 12 \\ 33 & 31 & 32  
 \end{array} \right] \ . 
 \]

 $V>0$ \'{e} uma matriz positiva definida. 
 O operador $\diag$, quando aplicado a uma matriz quadrada, 
 extrai o vetor na diagonal principal, e quando aplicado a um vetor, 
 produz a matriz diagonal correspondente. 
 \[ 
   \diag(A) = 
   \left[  \begin{array}{c} 
    A_1^1 \\ A_2^2 \\ \vdots \\ A_n^n 
   \end{array} \right] 
   \ \ , \ \ 
   \diag(a) = 
   \left[  \begin{array}{cccc} 
    a_1 & 0  & \ldots & 0 \\ 
    0 & a_2 & \ldots & 0 \\ 
    \vdots & \vdots & \ddots & \vdots \\ 
    0 & 0 & \ldots & a_n    
   \end{array} \right] 
 \]

 Uma lista de matrizes pode ser indexada por \'{\i}ndices 
 subscritos ou superscritos \`{a} esquerda. 
 No caso de matrizes blocadas, estes indices \`{a} esquerda indicam os 
 blocos de linhas (subscritos) e colunas (superscritos),  
 como por exemplo na matriz 
 \[ 
  A  = 
   \left[  \begin{array}{cccc} 
    {_1^1A} & {_1^2A} & \ldots & {_1^sA} \\ 
    {_2^1A} & {_2^2A} & \ldots & {_2^sA} \\ 
    \vdots & \vdots & \ddots & \vdots \\ 
    {_r^1A} & {_r^2A} & \ldots & {_r^sA}    
   \end{array} \right] 
 \] 
 Assim, ${_r^sA_i^j}$ \'{e} o elemento na 
 $i$-\'{e}sima linha e $j$-\'{e}sima coluna do bloco situado no 
 $r$-\'{e}simo bloco de linhas e $s$-\'{e}simo bloco de colunas 
 da matriz $A$.  
 Alternativamente, podemos escrever os \'{\i}ndices de bloco 
 entre chaves, i.e. podemos escrever ${_r^sA_i^j}$ como 
 $A\{r,s\}(i,j)$.     

 O operador $\Vec$ empilha as colunas da matriz argumento 
 em um \'{u}nico vetor. \\    
 O produto de Kronecker (ou produto direto, ou tensorial), 
 $\kron$ , \'{e} definido como segue: 
 \[ 
  \Vec(U^{1 \ate n}) = 
   \left[  \begin{array}{c} 
    u^1 \\ u^2 \\ \vdots \\ u^n 
   \end{array} \right] 
   \ \ , \ \ 
  A \kron B = 
   \left[  \begin{array}{cccc} 
    A_1^1 B & A_1^2 B & \ldots & A_1^n B \\ 
    A_2^1 B & A_2^2 B & \ldots & A_2^n B \\ 
    \vdots & \vdots & \ddots & \vdots \\ 
    A_m^1 B & A_m^2 B & \ldots & A_m^n B   
   \end{array} \right] 
 \]

 Uma {\it matriz de permuta\c{c}\~{a}o} \'{e} uma matriz obtida pela
permuta\c{c}\~{a}o de linhas ou colunas na matriz identidade.  Realizar,
na matriz identidade, uma dada permuta\c{c}\~{a}o de linhas, nos fornece
a correspondente matriz de permuta\c{c}\~{a}o de linhas; Analogamente,  
uma permuta\c{c}\~{a}o de colunas da identidade fornece a correspondente 
matriz de permuta\c{c}\~{a}o de colunas. 
 \index{Matriz!de permuta\c{c}\~{a}o} 

Dada uma (matriz de) permuta\c{c}\~{a}o de linhas, $P$ e uma (matriz de)
permuta\c{c}\~{a}o de colunas, $Q$, o correspondente vetor de
\'{\i}ndices de linha (coluna) permutados s\~{a}o
$$ p= (P \left[ \begin{array}{c} 1\\ 2\\ \vdots \\ m 
  \end{array} \right] )'$$
$$ q= \left[ \begin{array}{cccc} 1 & 2 & \ldots & n 
  \end{array} \right] Q $$

 Realizar uma permuta\c{c}\~{a}o de linhas (de colunas) numa
matriz qualquer $A$, de modo a obter a matriz permutada $\tilde A$,
equivale a multiplic\'{a}-la, \`{a} esquerda (\`{a} direita), pela
correspondente matriz de permuta\c{c}\~{a}o de linhas (de colunas). 
Ademais, se $p$ ($q$) \'{e} o correspondente vetor de \'{\i}ndices de
linha (de coluna) permutados,
 $$ {\tilde A}_i^j = (P A)_i^j = A_{p(i)}^j $$
$$ {\tilde A}_i^j = (A Q)_i^j = A_i^{q(j)} \ .$$

 Exemplo:
 Dadas as matrizes  
 $$
 A=\left[ \begin{array}{ccc} 11 & 12 & 13 \\ 21 & 22 & 23 \\
 31 & 32 & 33 \end{array} \right] \ , \  \ 
 P=\left[ \begin{array}{ccc} 0 & 0 & 1 \\ 1 & 0 & 0 \\ 0 & 1 & 0 
 \end{array} \right] \ , \  \ 
 Q=\left[ \begin{array}{ccc}0 & 1 & 0 \\ 0 & 0 & 1 \\ 1 & 0 & 0 
 \end{array} \right] \ ,$$ 
 $$ 
 p=q=\left[ \begin{array}{ccc} 3 & 1 & 2 \end{array} \right] \ , \ \ 
 PA=\left[ \begin{array}{ccc} 31 & 32 & 33 \\ 11 & 12 & 13 \\ 
 21 & 22 & 23 \end{array} \right] \ ,\ \ 
 AQ=\left[ \begin{array}{ccc} 13 & 11 & 12 \\ 23 & 21 & 22 \\ 
 33 & 31 & 32  \end{array} \right] \ . $$ 

 Uma matriz quadrada, A, \'{e} {\it sim\'{e}trica} sse for igual a
transposta, isto \'{e}, sse $A=A'$.  Uma {\it permuta\c{c}\~{a}o
sim\'{e}trica} de uma matriz quadrada $A$ \'{e} uma permuta\c{c}\~{a}o da
forma $\tilde A =PAP'$, onde $P$ \'{e} uma matriz de permuta\c{c}\~{a}o. 

 Uma matriz quadrada, $A$, \'{e} {\it ortogonal} sse sua inversa for
igual a sua transposta, isto \'{e}, sse $A^{-1}=A'$. 
  \index{Matriz!sim\'{e}trica} 
  \index{Matriz!ortogonal} 
(a) Matrizes de permuta\c{c}\~{a}o s\~{a}o ortogonais.
(b) Uma permuta\c{c}\~{a}o sim\'{e}trica de uma matriz 
sim\'{e}trica \'{e} ainda uma matriz sim\'{e}trica.

\section{Convexidade} 

Um ponto $y(l)$ \'{e} {\em combina\c{c}\~{a}o convexa} de $m$ 
pontos de $\Re ^n$, dados pelas colunas da matriz $X\ n\times m$, sse
 $$ \forall i \ , \ \  
    y(l)_i = \sum_{j=1}^{m} l_j * X_i^j \ , \ \ \  
    l_j\geq 0 \ \mid \ \sum_{j=1}^{m} l_j = 1 $$
ou, equivalentemente, em nota\c{c}\~{a}o matricial  
 $$ y(l) = \sum_{i=1}^m l_i * X^{j} \ , \ \ \ 
    l_j\geq 0 \ \mid \ \sum_{j=1}^{m} l_j = 1 $$
ou ainda, substituindo as somat\'{o}rias por produtos internos,     
 $$ y(l) = Xl \ ,\ \ l\geq 0 \mid {\bf 1}'l=1 $$

Em particular, um ponto $y(\lambda )$ \'{e} combina\c{c}\~{a}o convexa 
de 2 pontos, $z$ e $w$, se 
 $$ y(\lambda ) = (1-\lambda )z +\lambda w \ , \ \ 
    \lambda \in [0,\ 1] \ . 
 $$ 
 Geometricamente, estes s\~{a}o os pontos no segmento de reta que 
vai de $z$ a $w$. 
 
 Um conjunto, $C\in \Re ^n$, \'{e} {\em convexo} sse contiver qualquer
combina\c{c}\~{a}o convexa de dois quaisquer de seus pontos.  
 Um conjunto, $C\in \Re ^n$, \'{e} {\em limitado} sse a dist\^{a}ncia 
entre quaisquer dois de seus pontos \'{e} limitada:
 $$ \exists \delta \mid \forall x1 ,\ x2 \in C \ , \ \ 
    || x1 - x2 || \leq \delta 
 $$
Um conjunto n\~{a}o limitado \'{e} dito {\em ilimitado}.  As figuras 
1.1 e 1.2 
apresentam alguns exemplos de conjuntos conforme as defini\c{c}\~oes acima.

\begin{figure}[ht]
\[
\input{convexo.ptx}
\]
\caption{\label{fig1} Conjuntos convexos}
\end{figure}

\begin{figure}[ht]
\[
\input{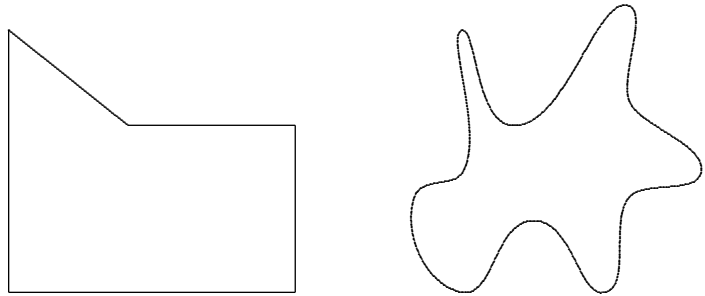}
\]
\caption{\label{fig2} Conjuntos n\~ao convexos}
\end{figure}

 Se denomina Ponto Extremal de um conjunto convexo $C$, a todo  ponto
$x$ que n\~{a}o pode ser representado como combina\c{c}\~{a}o  convexa
de dois pontos de $C$ distintos de $x$. O Perfil de $C$, $ext(C)$, 
\'{e} o conjunto de todos os pontos extremais de $C$.    

 Se denomina Casco Convexo (Fechado) de um conjunto $C$, $ch(C)$ 
($cch(C)$), \'{a} intersec\c{c}\~{a}o de todos os conjuntos convexos 
(fechados) que contem $C$. 

 {\bf Teorema:} Um conjunto $C$, convexo e compacto, i.e. convexo,
fechado e limitado, \'{e} igual ao casco convexo fechado de seu perfil, 
i.e. $C=cch(ext(C))$.  
  
{\bf Teorema:}  O casco convexo de um conjunto finito de pontos, $V$, 
\'{e} o conjunto de todas as combina\c{c}\~{o}es convexas de pontos de 
$V$, i.e. se $V=\{x_i, i=1\ldots n\}$, ent\~{a}o 
 $ch(V)=\{x \mid x=[x_1,\ldots x_n]l, l\geq 0, {\bf 1}'l=1\}$.   
 
 O {\em ep\'{\i}grafo} de uma curva em $\Re ^2$, $y=f(x)$, $x\in [a,b]$,
\'{e} o conjunto definido como 
  $ epig(f) \equiv \{ (x,y)\mid x\in [a,b] \wedge y\geq f(x) \}$.  
 Uma curva \'{e} convexa sse seu ep\'{\i}grafo \'{e} convexo.  
 Uma curva, $f(x)$, \'{e} {\em c\^{o}ncava} sse $-f(x)$ for convexa. 

\vspace{0.5cm}

{\bf Teorema:} Uma curva, $y=f(x)$, $\Re \mapsto \Re$, continuamente
diferenci\'{a}vel, e com primeira derivada sempre crescente, \'{e}
convexa.

\section{Poliedros}

 Uma {\em restri\c{c}\~{a}o} (n\~{a}o linear), em $\Re ^n$, \'{e} uma
inequa\c{c}\~{a}o da forma $g(x)\leq 0$, $g:\ \Re ^n \mapsto \Re $.  
 A {\em regi\~{a}o vi\'{a}vel} definida por $m$ restri\c{c}\~{o}es,
 $g(x)\leq 0$, $g:\ \Re ^n \mapsto \Re ^m$, 
\'{e} o conjuntos dos {\em pontos vi\'{a}veis} 
 $\{ x \mid g(x)\leq 0 \}$.  
 Dizemos que, num ponto vi\'{a}vel $x$, a restri\c{c}\~{a}o $g_i(x)$ \'{e} 
{\em justa} ou {\em ativa} sse valer a igualdade ($g_i(x)=0$), 
e {\em folgada} ou {\em inativa} caso contr\'{a}rio ($g_i(x)<0$). 

 Um {\em poliedro} em $\Re ^n$ \'{e} uma regi\~{a}o vi\'{a}vel definida por
{\em restri\c{c}\~{o}es lineares}: $Ax\leq d$.  Podemos sempre compor
uma restri\c{c}\~{a}o de igualdade, $a'x=\delta$, com um par de
restri\c{c}\~{o}es de desigualdade, 
 $a'x\leq \delta$ e $a'x\geq \delta$. 

\begin{figure}[ht]
\[
\input{poliedro.ptx}
\]
\caption{\label{fig3} Poliedros}
\end{figure}

A figura 1.3 acima mostra alguns poliedros.  
 
\vspace{0.5cm}

{\bf Teorema:} Poliedros s\~{a}o convexos, mas n\~{a}o necessariamente 
limitados.

\subsection*{Faces} 
 
 Uma {\em face} de dimens\~{a}o $k$, de um poliedro em $\Re ^n$ com $m$
restri\c{c}\~{o}es de igualdade, \'{e} uma regi\~{a}o vi\'{a}vel que
obedece justamente a $n-m-k$ das restri\c{c}\~{o}es de desigualdade do
poliedro. Em outras palavras: um ponto que obedece jstamente a $r$ 
restri\c{c}\~{o}es de desigualdade est\'{a} sobre uma face de 
dimens\~{a}o $k=n-m-r$.   
 Um {\em v\'{e}rtice} \'{e} uma face de dimens\~{a}o $0$. 
 Um {\em lado} \'{e} uma face de dimens\~{a}o $1$, um {\em ponto
interior} do poliedro tem todas as restri\c{c}\~{o}es de desigualdade
folgadas. 

 Eventualmente poder\'{\i}amos ter, num dado v\'{e}rtice, $n-m+1$
restri\c{c}\~{o}es de desigualdade justas.  Este v\'{e}rtice seria
super-determinado, pois obedeceria a $(n-m+1)+m=n+1$ equa\c{c}\~{o}es. 
 Todavia esta \'{e} uma situa\c{c}\~{a}o inst\'{a}vel: uma pequena
perturba\c{c}\~{a}o nos dados levaria este v\'{e}rtice {\em degenerado}
em 1 ou v\'{a}rios v\'{e}rtices n\~{a}o degenerados, vide figura 1.4 
 Doravante assumiremos sempre que situa\c{c}\~{o}es como esta n\~{a}o
ocorrem, i.e.  assumiremos a {\em hip\'{o}tese de n\~{a}o
degeneresc\^{e}ncia};

\begin{figure}[ht]
\[
\input{degenera.ptx}
\]
\caption{\label{fig4} V\'ertice degenerado}
\end{figure}

\subsection*{Forma Padr\~{a}o} 
 
Um poliedro na {\em forma padr\~{a}o}, em $\Re ^n$, \'{e} dado pelas $n$
{\em restri\c{c}\~{o}es de sinal} (i.e. $x_i\geq 0$ ) e $m<n$ 
{\em restri\c{c}\~{o}es de igualdade}:
 $$ P_A = \left\{ x\geq 0 \ \mid \ Ax=d \right\} 
    \ \ ,\ A\ m\times n 
 $$    
 
Podemos sempre reescrever um poliedro na forma padr\~{a}o, ainda que num
espa\c{c}o de dimens\~{a}o maior, usando os seguintes artif\'{\i}cios:
 \begin{itemize}  
 \item[1.] Substitua vari\'{a}veis irrestritas, pela diferen\c{c}a 
    de duas vari\'{a}veis positivas, $x^{+} - x^{-}$ onde 
    $x^{+} = max \{0,x\}$ e $x^{-} = max \{0,-x\}$.
     
 \item[2.] Acrescente uma vari\'{a}vel de folga $\chi \geq 0$ a cada 
 inequa\c{c}\~{a}o, 
    $$ a'x\leq \delta \ \Leftrightarrow \  
       \left[ \begin{array}{cc} a & 1 \end{array} \right] 
       \left[ \begin{array}{c} x \\ \chi \end{array} \right] 
       = \delta \ \ .  
    $$ 
 \end{itemize} 

Da defini\c{c}\~{a}o de v\'{e}rtice vemos que num poliedro padr\~{a}o,
$P_A$, um v\'{e}rtice \'{e} um ponto vi\'{a}vel com $n-m$
restri\c{c}\~{o}es de sinal justas, i.e.  $n-m$ vari\'{a}veis nulas;
estas s\~{a}o as {\em vari\'{a}veis residuais} do v\'{e}rtice. 
 Permutemos as linhas do vetor $x$ de modo a colocar as vari\'{a}veis
residuais nas $n-m$ \'{u}ltimas posi\c{c}\~{o}es, estando portanto as
vari\'{a}veis restantes, as {\em vari\'{a}veis b\'{a}sicas}, nas $m$
primeiras posi\c{c}\~{o}es.  Aplicando a mesma permuta\c{c}\~{a}o \`{a}s
colunas de $A$, as primeiras $m$ colunas da permuta\c{c}\~{a}o de $A$
formam a matriz $B$, ou {\em base} do v\'{e}rtice, e as demais colunas
de $A$ a {\em matriz residual}, $R$ com dimens\~ao $\ m\times (n-m)$. 
 Isto \'{e}, particionamos as colunas de $A$ como 
 \[ 
  A =  
 \left[ \begin{array}{cc} A^b & A^r \end{array} \right] = 
 \left[ \begin{array}{cc} B & R \end{array} \right] 
 \]

 Na forma blocada \'{e} f\'{a}cil explicitar as vari\'{a}veis n\~{a}o
nulas. Reescrevendo \\ 
 $x\geq 0\mid Ax=d$ como  
 \[ 
 \left[ \begin{array}{c} x_b \\ x_r \end{array} \right] 
 \geq 0 \mid 
 \left[ \begin{array}{cc} B & R \end{array} \right] 
 \left[ \begin{array}{c} x_b \\ x_r \end{array} \right] = d 
 \] 
temos 
 \[ 
    x_b = B^{-1} \left[ d- Rx_r \right] \ \ 
 \]  
 Anulando as vari\'aveis residuais, segue que  
 \[  
    x_b = B^{-1}d \ \ . 
 \]

 Da defini\c{c}\~{a}o de degenerec\^{e}ncia vemos que um v\'{e}rtice de um
poliedro padr\~{a}o \'{e} dege\-nerado sse tem, al\'{e}m das vari\'{a}veis
residuais, ao menos uma vari\'{a}vel b\'{a}sica nula.

\section{M\'{e}todo Simplex} 

 O problema de programa\c{c}\~{a}o linear (PPL) padr\~{a}o \'{e} o 
de minimizar uma fun\c{c}\~{a}o linear dentro de um poliedro padr\~{a}o: 
 $$ \mbox{min}\ \ cx, x\geq 0 \mid Ax=d \ \ .$$ 

 Suponha sabermos quais as vari\'{a}veis residuais (i.e.  as
restri\c{c}\~{o}es justas) de um dado v\'{e}rtice, e consequentemente 
conhecida uma matriz b\'asica B.  Permutando as
linhas de $x$ e as colunas de $c$ e $A$, para a forma blocada de
vari\'{a}veis b\'{a}sicas e residuais, reescrevemos o PPL como:
 $$ \mbox{min}\ \ 
 \left[ \begin{array}{cc} c^b & c^r \end{array} \right] 
 \left[ \begin{array}{c} x_b \\ x_r \end{array} \right] \ , \ \ 
 x \geq 0 \mid 
 \left[ \begin{array}{cc} B & R \end{array} \right] 
 \left[ \begin{array}{c} x_b \\ x_r \end{array} \right] = d \ \ .$$ 

Utilizaremos a nota\c{c}\~{a}o 
 $\tilde d \equiv B^{-1}d$ e $\tilde R \equiv B^{-1}R$.
Se mudarmos um \'{u}nico elemento, o $j$-\'{e}simo de $x_r$, 
permitindo que ele se torne positivo ( i.e. $x_{r(j)} > 0$), 
a {\em solu\c{c}\~{a}o b\'{a}sica}, $x_b$ se escreve como  
 \begin{eqnarray*} 
  x_b 
  &=& \tilde d - \tilde R x_r \\ 
  &=& \tilde d - x_{r(j)} \tilde R^j  
 \end{eqnarray*} 
 Esta solu\c{c}\~{a}o permanece vi\'{a}vel enquanto n\~{a}o negativa. 
Pela hip\'{o}tese de n\~{a}o degeneresc\^{e}ncia, $\tilde d >0$, e podemos
aumentar o valor de $x_{r(j)}$, mantendo a solu\c{c}\~{a}o b\'{a}sica
vi\'{a}vel, at\'{e} um patamar $\epsilon >0$, quando alguma 
vari\'avel b\'asica \'e anulada. 

O valor da fun\c{c}\~{a}o objetivo desta solu\c{c}\~{a}o b\'{a}sica \'{e}
 \begin{eqnarray*} 
   cx 
   &=&  c^b x_b + c^r x_r \\
   &=&  c^b B^{-1} [d- Rx_r] + c^r x_r \\  
   &=&  c^b \tilde d + (c^r - c^b \tilde R) x_r \\ 
   &\equiv &  \varphi - z x_r \\ 
   &=& \varphi - z^j x_{r(j)} 
 \end{eqnarray*}   
O vetor $z$ \'{e} denominado o {\em custo reduzido} na base. 

Os par\'{a}grafos precedentes sugerem o seguinte algoritmo para gerar
uma sequ\^{e}ncia de v\'{e}rtices vi\'{a}veis de valores
decrescentes, a partir de um v\'{e}rtice inicial:

\vspace{0.5cm}

\noindent
{\bf Algoritmo Simplex:} 

\vspace{0.5cm}

 \begin{enumerate}  
 \item Procure um \'{\i}ndice residual, $j$, tal que $z^j >0$. 
 \item Compute, para $k \in K\equiv \{l \mid \tilde R _l^j >0\}$\ , 
      \ \ $\epsilon _k= \tilde d _k / \tilde R _k^j$ \ ,  \\     
      e \ $i= \mbox{Argmin} _{k \in K} \ \  \epsilon_k$ \ ,  
      \ \ i.e. $\epsilon (i) = \min_k \epsilon_k$\ .  
             
 \item Fa\c{c}a a vari\'{a}vel $x_{r(j)}$ b\'{a}sica, e $x_{b(i)}$ residual. 
 \item Volte ao passo 1. 
 \end{enumerate} 

O Simplex n\~{a}o pode prosseguir se $z\leq 0$ no primeiro passo, ou se
no segundo passo o m\'{\i}nimo for tomado sobre um conjunto vazio.  O
segundo caso corresponde a termos um PPL ilimitado.  No primeiro caso o
v\'{e}rtice corrente \'{e} a solu\c{c}\~{a}o \'{o}tima do PPL!

Trocar o status b\'{a}sica/residual de um par de vari\'{a}veis \'{e} 
denominado {\em pivotar}.  Ap\'{o}s cada pivotamento necessitamos
recomputar a inversa da base, $B^{-1}$, o que pode ser feito com um 
m\'{\i}nimo de esfor\c{c}o.  


No restante desta se\c{c}\~{a}o usaremos o simplex para resolver alguns
problemas simples, interpretando geometricamente o processo de
solu\c{c}\~{a}o.  Na pr\'{o}xima se\c{c}\~{a}o provaremos formalmente a
corretude do algoritmo simplex.

\subsection*{Exemplo} 

 Consideremos o PPL 
 $\mbox{min} [-1,-1] x , \ 0 \leq x \leq 1$. \\ 
 Este PPL pode ser reescrito na forma padr\~{a}o com  
 $$
  c= \left[ \begin{array}{cccc} -1 & -1 & 0 & 0 \end{array} \right] \ \ \ 
  A= \left[ \begin{array}{cccc} 1 & 0 & 1 & 0 \\ 
                                0 & 1 & 0 & 1 \end{array} \right]    \ \ \ 
  d= \left[ \begin{array}{c} 1 \\ 1 \end{array} \right]
 $$ 
 Nos \'{e} dado um vertice inicial, $x=[0,0]$. \\ 
 Passo 1:  
 $r=[1,2]$, $b=[3,4]$, $B=A(:,b)=I$, $R=A(:,r)=I$, \\  
 $-z= c^r-c^b\tilde{R}= [-1,-1] -[0,0] \Rightarrow z= [1,1]$, 
 $j=1$, $r(j)=1$, \\ 
 $x_b= \tilde{d} -\epsilon \tilde{R}^j = 
  \left[ \begin{array}{c} 1 \\ 1 \end{array} \right]
  -\epsilon \left[ \begin{array}{c} 1 \\ 0 \end{array} \right] 
 \Rightarrow \epsilon^*=1 , \  i=1 , \ b(i)=3$ \\  
 Passo 2:  
 $r=[3,2]$, $b=[1,4]$, $B=A(:,b)=I$, $R=A(:,r)=I$, \\  
 $-z= c^r-c^b\tilde{R}= [0,-1] -[-1,0] \Rightarrow z= [-1,1]$, 
 $j=2$, $r(j)=2$, \\ 
 $x_b= \tilde{d} -\epsilon \tilde{R}^j = 
  \left[ \begin{array}{c} 1 \\ 1 \end{array} \right]
  -\epsilon \left[ \begin{array}{c} 0 \\ 1 \end{array} \right] 
 \Rightarrow \epsilon^*=1 , \  i=2 , \ b(i)=4$ \\  
 Passo 3:  
 $r=[3,4]$, $b=[1,2]$, $B=A(:,b)=I$, $R=A(:,r)=I$, \\  
 $-z= c^r-c^b\tilde{R}= [0,0] -[-1,-1] \Rightarrow z= [-1,-1] <0$

\subsection*{Obtendo um V\'{e}rtice Vi\'{a}vel} 

Usaremos o PPL auxiliar 
 $$ 
 \mbox{min} \ 
 \left[ \begin{array}{cc} 0 & 1 \end{array} \right] 
 \left[ \begin{array}{c} x \\ y \end{array} \right] 
 \ \ \mid \ \ 
 \left[ \begin{array}{c} x \\ y \end{array} \right] 
 \geq 0 \ \wedge \  
 \left[ \begin{array}{cc} A & 
        \mbox{\em diag}(\mbox{\em sign}(d))\end{array} \right] 
 \left[ \begin{array}{c} x \\ y \end{array} \right] 
 = d 
 $$ 

 Um v\'{e}rtice inicial para este PPL \'{e} 
 $ \left[ \begin{array}{cc} 0 & \mbox{\em abs}(d') \end{array} \right]'$.  
 Se o problema auxiliar tiver valor \'{o}timo zero, a solu\c{c}\~{a}o
\'{o}tima fornece um v\'{e}rtice vi\'{a}vel para o problema original;
caso contr\'{a}rio, o problema original \'{e} invi\'{a}vel.

\section{Dualidade} 

Dado um PPL qualquer, que chamaremos de PPL {\em Primal}, definiremos um
outro PPL, o PPL {\em Dual} (do PPL primal).  A teoria de dualidade
trata das rela\c{c}\~{o}es entre a solu\c{c}\~{a}o de um dado PPL, e a
solu\c{c}\~{a}o do seu dual. 

Dado um PPL na forma can\^{o}nica, o problema primal (PPLP):  
 $$ \mbox{min}\ \ cx \mid x\geq 0 \wedge Ax\geq d \ \ ,$$ 
definimos seu dual como (PPLD): 
 $$ \mbox{max}\ \ y'd \mid y\geq 0 \wedge y'A\leq c \ \ ,$$ 

O primal can\^{o}nico e seu dual tem uma interpreta\c{c}\~{a}o
econ\^{o}mica  intuitiva: O primal pode ser interpretado como o
cl\'{a}sico problema da ra\c{c}\~{a}o: $A_i^j$ \'{e} a quantidade de 
nutriente do tipo $j$ encontrada em uma unidade de alimento do tipo $i$.
$c^i$ \'{e} o custo de uma unidade de alimento do tipo $i$, e $d_j$ a 
necessidade di\'{a}ria m\'{\i}nima do nutriente $j$. A solu\c{c}\~{a}o 
\'{o}tima do primal, $x^*$ nos fornece a ra\c{c}\~{a}o nutricionalmente
vi\'{a}vel de custo m\'{\i}nimo. O dual pode ser interpretado como um 
fabricante de nutrientes sint\'{e}ticos, procurando o ``valor de
mercado''  para sua linha de nutrientes. A receita do fabricante por
ra\c{c}\~{a}o  \'{e} a fun\c{c}\~{a}o objetivo do dual a ser maximizada.
Para manter sua linha de produtos competitiva, nenhum alimento natural 
deve fornecer nutrientes mais barato que a correspondente mistura de 
nutrientes sint\'{e}ticos, estas s\~{a}o as restri\c{c}\~{o}es do dual. 
Os pre\c{c}os dos nutrientes, $y^*$ podem tamb\'{e}m ser interpretados 
como pre\c{c}os marginais: (dentro de um pequeno intervalo) de quanto 
aumentaria o valor do alimento $i$ se nele conseguissemos incrementar  a
concentra\c{c}\~{a}o do nutriente $j$. A corretude destas
interpreta\c{c}\~{o}es \'{e} demostrada pelas propriedades de dualidade 
discutidas a seguir.

{\bf Lema 1} O dual do dual \'{e} o pr\'{o}prio primal. 

Prova: 
Basta observar que o (PPLD) \'e equivalente a 
$$ \mbox{min}\ \ -y'd \mid y\geq 0 \wedge -y'A\geq -c \ \ ,$$

\vspace{0.5cm}

{\bf Teorema Fraco de Dualidade} 
 Se $x$ e $y$ s\~{a}o solu\c{c}\~{o}es vi\'{a}veis, respectivamente do
primal e do dual, ent\~{a}o existe um intervalo (gap) n\~{a}o negativo
entre o valor da solu\c{c}\~{a}o dual e o valor da solu\c{c}\~{a}o
primal:
 $$ cx \geq y'd \ \ .$$  

Prova: 
 Por viabilidade, $Ax\geq d$ e $y\geq 0$, portanto $y'Ax\geq y'd$. 
Analogamente, $y'A\leq c$ e $x\geq 0$ donde $y'Ax\leq cx$; 
 Portanto $cx\geq y'd$.

 QED. 

\vspace{0.5cm}

{\bf Corol\'{a}rio 1} 
 Se tivermos um par de solu\c{c}\~{o}es vi\'{a}veis, $x$ para o PPLP, e
$y$ para o PPLD, e se o valor das solu\c{c}\~{o}es primal e dual
coincidirem, i.e.  $cx^* =(y^*)'d$, ent\~{a}o ambas as solu\c{c}\~{o}es
s\~{a}o \'{o}timas. 

\vspace{0.5cm}
 
 {\bf Corol\'{a}rio 2} 
 
 Se o primal for um problema ilimitado, ent\~{a}o o dual \'{e}
invi\'{a}vel. 

\vspace{0.5cm}

Da mesma forma que podemos reescrever qualquer PPL na forma padr\~{a}o,
podemos reescrever qualquer PPL na forma da defini\c{c}\~{a}o do primal
ou dual. (Verifique).  
 Em vista do lema 1, vemos que a  rela\c{c}\~{a}o de dualidade est\'a
definida entre pares de problemas de PL, independentemente da forma com
que est\~{a}o escritos. 

 \vspace{0.5cm}

{\bf Lema 2}  

Dado o primal  na forma padr\~{a}o (PPLP): 
 $$ \mbox{min}\ \ cx \mid x\geq 0 \wedge Ax = d \ \ ,$$ 
o seu dual \'{e} (PPLD): 
 $$ \mbox{max}\ \ y'd \mid y\in \Re ^m \wedge y'A\leq c \ \ ,$$ 

\vspace{0.5cm}
  
 {\bf Teorema} (Prova de corretude do Simplex) 
  
 Provaremos que o simplex termina em um v\'{e}rtice \'{o}timo.  
Na parada do Simplex t\'{\i}nhamos 
 $z = -( c^r - c^b B^{-1}R ) \leq 0$. 
Consideremos $y'= c^bB^{-1}$ como candidato a solu\c{c}\~{a}o dual. 
 \begin{eqnarray*} 
  \lefteqn{ 
     \left[ \begin{array}{cc} c^b & c^r \end{array} \right] 
      - y' \left[ \begin{array}{cc} B & R \end{array} \right] } \\  
    &=& \left[ \begin{array}{cc} c^b & c^r \end{array} \right] 
     - c^b B^{-1}  \left[ \begin{array}{cc} B & R \end{array} \right] \\ 
    &=& \left[ \begin{array}{cc} c^b & c^r \end{array} \right] 
     - c^b  \left[ \begin{array}{cc} I & \tilde R \end{array} \right] \\ 
    &=&  \left[ \begin{array}{cc} c^b & c^r \end{array} \right] 
     - \left[ \begin{array}{cc} c^b & c^b \tilde R \end{array} \right] \\  
    &=& \left[ \begin{array}{cc} 0 & -z \end{array} \right] \geq 0 
 \end{eqnarray*}   
 Portanto $y$ \'{e} uma solu\c{c}\~{a}o dual vi\'{a}vel. 
 Ademais, seu valor (como solu\c{c}\~{a}o dual) \'{e}
 $y'd = c^b B^{-1} d = c^b \tilde d = \varphi$, 
e pelo corol\'{a}rio 1 ambas as solu\c{c}\~{o}es s\~{a}o \'{o}timas.  

\vspace{0.5cm}

 {\bf Teorema Forte de Dualidade} 
 Se o problema primal for vi\'{a}vel e limitado, assim ser\'{a} o dual;
al\'em disso, o valor das solu\c{c}\~{o}es \'{o}timas do primal e do dual
coincidem. 

 Prova: Construtiva, atrav\'{e}s do algoritmo simplex.  

\vspace{0.5cm}
 
 {\bf Teorema}  (Folgas Complementares) 
 
 Sejam $x$ e $y'$ solu\c{c}\~{o}es vi\'{a}veis de um PPL padr\~{a}o e de
seu dual. Estas solu\c{c}\~{o}es s\~{a}o \'{o}timas sse 
 $w'x=0$, onde $w=(c-y'A)$. 
 Os vetores $x$ e $w$ representam as folgas nas 
restri\c{c}\~{o}es de desigualdade do PPLP e do PPLD. 
 Como $x\geq 0$, $w\geq 0$ e $w'x=0$, cada termo do produto escalar se
anula, i.e.  $w_j x_j=0$, ou equivalentemente, se a j-\'{e}sima
restri\c{c}\~{a}o  de desigualdade \'{e} folgada no primal, ent\~{a}o a
correspondente  restri\c{c}\~{a}o no dual deve ser justa, e vice-versa;
dai o nome  folgas complementares.      

 Prova: Se as solu\c{c}\~{o}es s\~{a}o \'{o}timas, poder\'{\i}amos
t\^{e}-las obtido com o algoritmo Simplex.  Como na prova de corretude
do Simplex,
 $$ (c-y'A)x =   
   \left[ \begin{array}{cc} 0 & z \end{array} \right] 
   \left[ \begin{array}{c} x_b \\ 0 \end{array} \right] 
   = 0 
 $$ 
 Se $(c-y'A)x=0$, ent\~{a}o $y'(Ax)=cx$, ou $y'd=cx$, 
e pelo primeiro corol\'{a}rio do teorema fraco de dualidade, 
ambas as solu\c{c}\~oes s\~ao \'{o}timas.

\subsection*{Forma geral de Dualidade} 

Daremos a seguir o dual de um PPL em forma geral. 
Indicamos por um asterisco, ($*$), um vetor com vari\'{a}veis 
irrestritas.

PPLP: 
 \[  
 \min 
 \left[ \begin{array}{ccc} {^1c} & {^2c} & {^3c} \end{array} \right]
 \left[ \begin{array}{c} {_1x} \\ {_2x} \\ {_3x} \end{array} \right] \ \ \  
 \left. 
 \begin{array}{ccc} 
 {_1x} & \geq & 0 \\ {_2x} & * &  \\ {_3x} & \leq & 0  \end{array} \right.
 \ \ \mid \   
 \left[ \begin{array}{ccc} {^1_1A} & {^2_1A} & {^3_1A} \\  
   {^1_2A} & {^2_2A} & {^3_2A} \\ {^1_3A} & {^2_3A} & {^3_3A} \\\end{array} \right]
 \left[ \begin{array}{c} {_1x} \\ {_2x} \\ {_3x} \end{array} \right]
 \left. \begin{array}{cc} \leq & {_1d} \\ = & {_2d} \\ \geq & {_3d} 
  \end{array} 
 \right.  
 \] 

 PPLD: 
 \[ 
 \max 
 \left[ \begin{array}{ccc} {_1d}' & {_2d}' & {_3d}' \end{array} \right]
 \left[ \begin{array}{c} {_1y} \\ {_2y} \\ {_3y} \end{array} \right] \ \ \  
 \left. \begin{array}{ccc} 
 {_1y} & \leq & 0 \\ {_2y} & * &  \\ {_3y} & \geq & 0  \end{array} \right.
 \ \ \mid \   
 \left[ \begin{array}{ccc} {^1_1A} & {^2_1A} & {^3_1A} \\  
 {^1_2A} & {^2_2A} & {^3_2A} \\ {^1_3A} & {^2_3A} & {^3_3A} \\\end{array} \right]'
 \left[ \begin{array}{c} {_1y} \\ {_2y} \\ {_3y} \end{array} \right]
 \left. \begin{array}{cc} \leq & {^1c}' \\ = & {^2c}' \\ 
  \geq & {^3c}' \end{array} \right.
 \]  

Como corol\'{a}rio, ilustremos alguns  pares de problemas primal/dual 
$$
PPLP:\ \max cx \mid Ax \geq d \wedge x \geq 0 \ \ \ \ 
PPLD:\ \min y^{'} d  \mid y^{'} A \geq c \wedge y \leq 0 
$$ 
$$ 
PPLP:\ \max cx \mid Ax \leq d \wedge x \in \Re ^{n} \ \ \ \  
PPLD:\ \max y^{'} d \mid y^{'} A = c \wedge y \leq 0 
$$

\section{Problema Linear Param\'{e}trico}

 Consideremos agora o Problema Linear Param\'{e}trico (PLP), i.e. uma 
fam\'{\i}lia de PPLs em fun\c{c}\~{a}o de um par\^ametro, $\eta$.
 Como primeiro exemplo consideremos uma parametriza\c{c}\~{a}o linear  
do vetor do lado direito, 
 $$\min cx \ \ x \geq 0 \mid Ax=d \ \ , \ \ \ d= t + \eta p$$  
 Suponhamos ter resolvido o PPL acima, vi\'{a}vel e limitado, 
para um certo valor $\eta =\eta'$. 

 Na base \'otima $B$, a solu\c{c}\~{a}o b\'{a}sica $x_{b}$ 
pode ser escrita como 
 $$\tilde d = B^{-1}(t +\eta p) = \tilde t +\eta' \tilde p$$

 Analisemos agora o  comportamento deste problema quando variamos o 
par\^ametro  $\eta$, ou seja, consideremos o comportamento da
solu\c{c}\~{a}o do PPL para valores de $\eta$ pr\'{o}ximos de $\eta'$. 

 Consideremos $\eta$ crescente: Se $\tilde p \geq 0$, podemos aumentar o
par\^{a}metro $\eta$ sem jamais tornar $\tilde d$ invi\'{a}vel, i.e. 
sem jamais violar uma restri\c{c}\~{a}o de sinal.  Caso contr\'{a}rio,
podemos aumentar o par\^{a}metro apenas at\'{e} o valor cr\'{\i}tico
 $$
 {\eta}_j = {-\tilde t}_j / {\tilde p}_{j} 
 \ \ , \ \ \  
 k = arg\min_{j \mid {\tilde p}_{j}<0} \eta_j   
 \ \ , \ \ \  
 \eta_1 = \eta_k \ . 
 $$

Analogamente, consideremos o comportamento da solu\c{c}\~{a}o 
para $\eta$ decrescente.  Se $\tilde p \leq 0$, podemos diminuir o
par\^{a}metro $\eta$ sem jamais tornar $\tilde d$ invi\'{a}vel, i.e. 
sem jamais desobedecer uma restri\c{c}\~{a}o de sinal.  Caso contr\'{a}rio,
podemos diminuir o par\^{a}metro apenas at\'{e} o valor cr\'{\i}tico
 $$
 \eta_0 = \eta_k 
 \ \ , \ \ 
 \eta_{j} =  {-\tilde t}_j / {\tilde p}_j 
 \ \ , \ \ \  
 k = arg\max_{j \mid {\tilde p}_{j}>0} \eta_j 
 \ . 
 $$

Para $\eta$ entre este par de par\^{a}metros cr\'{\i}ticos, 
 $\eta_0 < \eta' < \eta_1$, 
 $\eta(\lambda) = (1-\lambda)\eta_0 +\lambda \eta_1$, 
vemos que a solu\c{c}\~{a}o do problema
ser\'{a} a correspondente combina\c{c}\~{a}o convexa das
solu\c{c}\~{o}es cr\'{\i}ticas:
 $$ x(\eta (\lambda )) 
    = (1-\lambda )x(\eta _0) +\lambda x(\eta _{1}) \ .$$
 Como $x(\eta )$ \'{e} linear entre um par de par\^{a}metros
cr\'{\i}ticos consecutivos, o valor \'{o}timo da
solu\c{c}\~{a}o do PPL com par\^{a}metro
 $\eta= (1-\lambda )\eta _0 +\lambda \eta _{1}$, $vopt(\eta )$, 
 \'{e} uma fun\c{c}\~{a}o linear em $\eta$.  Portanto o gr\'{a}fico de
  $vopt(\eta )$, para $\eta _0 \leq \eta < \eta_1$, 
 ser\'{a} um segmento de reta.

\section{Modelo de Sharpe}

 Devido a simplicidade da estrutura dos problemas de programa\c{c}\~ao 
linear, muitas vezes modelos n\~ao lineares s\~ao simplificados de forma
a  recair  no caso linear. Este artif\'{\i}cio tem sido bastante
empregado  em problemas de composi\c{c}\~ao de portf\'olios,  onde  o
objetivo  \'e  determinar uma  carteira de ativos que  minimize o risco 
e maximize o retorno do investidor. 

 Consideremos que o retorno de cada um dos ativos se escreva como 
 $$ r_{i} = a_{i} +e_{i} + b_{i} (a_{0} +e_0) , 
            \ i \in \{1\ldots n\} $$
 onde $a_0$ \'{e} o retorno esperado ``do mercado'', 
 $e_0$ um erro aleat\'{o}rio aferando este retorno,   
 $a_{i}$ \'e o retorno acima do mercado esperado do i-\'esimo ativo, 
 $e_i$ um erro afetando somente o retorno deste ativo, e 
 $b_{i}$ uma medida da rea\c{c}\~ao do ativo $i$ 
 \`{a}s mudan\c{c}as do \'{\i}ndice de mercado. 
 Admitamos ainda termos ruidos brancos n\~{a}o correlacionados, 
 i.e. $E(e_i)=0$, $Var(e_i)=s_{i,i}=s_i^2$, 
      $Cov(e_{i},e_{j})=s_{i,j}=0 , \ i \neq j$. 
 Em finan\c{c}as \'{e} usual a nota\c{c}\~{a}o $\alpha$, $\beta$ 
 e $\sigma$ para os vetores $a$, $b$ e $s$. 

 Seja $x_{i}$ a fra\c{c}\~{a}o do total de recursos que ser\'{a} alocado 
para o ativo $i$, $x\geq 0$, ${\bf 1}'x =1$. 
 O retorno do portf\'{o}lio, $r_P$, se escreve como 
 $$ r_p = r'x = (a +e)'x +(a_{0} +e_0) b' x $$ 
 definindo $r_P= r'x$, $a_P = a'x$, $b_P = b'x$,  
 a esper\^{a}n\c{c}a e vari\^ancia do portf\'{o}lio ser\~{a}o: 
 $$ E(r_{p}) =  a_P +a_0 b_P \ , \ \  
    Var(r_{p}) = s_{0}^{2} b_{P}^{2} + 
    \sum_{1}^{N} (s_{i} x_{i})^{2} .$$

 Definimos a fun\c{c}\~ao objetivo a ser maximizada, 
 como o retorno esperado do portf\'{o}lio poderado pela 
 toler\^{a}ncia ao risco do investidor, $\eta$, menos o 
 desvio padr\~{a}o do retorno:   
 $$f(x) = \eta E(r_{p})) -\mbox{std}(r_{p}).$$  
 Para linearizar o problema, assumimos que a carteira \'{e} bem
diversificada, i.e. que $x_i \approx 1/n$, e que nenhum ativo tem 
risco pr\'{o}prio muito maior que o risco de mercado, de modo que  
 $$s_0^2 \, b_P^2  > > \sum_{1}^{N} (s_{i}\, x_{i} )^{2}$$

 Recaimos assim em um problema de programa\c{c}\~{a}o linear, 
conhecido como o modelo de Sharpe: 
 $$max f(x) = \eta (a + a_0 b)' x  \  -s_0 b'x  
             \ , \ x\geq 0 \mid {\bf 1}' x =1$$
 a restri\c{c}\~{a}o do problema \'{e} conhecida como 
``restri\c{c}\~{a}o normal'', que simplesmente assegura a 
interpreta\c{c}\~{a}o do vetor $x$ como a fra\c{c}\~{a}o investida 
em cada ativo. Outras restri\c{c}\~{o}es lineares podem ser acrescentadas 
para modelar condi\c{c}\~{o}es impostas pelo investidor. 
 Por exemplo, a condi\c{c}\~{a}o de diversifica\c{c}\~{a}o pode ser 
escrita como $Ix \leq (\kappa/n){\bf 1}$, $\kappa << n$.

\section{Fluxos em Redes}

 Muitos problemas importantes podem ser formulados como o problema  de
encontrar o fluxo de minimo custo que satisfaz as equa\c{c}\~{o}es  de
oferta e demanda em uma rede. A Rede (ou Grafo) \'{e} descrita pelos
seus  V\'{e}rtices (ou N\'{o}s), numerados de $1$ a $n$, e pela Matriz
de Incid\^{e}ncia  de seus Arcos, $A$. Cada arco $(i,j)$, corresponde a
uma coluna de $A$, $A^{ij}$, tal que 
 $A^{ij}_i=-1$, $A^{ij}_j=+1$ e $A^{ij}_l=0,\; l\neq i,j$. 
 A demanda da mercadoria (ou commoditie) \'{e} representada pelo Vetor
de demanda, $b$. Uma demanda negativa, $b_i<0$, representa uma oferta,
ou disponibilidade da mercadoria no vertice $i$. Imporemos a
condi\c{c}\~{a}o  ${\bf 1}'b=0$ i.e., a oferta total iguala a demanda
total. 
 O fluxo de (mercadoria de) $i$ para $j$ \'{e} $x_{ij}$. 
 O custo de enviar uma unidade (de mercadoria) de $i$ para $j$ \'{e} 
dado pelo Vetor de Custos, $c^{ij}$. 
 Podemos ainda impor Restri\c{c}\~{o}es de Capacidade sobre os arcos, 
$l\leq x\leq k$. 
 Finalmente, o problema de Fluxo Capacitado de M\'{\i}nimo Custo  
pode ser formulado como: 
 $$ \mbox{min}\: c x\: , \; l\leq x \leq k \: \mid \: A x = b$$           
 Note que como a demanda total \'{e} nula, e o fluxo \'{e} conservativo,
poderiamos eliminar uma qualquer das equa\c{c}\~{o}es do sistema. 

 Um fluxo com $b=0$ denomina-se uma Circula\c{c}\~{a}o.    
 \'{E} facil transformar o problema de fluxo em um problema de 
circula\c{c}\~{a}o capacitado. Construimos a Rede de Circula\c{c}\~{a}o
acrescendo \`{a} rede original: 
 \begin{itemize} 
 \item  Dois v\'{e}rtices especiais: a Entrada (ou Fonte), 1, 
 e a Sa\'{\i}da (ou Sorvedouro, ou Dreno), n; 
 \item Arcos da entrada aos v\'{e}tices com oferta, e dos v\'{e}rtices
com demanda \`{a} sa\'{\i}da, com capacidades iguais \`{a} 
 quantidade absoluta de oferta ou demanda: \\ 
 $(1,j)$,\ $l_{1j}=0$,\ $k_{1j}=  b_j^+$,\ $c^{1j}=0$, \\   
 $(j,n)$,\ $l_{jn}=0$,\ $k_{jn}= (-b_j)^+$,\ $c^{jn}=0$; 
 \item O arco conectando a sa\'{\i}da \`{a} entrada: \\ 
 $(n,1)$,\ $l_{n1}= \beta = \sum b_j^+$, 
 $k_{n1}=+\infty$,\ $c^{n1}=0$. 
 \end{itemize} 
 Antes de estudar a resolu\c{c}\~{a}o do problema de circula\c{c}\~{a}o 
capacitada de m\'{\i}nimo custo, enunciaremos um lema auxiliar:  

 Teorma de Hoffman: Uma rede de circula\c{c}\~{a}o capacitada tem uma
circula\c{c}\~{a}o vi\'{a}vel sse 
 $l(V,\bar{V})\leq k(\bar{V},V),\; \forall\: V\subset N$.   
 Como o fluxo \'{e} conservativo, a condi\c{c}\~{a}o \'{e}
necess\'{a}ria. Verificaremos que a condi\c{c}\~{a}o \'{e} suficiente,
ao provar a corretude do algoritmao de calibragem. 
 
 Formulemos agora o problema de circula\c{c}\~{a}o capacitada de 
custo m\'{\i}nimo, como um problema de programa\c{c}\~{a}o linear. 

 Usando vari\'{a}veis de folga primais dadas pela dist\^{a}ncia
\`{a}s restri\c{c}\~{o}es de capacidade, 
 o problema, seu dual, e a condi\c{c}\~{a}o de folgas complementares, 
podem ser escritos na forma padr\~{a}o:  
 \begin{eqnarray*} 
 (P)\; \mbox{min}\: \bar{c}\,\bar{x} & &  \bar{x}\geq 0\: 
       \mid \: \bar{A} \bar{x} = \bar{b} \\ 
 \bar{x} &=& [x; \dot{x}; \ddot{x}] \\   
 \bar{c} &=& [c, 0, 0] \\ 
 \bar{b} &=& [0; l; k] \\    
 \bar{A} &=& 
   \left[ \begin{array}{ccc}  
    A & 0 & 0 \\ I & -I & 0 \\ I & 0 & I  
   \end{array} \right] \\ 
 (D)\; \mbox{max}\: \bar{y}\,\bar{b} & &  
       \bar{y} \bar{A} \leq \bar{c} \\ 
 \bar{y} &=& [y, \dot{y}, \ddot{y}] \\ 
 \mbox{Folg. Comp.} & & \bar{x}_p\, (\bar{c} -\bar{y}\,\bar{A})^p=0 \; 
 \Leftrightarrow     
 \end{eqnarray*} 
 $$ x_{ij}(c^{ij}-(-y^i+y^j+\dot{y}^{ij}+\ddot{y}^{ij}))=0\; , \; 
    \dot{x}_{ij}\dot{y}^{ij}=0 \; , \;  
   \ddot{x}_{ij}\ddot{y}^{ij}=0 ; \Leftrightarrow  $$  
 $$ x_{ij}(d^{ij} -\dot{y}^{ij} -\ddot{y}^{ij})=0 \; , \;  
    (x_{ij}-l_{ij})\dot{y}^{ij}=0 \; , \;  
    (k_{ij}-x_{ij})\ddot{y}^{ij}=0 . $$

 As vari\'{a}veis de folga para as restri\c{c}\~{o}es de capacidade 
s\~{a}o $\dot{x}$ e $\ddot{x}$.   
 As vari\'{a}veis $y^j$ s\~{a}o denominadas potencial, e interpretadas
como o pre\c{c}o de mercado  da commoditie na pra\c{c}a $j$. 
 Definimos o differencial de custo, ou diferen\c{c}a de potencial, 
 $d^{ij}= c^{ij} +y^i -y^j$. 
 Um diferencial de custo negativo representa uma possibilidade de
arbitragem (lucro): basta comprar o  produto na pra\c{c}a $i$,
transporta-lo, e vende-lo na pra\c{c}a $j$. 
 Esta interpreta\c{c}\~{a}o torna natural que o fluxo 
 $l_{ij}\leq x_{ij}\leq k_{ij}$ ,\ $0<l_{ij}<k_{ij}$, em um arco 
obedecendo \`{a}s condi\c{c}\~{o}es de viabilidade primal, viabilidade
dual e folgas complementares deva estar em uma das tr\^es 
codi\c{c}\~{o}es seguintes: 
  
 \begin{itemize} 
 \item[L:] $x_{ij}=l_{ij}$\ , $\ddot{y}^{ij}=0$\ , 
           $\dot{y}^{ij}=d^{ij}>0$. 
 \item[M:] $l_{ij}\leq x_{ij}\leq k_{ij}$\ , 
           $\dot{y}^{ij}=\ddot{y}^{ij}=d^{ij}=0$. 
 \item[K:] $x_{ij}=k_{ij}$\ , $\dot{y}^{ij}=0$\ , 
           $\ddot{y}^{ij}=d^{ij}<0$. 
 \item[ ] \ \ ou 
 \item[L:] $x_{ij}=l_{ij}$ e $d^{ij}\geq 0$. 
 \item[M:] $l_{ij}< x_{ij}< k_{ij}$ e $d^{ij}=0$. 
 \item[K:] $x_{ij}=k_{ij}$ e $d^{ij}\leq 0$. 
 \end{itemize} 
 $$\dot{y}^{ij}= \max(0, d^{ij})\; , \;   
  \ddot{y}^{ij}= \min(0, d^{ij})\ .$$

 O algoritmo de calibragem (out-of-kilter) resolve o problema de
circula\c{c}\~{a}o capacitada de custo m\'{\i}nimo com uma abordagem 
primal-dual. Nesta abordagem, iniciamos com uma circula\c{c}\~{a}o  e
potenciais quaisquer, e, a cada passo  do algoritmo, procuraremos
aproximar a circula\c{c}\~{a}o da viabilidade e condi\c{c}\~{o}es de
folgas complementares. Assumiremos que os dados do problema, 
 $c$, $l$ e $k$, s\~{a}o inteiros.  

 A seguir tabelamos as condi\c{c}\~{o}es poss\'{\i}veis de uma arco
descalibrado em uma circula\c{c}\~{a}o,  e algumas quantidades
associadas: 
 \begin{table}[ht]  
 \begin{center} 
 \begin{tabular}{|c|c|c|c|c|} 
 \hline  
  Cond. & $d_{ij}$ & $x_{ij}$ & desvio & Cap.Arco \\ 
 \hline 
  L' & $>0$ & $<l_{ij}$ & $l_{ij}-x_{ij}$         & $h_{ij}=l_{ij}-x_{ij}$\\  
  L''& $>0$ & $>l_{ij}$ & $d^{ij}(x_{ij}-l_{ij})$ & $h_{ji}=x_{ij}-l_{ij}$\\ 
  M' & $=0$ & $<l_{ij}$ & $l_{ij}-x_{ij}$         & $h_{ij}=l_{ij}-x_{ij}$\\  
  M''& $=0$ & $>k_{ij}$ & $x_{ij}-k_{ij}$         & $h_{ji}=x_{ij}-k_{ij}$\\ 
  K' & $<0$ & $<k_{ij}$ & $d^{ij}(x_{ij}-k_{ij})$ & $h_{ij}=k_{ij}-x_{ij}$\\  
  K''& $<0$ & $>k_{ij}$ & $x_{ij}-k_{ij}$         & $h_{ji}=x_{ij}-k_{ij}$\\   
 \hline 
 \end{tabular} 
 \end{center} 
 \end{table} 

 O desvio total, em rela\c{c}\~{a}o a ``situa\c{c}\~{a}o econ\^{o}mica
de equilibrio'', \'{e} a somatoria termos em arcos de dois tipos, 
correspondento a: \\  
 (1) Volume de fluxo violando restri\c{c}\~{o}es de capacidade, nas 
condi\c{c}\~{o}es L', M', M'' e K''. \\ 
 (2) Custos de fluxos anti-econ\^{o}micos em fluxos acima do limite
inferior com diferencial de custo positivo, na cond\c{c}\~{a}o L'', 
e perdas de oportunidade em fluxos abaixo do limite superior com 
deferencial de custo negativo, na condi\c{c}\~{a}o K'. 
  
 Dada uma circula\c{c}\~{a}o invi\'{a}vel ou n\~{a}o \'{o}tima,
encontraremos uma nova  circula\c{c}\~{a}o que mantenha as
condi\c{c}\~{o}es de folgas  complementares, e que esteja mais proxima
da viabilidade. 
 A nova circula\c{c}\~{a}o \'{e} obtida pela adi\c{c}\~{a}o de uma 
circula\c{c}\~{a}o em um circuito no grafo auxiliar, $H$.  

 Construimos $H$ com arcos descalibrados, tipos (1) e (2) acima, com 
capacidades m\'{a}ximas $h_{ij}$ na tabela. 
 Acrescentamos ainda a $H$ arcos correspondendo ao tipo \\ 
 (3) Arcos calibrados na condi\c{c}\~{a}o M, com capacidades 
 $h_{ij}=k_{ij}-x_{ij}$ e $h_{ji}=x_{ij}-l_{ij}$.   
 
  Escolhido um arco descalibrado, $(i,j)$, procuramos, por busca em
profundidade em $H$, um caminho de $j$ a $i$. Se um caminho for
encontrado, a  circula\c{c}\~{a}o com volume igual ao arco de
m\'{\i}nima capacidade $h_{ij}$ no circuito, pode ser somada \`{a}
circula\c{c}\~{a}o inicial, reduzindo seu desvio total. 

 Caso contrario, seja $J$ o conjunto dos v\'{e}rices alcan\c{c}ados na 
busca a partir de $j$. Seja 
 \begin{eqnarray*} 
 S &=& \{(v,w)\in G \mid v\in J, w\in \bar{J}, 
          d^{vw}>0, l_{vw}\leq x_{vw}<k_{vw} \} \\ 
 S'&=& \{(v,w)\in G \mid v\in \bar{J}, w\in J,  
          d^{vw}<0, l_{vw}< x_{vw}\leq k_{vw} \} \\ 
 \delta &=& \min(|d^{vw}|) , \; (v,w)\in S \cup S' 
 \end{eqnarray*}
 Se $\delta <\infty$, podemos aumentar o potencial dos v\'{e}rtices em
$\bar{J}$ de $\delta$, atualizar $H$, e continuar a busca, repetindo
ajustes de potencial at\'{e} alcan\c{c}ar $i$. Caso contr\'{a}rio,
encontramos um conjunto  que viola a condi\c{c}\~{a}o do teorema de
Hoffman.    

 A cada passo do algoritmo de calibragem reduzimos o desvio total da 
circula\c{c}\~{a}o corrente.  
 Assim, ap\'{o}s um n\'{u}mero finito de passos, ou verificaremos a
inviabilidade do problema, exibindo um  sub-conjunto de v'{e}rtices que
viola a condi\c{c}\~{a}o de Hoffman,  ou a circula\c{c}\~{a}o final
ser\'{a} viavel, obedecendo tamb\'{e}m a condi\c{c}\~{a}o de folgas 
complementares, i.e., a circula\c{c}\~{a}o final ser\'{a} \'{o}tima.

 \section{M\'{e}todos de Decomposi\c{c}\~{a}o} 

 Suponha termos um PPL na forma 
 $\min cx \geq 0 , Ax=b$  onde a matriz 
 $A=\left[ \begin{array}{c} \dot{A} \\ \ddot{A} \end{array} \right]$, 
 e o poli\'{e}dro descrito por $\ddot{A}x=\ddot{b}$ tem uma estrutura
``simples'', enquanto $\dot{A}x=\dot{b}$ implica em umas ``poucas''
restri\c{c}\~{o}es adicionais  que, infelizmente, complicam em muito o
problema. 
 Por exemplo: $\ddot{A}x=\ddot{b}$ \'{e} um problema de transporte, 
enquanto $\dot{A}x=\dot{b}$ imp\~{o}e restri\c{c}\~{o}es de capacidades
coletivas sobre conjuntos de arcos, ou  $\ddot{A}x=\ddot{b}$ um conjunto
de PPL's para grupos de vari\'{a}veis separadas, enquanto
$\dot{A}x=\dot{b}$ imp\~{o}e acoplamentos entre os PPL's (estrutura
diagonal blocada por linha).  

 Estudaremos agora o m\'{e}todo de Dantzig-Wolf, que nos permite
resolver  o PPL original, atravez de itera\c{c}\~{o}es entre um Problema
Mestre  pequeno, e um Subproblema simples. Suporemos que o poli\'{e}dro
simples \'{e}  limitado, sendo igual a combina\c{c}\~{a}o convexa de 
seus v\'{e}rtices. 
 $$\ddot{X}=\{x\geq 0 \mid \ddot{A}x=\ddot{b}\} = ch(V) 
   = Vl \; , \; l\geq 0 \mid {\bf 1}'l=1$$

 O PPL original equivale ao problema mestre:  
 $$ M:\;\; \min cVl \; , \; l\geq 0 \mid 
 \left[ \begin{array}{c} \dot{A}V \\ {\bf 1}' \end{array} \right] l = 
 \left[ \begin{array}{c} \dot{b} \\  1 \end{array} \right] $$ 
 obviamente esta \'{e} uma representa\c{c}\~{a}o que tem interesse 
te\'{o}rico, n\~{a}o sendo pr\'{a}tico encontrar os muitos v\'{e}rtices 
de $V$. Uma dada base $B$ \'{e} \'{o}tima sse 
 $$-z=[cV]_R -([cV]_B B^{-1})R \equiv [cV]_R - [y,\gamma]R \geq 0$$ 
 esta condi\c{c}\~{a}o equivale a termos, para todo indice residual, $j$,   
 $$cV^j - [y,\gamma] 
   \left[ \begin{array}{c} \dot{A}V^j \\ 1 \end{array} \right] \geq 0
   \; ,\; \mbox{ou}$$  
 $$ \gamma \leq cV^j - y \dot{A}V^j = (c - y \dot{A} )V^j \; , \; \mbox{ou}$$ 
 $$ \gamma \leq \min (c - y \dot{A})v \; , \; v \in \ddot{X}$$ 

 Definimos assim o sub-problema 
 $$S:\;\; \min (c - y \dot{A})v\; , \; 
   v\geq 0 \; \mid \; \ddot{A}v= \ddot{b}$$ 
 Se a solu\c{c}\~{a}o \'{o}tima de S, $v^*$ tem valor \'{o}timo 
 $(c - y\dot{A})v^*\geq \gamma$, a base $B$ \'{e} \'{o}tima para M. 
 Caso contrario, $v^*$ nos a pr\'{o}xima coluna para entrar na base: 
 $\left[ \begin{array}{c} \dot{A}v^* \\ 1 \end{array} \right]$.  

 A solu\c{c}\~{a}o \'{o}tima do problema auxiliar nos fornece ainda um
limite inferior para o problema original. Seja $x$ uma solu\c{c}\~{a}o
vi\'{a}vel qualquer do problema original, i.e. 
 $x\in \ddot{X} \mid \dot{A}x=\dot{b}$. Sendo $x$ mais restrito, 
 $(c-y\dot{A})x\geq (c-y\dot{A})v^*$, portanto, 
 $cx \geq y\dot{b} +(c-y\dot{A})v^*$. Note tamb\'{e}m que $y\dot{b}$ 
\'{e} o limite superior corrente. Note tamb\'{e}m que n\~{a}o \'{e}
necessario ter um crescimento monotonico do limite inferior, sendo 
portanto nescess\'{a}rio guardar o melhor limite inferior j\'{a} 
obtido.  

 Como vimos, a decomposi\c{c}\~{a}o de Dantzig-Wolf adapta-se muito  bem
a problemas com certo tipo de estrutura, como a estrutura diagonal 
bolcada por linhas. 
 Caso tivessemos uma estutura diagonal blocada por colunas, poderiamos
decompor o problema dual. Este \'{e}, essencialmente,  o m\'{e}todo de
decomposi\c{c}\~{a}o de Benders, cuja implementa\c{c}\~{a}o torna-se 
muito mais eficiente usando o algoritmo Simplex Dual, que estudaremos 
a seguir. M\'{e}todos de decomposi\c{c}\~{a}o s\~{a}o muito eficientes  
para obter boas (ainda que n\~{a}o precisamente \'{o}timas) 
solu\c{c}\~{o}es para problemas estruturados. 
 Algumas estruturas comumnmente encontradas na pr\'{a}tica de Pesquisa 
Operacional incluem as formas Angular Blocada por Linhas (ABL), 
Angular Blocada por Colunas (ABC), Angular Blocada por Linhas e Colunas 
(ABLC), Blocada Escalonada (BEs), etc. 
 Em algumas aplica\c{c}\~{o}es, como Programa\c{c}\~{a}o Estoc\'{a}stica 
Multiperiodo, encontramos a forma recursiva (ou aninhada) destas 
estruturas. A figura Fig.x ilustra algumas destas formas. 
 Para uma introdu\c{c}\~{a}o intuitiva, formal e unificada aos diversos 
m\'{e}todos de decomposi\c{c}\~{a}o recomendamos os artigos de 
A.M.Geoffrion referidos na literatura. 
 Para uma vis\~{a}o alternativa sobre decomposi\c{c}\~{a}o de problemas 
de grande porte, focada em m\'{e}todos computacionais de algebra linear 
para matrizes esparsas e estruturadas, veja Stern (1994).

 \section{Simplex Dual} 

 O algoritmo simplex dual \'{e} an\'{a}logo ao simplex, s\'{o} que 
trabalha com uma base corrente dual vi\'{a}vel, at\'{e} atingir 
viabilidade primal. O simplex dual \'{e} muito \'{u}til em diversas 
situa\c{c}\~{o}es em que resolvemos um PPL, e em seguida alteramos  
as restri\c{c}\~{o}es. 

 Trabalharemos com o PPL na forma padr\~{a}o e seu dual: 
 $$P:\; \min cx\; , \; x\geq 0 \; Ax=d \; \; \; \; \mbox{e} \; \; \; \; 
   D:\; \max y'd\; , \; y'A\leq c $$ 

 Em uma base dual vi\'{a}vel, $y=c^bB^{-1}$ \'{e} uma 
solu\c{c}\~{a}o dual, i.e. 
 \begin{eqnarray*} 
  \lefteqn{ 
     \left[ \begin{array}{cc} c^b & c^r \end{array} \right] 
      - y' \left[ \begin{array}{cc} B & R \end{array} \right] } \\  
    &=& \left[ \begin{array}{cc} c^b & c^r \end{array} \right] 
     - c^b B^{-1}  \left[ \begin{array}{cc} B & R \end{array} \right] \\ 
    &=& \left[ \begin{array}{cc} c^b & c^r \end{array} \right] 
     - c^b  \left[ \begin{array}{cc} I & \tilde R \end{array} \right] \\ 
    &=&  \left[ \begin{array}{cc} c^b & c^r \end{array} \right] 
     - \left[ \begin{array}{cc} c^b & c^b \tilde R \end{array} \right] \\  
    &=& \left[ \begin{array}{cc} 0 & -z \end{array} \right] \geq 0 
 \end{eqnarray*}   

 Queremos agora reescrever o dual em uma forma analoga \`{a} forma
padr\~{a}o, acrescentando vari\'{a}veis de folga, e  usando a
parti\c{c}\~{a}o $A=[B,R]$, como segue: 
 \begin{eqnarray*} 
  \max d'y  & &  A'y\leq c'  \\ 
  \max d'y  & & 
       \left[ \begin{array}{c} B' \\ R' \end{array} \right] y \leq 
       \left[ \begin{array}{c} {c^b}' \\ {c^r}' \end{array} \right] \\ 
  \max d'y  & & 
  \left[ \begin{array}{ccc} B' & I & 0 \\ R' & 0 & I \end{array} \right] 
  \left[ \begin{array}{c} y \\ w_b \\ w_r \end{array} \right] = 
  \left[ \begin{array}{c} {c^b}' \\ {c^r}' \end{array} \right] 
  \; , \; w\geq 0  
 \end{eqnarray*}  
 Nesta forma, a base dual, sua inversa e a correspondente
solu\c{c}\~{a}o b\'{a}sica s\~{a}o dadas por: 
 $$ \left[ \begin{array}{cc} B' & 0 \\ R' & I \end{array} \right] \; , \;    
 \left[ \begin{array}{cc} B^{-t} & 0 \\ -R'B^{-t} & I \end{array} \right] $$ 
 $$ \left[ \begin{array}{c} y \\ w_r \end{array} \right] =     
 \left[ \begin{array}{cc} B^{-t} & 0 \\ -R'B^{-t} & I \end{array} \right] 
 \left[ \begin{array}{c} {c^b}' \\ {c^r}' \end{array} \right]     
 -\left[ \begin{array}{cc} B^{-t} & 0 \\ -R'B^{-t} & I \end{array} \right] 
 \left[ \begin{array}{c} I \\ 0 \end{array} \right] w_b \; \; \mbox{i.e.}$$    
 $$  y= B^{-t}{c^b}' -B^{-t}w_b \; \; \; \; e \; \; \; \; 
   w_r= {c^r}' -R'B^{-t}{c^b}' +R'B^{-t}w_b $$ 

 Note que os \'{\i}ndices em $b$ e $r$ correspondem a \'{\i}ndices
b\'{a}sicos e  residuais no primal, sendo a situa\c{c}\~{a}o no dual
complementar. 
 Analogamente ao simplex na forma padr\~{a}o, podemos aumentar um elemento 
nulo do vetor residual, para melhorar o valor da solu\c{c}\~{a}o dual, 
 $$ d'y = d'B^{-t}({c^b}' -w_b) = \mbox{const} -{\tilde{d}}' w_b $$ 
 Se $\tilde{d}\geq 0$ a solu\c{c}\~{a}o b\'{a}sica primal \'{e} 
vi\'{a}vel, e temos a solu\c{c}\~{a}o \'{o}tima dual e primal.      
 Se houver um elemento $\tilde{d}_i < 0$, podemos aumentar o valor da 
solu\c{c}\~{a}o dual incrementando $w_{b(i)}$. 
 Podemos incrementar $w_{b(i)}=\nu$, sem perder viabilidade dual, 
enquanto mantivermos   
 $$w_r= {c^r}'-R'B^{-t}{c^b}' +\nu R'B^{-t}I^i \geq 0\;\;\mbox{transpondo}$$
 $$ c^r -\tilde{c^b}R +\nu B^{-1}_i R \geq 0 \; \; \mbox{i.e.}$$ 
 $$ -z +\nu \tilde{R}_i \geq 0 $$ 
 Fazendo 
 $j= \mbox{arg} \min \{\nu(j)=\: z^j / \tilde{R}_i^j 
      \; , \; j \mid \; \tilde{R}_i^j < 0 \}$, 
 temos o \'{\i}ndice que sai da base dual.  

 Podemos pois, na nova lista de \'{\i}ndices b\'{a}sicos do primal, $b$, 
excluir $b(i)$, incluir $r(j)$, atualizar a base, e prosseguir iteragindo 
at\'{e} a otimalidade dual, i.e. viabilidade primal.

\section{Programa\c{c}\~{a}o Inteira}

 Consiremos o PPL com restri\c{c}\~{o}es de caixa, onde algumas 
 vari\'{a}veis demem ser inteiras: \\ 
 $\min cx\; , \;  x \in N^p \times R^q 
  \: , \;  l\leq x\leq u  \: \mid \; Ax=d$. \\ 
 Se todas as vari\'{a}veis s\~{a}o inteiras o problema \'{e} dito um 
PPLI Puro, caso contr\'{a}rio \'{e} dito um PPLI Mixto.  A
condi\c{c}\~{a}o de integralidade destroi a convexidade do problema, 
dai sua grande dificuldade. 
 Estudaremos o m\'{e}todo de  Ramifica\c{c}\~{a}o e Poda (Brach and
Bound, Cut, Prune),  que pode ser aplicado a problemas  relativamente
grandes, mas com poucas restri\c{c}\~{o}es de integralidade. 

 Observemos inicialmente que, se $j<p$, sua solu\c{c}\~{a}o \'{o}tima 
ser\'{a} a solu\c{c}\~{a}o de um dos dois problemas seguintes, iguais 
ao PPLI original, a menos da restri\c{c}\~{a}o de caixa sobre a 
$j$-\'{e}sima varia\'{a}vel: 
 1- $l_j \leq x_j \leq h$     \; , \; 
 2- $h +1 \leq x_j \leq u_j$  \; , \; 
 $h \in N$ , \: $l_j \leq h \leq u_j -1$.   
 Desta forma formamos uma \'{a}rvore de busca por sucessiva divis\~{a}o 
(bifurca\c{c}\~{a}o, ramifica\c{c}\~{a}o) do PPLI original 
em dois PPLIs menores. Em cada n\'{o} resolvemos o problema relaxado  
(PPL comum, sem as condi\c{c}\~{e}es de integralidade). 
 As follhas  da \'{a}rvore s\~{a}o as solu\c{c}\~{o}es que obedecem \`{a}s
condi\c{c}\~{o}es  de integridade. 
 Note que pode ocorrer uma folha prematura, i.e. a solu\c{c}\~{a}o 
\'{o}tima de um problema relaxado pode obedecer a todas as
restri\c{c}\~{o}es de integridade (antes destas serem impostas por
cortes explicitos).   
 A melhor folha da \'{a}rvore (solu\c{c}\~{a}o  inteira) j\'{a}
encontrada, $x^*$, fornece um limite superior para a solu\c{c}\~{a}o 
\'{o}tima do PPLI, $cx^*$. 

 Sejam $PPL(v)$, $x(v)$, e $c x(v)$  o PPL relaxado no n\'{o} $v$, 
sua solu\c{c}\~{a}o \'{o}tima, e o valor \'{o}timo. 
 $cx(v)$ fornece um limite inferior para as solu\c{c}\~{a}o que podem 
ser obtidas por futuras ramifica\c{c}\~{o}es deste n\'{o}. 
 Um n\'{o} invi\'{a}vel recebe valor $cx(v)= +\infty$.   
 Sempre que $cx(v)\geq cx^*$, podemos eliminar (do processo de busca)
todas as ramifica\c{c}\~{o}es deste n\'{o},  i.e. podemos podar a
\'{a}rvore.  

 Se poss\'{\i}vel, terminamos o processo de ramifica\c{c}\~{a}o quando
encontramos uma solu\c{c}\~{a}o garantidamente  suficientemente
pr\'{o}xima do \'{o}timo.  Note que mesmo ap\'{o}s encontrar a 
solu\c{c}\~{a}o \'{o}tima, $x^*$ \'{e} poss\'{\i}vel que tenhamos que 
crescer muito a \'{a}rvore at\'{e} poder afirmar que $x^*$ \'{e} de fato 
a solu\c{c}\~{a}o \'{o}tima do PPLI. 
 Note ainda que o Simplex dual pode ser de grande ajuda para resolver 
os PPLs criados por ramifica\c{c}\~{a}o.       
 A efici\^{e}ncia do m\'{e}todo de ramifica\c{c}\~{a}o e poda depende 
da heur\'{\i}tica de constru\c{c}\~{a}o da \'{a}rvore,  i.e. do
processo das sucessivas escolhas de n\'{o}s, vari\'{a}veis e pontos de
corte, $PPL(v)$, $x_j$ e $h$. 

 Seja uma vari\'{a}vel b\'{a}sica, $x_j = \lfloor x_j \rfloor +f_j$, 
onde $\lfloor x_j \rfloor$ e $f_j$ indicam as partes inteira e 
fracion\'{a}ria de $x_j$.
 Ao ramificar o n\'{o} $v$, na varia\'{a}vel $x_j$, seus dois filhos 
tem a restri\c{c}\~{a}o de caixa $l_j \leq x_j \leq u_j$, substituida
por $l_j \leq x_j \leq h$ ou $h+1 \leq x_j \leq u_j$, onde o ponto de 
corte \'{e} $h= \lfloor x_j(v) \rfloor = x_j(v) -f_j$, ou 
$h+1= x_j(v) +(1-f_j)$.

 Uma heur\'{\i}stica de escolha de vari\'{a}vel \'{e} ramificar sempre a
vari\'{a}vel fracion\'{a}ria ``mais importante''. As prioridades  
poderiam vir de um vetor de pesos $p$, dados pela experi\^{e}ncia com o
problema. Poderiamos talves usar $p=|c|$.
 Definimos as penalidade associadas aos cortes  
 ``acima'' e ``abaixo'' de $x_j(v)$: 
 $P^l_j= f_j p_j$ e $P^u_j= (1-f_j) p_j$. 
 Estas penalidades pretendem prever o impacto dos cortes 
 nos valores \'{o}timos dos filhos de $v$, e nos permitem definir 
 duas heur\'{\i}sticas para escolha da vari\'{a}vel de corte: \\ 
 H1: \ Ramifique em uma vari\'{a}vel em 
 $\arg \min_j \min \{P^l_j, P^u_j \}$. 
 Esta heur\'{\i}stica visa nos levar a uma ``boa'' solu\c{c}\~{a}o 
 inteira. \\ 
 H2: \ Ramifique em uma vari\'{a}vel em   
 $\arg \max_j \max \{ P^l_j, P^u_j \}$.  
 Esta heur\'{\i}stica visa criar um n\'{o} que ser\'{a} rapidamente 
 podado, evitando um crescimento exagerado da \'{a}rvore.
 
 Analogamente, definimos duas heur\'{\i}sticas para escolha do n\'{o} a
 ser ramificado: \\ 
 BFS: \ Breadth First Serach, ou BEL, Busca Em Largura. 
 Dentre os n\'{o}s existentes na \'{a}rvore, ramifique o com melhor valor 
 relaxado, $cx(v)$. \\  
 DFS: Depth First Search, ou BEP, Busca Em Profundidade.  
 Dos n\'{o}s, ainda n\~{a}o podados, criados na mais recente
bifurca\c{c}\~{a}o, ramifique o com melhor valor relaxado, $cx(v)$. \\

\section{Exerc\'{\i}cios} 

 \begin{itemize}

 \item[1.] Geometria e lemas simples: 

 a- Desenhe o simplex, $S_n$,  e o cubo, $C_n$ de dimens\~{a}o 
 2 e 3. $S=\{ x\geq 0 \mid {\bf 1}'x \leq 1 \}$, 
 $C=\{ x\geq 0 \mid Ix \leq {\bf 1} \}$. 

 b- Reescreva $S_2$, $S_3$, $C_2$ e $C_3$ como poliedros padr\~{a}o em 
 $R^n$, onde $n= 3, 4, 4, 6$, respectivamente.   

 c- Prove que um poliedro (na forma padr\~{a}o) \'{e} convexo. 

 d- Prove os lemas 1 e 2 de dualidade. 

 e- Prove que um Poli\'{e}dro limitado \'{e} igual ao conjunto das 
combina\c{c}\~{o}es convexas de seus v\'{e}rtices.

 \item[2.] Escreva um programa para resolver um PPL na forma padr\~{a}o 
por enumera\c{c}\~{a}o exaustiva de todos os v\'{e}rtices. 
 Sugest\~{a}o: Use o programa combina.m para enumerar todos os grupos de $m$
colunas da matriz $A$, $m$ por $n$, $n>n$. Em seguida forme $B$, a
matriz  quadrada com estas $m$ colunas, verifique se $B$ \'{e} uma base
(\'{e} invers\'{\i}vel), e se a solu\c{c}\~{a}o b\'{a}sica \'{e} um
v\'{e}rtice, $\tilde d = B^{-1}d >0$. Calcule o valor das solu\c{c}\~{o}es
b\'{a}sicas vi\'{a}veis, e retorne a base \'{o}tima.

 \item[3.] Adapte e implemente o Simplex para usar a fatora\c{c}\~{a}o QR,
e atualizar, ao inv\'{e}s de reinverter, a base a cada passo do algoritmo. 
Incua uma forma de monitorar a qualidade da fatora\c{c}\~{a}o para, 
periodicamente reinverter a base, evitando a degrada\c{c}\~{a}o da 
inversa pelo ac\'{u}mulo de erro nas opera\c{c}\~{o}es aritm\'{e}ticas.  

 \item[4.]  Considere a solu\c{c}\~{a}o \'{o}tima do PPL padr\~{a}o com
vetor de custo parametrizado: \\ $c(\eta)=f +\eta g$.  

 a- Quanto podemos aumentar (ou diminuir) o par\^{a}metro $\eta$ antes
que a base que nos fornece a solu\c{c}\~{a}o \'{o}tima mude ?

 b- Ao passar por um ponto de mudan\c{c}a de base (em fun\c{c}\~{a}o de
aumento de $\eta$), i.e.  ao passar por um $\eta$ cr\'{\i}tico, qual a
pr\'{o}xima base \'{o}tima ?

 c- Conclua, usando o argumento de que existe um n\'{u}mero finito de
bases, que existe um n\'{u}mero finito de etas cr\'{\i}ticos.  

 d- Como poder\'{\i}amos utilizar o dual do problema parametrizado no
vetor de restri\c{c}\~{o}es para determinar a seq\"{u}\^{e}ncia de seus
etas ($\eta$) cr\'{\i}ticos ?

 e- Argumente que o valor \'{o}timo deste PPL param\'{e}trico, em
fun\c{c}\~{a}o do par\^{a}metro, \'{e} uma curva continua e c\^oncava, 
linear por trechos  i.e. composta de segmentos de reta. 

 f- Adapte e implemente o Simplex para listar os etas e as
solu\c{c}\~{o}es  cr\'{\i}ticas, e implemente um programa auxiliar para
calcular $x(\eta)$ por interpola\c{c}\~{a}o a partir da lista.

 \item[5.] Adapte e implemente o Simplex para restri\c{c}\~{o}es de
caixa, i.e., \\ 
 $\min cx\; , \;  l\leq x\leq u \: \mid \; Ax=d$. 
 
 Dica: Considere dada uma base viavel, $B$, e uma parti\c{c}\~{a}o 
 $\left[ \begin{array}{ccc} B & R & S \end{array} \right]$, onde \\  
 $l_b < x_b < u_b$, $x_r = l_r$, $x_s = u_s$, de modo que, \\ 
 $x_b = B^{-1}d -B^{-1}Rx_r -B^{-1}Sx_s$  \\      
 $cx = c^b B^{-1} d +(c^r -c^b B^{-1} R)x_r +(c^s -c^b B^{-1} S)x_s 
     = \varphi +z^r x_r +z^s x_s$    
 
 Se $z^{r(k)}<0$, podemos melhorar a solu\c{c}\~{a}o corrente aumentando 
 a variavel residual no limite inferior, 
  $x_{r(k)} = l_{r(k)} +\delta_{r(k)}$,      
 \ \ $x_b = B^{-1}d -B^{-1}R l_r -B^{-1}S u_s -\delta_{r(k)} B^{-1}R^k$. \\ 
 Todavia, $\delta_{r(k)}$ deve respeitar os seguintes limites: \\ 
 1- $x_{r(k)} = l_{r(k)} +\delta_{r(k)} \leq u_{r(k)}$, \ \   
 2- $x_b \geq l_b$, \ \ 
 3- $x_b \leq u_b$. \\ 
 Analogamente, se $z^{s(k)}>0$, podemos melhorar a solu\c{c}\~{a}o corrente 
 diminuindo a variavel residual no limite superior, 
 $x_{s(k)} = u_{s(k)} -\delta_{s(k)}$.

 \item[6.] Fa\c{c}a um estudo emp\'{\i}rico da complexidade esperada
(flops) do Simplex para  resolver um problema em fun\c{c}\~{a}o de:  
 1) Forma de atualiza\c{c}\~{a}o. 
 2) Crit\'{e}rio de entrada.    

 \item[6.] Implemente o algoritmo de ajuste. Compare seu desempenho 
com o Simplex em problemas de circula\c{c}\~{a}o capacitada. 

 \item[7.] Adapte e implemente o Simplex Dual para problemas com
restri\c{c}\~{o}es de caixa. 

 \item[8.] Implemente o algoritmo de poda e ramifica\c{c}\~{a}o.
 Compare empiricamente diferentes heur\'{\i}sticas de sele\c{c}\~{a}o
para o processo de ramifica\c{c}\~{a}o.    

 \item[9.] Implemente o m\'{e}todo de decomposi\c{c}\~{a}o de 
Dantzig-Wolf para problema na forma diagonal blocada. 

 \item[10.] Implemente o m\'{e}todo de decomposi\c{c}\~{a}o de 
Benders para o problema de programa\c{c}\~{a}o estoc\'{a}stica 
em dois est\'{a}gios. 

 \item[11.] Prepare um semin\'{a}rio apresentando um modelo da 
biblioteca de exemplos do GAMS.

\end{itemize}

%% file: cap2.tex
\chapter{Programa\c{c}\~{a}o N\~{a}o Linear}

\section{GRG: Gradiente Reduzido Generalizado} 

 Consideremos o problema de programa\c{c}\~{a}o n\~{a}o linear 
com restri\c{c}\~{o}es n\~{a}o lineares de igualdade, alem de 
restri\c{c}\~{o}es de caixa sobre as vari\'{a}veis, 
 \[ 
    \mbox{PPNL:} \ \ \min f(x) 
     \ \ , \ \ \  f: \Re^n \mapsto \Re 
 \] 
 \[ 
    l \leq x \leq u \ \g \ h(x)=0   
      \ \ , \ \ \ h: \Re^n \mapsto \Re^m  
 \] 

 O M\'{e}todo do Gradiente Reduzido Generalizado, ou GRG, 
imita os passos do Simplex, para uma lineariza\c{c}\~{a}o local 
do PPNL. Seja $x$ um ponto vi\'{a}vel inicial. 

 Como no Simplex, assumimos a hip\'{o}tese de N\~{a}o 
Degenerec\^{e}ncia, i.e., assumimos que no m\'{a}ximo $(n-m)$ 
das restri\c{c}\~{o}es de caixa sejam ativas em um ponto vi\'{a}vel. 
 Assim, podemos tomar $m$ das vari\'{a}veis com as restri\c{c}\~{o}es 
de caixa folgadas como sendo as vari\'{a}veis B\'{a}sicas 
(ou dependentes), e as $n-m$ vari\'{a}veis restantes como 
vari\'{a}veis Residuais (ou independentes). 
 Como no Simplex, particionamos todas as entidades vetoriais e 
matriciais reordenando as vari\'{a}veis de modo a agrupar 
as vari\'{a}veis b\'{a}sicas, $x_b$, e residuais, $x_r$, 
 \[ 
    x=   
 \left[ \begin{array}{c} x_b \\ x_r \end{array} \right] \ , \ \ 
    l=   
 \left[ \begin{array}{c} l_b \\ l_r \end{array} \right] \ , \ \ 
    u=   
 \left[ \begin{array}{c} u_b \\ u_r \end{array} \right] \ , \ \ 
    \nabla f(x)=   
 \left[ \begin{array}{cc} 
   \nabla^b f(x) & \nabla^r f(x) 
 \end{array} \right] 
 \]    
 \[ 
   J(x) \ = \ \ 
 \left[ \begin{array}{cc} 
   J^B(x) &  J^R(x) 
 \end{array} \right]  \ = \ \   
 \left[ \begin{array}{cc} 
   \nabla^b h_1(x) & \nabla^r h_1(x) \\  
   \nabla^b h_2(x) & \nabla^r h_2(x) \\  
   \vdots          & \vdots          \\ 
   \nabla^b h_m(x) & \nabla^r h_m(x) 
 \end{array} \right] 
 \] 

 Consideremos o efeito de uma pequena altera\c{c}\~{a}o no ponto 
vi\'{a}vel corrente, $x+\delta$, assumindo que as fun\c{c}\~{o}es 
$f$ e $h$ s\~{a}o cont\'{\i}nuas e diferenci\'{a}veis. 
 A correspondente altera\c{c}\~{a}o no  valor da 
solu\c{c}\~{a}o ser\'{a} 
 \[ 
    \Delta f \ = \ \ 
   f(x+\delta) -f(x) \ \ \approx \ \ 
   \nabla f(x) \; \delta \ \ = \ \ 
   \left[ \begin{array}{cc} 
     \nabla^b f(x) & \nabla^r f(x) 
   \end{array} \right] \ 
   \left[ \begin{array}{c} 
     \delta_b \\ \delta_r 
   \end{array} \right]       
 \] 
 Queremos tamb\'{e}m que o ponto alterado $x+\delta$ permane\c{c}a 
 (aproximadamente) vi\'{a}vel, i.e., 
 \[  
    \Delta h \ = \ \ 
   h(x+\delta) -h(x) \ \ \approx \ \ 
    J(x) \; \delta \ \ = \ \ 
   \left[ \begin{array}{cc} 
     J^b(x) & J^r(x) 
   \end{array} \right] \ 
   \left[ \begin{array}{c} 
     \delta_b \\ \delta_r 
   \end{array} \right] \ \ = \ \ 0       
 \] 
  Isolando $\delta_b$, e assumindo que a b\'{a}se 
 $J^b(x)$ \'{e} invers\'{\i}vel,  
 \begin{eqnarray*} 
   \delta_b  & = &  
       -\left( J^b(x) \right)^{-1} \; J^r(x) \; \delta_r \\         
   \Delta f & \approx & 
    \nabla^b f(x) \; \delta_b +\nabla^r f(x) \; \delta_r \\ 
            & = & 
  \left( \nabla^r f(x) -\nabla^b f(x) 
       \left( J^b(x) \right)^{-1} \; J^r(x)  \right) \; \delta_r 
    \ \ = \ \ \ z(x) \; \delta_r 
 \end{eqnarray*}

 Como o problema n\~{a}o \'{e} linear, n\~{a}o podemos 
garantir que no ponto \'{o}timo todas as vari\'{a}veis residuais 
tem uma restri\c{c}\~{a}o de caixa justa, o an\'{a}logo de um 
v\'{e}rtice no PPL padr\~{a}o.  
 Assim, n\~{a}o h\'{a} interesse em restringir $\delta_r$ 
a ter apenas uma componente n\~{a}o nula, como no Simplex. 
 Estas considera\c{c}\~{o}es sugerem mover o ponto vi\'{a}vel corrente 
na dire\c{c}\~{a}o (no espa\c{c}o das vari\'{a}veis residuais) 
do vetor $v_r$, oposta ao gradiente reduzido, sempre que 
a correspondente restri\c{c}\~{a}o de caixa for folgada, i.e., 
 \[  
    v_{r(i)} = \left\{ \begin{array}{cl} 
         -z^i & \mbox{se} \ z^i > 0 \ \mbox{e} \ x_{r(i)} > l_{r(i)} \\ 
         -z^i & \mbox{se} \ z^i < 0 \ \mbox{e} \ x_{r(i)} < u_{r(i)} \\ 
         0 & \mbox{caso contrario}  
          \end{array} \right. 
 \] 

  Na pr\'{o}xima sec\c{c}\~{a}o estudaremos condi\c{c}\~{o}es 
 gerais de converg\^{e}ncia para algoritmos de PNL, e veremos que 
 a descontinuidade do vetor $v_r$ em fun\c{c}\~{a}o da folga nas 
 restri\c{c}\~{o}es de caixa \'{e} indesej\'{a}vel. 
  Assim, usaremos uma forma cont\'{\i}nua do vetor de 
dire\c{c}\~{a}o, por exemplo, 
 \[  
    v_{r(i)} = \left\{ \begin{array}{cl} 
         -( x_{r(i)} -l_{r(i)} ) z^i & \mbox{se} \ z^i > 0 
              \ \mbox{e} \ x_{r(i)} > l_{r(i)} \\ 
         -( u_{r(i)} -x_{r(i)} ) z^i & \mbox{se} \ z^i < 0 
              \ \mbox{e} \ x_{r(i)} < u_{r(i)} \\ 
         0 & \mbox{caso contrario}  
          \end{array} \right. 
 \]

  A ideia do m\'{e}todo GRG \'{e} a de, a cada itera\c{c}\~{a}o do 
 algoritmo, movimentar o ponto vi\'{a}vel, dando um passo 
 $x +\delta$ com  $\delta=\eta v$, onde  
 $v_b= -\left( J^b(x) \right)^{-1}\;J^r(x)\;v_r$, 
 i.e., um passo de ``tamanho'' $\eta$ na dire\c{c}\~{a}o 
 (no espa\c{c}o das vari\'{a}veis residuais) $v_r$.  
  Para determinar o tamanho do passo, $\eta$, \'{e} necess\'{a}rio 
 fazer uma busca linear, respeitando as restri\c{c}\~{o}es de caixa. 
  Note ainda que $v_b$, a dire\c{c}\~{a}o no espa\c{c}o das vari\'{a}veis 
 b\'{a}sicas, foi escolhido de modo a que $x+ \eta v$, permane\c{c}a 
 um ponto aproximadamente vi\'{a}vel, pois estamos nos movendo no 
 hiperplano tangente a superf\'{\i}cie 
 (o espa\c{c}o tangente \`{a} variedade) 
 determinada por $h(x)=0$.           

  Um novo ponto $x$ dever\'{a} ser sempre acompanhado de uma 
 ``corre\c{c}\~{a}o'' $\Delta x$ para re-obter viabilidade exata 
 nas restri\c{c}\~{o}es n\~{a}o lineares, $h(x+\Delta x)=0$. 
  O ponto $x$ pode ser utilizado como ponto inicial de um m\'{e}todo 
 recursivo para encontrar um ponto exatamente vi\'{a}vel. 
  O m\'{e}todo de Newton-Raphson consiste de considerar Jacobiano
 $J^b(x)$ para calcular a corre\c{c}\~{a}o, 
 \[ 
    \Delta x_b= -\left( J^b(x) \right)^{-1} \; h(x_b, x_r) 
 \]

 \section{Busca Linear e Converg\^{e}ncia Local}

 Consideremos o problema de minimizar uma fun\c{c}\~{a}o, 
unidimensional, $f(x)$. Primeiramente, consideremos, como 
um problema auxiliar, o problema de encontrar a raiz (zero) 
de uma fun\c{c}\~{a}o derivavel, que aproximamos por sua  
s\'{e}rie de Taylor de primeira ordem, 
 $g(x) \approx q(x^k) +g'(x^k)(x-x^k)$. 
 Esta aproxima\c{c}\~{a}o implica que 
 $g(x^{k+1})\approx 0$, onde  
 \[ 
    x^{k+1} = x^k -g'(x^k)^{-1} g(x^k) 
 \] 
 Este \'{e} o M\'{e}todo de Newton para encontrar a raiz de 
uma fun\c{c}\~{a}o unidimensional.

 Mas se $f(x)$ \'{e} diferenciavel, encontrar seu ponto de 
m\'{i}nimo implica encontrar um ponto onde se anula a derivada 
primeira, assim, escrevemos o M\'{e}todo de Newton para 
minimiza\c{c}\~{a}o de um fun\c{c}\~{a}o unidimensional, 
 \[ 
    x^{k+1} = x^k -f''(x^k)^{-1} f'(x^k)  
 \] 

 Examinemos qu\~{a}o rapidamente a seq\"{u}\^{e}ncia gerada 
pelo m\'{e}todo de Newton se aproxima do ponto de m\'{\i}nimo, 
$x^*$, se a seq\"{u}\^{e}ncia j\'{a} estiver pr\'{o}xima do mesmo. 
 Assumindo diferenciabilidade at\'{e} terceira ordem, podemos 
escrever 
 \[ 
   0 = f'(x^*) = f'(x^k) + f''(x^k)(x^*-x^k) 
      +(1/2)f'''(y^k)(x^*-x^k)^2 \ , \ \mbox{ou} 
 \] 
 \[ 
    x^* = x^k -f''(x^k)^{-1}f'(x^k) 
         -(1/2)f''(x^k)^{-1}f'''(y^k)(x^*-x^k)^2 
 \] 
 Subtraindo da equa\c{c}\~{a}o que define o 
 m\'{e}todo de Newton, temos  
 \[ 
    (x^{k+1}-x^*) = 
      (1/2)f''(x^k)^{-1}f'''(y^k) \, (x^k-x^*)^2 
 \] 
   
 Como veremos adiante, este resultado implica que o m\'{e}todo de 
Newton converge rapidamente (quadraticamente) quando j\'{a} pr\'{o}ximo 
do ponto de \'{o}timo. Todavia, o m\'{e}todo de Newton necessita  muita
informa\c{c}\~{a}o diferencial sobre a fun\c{c}\~{a}o,  algo que pode
ser dif\'{\i}cil de obter. 
 Ademais, longe do ponto de \'{o}timo n\~{a}o temos nenhuma 
garantia sobre sua converg\^{e}ncia. A seguir, estudaremos 
m\'{e}todos que visam superar estas dificuldades.

 Examinemos agora o m\'{e}todo da Busca pela Raz\~{a}o \'{A}urea 
para minimizar uma fun\c{c}\~{a}o unidimensional e unimodal, 
 $ f(x)$, em um intervalo, $[x^1, x^4]$. 
 Se soubermos o valor da fun\c{c}\~{a}o em quatro pontos, os 
estremos do intervalo e mais dois pontos interiores, 
 $x^1<x^2<x^3<x^4$, a unimodalidade da fun\c{c}\~{a}o garante 
podermos determinar que o ponto de m\'{\i}nimo, $x^*$, est\'{a} 
em um subintervalo, i.e., 
 \[ 
    f(x^2)\leq f(x^3) \Rightarrow x^* \in [x^1, x^3] \ \ , \ \ 
    f(x^2) > f(x^3)   \Rightarrow x^* \in [x^2, x^4] \ \ . 
 \] 
 
 Sem perda de generalidade consideremos a forma de dividir o 
 intervalo $[0,1]$. Uma raz\~{a}o $r$ define uma divis\~{a}o 
 sim\'{e}trica da forma $0<1-r<r<1$. 
 Dividindo o subintervalo $[0,r]$, pela mesma raz\~{a}o 
 obtemos $0<r(1-r)<r^2<r$. 
 Para que apenas uma nova avalia\c{c}\~{a}o da fun\c{c}\~{a}o 
 seja necess\'{a}ria na pr\'{o}xima itera\c{c}\~{a}o, 
 devem coincidir os pontos $r^2$ e $1-r$, 
 i.e. $r^2+r-1=0$, de modo que $r=(\sqrt{5}-1)/2$, 
 a raz\~{a}o \'{a}urea $r\approx 0.6180340$.

 O m\'{e}todo da raz\~{a}o \'{a}urea \'{e} robusto, funcionando 
para qualquer fun\c{c}\~{a}o unimodal, utilizando apenas o 
valor da fun\c{c}\~{a}o nos pontos de avalia\c{c}\~{a}o. 
 No entanto, os extremos do intervalo de busca se aproximam
apenas linearmente com o n\'{u}mero de itera\c{c}\~{o}es.  
 M\'{e}todos polinomiais, estudados a seguir, tentam conciliar 
as melhores caracter\'{\i}sticas dos m\'{e}todos j\'{a} 
apresentados. 

 M\'{e}todos polinomiais para minimizar uma fun\c{c}\~{a}o 
unidimensional, $\min f(x +\eta)$, sobre $\eta \geq 0$, baseiam-se no
ajuste de um polin\^{o}mio que aproxima localmente $f(x)$, e a
subseq\"{u}ente  minimiza\c{c}\~{a}o anal\'{\i}tica do polin\^{o}mio
ajustado. 
 O mais simples destes m\'{e}todos \'{e} o do ajuste quadr\'{a}tico. 
 Consideremos conhecidos tr\^{e}s pontos, 
 $\eta_1, \eta_2, \eta_3$, com o 
respectivo valor da fun\c{c}\~{a}o, $f_i=f(x +\eta_i)$. 
 Considerando as equa\c{c}\~{o}es do o polin\^{o}mio interpolador 
 \[ 
    q(\eta)= a \eta^2 +b \eta + c  
     \ \ , \ \ 
    q(\eta_i) = f_i  
 \] 
 obtemos as constantes do polinomio,     
 \begin{eqnarray*}
  a &=& \frac{ 
  f_1(\eta_2 -\eta_3) +f_2(\eta_3 -\eta_1) +f_3(\eta_1 -\eta_2)
 }{ -(\eta_2 -\eta_1)(\eta_3 -\eta_2)(\eta_3 -\eta_1) } \\ 
  b &=& \frac{  
  f_1(\eta_3^2 -\eta_2^2) +f_2(\eta_1^2 -\eta_3^2) 
 +f_3(\eta_2^2 -\eta_1^2)
 }{ -(\eta_2 -\eta_1)(\eta_3 -\eta_2)(\eta_3 -\eta_1) } \\ 
  c &=& \frac{  
  f_1(\eta_2^2\eta_3 -\eta_3^2\eta_2) 
 +f_2(\eta_3^2\eta_1 -\eta_1^2\eta_3) 
 +f_3(\eta_1^2\eta_2 -\eta_2^2\eta_1)
 }{ -(\eta_2 -\eta_1)(\eta_3 -\eta_2)(\eta_3 -\eta_1) }   
 \end{eqnarray*} 

 Anulando a derivada do polin\^{o}mio interpolador, 
 $q'(\eta_4)= 2a\eta +b$, obtemos seu ponto de m\'{\i}nimo, 
 $\eta_4= a/2b$, ou diretamente, 
 \[ 
  \eta_4 = \frac{1}{2}  \frac{ 
  f_1(\eta_3^2 -\eta_2^2) +f_2(\eta_1^2 -\eta_3^2) 
 +f_3(\eta_2^2 -\eta_1^2) 
 }{ 
  f_1(\eta_3 -\eta_2) +f_2(\eta_1 -\eta_3) 
 +f_3(\eta_2 -\eta_1)
 } 
 \] 

  Procuraremos sempre utilizar os pontos iniciais no 
 ``padr\~{a}o de interpola\c{c}\~{a}o'',  
 $\eta_1 <\eta_2 <\eta_3$ e $f_1 \geq f_2 \leq f_3$, 
 i.e., tr\^{e}s pontos onde o ponto intermedi\'{a}rio tem o 
 menor valor da fun\c{c}\~{a}o. 
 Assim, garantimos que o m\'{\i}nimo do polin\^{o}mio  
 estar\'{a} dentro do intervalo inicial de busca, 
 $\eta_4\in [\eta_1, \eta_3]$, ou seja, que estamos 
 interpolando e n\~{a}o extrapolando a solu\c{c}\~{a}o. 

  Escolhendo $\eta_4$ e mais dois dentre os tr\^{e}s pontos 
 iniciais, temos uma nova trinca no padr\~{a}o desejado, e 
 estamos prontos para uma nova itera\c{c}\~{a}o. 
  Note ainda que, em geral, n\~{a}o podemos garantir que o 
 quarto ponto ser\'{a} o melhor da nova trinca. 
  Todavia, o quarto ponto sempre substitui um ponto da 
 trinca antiga que tinha valor maior, de modo que a soma 
 $z=f_1+f_2+f_3$ \'{e} monotonicamente decrescente no processo. 
  Como veremos a seguir, estas caracter\'{\i}sticas garantem a 
 converg\^{e}ncia global do algoritmo de busca linear por 
 ajuste quadr\'{a}tico.   

  Consideremos agora os erros em rela\c{c}\~{a}o ao ponto de 
m\'{\i}nimo, $\epsilon_i= x^* -x_i$.  
 Podemos expressar 
 $\epsilon_4 = g(\epsilon_1,\epsilon_2,\epsilon_3)$. 
 A fun\c{c}\~{a}o $g$ deve ser um polin\^{o}mio de segundo grau, 
pois o ajuste feito \'{e} quadr\'{a}tico, e sim\'{e}trico em seus 
argumentos, pois a ordem dos pontos \'{e} irrelevante. 
 Ademais, n\~{a}o \'{e} dif\'{\i}cil verificar que se $\epsilon_4$ 
se anula se dois dos tr\^{e}s erros iniciais se anulam. 
 Portanto, na proximidade do ponto de m\'{\i}nimo, $x^*$, 
temos a seguinte aproxima\c{c}\~{a}o do quarto erro: 
 \[ 
    \epsilon_4 = C \left( \epsilon_1 \epsilon_2 
     +\epsilon_1 \epsilon_3 +\epsilon_2 \epsilon_3 \right) 
 \]  

  Assumindo a convergencia do processo, o 
  $k$-\'{e}simo erro ser\'{a} approximadamente 
  $\epsilon_{k+4} = C \epsilon_{k+1} \epsilon_{k+2}$.   
  Tomando 
  $l_k= \log(C^{1/2}\epsilon_k)$, 
 temos 
  $l_{k+3} = l_{k+1} +l_{k}$, 
 cuja equa\c{c}\~{a}o caracter\'{\i}stica \'{e}    
  $\lambda^3 -\lambda -1 =0$. 
 A maior raiz desta equa\c{c}\~{a}o \'{e} 
  $\lambda \approx 1.3$.  
 Esta \'{e} a ordem de converg\^{e}ncia local do processo, 
 conforme definiremos a seguir.  

 Dizemos que uma seq\"{u}\^{e}ncia de n\'{u}meros reais 
 $r^k\rightarrow r^*$ converge ao menos em ordem $p>0$ se 
 \[  
    0 \leq \lim_{k\rightarrow \infty} 
    \frac{ | r^{k+1}-r^* | }{ | r^k-r^* |^p }  = \beta < \infty  
 \]  
 A ordem de converg\^{e}ncia da seq\"{u}\^{e}ncia \'{e} o 
 supremo das constantes $p>0$ nestas condi\c{c}\~{e}s.  
 Se $p=1$ e $\beta<1$, dizemos que a seq\"{u}\^{e}ncia tem 
 Converg\^{e}ncia Linear com Taxa $\beta$. 
 Se $\beta=0$, dizemos que a seq\"{u}\^{e}ncia tem 
 Converg\^{e}ncia Super-Linear.  

 Assim, para $c\geq 1$, $c$ \'{e} a ordem de converg\^{e}ncia 
 da seq\"{u}\^{e}ncia $a^{(c^k)}$. 
 Vemos tamb\'{e}m que $1/k$ converge em ordem $1$, embora n\~{a}o 
 seja linearmente convergente, pois $r^{k+1}/r^k\rightarrow 1$.   
 Finalmente, $(1/k)^k$ converge em ordem $1$, 
 pois  para qualquer $p>1$,   
 $r^{k+1}/(r^k)^p\rightarrow \infty$, 
 Todavia a converg\^{e}ncia \'{e} super-liner, pois        
 $r^{k+1}/r^k\rightarrow 0$.

 \pagebreak 

 \section{Partan} 

 Estudaremos agora o m\'{e}todo da Tangentes Paralelas, ou Partan, 
para resolver o problema de minimizar uma fun\c{c}\~{a}o convexa 
irrestrita. Este m\'{e}todo, como tantos outros, utiliza a id\'{e}ia 
de modelar a fun\c{c}\~{a}o a ser minimizada por uma fun\c{c}\~{a}o 
quadr\'{a}tica que, sem perda de generalidade, suporemos centrada 
na origem,  $f(x)=(1/2)x'Qx$. 

 Pela transforma\c{c}\~{a}o linear $y=Ux$, onde $U$ \'{e} o 
 fator de  Cholesky $U'U=Q$, podemos escrever $f$ com simetria 
 esf\'{e}rica, 
 \[ 
    f(y) = (1/2)(U^{-1}y)'Q(U^{-1}y)
        = (1/2)y'(U^{-t}(U'U)U^{-1})y = (1/2)y'Iy 
 \] 

 Lembremos as defini\c{c}\~{o}es de paralelismo e 
 ortogonalidade para planos de dimens\~{o}es arbitrarias: 
 Uma reta \'{e} paralela a um plano se for paralela 
 a uma reta deste plano. 
 Uma reta \'{e} ortogonal a um plano se for ortogonal  
 a toda reta deste plano. 
 Um plano \'{e} paralelo a outro plano se toda reta deste plano 
 for paralela ao outro plano. 
 Um plano \'{e} ortogonal a outro plano se toda reta deste plano 
 for ortogonal ao outro plano. 
 Note que estas defini\c{c}\~{o}es est\~{a}o de acordo com as 
 defini\c{c}\~{o}es usuais de espa\c{c}os vetoriais, mas s\~{a}o 
 diferentes das defini\c{c}\~{o}es usadas em geometria tridimensional.

 Uma propriedade fundamental de uma transforma\c{c}\~{a}o linear linear 
qualquer, $y=Ux$, \'{e} a de preservar as rela\c{c}\~{o}es de
Coplanaridade e Paralelismo, i.e.: 

 - Se os pontos $x^1\ldots x^k$ pertencem a um mesmo plano (reta), 
assim o \'{e} para os pontos $y^1\ldots y^k$, e vice-versa. 

 - Se os planos $\pi_{1,\ldots k}$ e $\pi_{k+1,\ldots k+h}$,  
 determinados, no antigo sistema de coordenadas pelos pontos 
 $x^1\ldots x^k$ e $x^{k+1}\ldots x^{k+h}$, s\~{a}o paralelos, 
 ent\~{a}o estes mesmos planos, determinados no novo sistema de 
 coordenadas pelos pontos $y^1\ldots y^k$ e $y^{k+1}\ldots y^{k+h}$, 
 tamb\'{e}m s\~{a}o paralelos.    

 Assim, se definirmos um bom algoritmo para o caso particular de 
fun\c{c}\~{o}es esf\'{e}ricas, usando apenas rela\c{c}\~{o}es de 
coplanaridade e paralelismo, este mesmo algoritmo permanecer\'{a} 
v\'{a}lido no caso geral. Este \'{e} o argumento b\'{a}sico no 
desenvolvimento do algoritmo Partan.

 Lembremos ainda que a rela\c{c}\~{a}o de ortogonalidade, $v'w=0$, 
 N\~{a}o \'{e} preservada por uma transforma\c{c}\~{a}o linear, 
 pois se $v=Ux$ e $w=Uy$, ent\~{a}o 
 $v'w = (Ux)'(Uy) = x'(U'U)y = x'Qy$.   
 Assim, utilizaremos a rela\c{c}\~{a}o de Q-Conjuga\c{c}\~{a}o, 
 $x'Qy=0$ , ou simplesmente conjuga\c{c}\~{a}o, se a forma quadr\'{a}tica 
 $Q$ estiver subentendida, como uma generaliza\c{c}\~{a}o do conceito de 
 ortogonalidade.

 Utilizaremos a seguinte nota\c{c}\~{a}o: 
 $p^k$ para um ponto qualquer;  
 $q^k$ para o gradiente de $f$ em $p^k$, i.e., $q^k= Qp^k$;   
 $\pi_k$ para o plano tangente a curva de n\'{\i}vel de $f$ 
 passando por $p^k$, i.e., ao plano passando por $p^k$ e ortogonal 
 a $q^k$; e   
 $\pi_{1,\ldots k}$ para o (hiper) plano de dimens\~{a}o 
 $k-1$ determinado pelos pontos $p_1\ldots p_k$. 
 Assim, $\pi_{1,2}$ \'{e} uma reta, $\pi_{1,2,3}$ \'{e} um plano 
 bidimensional, etc. 
 Finalmente, definimos os coeficientes de conjuga\c{c}\~{a}o, 
 $c_{k,j} \equiv (p^k)' Q p^j = (p^k)' q^j = c_{j,k}$.

 Examinemos inicialmente o problema bidimensional, do ponto de  vista
geom\'{e}trico. Dado uma fun\c{c}\~{a}o quadr\'{a}tica  bidimensional,
$f(x)=(1/2)x'Qx$, que sem perda de generalidade suporemos centrada na 
origem, temos o seguinte m\'{e}todo geom\'{e}trico para encontrar o ponto 
de m\'{\i}nimo, $p^*$, veja figura 2.1a.  

 Algoritmo das Cordas Paralelas: 

 1) Ao longo de duas retas paralelas, $r_0$ e $r_3$, encontre 
os pontos de m\'{\i}nimo $p^0$ e $p^3$. 

 2) Ao longo da reta $\pi_{0,3}$, encontre o ponto de m\'{\i}nimo, 
$p^4$.

 Pela simetria do problema esf\'{e}rico, fica claro que o centro $p^*$
pertence a reta $\pi_{0,3}$, de  modo que $p^4$ o ponto de m\'{\i}nimo
em $\pi_{0,3}$, \'{e} o  ponto de m\'{\i}nimo no plano, $p^*$.         

 Como $p^0$ e $p^3$ s\~{a}o pontos de m\'{\i}nimo de $f(x)$ nas 
retas $r_0$ e $r_3$, estas retas s\~{a}o tamb\'{e}m tangentes \`{a}s 
respectivas curvas de n\'{\i}vel de $f$ passando por $p^0$ e $p^3$, 
or seja, s\~{a}o $p^0$ e $p^3$ determinam ``tangentes paralelas''.  
 O algoritmo seguinte aproveita esta id\'{e}ia. 
 Uma reta passando por um ponto $p^k$ \'{e} dita Degenerada 
 se pertence ao plano $\pi_k$.

   \begin{figure}[hbt] 
   \centerline{\includegraphics[height=4.4in, width=6.4in,  
    angle=0, viewport= 80 500 420 740, clip
   ]{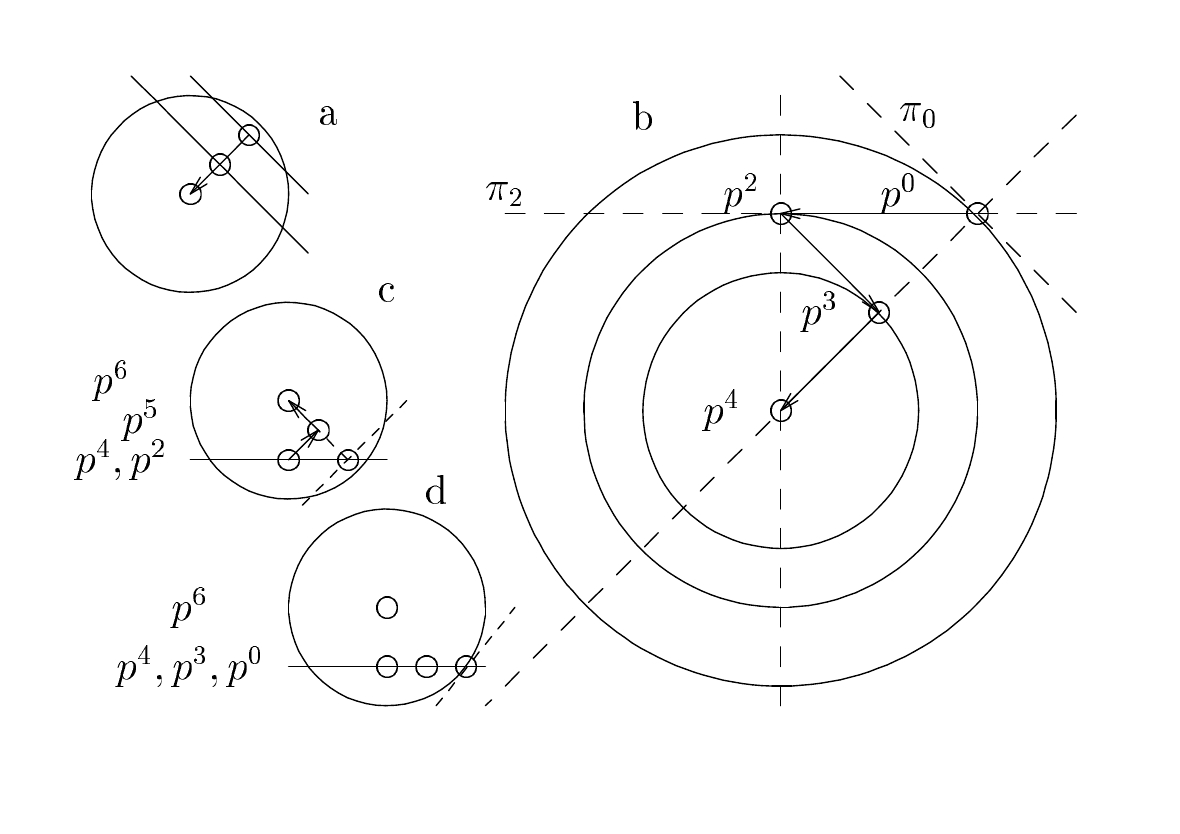}}
  \centerline{Figura 2.1 Partan na Esfera}
  \end{figure}

 Algoritmo Partan Bidimensional:

 1) Tome um ponto inicial qualquer, $p^0$. 

 2) Tome $p^2$ o ponto de m\'{\i}nimo sobre uma reta por $p^0$ 
    n\~{a}o degenerada.  

 3) Tome $p^3$ o ponto de m\'{\i}nimo na reta por $p^2$, 
    paralela a $\pi_0$. 

 4) Tome $p^4$ o ponto de m\'{\i}nimo na reta por $p_{0,3}$.

 Na figura 2.1b vemos que, para fun\c{c}\~{a}o esf\'{e}rica,  este
algoritmo nada mais \'{e} que uma  forma particular de implementar o
algoritmo das cordas paralelas,  sendo portanto correto. Como as
rela\c{c}\~{o}es de coplanaridade e paralelismo (usadas para especificar
o algoritmo) s\~{a}o invariantes  por transforma\c{c}\~{o}es lineares,
vemos tamb\'{e}m que o algoritmo \'{e} correto para fun\c{c}\~{o}es 
quadr\'{a}ticas quaisquer.     
 Repare que os vetores $w^4=p^4-p^2$ e $w^2=p^2-p^0$ s\~{a}o ortogonais. 
 Vamos agora estender este algoritmo para o espa\c{c}o tridimensional.

 Algoritmo Partan Tridimensional:   
 
 1) Tome um ponto inicial qualquer, $p^0$. 

 2) Tome $p^2$ o ponto de m\'{\i}nimo sobre uma reta por $p^0$ 
    n\~{a}o degenerada.

 3) Tome $p^3$ o ponto de m\'{\i}nimo em uma reta por $p^2$, 
    paralela a $\pi_0$ e n\~{a}o degenerada. 

 4) Tome $p^4$ o ponto de m\'{\i}nimo na reta por $p_{0,3}$. 

 5) Tome $p^5$ o ponto de m\'{\i}nimo na reta por $p^4$ 
    paralela a intersec\c{c}\~{a}o de $\pi_0$ e $\pi_2$. 

 6) Tome $p^6$ o ponto de m\'{\i}nimo na reta $\pi_{2,5}$.

 Vejamos que $p^6$ \'{e} realmente o ponto de \'{o}timo, 
olhando o problema no sistema de coordenadas com simetria 
esf\'{e}rica, veja a figura 2.1. 
 Para tanto basta verificar que: \\ 
 1) $p^5 -p^4 \para \pi_0 \cap \pi_2$ , \    
 2) $p^* -p^0 \perp \pi_0$  e   
    $p^* -p^2 \perp \pi_2$ , \      
 3) $p^5 -p^4 \perp \pi_{0,2,*}$ , \\  
 4) $p^5 -p^4 \perp \pi_{0,2}$ , \  
 5) $p^5 -p^4 \in \pi_{2,4,*}$ , \ 
 6) $p^6 = p^*$. 

 As afirma\c{c}\~{o}es acima s\~{a}o v\'{a}lidas: 
 1) por constru\~{a}o, 
 2) por s\'{\i}metria, 
 3) por 1 e 2, \\ 
 4) pois $\pi_{0,2} \in \pi_{0,2,*}$, 
 5) por simetria, e 
 6) segue da corretude do Partan 2D.

 Podemos ainda fazer algumas observa\c{c}\~{o}es a respeito da 
 geometria do Partan 3D com simetria esf\'{e}rica. 
 Primeiro, vemos de $p^{2k}$ \'{e} o ponto de m\'{\i}nimo do plano 
 $\pi_{0,2,\ldots 2k}$. 
 Segundo vemos de o vetor 
 $w^{2k}\equiv p^{2k}-p^{2k-2}$ \'{e} perpendicular a este mesmo plano. 
 Assim, os vetores $w^{2k}$ s\~{a}o mutuamente ortogonais. 

 Estas observa\c{c}\~{o}es s\~{a}o a b\'{a}se para a generaliza\c{c}\~{a}o 
 do Partan $n$-dimensional, e tamb\'{e}m para a prova de sua corretude.

 Algoritmo Partan N-dimensional:   
 
 1) Tome um ponto inicial qualquer, $p^0$. 

 2) Tome $p^2$ o ponto de m\'{\i}nimo sobre uma reta por $p^0$ 
    n\~{a}o degenerada.

 3) Para $k=1,2,\ldots n-1$: 

    3a) Tome $p^{2k+1}$ o ponto de m\'{\i}nimo em uma reta 
    por $p^{2k}$ n\~{a}o degenerada e paralela  \\ 
    \mbox{} \ \ \ \ \ \  aos planos  
    $\pi_0, \pi_2, \ldots \pi_{2k-2}$.   

    3b) Tome $p^{2k+2}$ o ponto de m\'{\i}nimo na reta por 
    $p^{2k-2}$ e $p^{2k+1}$.

 Teorema de Corretude do Partan $n$ Dimensional: 
 O algoritmo Partan  $n$ dimensional encontra o ponto de m\'{\i}nimo
 $p^*=p^{2n}$. 

 Para provar o teorema de corretude, provemos o seguinte teorema, 

 Teorema de Conjuga\c{c}\~{a}o do Partan: 
 No algoritmo Partan $n$ dimensional, os vetores 
 $w^{2k}\equiv p^{2k}-p^{2k-2}$, $k=1\ate n$, 
 s\~{a}o mutuamente conjugados.  

 O teorema de conjuga\c{c}\~{a}o do Partan implica imediatamente no 
teorema de corretudo do Partan, pois em $n$ dimens\~{o}es, n\~{a}o 
pode haver mais que $n$ vetores n\~{a}o nulos $Q$-conjugados, 
com $Q$ n\~{a}o singular. 
 O teorema de conjuga\c{c}\~{a}o decorre imediatamente do seguinte 
lema, devido a B.V.Shah, J.Buehler e O.Kempthorne.

 Lema SBK:  
 Para $k=1\ate n$, \ \ $c_{2k,0} = c_{2k,2} = \ldots = c_{2k,2k}$.

 Para provar o Lema SBK, notemos inicialmente que as equa\c{c}\~{o}es de
constru\c{c}\~{a}o  dos pontos $p^k$ do Partan implicam em v\'{a}rias
identidades  entre os coeficientes de conjuga\c{c}\~{a}o:

 (a) Pela condi\c{c}\~{a}o de paralelismo,  
 $(p^{2k+1}-p^{2k}) \perp q^{2j}$ ou, equivalentemente, 
 para $k=1\ate n-1$, $j=0\ate k-1$, \ \    
 $c_{2k+1,2j}=c_{2k,2j}$.

 (b) Pela condi\c{c}\~{a}o de convexidade, 
 $p^{2k+2} = \lambda_{k+1} p^{2k+1} +\lambdab_{k+1} p^{2k-2}$, 
 onde $\lambdab=(1-\lambda)$, e portanto, 
 para $k=1\ate n-1$, $j=0\ate k-1$, \ \  
 $c_{2k+2,j} = \lambda_{k+1} c_{2k+1,j} +\lambdab_{k+1} c_{2k-2,j}$.

 (c) Pela condi\c{c}\~{a}o de minimiza\c{c}\~{a}o, 
 $(p^2-p^0) \perp q^2=0$ e $(p^k-p^{k-1}) \perp q^k=0$, ou   
 para $k=3\ate 2n$, \ \   
 $c_{2,2}=c_{2,0}$ e $c_{k,k}=c_{k,k-1}$. 

 (d) Pela condi\c{c}\~{a}o de colinearidade de  
 $p^{2k+2}-p^{2k+1}$ e $p^{2k+1}-p^{2k-2}$,  
 para $k=1\ate n-1$,  \\ 
 $c_{2k+2,2k+2} = c_{2k+1,2k+2} = c_{2k-2,2k+2}$.

 Por (c) o Lema SBK \'{e} verdadeiro para $k=1$. 
 Como hip\'{o}tese indu\c{c}\~{a}o, assumamos que o Lema SBK \'{e} 
 verdadeiro at\'{e} $k$, e provemos que \'{e} verdadeiro para 
 $k+1$, i.e., que \\ 
 $c_{2k+2,0} = c_{2k+2,2} = \ldots = 
  c_{2k+2,2k-2} = c_{2k+2,2k} = c_{2k+2,2k+2}$.

 Usando (c), (b), (a), e novamente (c), temos a identidade (e): 
 \[ 
    c_{2k+2,2k+2} -c_{2k+2,2k} = 
    c_{2k+2,2k+1} -c_{2k+2,2k} =  
 \] 
 \[ 
    \lambda_{k+1}(c_{2k+1,2k+1} -c_{2k+1,2k}) 
    +\lambdab_{k+1}(c_{2k-2,2k+1} -c_{2k-2,2k}) 
    =0   
 \] 
 e, usando (d), temos a identidade (f): 
 \[ 
    c_{2k+2,2k+2} = c_{2k-2,2k+2}
 \] 
 De (e) e (f), temos as tres \'{u}ltimas igualdades do Lema SBK. 
 As demais igualdades s\~{a}o obtidas tomando $j\leq k-1$, 
 e escrevendo a identidade (g):  
 \[ 
    c_{2k+2,2k-2} -c_{2k+2,2j} = 
    \lambda_{k+1}(c_{2k+1,2k-2} -c_{2k+1,2j}) 
    +\lambdab_{k+1}(c_{2k-2,2k-2} -c_{2k-2,2j}) 
 \] 
 O fator da combina\c{c}\~{a}o convexa $\lambda_{k+1}=1$, 
 pela hip\'{o}tese de indu\c{c}\~{a}o. 
 Usando (a) em ambos os termos com fator $\lambda_{k+1}$ em (g), 
 e usando novamente a hip\'{o}tese de indu\c{c}\~{a}o, temos 
 \[ 
   c_{2k+1,2k-2} -c_{2k+1,2j} = c_{2k,2k-2} -c_{2k,2j} =0  
 \] 
 de forma que (g) \'{e} zero, e o lema SBK \'{e} satisfeito 
 para $2k+2$, Q.E.D.

 Uma das observa\c{c}\~{o}es que nos levaram ao Partan era que 
 $p^{2k}$ \'{e} o ponto de m\'{\i}nimo do plano $\pi_{0,2,\ldots 2k}$. 
 Esta mesma observa\c{c}\~{a}o nos fornece uma forma simples e 
 eficiente de escolher a dire\c{c}\~{a}o de $p^{2k+1}-p^{2k}$ 
 respeitando a condi\c{c}\~{a}o de paralelismo com os planos 
 $\pi_0,\pi_2,\ldots \pi_{2k-2}$. 
 Basta tomar (sempre) a dire\c{c}\~{a}o $p^{2k+1}-p^{2k}$ 
 perpendicular a $\pi_{2k}$, isto \'{e}, na dire\c{c}\~{a}o do gradiente. 
 Isto acontece pois s\~{a}o paralelos os tr\^{e}s planos, 
 $\pi_{2k}$, $\pi_{0,2,\ldots 2k}$ e $\pi'_{0,2,\ldots 2k}$, 
 onde o \'{u}ltimo plano \'{e} o paralelo ao pen\'{u}ltimo 
 passando pela origem. Todos os gradientes 
 $q^0,q^2,\ldots q^{2k-2}$ est\~{a}o em $\pi'_{0,2,\ldots 2k}$, 
 de modo que tomar a dire\c{c}\~{a}o do gradiente $q^{2k}$ 
 garante a perpendicularidade com os gradientes anteriores, 
 $q^0,q^2,\ldots q^{2k-2}$ ou, equivalentemente, a condi\c{c}\~{a}o       
 de paralelismo. Este \'{e} o algoritmo Partan Gradiente.

 Como Partan Gradiente \'{e} um caso particular do Partan, 
 ele encontra a solu\c{c}\~{a}o \'{o}tima de uma fun\c{c}\~{a}o 
 quadr\'{a}tica em um numero finito ($2n$) de passos. 
 Todavia, mesmo se o modelo quadr\'{a}tico para a fun\c{c}\~{a}o 
 objetivo for pobre, os passos impares s\~{a}o nada mais que 
 minimiza\c{c}\~{o}es na dire\c{c}\~{a}o de m\'{a}ximo declive. 
 Isto explica porque o Partan Gradiente \'{e} um algoritmo de 
 otimiza\c{c}\~{a}o irrestrita com as melhores caracter\'{\i}sticas 
 de dois mundos, tendo a robustes do m\'{e}todo de Cauchy 
 quando longe do ponto de \'{o}timo, e convergencia quadr\'{a}tica      
 quando perto da solu\c{c}\~{a}o \'{o}tima. 

 O Partan necessita de duas buscas lineares (passo impar e par)
 para cada dimens\~{a}o do problema. Longe do ponto de \'{o}timo 
 o M\'{e}todo de Cauchy usaria apenas uma busca. 
 Perto da solu\c{c}\~{a}o \'{o}tima poderiamos usar tamb\'{e}m apenas 
 uma busca linear por dimens\~{a}o, caso soubessemos gerar facilmente 
 as $n$ dire\c{c}\~{o}es conjugadas. 
 Esta \'{e} a id\'{e}ia b\'{a}sica do(s) algoritmo(s) tipo 
 Gradientes Conjugados. Todavia este corte pela metade o n\'{u}mero 
 de buscas linears tem um custo impl\'{\i}cito: 
 A implementa\c{c}\~{a}o de um mecanismo de monitoramento do comportamento 
 do algoritmo, para decidir em que momento fazer a tranzi\c{c}\~{a}o de 
 Cauchy para Gradientes Conjugados.

 \section{Converg\^{e}ncia Global}

 Estudaremos nesta se\c{c}\~{a}o as condi\c{c}\~{o}es de 
converg\^{e}ncia de um algoritmo para a solu\c{c}\~{a}o \'{o}tima 
de um problema de otimiza\c{c}\~{a}o n\~{a}o linear. 
 Seguiremos de perto a apresenta\c{c}\~{a}o desenvolvida por 
W.I.Zangwill.  

 Definimos Algoritmo como um processo iterativo gerando uma 
seq\"{u}\^{e}ncia de pontos, $x^0,x^1,x^2\ldots$ que obedece a uma  
equa\c{c}\~{a}o de recurs\~{a}o da forma  
 $x^{k+1} \in A_k(x^k)$, onde o Mapa Ponto a Conjunto $A_k(x^k)$ define 
os poss\'{i}veis sucessores de $x^k$ na seq\"{u}\^{e}ncia. 
 
 A id\'{e}ia de usar um mapa ponto-a-congunto, ao inv\'{e}s de uma 
fun\c{c}\~{a}o, ou mapa ponto-a-ponto, nos permite estudar de forma 
unificada classes de algoritmos, incluindo diversas
implementa\c{c}\~{o}es de v\'{a}rios detalhes, c\'{a}lculos aproximados
ou inexatos, vari\'{a}veis randomizadas, etc.     
 A propriedade b\'{a}sica que queremos dos mapas definindo os algoritmos 
\'{e} a propriedade de Fechamento, definida a seguir. 

 Um mapa ponto-a-conjunto do espa\c{c}o $X$ no espa\c{c}o $Y$, \'{e} 
fechado em $x$ na seguinte condi\c{c}\~{a}o: 
 Se a seq\"{u}\^{e}ncia $x^k$ converge para $x$, 
  e a seq\"{u}\^{e}ncia $y^k$ converge para $y$, 
  onde $y^k \in A(x)$, ent\~{a}o o ponto limite na imagem, $y$, 
  pertence a imagem por $A$ do ponto limite no dom\'{\i}nio, $x$, i.e.  
  \[ 
     x^k \rightarrow x \ , \ 
     y^k \rightarrow y \ , \ 
     y^k \in A(x^k) \ \Rightarrow \ 
     y \in A(x) \ . 
  \] 
 O mapa \'{e} fechado em $C\subseteq X$ se for fechado em qualquer ponto de 
$C$. Note que se trocarmos na defini\c{c}\~{a}o de fechamento a 
rela\c{c}\~{a}o de pertinencia pela rela\c{c}\~{a}o de igualdade, 
obtemos a defini\c{c}\~{a}o de continuidade para mapas ponto-a-ponto. 
 Assim, a propriedade de fechamento generaliza a propriedade de 
continuidade. Com efeito, uma fun\c{c}\~{a}o cont\'{\i}nua \'{e} 
fechada, embora o inverso n\~{a}o seja necessariamente verdadeiro. 

 A id\'{e}ia b\'{a}sica do teorema de converg\^{e}ncia global de 
Zangwill \'{e} procurar alguma caracter\'{\i}stica que ``melhore''  
continuamente'' a cada itera\c{c}\~{a}o do algoritmo. 
 Esta caracter\'{\i}stica \'{e} representada pelo conceito de 
fun\c{c}\~{a}o de descend\^{e}ncia. 

 Seja $A$ um algoritmo em $X$ para resolver o problema $P$, e seja 
 $S\subset X$ o conjunto de solu\c{c}\~{o}es de $P$. 
 Uma fu\c{c}\~{a}o $Z(x)$ \'{e} uma fun\c{c}\~{a}o de 
 descend\^{e}ncia para $(X,A,S)$ se a composi\c{c}\~{a}o de 
 $Z$ e $A$ sempre decresce fora do conjunto solu\c{c}\~{a}o, 
 e n\~{a}o aumenta dentro dele, i.e.:  \\ 
 $x\notin S \wedge y\in A(x) \Rightarrow Z(y) < Z(x)$ 
 \ \ e \ \  
 $x\in S \wedge y\in A(x) \Rightarrow Z(y) \leq Z(x)$ \ .

 Em problemas de otimiza\c{c}\~{a}o, muitas vezes uma boa 
fun\c{c}\~{a}o de descend\^{e}ncia \'{e} a pr\'{o}pria fun\c{c}\~{a}o 
objetivo. Outras vezes, fun\c{c}\~{o}es de descend\^{e}ncia 
mais complexas tem de ser utilizadas, como por exemplo a soma da 
fun\c{c}\~{a}o objetivo com termos auxiliares, como penalidades para a 
viola\c{c}\~{a}o de restri\c{c}\~{o}es, ou uma fun\c{c}\~{a}o 
decrescente de um contador de itera\c{c}\~{o}es sem melhora no objetivo 
para certos tipos de problemas degenerados.  

 Antes de enunciar o Teorema de Zangwill, recordemos dois conceitos 
de elementares de topologia de conjuntos:  
 Um Ponto De Acumula\c{c}\~{a}o de uma seq\"{u}\^{e}ncia \'{e} 
um ponto limite para uma se suas sub-seq\"{u}\^{e}ncias.    
 Um conjunto \'{e} Compacto sse qualquer seq\"{u}\^{e}ncia (infinita) 
tem um ponto de acumula\c{c}\~{a}o dentro do conjunto. 
 Em $R^n$, um conjunto \'{e} compacto sse \'{e} fechado e limitado.

 Teorema de Converg\^{e}ncia Global (Zangwill): 

 Seja $Z$ uma fun\c{c}\~{a}o de descend\^{e}ncia para o algoritmo 
 $A$ definido em $X$ com conjunto de solu\c{c}\~{o}es $S$, e 
 $x^0,x^1,x^2,\ldots$ uma seq\"{u}\^{e}ncia gerada pelo algoritmo 
 tal que: \\  
 A) O mapa $A$ \'{e} fechado em qualquer ponto fora do conjunto 
 solu\c{c}\~{a}o, \\  
 B) Todos os pontos da seq\"{u}\^{e}ncia permanecem dentro de 
    um compacto, $C\subseteq X$, e \\ 
 C) $Z$ \'{e} cont\'{\i}nua. \\ 
 Ent\~{a}o qualquer ponto de acumula\c{c}\~{a}o da seq\"{u}\^{e}ncia 
 estar\'{a} no conjunto solu\c{c}\~{a}o.  
 
 Pela compacidade de $C$, a seq\"{u}\^{e}ncia gerada pelo algoritmo 
 tem um ponto de acumula\c{c}\~{a}o, $x\in C\subseteq X$, para uma 
 subseq\"{u}\^{e}ncia de acumula\c{c}\~{a}o, $x^{h(k)}$. 
 Pela continuidade de $Z$ em $X$, o limite do valor de $Z$ na   
 subseq\"{u}\^{e}ncia de acumula\c{c}\~{a}o coincide com o valor de $Z$ 
 no ponto de acumula\c{c}\~{a}o, i.e., 
 $Z(x^{h(k)})\rightarrow Z(x)$.  
  Mas a seq\"{u}\^{e}ncia completa, $Z(x^k)$ \'{e} monotonicamente 
 n\~{a}o crescente, de modo que se 
 $h(k) \leq j \leq h(k+1)$ ent\~{a}o 
 $Z(x^{h(k)}) \geq Z(x^j) \geq Z(x^{h(k+1)})$, 
 de modo que o valor de $Z$ na 
 seq\"{u}\^{e}ncia completa tamb\'{e}m converge para o valor de 
 $Z$ no ponto de acumula\c{c}\~{a}o i.e., 
 $Z(x^k)\rightarrow Z(x)$.

 Consideremos agora, por absurdo, que $x$ n\~{a}o seja uma 
 solu\c{c}\~{a}o, de modo que $Z(A(x))<Z(x)$. 
 Consideremos a subseq\"{u}\^{e}ncia dos sucessores dos pontos na 
 primeira subseq\"{u}\^{e}ncia de acumula\c{c}\~{a}o, 
 $x^{h(k)+1}$, seq\"{u}\^{e}ncia esta que, por compacidade, tem 
 tam\'{e}m um ponto de acumula\c{c}\~{a}o, $x'$. 
 Mas pelo resultado no par\'{a}grafo anterior, o valor da 
 fun\c{c}\~{a}o de descend\^{e}ncia em ambas as subseq\"{u}\^{e}ncias 
 converge para o valor limite da seq\"{u}\^{e}ncia completa, i.e.,  
 $\lim Z(x^{h(k)+1}) = \lim Z(x^k) = \lim Z(x^{h(k)})$. 
 Fica demonstrada assim a impossibilidade de que $x$ n\~{a}o seja 
 uma solu\c{c}\~{a}o.

 A formula\c{c}\~{a}o de muitos algoritmos \'{e} feita pela 
 composi\c{c}\~{a}o de v\'{a}rios passos. Assim, o mapa 
 descrevendo o algoritmo completo \'{e} a composi\c{c}\~{a}o  
 de v\'{a}rios mapas, um para cada passo. 
 Um exemplo t\'{\i}pico \'{e} termos um passo correspondente a escolha 
 de uma dire\c{c}\~{a}o de busca, e o passo seguinte correspondente 
 a uma busca linear. 
 Os lemas apresentados a seguir s\~{a}o \'{u}teis na 
 constru\c{c}\~{a}o de mapas compostos. 
  
 Primeiro Lema de Composi\c{c}\~{a}o: \\ 
 Seja $A$ de $X$ em $Y$, e $B$ de $Y$ em $Z$, mapas ponto a conjunto, 
 $A$ fechado em $x\in X$, $B$ fechado em $A(x)$. 
 Se para qualquer seq\"{u}\^{e}ncia $x^k$ convergindo para $x$, 
 $y^k \in A(x^k)$ tem um ponto de acumula\c{c}\~{a}o $y$, 
 ent\~{a}o o mapa composto $B\circ A$ \'{e} fechado em $x$. 

 Demonstra\c{c}\~{a}o:   \\  
 Como $A$ \'{e} fechado em $x$, $y\in A(x)$. 
 Como $y^h(k) \rightarrow y$ e $B$ \'{e} fechado em $y$, 
 uma seq\"{u}\^{e}ncia $z^{h(k)}\in B(y^{h(k)})$ converge para 
 um ponto $z\in B(y)\subseteq B(A(x))$.

 Segundo Lema de Composi\c{c}\~{a}o: \\ 
 Seja $A$ de $X$ em $Y$, e $B$ de $Y$ em $Z$, mapas ponto a conjunto, 
 $A$ fechado em $x\in X$, $B$ fechado em $A(x)$. 
 Se $Y$ for compacto, ent\~{a}o o mapa composto 
 $B\circ A$ \'{e} fechado em $x$.

 Terceiro Lema de Composi\c{c}\~{a}o:  \\ 
 Seja $A$ um mapa ponto a ponto de $X$ em $Y$, 
 e $B$ um mapa ponto a conjunto de $Y$ em $Z$.  
 Se $A$ for cont\'{\i}nuo em $x$, e $B$ for fechado em $A(x)$. 
 ent\~{a}o o mapa composto 
 $B\circ A$ \'{e} fechado em $x$.


%% file: cap3.tex
\chapter{Programa\c{c}\~{a}o Quadr\'{a}tica}

 No cap\'{\i}tulo anterior vimos um modelo relativamente simples de 
composi\c{c}\~ao de carteiras que recaiu em um problema com 
fun\c{c}\~ao objetivo quadr\'atica. Para esta categoria de problemas 
uma alternativa \'e, realizar algum tipo de lineariza\c{c}\~ao 
que nos permita empregar o algoritmo simplex. Todavia, \'e poss\'{\i}vel 
explorar a estrutura da fun\c{c}\~ao objetivo no caso em que  as 
restri\c{c}\~oes do problema s\~ao lineares, conseguindo construir 
algoritmos para resolver o problema quadr\'{a}tico.  

Estudaremos  aqui este caso, assumindo que a fun\c{c}\~ao 
sendo minimizada \'e quadr\'atica, isto \'e, da forma 
$x^{'} Q x + \eta p^{'} x$ e que as 
restri\c{c}\~oes s\~ao lineares. Para resolv\^e-lo iremos construir 
um algoritmo similar ao  simplex, baseado nas condi\c{c}\~oes de 
otimalidade de Lagrange.

\section{Multiplicadores de Lagrange} 

Um problema de programa\c{c}\~{a}o n\~{a}o linear tem a forma:
 $$\min f(x),\ x \mid g(x)\leq 0 \wedge h(x)=0\ , \  
   f:\Re ^n\mapsto \Re\ , \ g:\Re ^n\mapsto \Re ^m\ , \ h:\Re ^n\mapsto \Re ^k \ . 
 $$  
 Podemos imaginar $f$ como um {\em potencial}, ou como a ``altura'' de
uma superf\'{\i}cie.  {\em Equipotenciais}, ou {\em curvas de n\'{\i}vel},
s\~{a}o curvas nas quais $f(x)$ mant\'{e}m um valor constante.  O gradiente,
 $$\nabla f \equiv \partial f /\partial x = \left[ \begin{array}{cccc} 
   \partial f /\partial x_1, & \partial f /\partial x_2, & \ldots & 
   \partial f /\partial x_n 
   \end{array} \right] 
 $$ 
 nos d\'{a} a dire\c{c}\~{a}o de ``maior inclina\c{c}\~{a}o'' da
superf\'{\i}cie $f(x)$.  Temos tamb\'{e}m que o gradiente de $f$ num ponto
$x$, \'{e} ortogonal (ou perpendicular) \`{a} curva de n\'{\i}vel de $f$
que passa por $x$.

 Daremos a seguir uma explica\c{c}\~{a}o intuitiva para uma
condi\c{c}\~{a}o necess\'{a}ria de otima\-lidade.  Podemos
imaginar uma part\'{\i}cula puxada ``para baixo'' por uma ``for\c{c}a''
$-\nabla f(x)$.  Um ponto de m\'{\i}nimo ser\'{a} portanto um {\em ponto
de equil\'{\i}brio} para a part\'{\i}cula, i.e.  um ponto onde a for\c{c}a
que puxa a part\'{\i}cula para baixo se anula, ou ent\~{a}o \'{e}
equilibrada por ``for\c{c}as de rea\c{c}\~{a}o'' exercidas pelas
restri\c{c}\~{o}es. 
 A for\c{c}a de rea\c{c}\~{a}o de uma restri\c{c}\~{a}o de desigualdade,
$g_i(x)\leq 0$, deve ser:
 \begin{itemize}
 \item[a)] perpendicular \`{a} curva de n\'{\i}vel desta restri\c{c}\~{a}o
(pois a part\'{\i}cula pode mover-se livremente ao longo da curva de
n\'{\i}vel),
 \item[b)] uma for\c{c}a de rea\c{c}\~{a}o ``para dentro'' da regi\~{a}o
vi\'{a}vel (i.e.  impedindo que a part\'{\i}cula saia para fora da
regi\~{a}o vi\'{a}vel). 
 \item[c)] Ademais esta restri\c{c}\~{a}o de desigualdade s\'{o} pode exercer
uma for\c{c}a de rea\c{c}\~{a}o quando estiver ativa, algo como: ``a
part\'{\i}cula n\~{a}o pode apoiar-se numa parede (restri\c{c}\~{a}o) em
que n\~{a}o esteja encostada''. 
 \end{itemize} Uma restri\c{c}\~{a}o de igualdade, $h_i(x)=0$, pode ser
vista como um par de restri\c{c}\~{o}es de desigualdade, $h_i(x)\leq 0$
e $h_i(x)\geq 0$; Ademais, dentro da regi\~{a}o vi\'{a}vel, este par de
restri\c{c}\~{o}es ser\'{a} sempre ativo. 

 A nossa discuss\~{a}o intuitiva pode ser resumida analiticamente nas 
 {\em Condi\c{c}\~{o}es de Lagrange}:
 Se $\hat x\in \Re ^n$ \'{e} um ponto \'{o}timo ent\~{a}o:
 $$\exists u\in \Re ^m\ , \ v\in \Re ^k\ \mid 
   u\nabla g +v\nabla h -\nabla f =0 
   \ ,\ \mbox{onde}\ \ u\leq 0 \wedge ug=0 
 $$ 
 A condi\c{c}\~{a}o $u\leq 0$ implica que a rea\c{c}\~{a}o das
restri\c{c}\~{o}es de desigualdade aponta para dentro da regi\~{a}o
vi\'{a}vel, enquanto a {\em condi\c{c}\~{a}o de complementaridade},
$ug=0$, implica que apenas restri\c{c}\~{o}es ativas podem exercer
for\c{c}as de rea\c{c}\~{a}o.  Os vetores $u$ e $v$ s\~{a}o conhecidos
como {\em multiplicadores de Lagrange}.

\section{Condi\c{c}\~{a}o de Otimalidade} 

O Problema de Programa\c{c}\~{a}o Quadr\'{a}tica (QP) \'{e}:
 $$ min f(x)\equiv (1/2)x'Qx -\eta p'x  
\mid x\geq 0 \wedge Te*x=te \wedge Tl*x\leq tl 
   $$
 onde as dimens\~{o}es da matrizes s\~{a}o:  
 $Te \ me\times n,\ me<n$, 
 $Tl \ ml\times n,\ ml<n$,  
 $Ml={1,2,\ldots ml},\ Me={1,2,\ldots me},\ N={1,2,\ldots n}$. 
 e, por hip\'{o}tese, a forma quadr\'{a}tica \'{e} sim\'{e}trica e 
 positiva definida, isto \'e $Q=Q',\ Q>0$.

 No QP, o gradiente da fun\c{c}\~{a}o objetivo \'{e}: 
 $$ \nabla f = x'Q -\eta p'\ , $$
 e os gradientes das restri\c{c}\~{o}es s\~{a}o: 
 $$ g_i(x) =  T_{i}x \leq t_{i} \Rightarrow \nabla g_{i} = T_{i}\ , $$
 gerando as condi\c{c}\~{o}es de equil\'{\i}brio de Lagrange, 
 ou condi\c{c}\~{o}es de otimalidade:
 $$ x\in R_+^n, s\in R_+^n, l\in R_+^{ml}, e\in \Re ^{me},\   \ \mid \  
    -(x'Q -\eta p')  +s' -l'Tl +e'Te =0  $$  
 $$   \ \wedge \ \forall i\in N \ , \ x_{i}s_{i}=0  
    \ \wedge \ \forall k\in Ml \ , \  (Tl*x -tl)_{k}l_{k}=0 $$
 ou
 $$ x\in R_+^n, s\in R_+^n, l\in R_+^{ml}, e\in \Re ^{me}, y\in R_+^{ml}\   
    \ \mid \ Qx -s' +Tl'*l +Te'e  = \eta p  $$ 
 $$   \ \wedge \ \forall i\in N \ , \  x_{i}s_{i}=0  
    \ \wedge \ \forall k\in Ml \ , \  yl_{k}l_{k}=0 \   
    \mbox{onde}\ yl=(tl -Tl*x)$$
 
 As condi\c{c}\~{o}es de complementaridade (CC), $x's=0$ e $yl'l=0$,
indicam que s\'{o} restri\c{c}\~{o}es justas podem equilibrar
componentes negativas do gradiente da fun\c{c}\~{a}o objetivo. 
Com a mudan\c{c}a de vari\'{a}veis $e=ep-em$, $ep,\ em \geq 0$, o  
ponto \'{o}timo dado pelas  ``Equa\c{c}\~{o}es de Viabilidade e 
Otimalidade'' ou EVO:
 $$ 
 \left[ \begin{array}{c}  
  x\\ l\\ ep\\ en\\ s\\ yl \end{array} \right] \geq 0 
 \mid 
 \left[ \begin{array}{rrrrrr} 
  Tl & 0   & 0   & 0    & 0  & I \\ 
  Te & 0   & 0   & 0    & 0  & 0 \\ 
  Q  & Tl' & Te' & -Te' & -I & 0   
 \end{array} \right]
 \left[ \begin{array}{c}  
  x\\ l\\ ep\\ en\\ s\\ yl \end{array} \right]
 = \left[ \begin{array}{c} tl\\ te\\ \eta p \end{array} \right] $$ 
 $$ x's=0,\ yl'l=0. $$

\section{Complementaridade Linear} 
 
Em EVO as CC implicam que, na solu\c{c}\~{a}o, s\~{a}o nulos: pelo
menos, $n$ elementos dentre $x$ e $s$; e pelo menos $ml$ elementos
dentre $yl$ e $l$.  $ep$ e $en$ s\~{a}o, respectivamente, a parte
positiva e negativa do vetor irrestrito $e$, de modo que est\'{a}
impl\'{\i}cita na formula\c{c}\~{a}o do problema que, na
solu\c{c}\~{a}o, s\~{a}o nulos pelo menos $me$ elementos dentre $ep$ e
$en$.  Assim, na solu\c{c}\~{a}o, n\~{a}o h\'{a} mais que $ml+me+n$ 
vari\'{a}veis n\~{a}o nulas, que podem ser escritas como uma 
uma solu\c{c}\~{a}o b\'{a}sica do sistema EVO.  Isto sugere usarmos o
algoritmo Simplex para resolver o QP.  Assumiremos agora, por mera
conveni\^{e}ncia para as aplica\c{c}\~{o}es que se seguem, que $tl\geq 0$
(alterar o material que se segue de modo a prescindir desta hip\'{o}tese
\'{e} um exerc\'{\i}cio trivial).  Formularemos EVO como um PPL na forma
padr\~{a}o, mais as CC.  Este \'e o problema de complementaridade
linear, PCL:
 $$ 
 \min 
 \left[ \begin{array}{cccccccc} 0 & 0 & 0 & 0 & 0 & 0 & 1 & 1 
 \end{array} \right]   
 \left[ \begin{array}{c}  
  x\\ l\\ ep\\ en\\ s\\ yl\\ ye\\ yq \end{array} \right]  
 \ \ \left[ \begin{array}{c}  
  x\\ l\\ ep\\ en\\ s\\ yl\\ ye\\ yq \end{array} \right] \geq 0  
 \mid 
 $$ 
 $$   
 \left[ \begin{array}{rrrrrrrr} 
  Tl & 0 & 0 & 0 & 0 & I & 0 & 0 \\ 
  Te & 0 & 0 & 0 & 0 & 0 & De & 0 \\ 
  Q & Tl' & Te' & -Te' & -I & 0 & 0 & Dq  
 \end{array} \right]
 \left[ \begin{array}{c}  
  x\\ l\\ ep\\ en\\ s\\ yl\\ ye\\ yq \end{array} \right]
 = \left[ \begin{array}{c} tl\\ te\\ \eta p \end{array} \right] 
 $$
 onde temos as CC, os novos blocos, e o v\'{e}rtice inicial, dados por,
respectivamente
 $$
 x's=0,\ yl'l=0,\ \  
 Dq=diag( sign( \eta p )),\   
 De=diag( sign( te )),\ \  
 tl\geq 0.$$ 
 $$  
 { \left[ \begin{array}{c}  
  x\\ l\\ ep\\ en\\ s\\ yl\\ ye\\ yq \end{array} \right] }_0 =
 \left[ \begin{array}{c} 0\\ 0\\ 0\\ 0\\ 0\\ tl\\ 
         |\ te\ | \\ |\ p\ | \end{array} \right] \ . $$ 
 

Na solu\c{c}\~{a}o inicial as CC s\~{a}o satisfeitas, pois $x's=0'0=0$,
e $yl'l=tl'0=0$.  Para assegurar que as CC continuem sendo satisfeitas,
usaremos uma regra adicional no simplex: N\~{a}o permitiremos que
 \begin{itemize} 
  \item $x_i$ ou se torne b\'{a}sica se $s_i$ for uma vari\'{a}vel 
   b\'{a}sica, e vice-versa,  
  \item $yl_i$ ou se torne b\'{a}sica se $l_i$ for uma vari\'{a}vel 
   b\'{a}sica, e vice-versa. 
 \end{itemize} 
  Regras proibitivas como esta s\~{a}o denominadas regras `Tabu'.  
  Precisamos agora demonstrar que, mesmo com esta regra tabu, o
algoritmo termina num v\'{e}rtice \'{o}timo do PCL de valor 0, i.e., com
$ye=0$, $yq=0$ e $x$ a solu\c{c}\~{a}o \'{o}tima do QP. 

Assumiremos que $Tex=te$.  Para tanto minimizaremos primeiro a soma das
vari\'{a}veis artificiais ($ye$) correspondentes a estas
restri\c{c}\~{o}es, i.e.  levaremos as vari\'{a}veis
artificiais $ye$ para fora da base.  Em seguida, mantendo $ye$ fora da base, 
tentaremos levar as vari\'{a}veis artificiais $yq$ para fora da base. 

Consideremos a possibilidade do algoritmo terminar num v\'{e}rtice de
valor positivo, sem nenhuma vari\'{a}vel residual de custo reduzido
negativo cuja vari\'{a}vel complementar j\'{a} n\~{a}o seja b\'{a}sica. 
Derivaremos desta possibilidade uma contradi\c{c}\~{a}o, provando 
assim a corretude do algoritmo PCL para resolu\c{c}\~{a}o do QP.  
Particionemos cada um dos vetores $x$, $s$, $yl$ e $l$ em tr\^{e}s 
blocos, de modo que as vari\'{a}veis: 
 \begin{itemize} 
 \item em $x_1$ sejam b\'{a}sicas, e portanto suas
complementares em $s_1$ sejam residuais,
 \item em $s_2$ sejam b\'{a}sicas, e portanto suas
complementares em $x_2$ sejam residuais,
 \item em $yl_1$ sejam b\'{a}sicas, e portanto suas
complementares em $l_1$ sejam residuais,
 \item em $l_2$ sejam b\'{a}sicas, e portanto suas
complementares em $yl_2$ sejam residuais. 
 \end{itemize} 
  Esta solu\c{c}\~{a}o \'e a solu\c{c}\~{a}o \'{o}tima do seguinte PPL
(as vari\'{a}veis omitidas s\~{a}o as proibidas de entrar na base pela
regra de complementariedade), PAUX:
 $$
 \min 
 \left[ \begin{array}{cccccccccc} 
  0 & 0 & 0 & 0 & 0 & 0 & 0 & 0 & 0 & 1 \end{array} \right]
 \left[ \begin{array}{c} 
 x_1 \\ x_3 \\ l_2 \\ l_3 \\ e \\ s_2 \\ s_3 \\ 
 yl_1 \\ yl_3 \\ yq \end{array} \right]
 \ , \ \   
 \left. \begin{array}{ccc} 
 x_1 & \geq & 0 \\ x_3 & \geq & 0 \\ l_2 & \geq & 0 \\ l_3 & \geq & 0 \\ 
 e & * & \\ u_2 & \geq & 0 \\ u_3 & \geq & 0 \\ 
 yl_1 & \geq & 0 \\ yl_3 & \geq & 0 \\ yq & \geq & 0  \end{array} \right.
 \ \ \mid \   
 $$ 
 $$ 
 \left[ \begin{array}{cccccccccc} 
 Tl_1^1 & Tl_1^3 & 0 & 0 & 0 & 0 & 0 & I & 0 & 0 \\ 
 Tl_2^1 & Tl_2^3 & 0 & 0 & 0 & 0 & 0 & 0 & I & 0 \\ 
 Te^1 & Te^3 & 0 & 0 & 0 & 0 & 0 & 0 & 0 & 0 \\ 
 Q^1_1 & Q_1^3 & {Tl_2^1}' & {Tl_3^1}' & {Te^1}' & 0 & 0 & 0 & 0 & Dq_1 \\ 
 Q_2^1 & Q_2^3 & {Tl_2^2}' & {Tl_3^2}' & {Te^2}' & -I & 0 & 0 & 0 & Dq_2 \\ 
 Q_3^1 & Q^3_3 & {Tl_2^3}' & {Tl_3^3}' & {Te^3}' & 0 & -I & 0 & 0 & Dq_3  
 \end{array} \right]
 \left[ \begin{array}{c}  
  x_1 \\ x_3 \\ l_2 \\ l_3 \\ e \\ s_2 \\ s_3 \\ yl_1 \\ yl_3 \\ yq 
 \end{array} \right] = 
 \left[ \begin{array}{c} tl_1 \\ tl_2 \\ te \\ 
        \eta p_1 \\ \eta p_2 \\ \eta p_3 \\ \end{array} \right]
 $$
 e consideremos tamb\'{e}m seu dual, DAUX: 
 $$
 \max 
 \left[ \begin{array}{cccccc} tl_1' & tl_2' & te' & 
       \eta {p_1}' & \eta {p_2}' & \eta {p_3}' 
 \end{array} \right]
 \left[ \begin{array}{c} ul_1 \\ ul_2 \\ ue \\ uq_1 \\ uq_2 \\ uq_3 
 \end{array} \right] 
 $$

 $$
 \left[ \begin{array}{cccccc} 
 {Tl_1^1}' & {Tl_2^1}' & {Te^1}' & {Q^1_1}' & {Q^1_2}' & {Q^1_3}' \\    
 {Tl_1^3}' & {Tl_2^3}' & {Te^3}' & {Q^3_1}' & {Q^3_2}' & {Q^3_3}' \\    
 0 & 0 & 0 & Tl_2^1 & Tl_2^2 & Tl_2^3 \\ 
 0 & 0 & 0 & Tl_3^1 & Tl_3^2 & Tl_3^3 \\ 
 0 & 0 & 0 & Te^1 & Te^2 & Te^3 \\ 
 0 & 0 & 0 & 0 & -I & 0 \\ 
 0 & 0 & 0 & 0 & 0 & -I \\ 
 I & 0 & 0 & 0 & 0 & 0 \\ 
 0 & I & 0 & 0 & 0 & 0 \\ 
 0 & 0 & 0 & Dq_1' & Dq_2' & Dq_3'  
 \end{array} \right] 
 \left[ \begin{array}{c} 
 ul_1 \\ ul_2 \\ ue \\ uq_1 \\ uq_2 \\ uq_3     
 \end{array} \right] 
 \ \   
 \left. \begin{array}{ccl}  
     \geq & 0 & \mbox{1a},\ \mbox{justa},\  x_1>0 \\   
     \geq & 0 & \mbox{2a},\ \mbox{folgada}, \\ 
     \geq & 0 & \mbox{3a},\ \mbox{justa},\ l_2>0 \\   
     \geq & 0 & \mbox{4a},\ \mbox{folgada}, \\  
     =    & 0 & \mbox{5a},\ e\ \mbox{irrestrito} \\   
     \geq & 0 & \mbox{6a},\ \mbox{justa},\ s_2>0 \\  
     \geq & 0 & \mbox{7a},\ \mbox{folgada}, \\   
     \geq & 0 & \mbox{8a},\ \mbox{justa},\ yl_1>0 \\  
     \geq & 0 & \mbox{9a},\ \mbox{folgada}, \\   
     \geq & 1 & \mbox{10a}       
  \end{array} \right.  
 $$ 
 
 No dual apresentamos a solu\c{c}\~{a}o \'{o}tima, particionada conforme
o primal.  Nas restri\c{c}\~{o}es do DAUX temos igualdade na 5a linha 
devido a termos $e$ sem restri\c{c}\~{a}o de sinal, e igualdades nas 
1a, 3a, 6a, e 8a linhas pelo teorema de folgas complementares, pois  
as vari\'{a}veis respectivamente indicadas s\~{a}o b\'{a}sicas.  

 Da 6a linha blocada das restri\c{c}\~{o}es do DAUX vemos que 
 $uq_2 =0$, e da 8a linha vemos que $ul_1 =0$. 
 Podemos pois escrever a equa\c{c}\~{a}o (inequa\c{c}\~{a}o justa) da 
 1a linha descartando os termos em $ul_1$ e $uq_2$. 
 Multiplicando esta equa\c{c}\~{a}o por $uq_1'$ temos  
 $$ uq_1' 
    \left[ \begin{array}{cccc} {Tl_2^1}' & {Te^1}' & {Q_1^1}' & {Q_3^1}'  
    \end{array} \right] 
    \left[ \begin{array}{c} ul_2 \\ u_e \\ uq_1 \\ uq_3  
    \end{array} \right] 
    = uq_1' {\bf 0} = 0 
 $$ 
 analogamente, multiplicando a inequa\c{c}\~{a}o da 2a linha por 
 $uq_3'$, e notando que pela 7a linha $uq_3'\leq 0$, temos 
 $$ uq_3' 
    \left[ \begin{array}{cccc} {Tl_2^3}' & {Te^3}' & {Q_1^3}' & {Q_3^3}'  
    \end{array} \right] 
    \left[ \begin{array}{c} ul_2 \\ u_e \\ uq_1 \\ uq_3'  
    \end{array} \right] 
    \leq 0 
 $$ 
    
Somando e transpondo as duas \'ultimas desigualdades temos 
  \begin{eqnarray*} 
    \left[ \begin{array}{cccc} ul_2' & u_e' & uq_1' & uq_3'  
    \end{array} \right] 
    \left[ \begin{array}{c} Tl_2^1 \\ Te^1 \\ Q_1^1 \\ Q_3^1
    \end{array} \right] uq_1  & & \\ 
    +  
    \left[ \begin{array}{cccc} ul_2' & u_e' & uq_1' & uq_3'  
    \end{array} \right] 
    \left[ \begin{array}{c} Tl_2^3 \\ Te^3 \\ Q_1^3 \\ Q_3^3  
    \end{array} \right] 
    uq_3  & \leq 0 &    
 \end{eqnarray*} 
 ou 
 \begin{eqnarray*} 
 ul_2' 
 \left[ \begin{array}{cc} Tl_2^1 & Tl_2^3 \end{array} \right]  
 \left[ \begin{array}{c} uq_1 \\ uq_3 \end{array} \right] & & \\ 
 \ +\   
 ue' 
 \left[ \begin{array}{cc} Te^1 & Te^3 \end{array} \right]  
 \left[ \begin{array}{c} uq_1 \\ uq_3 \end{array} \right] & & \\ 
 \ +\ 
 \left[ \begin{array}{cc} uq_1' & uq_3' \end{array} \right] 
 \left[ \begin{array}{cc} Q_1^1 & Q_1^3 \\ Q_3^1 & Q_3^3 
 \end{array} \right] 
 \left[ \begin{array}{c} uq_1 \\ uq_3 \end{array} \right] 
 & \leq 0 &  
 \end{eqnarray*}  

 Notemos agora que os dois primeiros termos desta \'{u}ltima
inequa\c{c}\~{a}o s\~{a}o n\~{a}o negativos:
 \begin{itemize} 
 \item Pela 9a linha das restri\c{c}\~{o}es do DAUX vemos que
$ul_2\geq 0$, e pela 6a linha que $uq_2 = 0$, donde, usando a 3a linha, 
concluimos que o primeiro termo \'{e} n\~{a}o negativo. 
 \item Pela 5a linha o segundo termo \'{e} nulo. 
 \end{itemize}   
Obtivemos assim uma contradi\c{c}\~{a}o com a positividade de $Q$.

\section{Problema Quadr\'{a}tico Param\'{e}trico} 

Consideremos agora o Problema Quadr\'{a}tico Param\'{e}trico (QPP) 
que consiste de encontrar todas as solu\c{c}\~{o}es \'{o}timas do 
QP em fun\c{c}\~{a}o do par\^{a}metro $\eta$: estas s\~{a}o as 
``solu\c{c}\~{o}es eficientes'' do QPP.  
 Da mesma forma que reduzimos o problema de programa\c{c}\~{a}o
quadr\'{a}tica (QP) a um problema de programa\c{c}\~{a}o linear (PPL),
o problema de complementariedade linear (PCL), reduziremos o problema
quadr\'{a}tico param\'{e}trico (QPP) a um problema linear
param\'{e}trico (PLP). 

Consideremos no QP, uma solu\c{c}\~{a}o b\'{a}sica das EVO,
 $$ 
 \left[ \begin{array}{c}  
  x\\ l\\ ep\\ en\\ s\\ yl \end{array} \right] \geq 0 
 \mid 
 \left[ \begin{array}{rrrrrr} 
  Tl & 0   & 0   & 0    & 0  & I \\ 
  Te & 0   & 0   & 0    & 0  & 0 \\ 
  Q  & Tl' & Te' & -Te' & -I & 0   
 \end{array} \right]
 \left[ \begin{array}{c}  
  x\\ l\\ ep\\ en\\ s\\ yl \end{array} \right]
 = \left[ \begin{array}{c} tl\\ te\\ \eta p \end{array} \right] $$ 
 $$ 
 x's=0\ ,\ yl'l=0. 
 $$
 ap\'{o}s determinar um primeiro par\^{a}metro cr\'{\i}tico, podemos
reescrever o problema param\'{e}trico introduzindo o termo parametrizado
do vetor do lado direito como a \'{u}ltima coluna da matriz de
coeficientes de um novo PPL:
 $$ 
 \min \pm \eta $$ 
 $$   
 \left[ \begin{array}{c}  
  x\\ l\\ ep\\ en\\ s\\ yl \\ \eta \end{array} \right] \geq 0 
 \mid 
 \left[ \begin{array}{rrrrrrr} 
  Tl & 0   & 0   & 0    & 0  & I & 0 \\ 
  Te & 0   & 0   & 0    & 0  & 0 & 0 \\ 
  Q  & Tl' & Te' & -Te' & -I & 0 & -p   
 \end{array} \right]
 \left[ \begin{array}{c}  
  x\\ l\\ ep\\ en\\ s\\ yl \\ \eta \end{array} \right]
 = \left[ \begin{array}{c} tl\\ te\\ 0 \end{array} \right] $$ 
 $$ 
 x's=0\ ,\ yl'l=0   
 $$  

Como no PLP, $x(\eta )$, a curva das {\em solu\c{c}\~{o}es eficientes}
do QPP \'{e} linear entre um par de par\^{a}metros cr\'{\i}ticos
consecutivos.  Vemos pois que o valor \'{o}timo da solu\c{c}\~{a}o do
problema quadr\'{a}tico com o correspondente par\^{a}metro
 $\eta = (1-\lambda )\eta _k +\lambda \eta _{k+1}$, $vopt(\eta )$, 
 \'{e} uma fun\c{c}\~{a}o quadr\'{a}tica em $\eta$:  
 \begin{eqnarray*}
 vopt(\eta ) &=& (1-\lambda )p'x(\eta_k) +\lambda p'x(\eta_{k+1}) \\  
  & & +(1-\lambda )^2 {x(\eta_k)}'Qx(\eta_k)          
      +{\lambda}^2 {x(\eta_{k+1})}'Qx(\eta_{k+1})    \\
  & & + 2 \lambda (1-\lambda ) {x(\eta_k)}'Qx(\eta_{k+1}) \ . 
 \end{eqnarray*}  
 Portanto o gr\'{a}fico de da parte linear versus a parte quadr\'{a}tica 
 de $vopt(\eta )$, para $-\infty <\eta < \infty$, ser\'{a} uma curva 
 cont\'{\i}nua composta de arcos de quadr\'{a}ticos. 
 
\section{Implementa\c{c}\~{a}o Computacional} 

 A implementa\c{c}\~{a}o dos algoritmos para QP e 
QPP deve contemplar v\'{a}rios aspectos computacionais, 
entre eles: 
 \begin{itemize} 
   \item[a.] ser numericamente est\'{a}vel, 
   \item[b.] preservar a estrutura blocada nas sucessivas bases do simplex, 
   \item[c.] manter a esparsidade das sub-matrizes das bases, 
   \item[d.] utilizar estruturas de dados eficientes para matrizes esparsas, 
   \item[e.] controlar dinamicamente os erros com  opera\c{c}\~{o}es de 
    ponto flutuante. 
 \end{itemize} 
 Estes aspectos, comuns a algoritmos de otimiza\c{c}\~{a}o envolvendo
\'{a}lgebra linear computacional, fogem ao escopo introdut\'{o}rio
deste trabalho, sendo tratados em algumas das refer\^{e}ncias
bibliogr\'{a}ficas citadas. 
 
 \'{E} importante ressaltar que, mesmo usando as mais avan\c{c}adas 
t\'{e}cnicas de estabiliza\c{c}\~{a}o, os algoritmos dependem 
fortemente de termos: 
 \begin{enumerate} 
 \item A matriz de covari\^{a}ncia positiva definida. 
 \item As restri\c{c}\~{o}es linearmente independentes. 
 \end{enumerate} 
 Ao tratarmos um problema ``malposto'', i.e.  um problema pr\'{o}ximo
de violar qualquer das duas condi\c{c}\~{o}es acima, o algoritmo
torna-se inst\'{a}vel, produzindo resultados incorretos. Para tentar sanar 
estes problemas procura-se tomar alguns cuidados no momento de 
modelar o problema. No cap\'{\i}tulo 3 apresentaremos algumas sugest\~oes 
nesta dire\c{c}\~ao para o problema de sele\c{c}\~ao de carteiras.


%% file: cap4.tex
\chapter{Modelo de Markowitz}

\section{An\'{a}lise de M\'{e}dia e Vari\^{a}ncia} 

O modelo de Markowitz \'{e} um modelo para forma\c{c}\~{a}o de
portf\'{o}lios, que visa maximizar a utilidade de um investidor, que 
deve escolher um conjunto de ativos para compor uma carteira, obedecendo 
a restri\c{c}\~oes de disponibilidade de recursos ou outra natureza.  
Consideram-se $n$ ativos, com taxas de retorno
$r_1,\ldots ,r_n$, sendo que um portf\'{o}lio \'{e} especificado pela quantidade
investida em cada ativo, $x_1,\ldots ,x_n$.  As taxas de retorno s\~{a}o
consideradas vari\'{a}veis aleat\'{o}rias.  Os dados do modelo (que
sup\~{o}em-se conhecidos) s\~{a}o o vetor das taxas de retorno
esperadas, $E(r)$, e a matriz de covari\^{a}ncia das taxas de retorno,
$Cov(r)$.  \'{E} um pressuposto do modelo termos a matriz de
covari\^{a}ncia positiva definida. 

 O modelo de Markowitz usa uma fun\c{c}\~{a}o utilidade do
tipo m\'{e}dia-e-vari\^{a}ncia vide ap\^endice B), e define como portf\'{o}lio \'{o}timo um
portf\'{o}lio  $x$ que maximiza a fun\c{c}\~{a}o utilidade estando 
o mesmo  sujeito a restri\c{c}\~{o}es lineares de igualdade, 
desigualdade e sinal: 
 $$\max_x\ U(x)= \eta x'E(r) -x'Cov(r)x \ , \ \ 
   x\geq 0\ \mid \ Te*x=te\ \wedge \ Tl*x\leq tl $$   
 O par\^{a}metro $1/\eta$ \'{e} a avers\~{a}o ao risco do investidor. 

 Nos cap\'{\i}tulos anteriores, estudamos algoritmos para resolver este
problema de otimiza\-\c{c}\~{a}o:  dado valor particular de $\eta$
este era o QP - Problema de Programa\c{c}\~{a}o Quadr\'{a}tica. 
Encontrar a curva das solu\c{c}\~{o}es \'{o}timas em fun\c{c}\~{a}o do
par\^{a}metro $\eta$, $x(\eta )$, (as solu\c{c}\~{o}es eficientes) era o
QPP - Problema Quadr\'{a}tico Param\'{e}trico. 

Para n\~{a}o sobrecarregar a nota\c{c}\~{a}o, ao falarmos dos
portf\'{o}lios $x$ e $y$, denotaremos $x'E(r)$ ou $Cov(x'r,y'r)$ por,
respectivamente, $e(x)$ ou $e_x$, e $\sigma(x,y)$ ou $\sigma_{x,y}$.  
 No caso de uma fam\'{\i}lia de portf\'{o}lios parametrizada por um
escalar, $x(\lambda ),\ \lambda \in \Re$ denotamos $x'(\lambda )E(r)$ por
$e(\lambda )$ ou $e_\lambda$, e $Cov(x(\alpha )'r, x(\beta )'r)$ por
$\sigma(\alpha ,\beta)$ ou $\sigma_{\alpha ,\beta}$. 

\section{Distribui\c{c}\~{a}o das Taxas de Retorno} 

Dados, em dois instantes de tempo $0$ e $t$, o valor ou pre\c{c}o de
um ativo, $v_0$ e $v_t$, existem v\'arias defini\c{c}\~{o}es de ``taxa de
retorno'' do ativo neste per\'{\i}odo.  As mais comuns s\~{a}o:
 \begin{itemize} 
 \item A {\em taxa de retorno simples},   
  $rs \mid v_t= (1+rs)v_0$ donde $rs= (v_t/v_0) -1$ 
 \item A {\em taxa de retorno $t$ vezes composta},  
  $rc \mid v_t= (1+rc)^t v_0$ donde $rc= (v_t/v_0)^{(1/t)} -1$ 
 \item A {\em taxa de retorno continuamente composta},
$v_{t} = v_0 \exp(r*t)$, isto \'e, dada por  
$r = (1/t)\ln(v_t/v_0)$ = $ (1/t)(\ln(v_t) -\ln(v_0))$
  A taxa de retorno continuamente composta pode  ser derivada 
como o limite de Nepper, i.e.  o limite da taxa de retorno $t$ vezes 
composta quando $t\rightarrow \infty$, ou atrav\'es da 
equa\c{c}\~ao diferencial 
$r \mid \frac{ \partial v(t) ) }{ \partial t} = r v(t)$,  
 com condi\c{c}\~{a}o de contorno $v(0) = v_0$ e $v(t)=v_t$.

 \end{itemize} 

Para manter a coer\^{e}ncia de algumas interpreta\c{c}\~{o}es, e para
facilitar o uso de m\'{e}todos estat\'{\i}sticos de estima\c{c}\~{a}o,
gostar\'{\i}amos que a distribui\c{c}\~{a}o da taxa de retorno utilizada
tivesse algumas propriedades: minimamente gostar\'{\i}amos de ter uma
distribui\c{c}\~{a}o sim\'{e}trica, idealmente, queremos uma
distribui\c{c}\~{a}o Normal.  Na maioria das situa\c{c}\~{o}es, a taxa
de retorno continuamente composta \'{e} a que melhor se adapta a estes
quesitos.

\section{Fronteira Eficiente}

Consideremos a curva das solu\c{c}\~{o}es eficientes do QPP, 
 $x(\eta ), 0\leq \eta \leq \infty$.  O gr\'{a}fico do valor da parte
linear versus a parte quadr\'{a}tica do valor das solu\c{c}\~{o}es
eficiente,
 $$e(\eta )\equiv E(r)'x(\eta) 
   \ \ \mbox{contra} \ \   
   v(\eta )\equiv x'(\eta)Cov(r)x(\eta)\ , 
 $$  
 \'{e} denominado {\em fronteira eficiente}.  Podemos interpretar o
par\^{a}metro $\eta$ como uma medida de quanto maximizar o retorno
esperado, $e(\eta )$, \'{e} prefer\'{\i}vel a minimizar o risco do
investimento, $x'Sx$.  No extremo $\eta =0$ queremos apenas minimizar o
risco; no extremo oposto, $\eta =\infty$, queremos apenas maximizar o
retorno esperado. 

Provemos que a fronteira eficiente \'{e} uma curva c\^{o}ncava. 
Usaremos para tanto o fato (demonstrado no cap\'{\i}tulo 2) de que a
solu\c{c}\~{a}o eficiente do QPP entre dois par\^{a}metros
cr\'{\i}ticos, digamos $\eta_0$ e $\eta_1$, $\eta_1 >\eta_0$, \'{e} a
correspondente combina\c{c}\~{a}o convexa das solu\c{c}\~{o}es
eficientes nestes dois par\^{a}metros cr\'{\i}ticos, i.e., escrevendo um
par\^{a}metro intermedi\'{a}rio como 
 $$ \eta (\lambda )= (1-\lambda )\eta_0 +\lambda \eta_1$$ 
 temos a solu\c{c}\~{a}o eficiente: 
 $$ x(\eta (\lambda )) = (1-\lambda )x(\eta_0) +\lambda x(\eta_1)$$    
 de modo que 
 $$ e(x(\eta (\lambda ))) = 
    (1-\lambda )E(r)'x(\eta_0) +\lambda E(r)'x(\eta_1)$$  
 ou 
 $$e(\eta (\lambda )) = (1-\lambda )e(\eta_0) +\lambda e(\eta_1)$$        
Numa nota\c{c}\~ao mais compacta, 
 $$e_\lambda = (1 -\lambda)e_0 +\lambda e_1$$ 
Analogamente,  
 \begin{eqnarray*} 
  \lefteqn{ v(x(\eta (\lambda ))) =\    
     (1-\lambda )^2 Cov(r'x(\eta_0), r'x(\eta_0)) }\\   
  & &  +2\lambda (1-\lambda ) Cov(r'x(\eta_0), r'x(\eta_1)) 
       +\lambda ^2 Cov(r'x(\eta_1), r'x(\eta_1))  
 \end{eqnarray*}
 ou 
 $$ v(\eta (\lambda )) = 
    (1-\lambda )^2 \sigma (\eta_0, \eta_0)  
    +2\lambda (1-\lambda ) \sigma (\eta_0, \eta_1)
    +\lambda^2 \sigma (\eta_1, \eta_1) $$ 
 ou ainda 
 $$ v_\lambda = (1-\lambda )^2 \sigma_{0,0}   
    +2\lambda (1-\lambda ) \sigma_{0,1} +\lambda^2 \sigma_{1,1}$$  
 portanto 
 $$ \frac{ \partial e_\lambda }{ \partial \lambda } = e_1 -e_0$$ 
 $$ \frac{ \partial v_\lambda }{ \partial \lambda } = 
    -2(1 -\lambda)\sigma_{0,0} +2(1 -2\lambda )\sigma_{0,1} 
   +2\lambda \sigma_{1,1}$$  
 $$ \frac{ \partial ^2 v_\lambda }{ \partial \lambda^2} =  
   2(\ \sigma_{0,0} -2\sigma_{0,1} +\sigma_{1,1}\ )$$ 
 Como assumimos que a matriz de covari\^{a}ncia \'e positiva definida,
n\~{a}o podemos ter $\rho_{0,1}=1$.  Assim, sabemos que ( vide ap\^endice ) 
a  \'{u}ltima express\~{a}o \'{e} positiva, e portanto, entre dois pontos
cr\'{\i}ticos, a fronteira eficiente \'{e} c\^{o}ncava (note que
definimos a fronteira eficiente como o gr\'{a}fico $e(\eta )\times
v(\eta )$, e n\~{a}o $v(\eta)\times e(\eta)$ que seria uma curva convexa). 

\begin{figure}[ht]
\[
\input{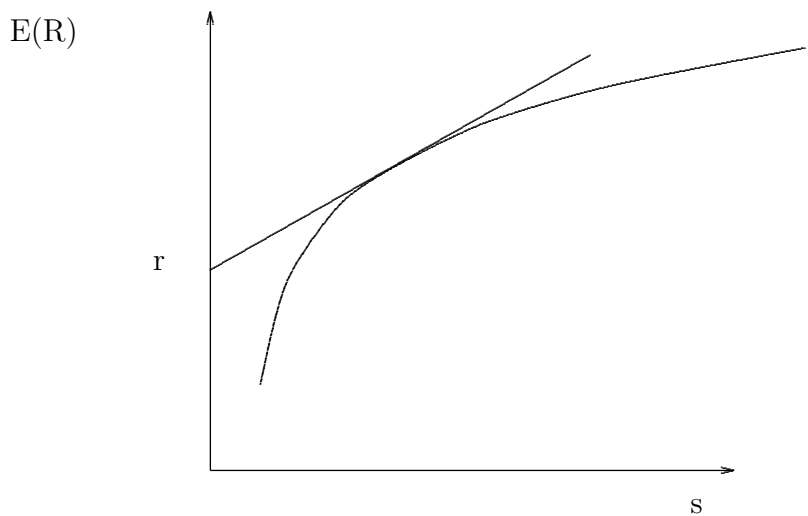}
\]
\caption{\label{fig50} O gr\'afico risco X retorno}
\end{figure}

Nos pontos cr\'{\i}ticos a derivada $\partial \sigma / \partial e$ \'{e}
descont\'{\i}nua.  Para provar que a fronteira eficiente, considerada
integralmente, \'{e} c\^{o}ncava, basta provar que, num ponto
cr\'{\i}tico, esta derivada n\~{a}o pode aumentar.  Consideremos a
possibilidade de que tal ocorresse: Neste caso ter\'{\i}amos uma corda
entre portf\'{o}lios $x$ e $y$, situados numa vizinhan\c{c}a
suficientemente pr\'{o}xima do ponto cr\'{\i}tico, cuja utilidade domina
a fronteira eficiente.  Mas como a correla\c{c}\~{a}o entre
estes dois portf\'{o}lios, $\rho_{x,y}<1$ esta corda \'{e} por sua vez
dominada pela utilidade das combina\c{c}\~{o}es convexas destes
portf\'{o}lios, $(1-\lambda)x +\lambda y$; contradizendo a hip\'{o}tese
de uma fronteira como a considerada ser realmente eficiente.

No cap\'{\i}tulo precedente discutimos brevemente a quest\~ao de 
problemas mal postos, nos quais a matriz de covari\^ancia podia 
deixar de ser positiva definida ou as restri\c{c}\~oes poderiam ser 
LD. Nos problemas de composi\c{c}\~ao de carteiras as seeguintes sugest\~oes 
s\~ao \'uteis para evitar problemas mal postos:

 As seguintes sugest\~{o}es s\~{a}o \'{u}teis para evitar problemas 
malpostos: 
 \begin{enumerate} 

 \item Adi\c{c}\~{a}o gradual de ativos ao portf\'{o}lio:  
  \begin{enumerate}  
  \item Ao inv\'{e}s de considerar de partida todos os ativos
dispon\'{\i}veis no mercado, comece com um subgrupo b\'{a}sico de
tamanho mais modesto, como por exemplo os ativos j\'{a} presentes no
portf\'{o}lio, ou os mais l\'{\i}quidos no mercado. 
  \item A seguir acrescente novos subgrupos ao modelo, agrupando sempre
ativos semelhantes, i.e.  altamente correlacionados.  Crit\'{e}rios
\'{u}teis de semelhan\c{c}a incluem o pa\'{\i}s sede da empresa, seu
setor de atividade, etc. 
  \item Mantenha na modelagem apenas os ativos em cada subgrupo
adicional que efetivamente entram na composi\c{c}\~{a}o de carteiras
eficientes. 
  \end{enumerate} 

 \item Adi\c{c}\~{a}o gradual de restri\c{c}\~{o}es: 
  \begin{enumerate} 
  \item Ao inv\'{e}s de considerar de partida todas as
restri\c{c}\~{o}es desejadas, comece com um subgrupo de tamanho mais
modesto, como as restri\c{c}\~{o}es de igualdade.  A seguir acrescente,
dentre as condi\c{c}\~{o}es de desigualdade desejadas, apenas as que se
mostrarem violadas. 
  \item Tome cuidado para n\~{a}o usar um restri\c{c}\~{a}o de igualdade
quando o que se deseja no modelo \'{e} apenas uma restri\c{c}\~{a}o de
desigualdade. 
  \item Tome muito cuidado para n\~{a}o escrever restri\c{c}\~{o}es
redundantes.  Os exemplos abaixo mostram alguns exemplos de
restri\c{c}\~{o}es redundantes:
   \begin{enumerate} 
   \item $$x_1+x_2\leq 1 \ , \ 0 \leq  x_1\leq 0.5 \ , \ 0 \leq x_2\leq 0.5 \ .$$  
   \item $$x_1+x_2= 1 \ , \ 0 \leq x_1\leq 0.5 \ , \ 0 \leq x_2\leq 0.5 \ .$$  
   \end{enumerate} 
  \end{enumerate}

 \end{enumerate}

\section{Modelos de Tobin e Brennan} 

O modelo de Tobin considera, al\'{e}m dos ativos de risco, um ativo sem
risco de taxa de retorno $e_0$, com desvio padr\~{a}o $\sigma_0=0$. 
Consideremos uma combina\c{c}\~{a}o convexa entre o ativo sem risco e um
portf\'{o}lio eficiente, $x(\eta)$, o portf\'{o}lio  
 $${x}(\lambda)=(1-\lambda)x_0 +\lambda x(\eta)$$   
 A covari\^{a}ncia de uma vari\'{a}vel aleat\'{o}ria com uma
constante \'{e} sempre zero, portanto o retorno esperado e o desvio
padr\~{a}o deste portf\'{o}lio s\~{a}o dados por:
 $$ {e}(\lambda)= (1-\lambda)e_0 +\lambda e(\eta)$$ 
 $$ {\sigma}(\lambda) = \lambda \sigma(\eta)$$ 
 Estes portf\'{o}lios est\~{a}o representados, pela
semi-reta secante \`{a} fronteira eficiente.  Vemos todavia que a
utilidade de qualquer ponto na reta secante \'{e} dominada pela
utilidade de um ponto na semi-reta tangente a fronteira eficiente (em
virtude da concavidade da fronteira eficiente, esta semi-reta \'{e}
\'{u}nica). 

 Par\^{a}metros $\lambda >1$ podem ser interpretados como
empr\'{e}stimo (possuir uma quantidade negativa) do ativo sem risco
aplicados ao portf\'{o}lio tangente, $x(\eta)$.  O par\^{a}metro
correspondente ao ponto de tang\^{e}ncia, $\hat \eta$, \'{e} a raiz de
 $$ h(\eta) = \ 
    \frac{ \partial e(\eta ) }{ \partial \eta} \ / \  
    \frac{ \partial \sigma(\eta ) }{ \partial \eta} 
    \ \ - \  
    ( e(\eta ) -e_0 ) \ / \ \sigma(\eta ) 
 $$   
 Se o ponto de tang\^{e}ncia n\~{a}o \'{e} cr\'{\i}tico, $h(\eta)=0$ traduz  
a igualdade entre as inclina\c{c}\~{o}es (derivadas) da fronteira eficiente 
e da reta secante.  
 A fun\c{c}\~{a}o $h(\eta)$ \'{e} mon\'{o}tona, mas descont\'{\i}nua nos
pontos cr\'{\i}ticos.  Se $h(\eta )$ muda de sinal num ponto de
descontinuidade, ent\~{a}o o par\^{a}metro de tang\^{e}ncia coincide com
um par\^{a}metro cr\'{\i}tico.  

O modelo de Brennan assume dois ativos sem risco, a
interpreta\c{c}\~{a}o deste modelo \'{e} que geralmente pagamos por um
empr\'{e}stimo uma taxa maior do que a que recebemos por um dep\'{o}sito
no ativo sem risco.  Nesta situa\c{c}\~{a}o 
temos dois pontos de tang\^{e}ncia, e uma fronteira
eficiente que est\'{a} na primeira semi-reta para $0\leq \lambda < 1$,
coincide com a fronteira eficiente do modelo de Markowitz para
$\lambda=1$, e esta sobre a segunda semi-reta para $\lambda>1$. 

\mbox{}\\ 

\input{ef1.tex}

\section{Modelos de \'{I}ndices} 

O modelo de Cohen e Pogue considera {\em taxas de crescimento
setoriais}, $c_k$, para os setores da economia $k=1,\ldots K$.  A taxa
de retorno de cada ativo, $r_i$ $i=1\ldots N$ \'{e} a soma de uma {\em
taxa pr\'{o}pria}, $a_i$, mais as taxas de crescimento setoriais
ponderadas por {\em fatores de sensibilidade}, $B_{i,k}$.  Neste tipo de
modelo temos sempre que $k<<n$.  Ademais, o modelo sup\~{o}e que as
taxas pr\'{o}prias s\~{a}o vari\'{a}veis aleat\'{o}rias independentes
entre si, e tamb\'{e}m independentes das taxas setoriais:
 $$ r_i = a_i +\sum_k B_{i,k}c_k \ , \ \  
    Cov(a_i,a_j)=0,\ Cov(a_i,c_k)=0
 $$ 

Usando as leis de transforma\c{c}\~{a}o, podemos calcular a
esperan\c{c}a e covari\^{a}ncia do vetor dos retornos:
 $$ E(r) = E(a) +B\, E(c) \ , \ \  
    Cov(r) = diag([Var(a_1)\ldots Var(a_n)]) \   
    + B\, Cov(c)\, B' 
 $$

Para usar este modelo temos que estimar $E(a_i)$, $Var(a_i)$, $B_{i,k}$,
$E(c)$ e $Cov(c)$, i.e.  da ordem de $2n +k*n +k^2$, ao inv\'{e}s de
$n^2$ par\^{a}metros.  Para fazer estas estimativas s\~{a}o usadas uma
s\'{e}rie de t\'{e}cnicas que integram estat\'{\i}stica e
otimiza\c{c}\~{a}o. 
A partir de s\'{e}ries hist\'{o}ricas de valores de $r_i$ e $c_{i,k}$, 
costumamos estimar o vetor $a$ e a matriz $B$, por: 
  \begin{enumerate}
   \item Modelos Lineares, como por exemplo: 
    \begin{enumerate} 
     \item An\'{a}lise de regress\~{a}o,  
     \item Filtro de Kalman; ou 
    \end{enumerate}
   \item Modelos n\~{a}o Lineares, como por exemplo: 
    \begin{enumerate} 
     \item Regress\~{a}o n\~{a}o Linear, 
     \item Redes Neurais, 
     \item Filtros Adaptativos. 
    \end{enumerate} 
  \end{enumerate}   

A an\'{a}lise e previs\~{a}o de \'{\i}ndices \'{e} feita utilizando
todos os instrumentos listados acima, e ainda t\'{e}cnicas
econom\'{e}tricas como An\'{a}lise de Clusters, 
An\'{a}lise de Box-Jenkins, An\'{a}lise Espectral, 
An\'{a}lise de Insumo-Produto, etc.

\subsection*{Modelos Diagonais}

Um modelo de \'{\i}ndices est\'{a} em forma {\em diagonal}, se a matriz
de covari\^{a}ncia dos \'{\i}ndices \'{e} uma matriz diagonal,
 $$Cov(c)= \mbox{diag}( [ \sigma_1^2, \ldots \sigma_n^2 ] ) 
   = \left[ \begin{array}{ccc} 
   \sigma_{1,1} & & 0 \\ & \ddots & \\ 0 & & \sigma_{k,k} 
   \end{array} \right] 
 $$ 
 Modelos diagonais permitem uma interpreta\c{c}\~{a}o mais simples, 
pois $i\neq j\Rightarrow Cov(c_i,c_j)=0$. 
 Na verdade, qualquer modelo de \'{\i}ndices pode ser posto na 
forma diagonal, como veremos a seguir. 

\'{E} um teorema b\'{a}sico de An\'{a}lise Num\'{e}rica que, dada uma
matriz sim\'{e}trica e positiva definida, $S$, podemos computar sua {\em
fatora\c{c}\~{a}o de Cholesky}, i.e.  podemos encontrar uma matriz
triangular inferior, $L$, tal que $S=LL'$. 
 
Usaremos agora a fatora\c{c}\~{a}o de Cholesky para transformar um
modelo de \'{\i}ndices geral, cuja matriz de covari\^{a}ncia dos
\'{\i}ndices n\~{a}o \'e diagonal, num modelo diagonal equivalente. 
Consideremos o novo vetor de \'{\i}ndices
 $$ d = L^{-1}c \ , \ \mbox{onde}\ \ Cov(c)= S= LL' \ .$$ 

A matriz de covari\^{a}ncia dos novos \'{\i}ndices \'{e} diagonal, pois 
 $$ Cov(d) = L^{-1}Cov(c)L^{-t} = L^{-1}(LL')L^{-t} 
    = (L^{-1}L)(L'L^{-t})= I 
 $$ 
 O modelo originalmente escrito em fun\c{c}\~{a}o dos \'{\i}ndices $c$,
pode ser facilmente reescrito em fun\c{c}\~{a}o dos novos \'{\i}ndices,
$d$:
 $$ r = a +Bc \Rightarrow 
   r = a +B LL^{-1}c = a +\tilde B d 
 $$

\section{Modelos de Equil\'{\i}brio} 

Apresentaremos agora alguns modelos de equil\'{\i}brio.  Estes modelos
tem uma natureza essencialmente diferente dos modelos que estudamos
at\'{e} agora.  Os modelos que vimos at\'{e} agora s\~{a}o modelos de
decis\~{a}o: eles sup\~{o}em que um dado investidor tem certas
prefer\^{e}ncias, e que este investidor foi a campo e ``mediu'' algumas
caracter\'{\i}sticas dos ativos no mercado.  A partir da\'{\i} os
modelos de decis\~{a}o fornecem algoritmos que encontram as decis\~{o}es
``\'{o}timas'' (ou racionais) para este investidor. 

Modelos de equil\'{\i}brio assumem que todos os investidores no mercado
se comportam racionalmente usando um mesmo modelo de decis\~{a}o. 
Geralmente estes modelos tamb\'{e}m assumem que todos eles est\~{a}o de
{\em acordo} quanto \`{a}s entradas do modelo (suas medidas e
previs\~{o}es sobre as caracter\'{\i}sticas relevantes dos ativos no
mercado s\~{a}o id\^{e}nticas).  A partir destas hip\'{o}teses estes
modelos calculam propriedades de {\em pontos de equil\'{\i}brio} (ou
pontos estacion\'{a}rios) do mercado.

\subsection*{Modelo CAPM} 

O modelo CAPM assume que todos os investidores usam o modelo de
Markowitz para tomada de decis\~{a}o, que existe um ativo sem risco
acess\'{\i}vel a todos os investidores, que podem tom\'{a}-lo emprestado
ou nele investir com uma mesma taxa de retorno $r_0$, e que todos os
investidores est\~{a}o de acordo quanto ao retorno esperado e a matriz
de covari\^{a}ncia dos retornos dos ativos de risco no mercado. 

A primeira conclus\~{a}o a que chegamos neste cen\'{a}rio \'{e} que
todos os investidores concordam sobre a fronteira eficiente do mercado. 
Ademais, como no modelo de Tobin, todos os investidores formam seu
portf\'{o}lio sobre a {\em linha de mercado}, a partir do ativo sem
risco e de um mesmo portf\'{o}lio tangente, o {\em portf\'{o}lio de
mercado}, $xm$. 

Sob condi\c{c}\~{o}es de equil\'{\i}brio, todas os ativos deveriam estar
presentes no portf\'{o}lio de mercado (pois ningu\'{e}m teria interesse
num ativo fora de $xm$).  Examinemos agora a combina\c{c}\~{a}o convexa
de $xm$ com o portf\'{o}lio $xi$ constitu\'{\i}do apenas do $i$-\'{e}simo
ativo, $x_i$:
 $$ xc(\lambda ) = (1-\lambda )xi +\lambda xm   
    \ ,\ \ \mbox{com} \  
    e(\lambda)\equiv e(xc(\lambda )'r) \ \mbox{e} \ 
    \sigma_{\lambda ,\lambda}\equiv Var(xc(\lambda )'r)
 $$
 $xc(1)$ coincide com $xm$, e para todos os demais valores de $\lambda
$, a solu\c{c}\~{a}o $xc(\lambda )$ \'{e} dominada pela fronteira
eficiente; i.e., a curva
 $e(xc(\lambda ))\times \sigma (xc(\lambda ))$ 
tangencia a fronteira eficiente. 

Mas 
 \begin{eqnarray*} 
 e(\lambda) &=& 
 (1-\lambda )e(xi) +\lambda e(xm) \\ 
 \frac{ \partial e(\lambda ) }{ \partial \lambda } &=& 
 e(xm) -e(xi) \\ 
 \sigma_{\lambda ,\lambda} &=& 
 (1-\lambda )^2\sigma_{xi,xi} +2\lambda (1- \lambda )\sigma_{xi,xm} 
 +\lambda^2 \sigma_{xm,xm} \\ 
 \frac{ \partial \sigma_{\lambda} }{ \partial \lambda } &=& 
 \frac{ -(1 -\lambda )\sigma_{xi,xi} +(1-2\lambda )\sigma_{xi,xm} 
 +\lambda \sigma_{xm,xm} }{ \sigma_{\lambda } }   
 \end{eqnarray*} 

 Igualando, em $\lambda =1$ a derivada, 
 $\partial e(xc) / \partial \sigma(xc)$ 
 \`{a} inclina\c{c}\~{a}o da linha de mercado temos 
 $$ \frac{ (e(xm) -e(xi) ) \sigma_{xm} } 
    { \sigma_{xm,xm} - \sigma_{xi,xm} } = 
    \frac{ e(xm) -r_0 }{ \sigma_{xm} } 
 $$ 
 simplificando, 
 $$e(xi) = r_0 +(e(xm)-r_0)\sigma_{xi,xm}/\sigma_{xm,xm}$$    

No modelo de \'{\i}ndices, com um \'{u}nico \'{\i}ndice, tinhamos 
 $$ r_i = \alpha_i +\beta_i c $$ Nesta situa\c{c}\~{a}o as
condi\c{c}\~{o}es de equil\'{\i}brio se traduzem como
 $$ \alpha_i=r_0 \ , \ \ 
    c= e(xm) - r_0 \ , \ \  
    \beta_i = \frac{ \sigma_{xi,xm}}{\sigma_{xm,xm}} 
            = \frac{e(xi)}{e(xm) -r_0}  
 $$ 
Vemos portanto que o modelo CAPM espera de cada ativo um retorno 
igual ao do ativo sem risco, mais um retorno proporcional a taxa de 
crecimento da economia; este fator de proporcionalidade \'{e} por sua vez 
proporcional a covari\^{a}ncia do ativo com o portf\'{o}lio de mercado. 

\subsection*{Modelo APT} 

O modelo APT \'{e} um modelo de equil\'{\i}brio baseado no modelo de
\'{\i}ndices, que sup\~{o}em ser poss\'{\i}vel construir
``portf\'{o}lios de arbitragem'' com as seguintes caracter\'{\i}sticas:
 \begin{enumerate}  
  \item $ x' {\bf 1}=0$,  
  \item $ x' B = 0 $, 
  \item $ Var(x' a) \approx 0$.   
 \end{enumerate} 
 Isto \'{e}, portf\'{o}lios de valor zero, 
{\em imunes} a todas as taxas de crescimento setorial. 
e quase livres de risco pr\'{o}prio.  

A condi\c{c}\~{a}o de equil\'{\i}brio do modelo \'{e}: ``As
condi\c{c}\~{o}es acima implicam que a taxa esperada de retorno de um
portf\'{o}lio de arbitragem deve ser nula, $x'E(a)=0$''.  Esta
condi\c{c}\~{a}o de equil\'{\i}brio \'{e} motivada pela possibilidade de
ganhar dinheiro sem risco, caso a condi\c{c}\~{a}o n\~{a}o seja
satisfeita.  Matematicamente, para qualquer portf\'{o}lio de arbitragem, a
condi\c{c}\~{a}o de equil\'{\i}brio se expressa como:
 $$ x' \left[ \begin{array}{cc} 1 & B \end{array} \right] = 0 
  \Rightarrow x' E(r) = 0 
 $$  
 Mas a implica\c{c}\~{a}o acima \'{e} equivalente a termos 
 $$ E(r) = \left[ \begin{array}{cc} 1 & B \end{array} \right] 
           \left[ \begin{array}{cccc} 
            l_0 & l_1 & \ldots l_k \end{array} \right] '  
 $$ 
Assim o modelo APT conclui que, em condi\c{c}\~{o}es de equil\'{\i}brio,
o retorno esperado de um dado ativo, $E(r_i)$, \'{e} uma constante,
$l_0$, mais termos proporcionais aos {\em pr\^{e}mios de risco} de cada
setor, $l_k$ ponderados pelo fator de sensibilidade, $B_{i,k}$. 
Finalmente, usando novamente um argumento de arbitragem,
conclu\'{\i}mos que a constante $l_0$ deve ser a taxa de retorno do
ativo sem risco, $r_0$.

%% file: ef1.tex
\unitlength=1.10mm
\special{em:linewidth 0.4pt}
\linethickness{0.4pt}
\begin{picture}(100.00,157.00)
\put(1.00,1.00){\vector(0,1){156.00}}
\put(1.00,1.00){\vector(1,0){99.00}}
\put(1.00,50.00){\vector(1,0){99.00}}
\put(1.00,100.00){\vector(1,0){99.00}}
\bezier{228}(46.00,124.00)(57.00,147.00)(88.00,148.00)
\bezier{240}(21.00,117.00)(27.00,141.00)(62.00,142.00)
\put(21.00,117.00){\line(5,3){41.00}}
\bezier{476}(5.00,53.00)(6.00,84.00)(94.00,90.00)
\put(48.00,84.00){\circle{2.00}}
\bezier{480}(4.00,4.00)(8.00,38.00)(94.00,40.00)
\bezier{216}(21.00,11.00)(24.00,41.00)(41.00,24.00)
\put(92.00,95.00){\line(-4,-1){91.00}}
\bezier{204}(46.00,124.00)(14.00,119.00)(4.00,103.00)
\put(62.00,142.00){\circle{2.00}}
\put(21.00,117.00){\circle{2.00}}
\put(28.00,30.00){\circle{2.00}}
\put(21.00,11.00){\circle{2.00}}
\put(41.00,24.00){\circle{2.00}}
\put(1.00,57.00){\line(1,1){40.00}}
\put(12.00,68.00){\circle{2.00}}
\put(1.00,19.00){\line(5,2){70.00}}
\put(24.00,11.00){\makebox(0,0)[lc]{xi=xc(0)}}
\put(26.00,32.00){\makebox(0,0)[rc]{xm=xc(1)}}
\put(44.00,24.00){\makebox(0,0)[lc]{xc(2)}}
\put(48.00,123.00){\makebox(0,0)[lc]{Critical Point}}
\put(23.00,115.00){\makebox(0,0)[lc]{x}}
\put(64.00,140.00){\makebox(0,0)[lc]{y}}
\end{picture}

%% file: cap5.tex
\chapter{Programa\c{c}\~{a}o Din\^{a}mica}

Neste cap\'{\i }tulo procuraremos introduzir conceitos b\'{a}sicos de 
programa\c{c}\~{a}o din\^{a}mica e a\-presentar exemplos de resolu\c{c}\~{a}o 
de problemas. Programa\c{c}\~{a}o din\^{a}mica \'{e} uma metodologia que 
pode ser empregada para resolver uma extensa classe de problemas. Mais que uma 
t\'{e}cnica, \'{e} uma abordagem conceitual que procura subdividir o problema de 
interesse em subproblemas de mais simples resolu\c{c}\~{a}o, estabelecendo uma 
rela\c{c}\~{a}o de depend\^{e}ncia entre os diversos subproblemas. Essa rela\c{c}\~{a}o 
possibilita que a solu\c{c}\~{a}o \'{o}tima do problema original seja obtida 
recursivamente: em cada est\'{a}gio de resolu\c{c}\~{a}o busca-se a solu\c{c}\~{a}o 
\'{o}tima de um subproblema que depende apenas da solu\c{c}\~{a}o \'{o}tima do 
subproblema anteriormente resolvido.

 O objetivo de um problema de programa\c{c}\~ao din\^amica 
ou controle \'{o}timo \'{e} otimizar as a\c{c}\~oes de um operador capaz de 
influenciar o sistema em estudo. Procura-se  determinar uma regra 
de decis\~{a}o \'otima, ou seja, que minimize o custo total associado \`{a} 
tomada de decis\~{a}o. Discutiremos a quest\~ao assumindo dois agentes: 
o operador que tem \`{a} sua disposi\c{c}\~{a}o um conjunto de controles, 
ou a\c{c}\~oes, para atuar sobre o sistema, e o administrador 
do sistema a quem cabe estabelecer uma pol\'{\i}tica que determine o 
controle a ser usado pelo operador em fun\c{c}\~ao do estado do sistema. 

A evolu\c{c}\~{a}o do sistema ao longo do tempo \'{e} descrita por uma 
equa\c{c}\~{a}o relacionando o estado em que o mesmo se encontrar\'{a} no instante 
t+1 e seu hist\'{o}rico passado e presente; tamb\'{e}m nesta descri\c{c}\~{a}o 
considera-se o conjunto de a\c{c}\~{o}es tomadas pelo operador dentro do horizonte de 
tempo em quest\~{a}o, i.e., os controles exercidos nos instantes 
presente e passado.

\section{Conceitos B\'{a}sicos}

Os conceitos apresentados a seguir ser\~{a}o utilizados neste
cap\'{\i}tulo  e sintetizam alguns dos principais aspectos  de
programa\c{c}\~{a}o din\^{a}mica.

\begin{itemize}
\item[1]
{\bf Tempo, est\'{a}gio, per\'{\i }odo, \'{e}poca, instante}: 
 \'{e} um par\^{a}metro de  evolu\c{c}\~{a}o do sistema. Muitas vezes
representa tempo, mas nem sempre esta  interpreta\c{c}\~{a}o \'{e} a
mais adequada. Pode ser inerente \`{a} natureza do  problema mas com
freq\"u\^{e}ncia \'{e} introduzido artificialmente para subdividir o
problema original em subproblemas.

\item [2] 
{\bf Horizonte:} 
 um limitante superior para o \'{\i }ndice de tempo $t$.  Neste
trabalho, o dom\'{\i}nio do tempo $t$ ser\'{a} sempre discreto, uma
consequ\^{e}ncia da  t\^{o}nica algor\'{\i }tmica e computacional
adotada. Geralmente tem a  forma $t\in \{0,1,2,\ldots ,h\}$ ou $t\in
\{1,2,\ldots,h\}$, onde $h$ \'{e} o horizonte (de planejamento). 
Podemos ter $h=+\infty$. 

\item [3]
{\bf  Estado:} 
descri\c{c}\~{a}o das condi\c{c}\~{o}es do sistema no instante $t$, 
denotada por $x(t)$. Quando n\~{a}o houver ambig\"uidade, o instante $t$ 
em discuss\~{a}o ser\'{a} omitido. 

\item [4]
{\bf  A\c{c}\~{a}o, decis\~{a}o, controle:} 
 $u(t)$. Descreve a influ\^{e}ncia do operador sobre o  sistema.
Otimizar as a\c{c}\~{o}es do operador \'{e} o objetivo de um problema de
 programa\c{c}\~{a}o din\^{a}mica ou de controle \'{o}timo.
 
 Neste livro veremos exemplos de espa\c{c}os de estado e controle
discretos e cont\'{\i }nuos, ainda  que utilizemos somente par\^{a}metro
de tempo discreto.

\item [5] 
{\bf Equa\c{c}\~{a}o de Evolu\c{c}\~{a}o Din\^{a}mica:}
\'E a equa\c{c}\~ao que nos d\'{a} o 
pr\'{o}ximo estado do sistema em fun\c{c}\~{a}o dos estados e controles
presentes e passados,  
$x(t+1)=g(x(t),u(t),x(t-1),u(t-1),\ldots),x(1),u(1)$. 

\item[6] 
{\bf Trajet\'{o}ria:} 
uma poss\'{\i}vel seq\"{u}\^{e}ncia de estados e controles, 
do instante inicial ao horizonte, 
$x(1), u(1), x(2), u(2), \ldots x(h)$.

\item [7]
{\bf  Evolu\c{c}\~{a}o sem mem\'{o}ria, Sistema Markoviano:} 
$x(t+1) = g(x(t),u(t))$ \'{e} fun\c{c}\~{a}o apenas do estado e 
controles presentes. Muito conveniente em programa\c{c}\~{a}o din\^{a}mica, 
sendo que muitas vezes adotamos uma descri\c{c}\~{a}o redundante de estados 
apenas para obter markovianidade. A grande maioria dos problemas de 
programa\c{c}\~{a}o 
din\^{a}mica s\~{a}o markovianos ou s\~{a}o tais que a fun\c{c}\~{a}o 
$g(.)$ depende de um n\'{u}mero finito de estados passados. Neste \'{u}ltimo 
caso $x(t+1)=g(x(t),u(t),x(t-1),u(t-1),\ldots,x(t-n),u(t-n))$, 
isto \'{e}, a equa\c{c}\~{a}o de evolu\c{c}\~{a}o \'{e} 
uma equa\c{c}\~{a}o de diferen\c{c}as finitas de ordem n+1. 

\item [8]
{\bf  Restri\c{c}\~{o}es:} condi\c{c}\~{o}es especificando as
trajet\'{o}rias vi\'{a}veis (admiss\'{\i }veis) para o sistema, geralmente
expressas na forma $R(x(0),u(0),\ldots,x(h),u(h))\geq 0.$ Geralmente
estudaremos restri\c{c}\~{o}es independentes do passado, 
$R(t,x(t),u(t))\geq 0$.

\item [9]
{\bf Custo instant\^{a}neo:} $c_t = c(t,x(t),u(t))$, o custo do
controle $u(t)$ estando o sistema no estado $x(t)$. 
 
\item [10]
{\bf Custo aditivo, custo descontado, custo m\'{e}dio:}
 a soma, a soma
descontada, ou a m\'{e}dia dos custos instant\^{a}neos ao longo de uma
trajet\'{o}ria (caminho de evolu\c{c}\~{a}o) do sistema.  
 $S= {\sum}_{t=1}^{h} c_t$, 
 $S= {\sum}_{t=1}^{h} {\beta}^{t}*c_t$, ou 
 $S= (1/h){\sum}_{t=1}^{h} c_t$, onde $c_t=c(t,x(t),u(t))$. 

\item [11]
{\bf Fun\c{c}\~{a}o de custo:} 
uma fun\c{c}\~{a}o\
$f(x(0),u(0),x(1),u(1),\ldots ,x(h),u(h))$ que pretendemos minimizar. 
Geralmente $f()$ \'{e} uma fun\c{c}\~{a}o simples do custo aditivo ou
descontado.  Muitas vezes utilizamos truncamentos da fun\c{c}\~{a}o de
custo (fun\c{c}\~{o}es de custo futuro),
 $f(x(t),u(t),x(t+1),u(t+1),\ldots ,x(h),u(h))$, para sub-dividir o
problema em problemas auxiliares. 

\item [12]
{\bf Pol\'{\i}tica:} 
Uma regra para tomada de decis\~{a}o. Se o sistema se encontra 
em um estado $x(t)$, a ado\c{c}\~{a}o de uma pol\'{i}tica $\Pi$
gera uma a\c{c}\~ao ou controle $u(t)$, dado por 
$u(t)={\Pi}(t,x(t),u(t-1),x(t-1),\ldots,u(0),x(0))$.  Geralmente
estudaremos pol\'{\i}ticas que dependem apenas do estado presente,
$u(t)={\Pi}(t,x(t))$.  Uma pol\'{\i}tica \'{e} vi\'{a}vel se respeita as
restri\c{c}\~{o}es do sistema.

\item [13]
{\bf  Pol\'{\i }tica \'{o}tima:} 
a pol\'{\i }tica vi\'{a}vel \'{o}tima \'{e} 
aquela que minimiza a fun\c{c}\~{a}o de custo

\item [14]
{\bf Princ\'{\i }pio de otimalidade:} 
 uma prova (argumento) de que a pol\'{\i }tica  \'{o}tima \'{e}
fun\c{c}\~{a}o apenas do estado presente. Note que a  markovianidade do
sistema n\~{a}o implica num princ\'{\i }pio de otimalidade, sendo 
f\'{a}cil apresentar contra-exemplos com fun\c{c}\~{o}es de custo
n\~{a}o aditivas.
 
\item[15]
{\bf Condi\c{c}\~{a}o de contorno:} 
base para a aplica\c{c}\~{a}o recursiva da equa\c{c}\~{a}o de Bellman.
Estabelece o valor da fun\c{c}\~ao custo no horizonte de planejamento. 

\item [16]
{\bf  Equa\c{c}\~{a}o de Bellman ou equa\c{c}\~{a}o de otimalidade:} uma 
afirma\c{c}\~{a}o sobre a pol\'{\i }tica \'{o}tima, 
baseada no princ\'{\i }pio de otimalidade e que nos permite calcular a  
solu\c{c}\~{a}o \'{o}tima do problema recursivamente. Muitas vezes a 
equa\c{c}\~{a}o de Bellman n\~{a}o \'{e} montada diretamente sobre 
a pol\'{\i }tica \'{o}tima, mas sobre o \'{o}timo da fun\c{c}\~{a}o 
custo numa s\'{e}rie de problemas auxiliares.
 
\item [17]
{\bf Recupera\c{c}\~{a}o da pol\'{\i}tica \'{o}tima ou backtracking}:
recupera\c{c}\~{a}o da pol\'{\i}tica \'{o}tima a partir da tabela de
valores calculados com a equa\c{c}\~{a}o de Bellman.

\item [18]
{\bf  Equa\c{c}\~{a}o de evolu\c{c}\~{a}o em sistemas estoc\'{a}sticos}: a 
evolu\c{c}\~{a}o de um sistema estoc\'{a}stico markoviano \'{e} dada pela 
probabilidade (distribui\c{c}\~{a}o de probabilidade) de transi\c{c}\~{a}o 
do sistema para o estado $x(t+1)$, a partir do estado $x(t)$ com controle 
$u(t)$, descrita como $pr(t,x(t),u(t),x(t+1))$.
 
\item [19]
{\bf Fun\c{c}\~{a}o de custo em sistemas estoc\'{a}sticos}: nestes sistemas 
adota-se ao inv\'es da fun\c{c}\~ao custo, a esperan\c{c}a das 
mesmas. Assim, as fun\c{c}\~{o}es mais utilizadas em sistemas 
estoc\'{a}sticos s\~{a}o:

 \begin{enumerate}
 \item O valor esperado do custo aditivo (ou descontado):
  $f_{\Pi} = E(S= {\sum}_{t=0}^{h} c(t,x,u))$.
 \item Uma combina\c{c}\~{a}o linear do valor esperado e da
 vari\^{a}ncia do custo aditivo:
  $f_{\Pi} = E(S) - Var(S)$.
\item A esperan\c{c}a de uma fun\c{c}\~{a}o do custo aditivo ou
descontado, como por exemplo no controle quadr\'{a}tico Gaussiano
sens\'{\i}vel a risco (ELQG):

  $$f_{\Pi} = (-2/\theta) \log E( \exp (-\theta S /2) )$$
        $$  = E(S) + (\theta /4)Var(S) + O({\theta}^2)$$
\end{enumerate}

\end{itemize}

\section{Dist\^{a}ncia M\'{\i }nima em um Grafo}

Um problema cl\'{a}ssico em teoria dos grafos, denominado 
problema do caminho m\'{\i}nimo, possui diversas aplica\c{c}\~oes 
tanto em engenharia como em finan\c{c}as. Iremos resolv\^{e}-lo 
dentro do contexto de programa\c{c}\~{a}o din\^{a}mica.

Primeiramente apesentaremos algumas defini\c{c}\~{o}es elementares:

\begin{itemize}

\item[]
Um {\bf grafo}, $G=(X,U)$, \'{e} uma estrutura composta de um conjunto de 
{\bf v\'{e}rtices},
$X=\{1,\ldots , n\}$, e um conjunto de arestas $U \subseteq X \times X$. 
Imagine v\'{e}rtices como sendo cidades, e uma aresta $u=(x,y)$ como uma
estrada de m\~{a}o \'{u}nica indo de $x$ para $y$.  

\item[]
Um grafo \'{e} {\bf completo} se cont\'{e}m todas as $n^2$ arestas 
poss\'{\i}veis. 

\item[]
Um {\bf caminho} \'{e} uma sequ\^{e}ncia de arestas consecutivas, 
 $$p = ( (v_0, v_1), (v_1,v_2),(v_2,v_3), \ldots , (v_{t-1}, v_t) ).$$

\item[]
Um {\bf ciclo} \'{e} um caminho
que come\c{c}a e termina no mesmo v\'{e}rtice, $v_0=v_t$, sem repetir
nenhum outro v\'{e}rtice.  

\item[]
O {\bf tamanho} de um caminho, $|p|$ , \'{e} o
n\'{u}mero de arestas que o comp\~{o}em.  Denotamos por $p(s)=v_s$ o
$s$-\'{e}simo v\'{e}rtice em $p$. 

\item[]
O {\bf custo} \'{e} uma fun\c{c}\~{a}o $c(u)$, $U \mapsto {\Re}$. 
O custo aditivo de um caminho \'{e} a soma dos custos das arestas
que o comp\~{o}em, i.e. $c(p) = \sum_{1}^{t} c((v_{t-1},v_t))$.
\end{itemize}

\noindent
O problema que queremos resolver \'{e} 
determinar $p^*(1,y)$, o caminho de m\'{\i}nimo custo de
$1$ at\'{e} $y$.  Assumiremos v\'{a}lidas as seguintes hip\'{o}teses:
 \begin{itemize}
 \item O custo de uma aresta nunca \'{e} negativo. 
 \item O grafo \'{e} completo.
 \end{itemize}

A primeira hip\'{o}tese simplifica grandemente o problema. J\'{a} a segunda
\'{e} apenas uma conveni\^{e}ncia de modelagem e pode ser assumida 
sem perda de generalidade uma vez que arestas inexistentes podem ser 
modeladas por arestas de custo muito alto, ou $+\infty$. 

 Para resolver o problema do caminho m\'{\i }nimo iremos utilizar o
princ\'{\i }pio de  programa\c{c}\~{a}o din\^{a}mica, construindo
problemas auxiliares.  No t-\'esimo subproblema queremos determinar um
caminho de no m\'aximo  $t$ arestas que minimize a dist\^ancia entre o
v\'ertice (estado) origem, 1,   e o destino y. O \'\i ndice de ``tempo''
$t$ neste caso indicar\'a os  subproblemas construidos  e estar\'a
associado ao tamanho dos caminhos  construidos. A decis\~ao a ser tomada
em cada estado (cada v\'ertice do  grafo) consiste em determinar a
pr\'oxima aresta que ir\'a compor o  caminho m\'\i nimo.

Vamos subidividir o problema considerando v\'arios  subproblemas nos 
quais iremos determinar um caminho de custo m\'{\i }nimo entre  o v\'ertice 
(estado) 1 e o v\'ertice  y com a restri\c{c}\~{a}o de conter no m\'{a}ximo 
$t$  arestas.

Ou seja, para $t \in \{0,1,2,\ldots\}$ consideramos:

{\bf Problema t:}  Encontre o m\'{\i}nimo custo aditivo, para um caminho 
de $1$ at\'{e} $y$ de tamanho menor ou igual a $t$, i.e. determine a 
fun\c{c}\~{a}o de custo
 $$f^*(t,1,y)= \min _{p\in \bar{P}(t,1,y)} c(p)$$ onde
 $$P(t,1,y)=\{p  |\ |p| = t, p(0)=1, p(t)=y\} \ \ 
 {\rm e}\ \  \bar{P}(t,1,y)={\cup}_{s=0}^{t} P(s,1,y)$$
 Caso n\~{a}o exista uma caminho de $1$ a $y$ de tamanho menor ou igual
a $t$ temos, pela defini\c{c}\~{a}o de $\min$, $f^*(t,1,y)=+\infty$. 

\vspace{0.5cm}

\noindent
O primeiro problema auxiliar tem solu\c{c}\~{a}o trivial:
 $$f^*(t=0,1,1)=0 \ \ e \ \ f^*(t=0,1,y)=+\infty \ , \ \ \forall y\neq 1$$.

Temos tamb\'{e}m a rela\c{c}\~{a}o de recorr\^{e}ncia:
 $$f^*(t+1,1,y) = {\min}_{u=(x,y)} \{ f^*(t,1,y) , (f^*(t,1,x) +c(x,y))\}.$$
Esta recorr\^{e}ncia se explica da seguinte forma: Qualquer caminho
 $p\in P(t,1,x)$  seguido da aresta $u=(x,y)$ resulta num caminho em 
 $P(t+1,1,y)$, e qualquer caminho em $P(t+1,1,y)$
pode ser obtido exatamente desta forma (por que?).

Estabelecida a recorr\^{e}ncia podemos 
usar a solu\c{c}\~{a}o do problema auxiliar com $t=0$ e a
rela\c{c}\~{a}o de recorr\^{e}ncia para determinar os \'{o}timos da
fun\c{c}\~{a}o de custo, $f^*(t,1,y)$, para $t\in \{1, 2, \ldots\}$.  Em
particular, a primeira aplica\c{c}\~{a}o da rela\c{c}\~{a}o de
recorr\^{e}ncia nos fornece:
 $$ f^*(1,1,y) = c((1,y)) , \forall y\in X \ .$$

\begin{figure}[ht]
\[
\input{grafo.ptx}
\]
\caption{\label{fig6} Um grafo orientado}
\end{figure}

Na figura 4.1 temos um grafo com 5 v\'{e}rtices.
 Seus custos 
s\~ao apresentados na Tabela I abaixo:

\begin{center}
{\bf Tabela I}
\end{center}
 $$  
\begin{array}{c|ccccc} 
 C(i,j)  & 1 & 2 & 3 & 4 & 5  \\ 
\hline 
 1 & 0 & 2 & \infty & 1 & 2 \\ 
 2 & 2 & 0 & 2 & \infty & 0 \\ 
 3 & \infty & 2 & 0 & 3 & 4 \\ 
 4 & 1 & \infty & 3 & 0 & 0 \\ 
 5 & 2 & 0 & 4 & 0 & 0  
\end{array} 
 $$ 

\vspace{0.5cm}

Na tabela  abaixo temos o exemplo de aplica\c{c}\~{a}o da recorr\^{e}ncia
para $t \in \{0, 1,\ldots, n-1\}$ no grafo da figura 4.1. 

 $$  
\begin{array}{c|ccccc} 
 t \backslash y & 1 & 2 & 3 & 4 & 5 \\ 
\hline 
 0 & {\bf 0} & \infty & \infty & \infty & \infty \\ 
 1 & 0 & 2 &  \infty & {\bf 1} & 2 \\ 
 2 & 0 & 2 & 4 & 1 & {\bf 1} \\ 
 3 & 0 & {\bf 1} & 4 & 1 & 1 \\ 
 4 & 0 & 1 & {\bf 3} & 1 & 1 
\end{array} 
 $$  
\begin{center}
{\bf Tabela II}
\end{center}
\vspace{0.5cm}

{\bf Lema 1:} 
$\forall t\geq n,  f^*(t,x,y)=f^*(n-1,x,y)$

{\bf Prova: } 
Considere $p^*_t$ o caminho de m\'{\i}nimo custo de $x$ a $y$ com
tamanho $t\geq n$.  Provemos que existe um caminho de tamanho $t-1$ com
custo menor ou igual ao custo de $p^*_t$.  Como $t\geq n$, e s\'{o} ha $n$
v\'{e}rtices no grafo, $pt$ cont\^{e}m ao menos um ciclo.  Como o custo
deste ciclo \'{e} n\~{a}o negativo, basta remov\^{e}-lo para obter um
caminho como o desejado.  Disto segue trivialmente o lema (por que?). 

\noindent
{\bf QED}

\vspace{0.5cm}

 Pelo {\bf Lema 1} sabemos que, $f^*(n-1,1,y)$, calculado na \'{u}ltima
linha da Tabela II, \'{e} $f^*(1,y)$, o custo da solu\c{c}\~{a}o
\'{o}tima do problema original! De posse da Tabela I, e de uma
calculadora de m\~{a}o, \'{e} f\'{a}cil {\bf recuperar} (backtrack)
 $$ p^* = {\rm argmin}_{p\in \bar{P}(n,1,y)} {c(p)} ,$$
 o caminho que realiza este m\'{\i}nimo.  Na  Tabela I temos o caminho
\'{o}timo para um dado destino tra\c{c}ado nos val\^{o}res \'{o}timos 
do custo futuro, $f^*$, em {\bf negrito}. 

Explique exatamente como foi poss\'{\i}vel tra\c{c}ar este caminho na
tabela pronta.  Seria poss\'{\i}vel ir tra\c{c}ando o caminho ao longo
da constru\c{c}\~{a}o da tabela, i.e.  ap\'{o}s o c\'{a}lculo de cada
linha? Como este tra\c{c}ado na tabela efetivamente nos d\'{a} o caminho
\'{o}timo, assinalado na figura-1?

O programa mindist.m implementa em Matlab algoritmo de m\'{\i}nima 
dist\^{a}ncia. Estude-o e escreva o algoritmo de backtraking.

\subsection*{Algoritmo de Dijkstra} 

O programa dijk.m implementa em MATLAB o c\'{a}lculo da tabela de custos
\'{o}timos.  O algoritmo de Dijkstra \'{e} uma formula\c{c}\~{a}o\
mais eficiente, requerendo apenas da ordem de $n^2$ opera\c{c}\~{o}es
aritm\'{e}ticas e espa\c{c}o de mem\'{o}ria, contra
respectivamente $n^3$ e $n^2$ no primeiro programa. 

Tente explicar o algoritmo de Dijkstra em dijk.m e justificar sua
corretude.  Dica: Ao t\'{e}rmino da fase $t=0,1,\ldots ,n-1$ temos ao
menos $t+1$ v\'{e}rtices para os quais o custo do caminho \'{o}timo
j\'{a} foi determinado corretamente.  Estes s\~{a}o os \'{\i}ndices
n\~{a}o mais  presentes no conjunto {\it inder}.

 \section{Cadeias de Markov e Custo Esperado}

A natureza dos problemas de finan\c{c}as nos leva a considerar modelos 
estoc\'{a}sticos: sejam pre\c{c}os, taxas, volumes ou outras as 
vari\'{a}veis de decis\~{a}o, devemos levar em conta a incerteza inerente 
a elas. H\'{a} v\'{a}rias maneiras de abordar esta quest\~{a}o: uma delas \'e 
interpretar algumas vari\'{a}veis de decis\~{a}o como vari\'{a}veis 
aleat\'{o}rias, estudando sua distribui\c{c}\~{a}o de probabilidade. Outra 
forma de  inserir aleatoriedade nos modelos \'{e} considerar que h\'{a} 
uma probabilidade conhecida de  mudan\c{c}a de estado no sistema em quest\~ao. 
Essa segunda abordagem nos permite utilizar os princ\'\i pios de 
programa\c{c}\~{a}o din\^{a}mica apresentados anteriormente. 

Muitos problemas em finan\c{c}as procuram determinar uma sequ\^encia 
de decis\~{o}es \'{o}timas ao longo de um horizonte de planejamento. 
Em geral a tomada de decis\~{a}o em um 
dado instante depende do estado em que  o sistema se encontra naquele particular 
momento e da a\c{c}\~{a}o, ou controle, que ser\'{a} exercido no sistema a 
partir de ent\~{a}o. Os modelos de programa\c{c}\~{a}o din\^{a}mica s\~{a}o 
facilmente empreg\'{a}veis nestes casos e possibilitam que seja introduzida 
aleatoriedade no problema. 
 
Neste cap\'\i tulo apresentaremos uma classe de modelos denominada 
cadeias de Markov, procurando inser\'{\i}-los no contexto de 
programa\c{c}\~ao din\^amica. Como no cap\'\i tulo precedente iremos 
procurar construir uma express\~{a}o 
recursiva que forne\c{c}a a a\c{c}\~{a}o \'{o}tima a ser tomada em cada 
per\'\i odo, 
considerando que h\'{a} um custo associado a cada uma das poss\'\i veis 
a\c{c}\~{o}es. Nos exemplos que ser\~{a}o  apresentados, teremos como 
meta minimizar o custo esperado.

 \subsection*{Decis\~{a}o por custo esperado}

Consideremos um sistema com estados $x \in X=\{1,2,\ldots \}$, 
evoluindo num par\^{a}metro discreto $t\in T=\{1,2,\ldots ,h\}$, 
ou seja, $x(t) \in X$. A cada instante  t tomamos uma 
a\c{c}\~{a}o $u(t) \in U=\{0,1,2,\ldots\}$.
A a\c{c}\~{a}o tomada \'e fun\c{c}\~ao da pol\'{\i}tica $\Pi$ adotada, isto \'{e}, 
de uma regra de decis\~{a}o previamente estabelecida. 

A cada controle $u(t)$, associamos um custo determin\'{\i}stico 
$c_{x}^{u}(t)$, que \'{e} fun\c{c}\~{a}o do estado, do controle 
e do instante de tempo considerado.  Consequentemente podemos 
atribuir a cada trajet\'{o}ria $(x(1),u(1),x(2),u(2), \dots, x(h),u(h))$ 
um custo aditivo  dado por $\sum_{t=1}^h c_{x}^{u}(t)$

A aleatoriedade no sistema \'{e} introduzida atrav\'{e}s de 
probabilidades de mudan\c{c}a de estado. Assim, definimos 
$P_{x,y}^{u}(t)$ como a probabilidade de transi\c{c}\~{a}o do
estado $x(t)$ para o estado $y=x(t+1)$, e obviamente tomamos 
 $\sum_{y} P_{x,y}^{u}(t) = 1 , \forall x,u,t $. 
Com isto podemos falar no custo esperado associado a uma pol\'{\i}tica, 
pois fixar uma  pol\'\i tica determina a probabilidade de cada trajet\'oria 
poss\'\i vel. 

Assim, definimos o custo associado a uma pol\'\i tica $\Pi$, como 
sendo o valor esperado do custo aditivo, a partir de um estado inicial 
$x(t)$: $$f_{\Pi}(1,x(1))=E (\sum_{t=1}^h c_{x}^{u}(t))$$
Igualmente podemos construir, para uma fixada pol\'\i tica e um dado 
instante $t$, a fun\c{c}\~{a}o  de custo futuro, 
$$f_{\Pi}(t,x(t))=E (\sum_{s=t}^h c_{x}^{u}(s)),$$ o custo esperado 
futuro a partir do estado $x(t)$. 
Definimos tamb\'{e}m o \'{o}timo da fun\c{c}\~{a}o de custo futuro como 
$f^*(t,x)= \min_{\Pi} f_{\Pi}(t,x)$. Queremos determinar a pol\'\i tica 
\'otima  $\Pi^{*}$  que minimize a fun\c{c}\~ao de custo. 

Uma express\~{a}o recursiva para o c\'{a}lculo do \'{o}timo da 
fun\c{c}\~{a}o \'e  facilmente obtida: 

\begin{itemize}
\item 
Condi\c{c}\~{a}o de contorno: para $t$ no horizonte h, temos 
 $f^*(h,x)= \min_u c_{x}^{u}(h)$.

\item
Para $t\in \{ h-1,h-2,\ldots 1\}$ vale a rela\c{c}\~{a}o de recorr\^{e}ncia
 $$f^*(t,x)=\min_{u} \{c_{x}^{u}(t) +
 \sum_{y\in X} P_{x,y}^{u}(t)*f^{*}(t+1,y)\}$$
\end{itemize}

Como no problema de m\'\i nima dist\^{a}ncia num grafo, a partir da 
condi\c{c}\~{a}o de contorno (neste caso no horizonte h) e da equa\c{c}\~{a}o 
de Bellman (recorr\^{e}ncia sobre a  fun\c{c}\~{a}o f()), \'{e} poss\'\i vel 
resolver todos os problemas auxiliares, e a partir destes recuperar a 
solu\c{c}\~{a}o \'{o}tima do problema original.

\section{Hedging}

Ilustremos a t\'{e}cnica de controle em cadeias de markov 
em um problema concreto. O problema em quest\~ao \'{e} aned\'otico, 
mas ilustra o uso de estoques reguladores como forma de Hedging, 
ou prote\c{c}\~{a}o contra as ocila\c{c}\~{o}es de pre\c{c}o ou 
demanda no mercado. 

O Bigode, que vende cachorro-quente na porta da faculdade de economia,
tem uma  demanda constante de uma caixa de salsicha (uma unidade) por
dia.  O Bigode vende esta unidade a um pre\c{c}o constante de US\$ 120,
todo dia. Ele  compra esta unidade no mercado a um pre\c{c}o que varia
aleatoriamente, sendo que o mercado abre cada dia no estado $m$, onde a
unidade custa $v(m)$, com probabilidade $p(m)$ conforme descrito na
tabela abaixo. 

$$
\begin{array}{|c|c|c|c|}
\hline
$m$  & 1  &  2  & 3   \\
\hline
$v(m)$ & 90 & 100 & 110 \\
\hline
$p(m)$ & 0.2 & 0.7 & 0.1 \\
\hline
\end{array}
$$

Al\'{e}m da unidade sendo vendida no dia, o Bigode tem em sua casa um
estoque de $l(t)$ unidades. O semestre letivo tem $h+1$ dias consecutivos e, 
exceto no \'{u}ltimo dia
$t=h+1$, o Bigode tem que decidir quantas unidades comprar, $u(t)$. 
Deste modo, $l(t+1)=l(t)+u(t)-1$.
A decis\~{a}o sobre quanto comprar \'{e} tomada estando ciente do estado
$x(t)=(l(t),m(t))$. 

Como o Bigode n\~{a}o pode faltar com seus clientes, nem guardar este
produto perec\'{\i}vel por mais de 4 dias, impomos as
restri\c{c}\~{o}es $0\leq l(t)\leq 3 , \forall 1\leq t\leq h$.  Para
maximizar seus lucros, o Bigode quer uma pol\'{\i}tica (vi\'{a}vel)
$u(t)={\Pi}^{*} (t,l(t),m(t))$ que minimize a esperan\c{c}a de seu custo
operacional durante o semestre.  

Com a pol\'{\i}tica \'{o}tima, ${\Pi}^*$, o custo operacional esperado,
do dia $t$ at\'{e} o final do semestre, \'{e} (equa\c{c}\~{a}o de Bellman):
 $$f^*(t,l(t),m(t))=  {\min}_{u | 0\leq l+u-1 \leq 3}
 ( u*v(m) + {\sum}_{n=1}^{3} p(n)*f^{*}(t+1,l+u-1,n) )$$

O resultado da pol\'{\i}tica \'{o}tima no \'{u}ltimo dia de compra do
semestre, $t=h$, \'{e} (condi\c{c}\~{a}o de contorno):
 $$f^*(h,l,m)=\ \   0,\ \  {\rm se}\ \  1\leq l\leq 3;
             \ \ \ v(m),\ \  {\rm se} \ \ l=0.
 $$

\section{Custo Descontado}

Em muitas situa\c{c}\~{o}es reais existem oportunidades de investimento 
em cada per\'{\i}odo com uma taxa de juro $r(t)$. Na presen\c{c}a desta 
oportunidade, para pagar o custo $ c_x^u(t)$ do controle $u(t)$ no instante 
$t$ bastaria tomar no instante $t-1$ o valor presente 
deste custo, ou seja, $\delta (t-1) c_x^u(t)$, onde 
$\delta (t-1) = \frac{1}{1+ r(t-1)}$

Nestas condi\c{c}\~{o}es gostar\'{\i}amos de 
substituir, como crit\'{e}rio de avalia\c{c}\~{a}o de uma pol\'{\i}tica, 
o custo esperado pelo custo descontado esperado: 
        
$$f_{\Pi}(1,x(1)) = 
  E (\sum_{t=1}^h \left[ \prod_{s=1}^{(t-1)} \delta (s) \right] 
  c_{x}^{u}(t))$$

A equa\c{c}\~{a}o de Bellman (rela\c{c}\~{a}o de recorr\^{e}ncia) \'{e}
facilmente adaptada para  incorporar o fator de desconto: 

$$f^*(t,x)=\min_{u} \{c_{x}^{u}(t) +
 \delta (t) \sum_{y\in X} P_{x,y}^{u}(t)*f^{*}(t+1,y)\}$$
No caso de uma taxa de desconto constante estas equa\c{c}\~{o}es 
se reduzem respectivamente a: 

$$f_{\Pi}(1,x(1))=E (\sum_{t=1}^h \delta^{(t-1)} c_{x}^{u}(t))$$
 
$$f^*(t,x)=\min_{u} \{c_{x}^{u}(t) +
 \delta \sum_{y\in X} P_{x,y}^{u}(t)*f^{*}(t+1,y)\}$$

Como o custo presente dos controles no horizonte de planejamento
diminuem exponencialmente com $h$, a pol\'{\i}tica \'{o}tima inicial,
$\Pi (1,x)$ $u(1)$, tende a convergir para um controle $u(1)$ a medida
que o horizonte se distancia (i.e. que $h$ aumenta). Pela mesma
raz\~{a}o, para horizontes long\'\i nquos, os controles \'{o}timos iniciais
tendem a ser pouco sens\'{\i}veis \`{a}s condi\c{c}\~{o}es de contorno.
Ambas as  caracter\'{\i}sticas tornam  geralmente mais robustas as
modelagens pelo  crit\'{e}rio de custo descontado.     

\section{Precifica\c{c}\~{a}o de Contratos Derivativos} 

Em um contrato de op\c{c}\~{a}o de compra (venda) o contratante adquire
do contratado o direito de comprar (vender) uma mercadoria do (ao)
contratado  por um pre\c{c}o de exerc\'{\i}cio pr\'{e}-estabelecido,
$K$. Note que o contratante adquire o direito de fazer uma compra
(venda), sem ter a obriga\c{c}\~{a}o de faze-la (ao contr\'ario de um
simples contrato  futuro de compra ou venda), da\'{\i} o nome contrato de
op\c{c}\~{a}o. 
   
Existem duas modalidades b\'{a}sicas de contrato de op\c{c}\~{a}o: em um
contrato de tipo Americano o contratante tem o direito de exercer sua 
op\c{c}\~{a}o a qualquer instante $t$ desde a assinatura do contrato
at\'{e} at\'{e} sua data de vencimento, $0\leq t\leq h$. No contrato de
tipo Europeu a op\c{c}\~{a}o pode somente ser exercida na data de
vencimento do contrato. Um contrato  op\c{c}\~{a}o de compra (venda)
\'{e} tambem denominado um call (put). O pre\c{c}o pelo qual o
contratante  adquire o call (put) \'{e} o pre\c{c}o do contrato, 
$C$ ($P$).  

 A mercadoria especificada no contrato \'{e} chamada ativo fundamental
do contrato.  O contratado recebe uma remunera\c{c}\~{a}o para absorver
os riscos decorrente das varia\c{c}\~{o}es de pre\c{c}o  do ativo
fundamental, e \'{e} por isto denominado especulador.  O contratante
paga para se proteger contra os mesmos riscos, e \'{e} por  isto
denominado hedger. Contratos de op\c{c}\~{a}o s\~{a}o denominados 
ativos derivativos, pois os ganhos e perdas de ambas as partes  resultam
(derivam) da evolu\c{c}\~{a}o do pre\c{c}o do ativo fundamental.    

 O call ou put deve especificar a mercadoria em termos de  quantidade,
local de entrega (pra\c{c}a) e qualidade.  Ambas as partes  estabelecem
um intermediador id\^{o}neo, por exemplo uma bolsa de mercadorias, para
dirimir quaisquer duvidas.   No exerc\'{\i }cio de uma op\c{c}\~{a}o
teriamos ent\~{a}o a entrega f\'{\i}sica da mercadoria na pra\c{c}a
especificada com verifica\c{c}\~{a}o, por parte da bolsa, da quantidade
e qualidade.  

A mercadoria especificada no contrato \'{e} muitas vezes uma mercadoria
amplamente comercializada, n\~{a}o importando o fornecedor (uma
commoditie), e cujo pre\c{c}o de mercado, $S(t)$, \'{e} tambem  divulgado
pela bolsa. Nestas condi\c{c}\~{o}es, \'{e} geralmente mais c\^omodo para
ambas as partes  substituir a entrega f\'{\i}sica por um acerto de
contas em fun\c{c}\~{a}o da diferen\c{c}a entre o pre\c{c}o de
exerc\'{\i}cio e o pre\c{c}o de mercado da comoditie na data do
exerc\'{\i}cio.    Por exemplo, num call, o vendedor da
op\c{c}\~{a}o (contratado ou especulador) simplesmente pagaria ao
comprador da op\c{c}\~{a}o (contratante ou hedger) o valor $max\{0,
(S(t) -K)\}$.  No caso de uma put este valor seria $max\{0, (K -S(t))\}$.

 Para colocar os contratos de op\c{c}\~{o}es no contexto de
programa\c{c}\~{a}o din\^{a}mica em cadeias de programa\c{c}\~ao  
din\^amica \'{e} preciso 
modelar a evolu\c{c}\~{a}o do pre\c{c}o $S(t)$ por um processo
estoc\'{a}stico discreto. No modelo de passeio aleat\'{o}rio trinomial
temos que  o pre\c{c}o poderia assumir apenas tr\^es valores:
 $Pr(S(t+1)=fu(t)*S(t))=pu(t)$, $Pr(S(t+1)=fd(t)*S(t))=pd(t)$,
 $Pr(S(t+1)=S(t))=1-pu(t)-pd(t)$, onde $fu(t)$, $fd(t)$, $pu(t)$ e 
 $pd(t)$ s\~{a}o, respectivamente, os fatores de subida e descida do
pre\c{c}o fundamental entre os instantes $t$ e $t+1$, e suas
probabilidades. No modelo binomial temos $pu(t)+pd(t)=1$. No modelo
geom\'{e}trico h\'{a} uma  simetria entre os fatores de subida e
descida, $fd(t)=1/fu(t)$.  
 Esta hip\'{o}tese de simetria, que \'{e} motivada por certas
considera\c{c}\~{o}es econ\^{o}micas e emp\'{\i}ricas,  em muito
facilita a representa\c{c}\~{a}o do espa\c{c}o de estados. No modelo
estacion\'{a}rio os fatores e suas probabilidades s\~{a}o  constantes no
tempo.  Assim no modelo binomial geom\'{e}trico  temos apenas os
par\^{a}metros $fu(t)$ e $pu(t)$. Mesmo num modelo t\~{a}o simples
podemos  retratar um processo onde a taxa\\
 $r(t)=ln(S(t+1)/S(t))$ tem valor esperado e  desvio padr\~{a}o 
 (tend\^{e}ncia e volatilidade)  
 $$\mu(t)=pu(t)*fu(t)+(1-pu(t))/fu(t)\ \ \mbox{e} \ \ 
   \sigma(t)=\sqrt{pu(t)*(1-pu(t))*(fu(t)-1/fu(t))^2}$$     

 Estes conceitos podem ser colocados dentro da linguagem de 
programa\c{c}\~ao din\^amica, como pode ser visto nos exerc\'{\i}cios 
abaixo:

\begin{itemize}
 \item[1.] 
 Escreva um programa para ajudar o bigode a gerenciar sua empresa. 

 \item[2.]
 Call Americano de salchicha: Modifique o programa do Bigode para  um
mercado de salchicha onde o pre\c{c}o segue um passeio aleat\'{o}rio
trinomial geom\'{e}trico sim\'{e}trico e estacion\'{a}rio, 
definido pela tend\^{e}ncia e volatilidade di\'{a}rias $ms$ e $ss$. 
 Admita que um especulador
ofere\c{c}a ao Bigode no primeiro dia do semestre a oportunidade de
comprar por  $C$ um (1) call americano para $Q$ caixas, com pre\c{c}o de
exerc\'{\i}cio $K$, expirando no final do semestre. 
 Considere que as  caixas a serem entregues no eventual exerc\'{\i}cio
da op\c{c}\~{a}o s\~{a}o  do tipo longa-vida, que podem ser armazenadas
at\'{e} o final do  semestre, independentemente do estoque regular. O
espa\c{c}o  de estados do modelo deve incluir, cotidianamente, o
pre\c{c}o do mercado,  o estoque regular, o estoque longa-vida, e a
posse ou n\~{a}o do  call n\~{a}o exercido. 
 O espa\c{c}o de controle deve incuir, cotidianamente, a possibilidade 
do Bigode se abastecer no mercado at\'{e} o limite do estoque regular, e
a possibilidade de exercitar o call caso o Bigode o possua. No primeiro
dia o Bigode deve decidir tamb\'em pela compra ou n\~{a}o  do call.
Queremos como sa\'{\i}da deste programa de  programa\c{c}\~ao din\^amica
a decis\~{a}o \'{o}tima no primeiro dia pelo crit\'{e}rio do custo
descontado com  taxa de desconto di\'{a}ria $\delta$, e tamb\'em a
rendabilidade esperada do neg\'{o}cio do Bigode. 

 \item[3.]
 Escreva um programa de precifica\c{c}\~{a}o, que utilizando o m\'{e}todo
da bisec\c{c}\~{a}o e invocando o programa de programa\c{c}\~ao 
din\^amica, determine  valor $C$  a partir do qual a call deixa de ser 
interessante para o Bigode. 

 \item[4.]
 Adapte o programa de  programa\c{c}\~ao 
din\^amica, para uma taxa de desconto $\delta (t)$, que tem
valor  inicial $\delta (1)$ e segue um passeio aleat\'{o}rio trinomial
geom\'{e}trico estacion\'{a}rio, definido pela tend\^{e}ncia e 
volatilidade di\'{a}rias $md$ e $sd$. 

\item[]
O programa Bigopt.m, listado em ap\^{e}ndice, resolve parcialmente o 
exerc\'{\i}cio 1. 

\end{itemize}

 \section{Pol\'{\i}ticas de Scarf  (s,S)} 

 Diferentes caracter\'{\i}sticas ou objetivos parciais de um sistema 
produtivo t\'{\i}pico sugerem tend\^{e}ncias divergentes para a
pol\'{\i}tica de estoques. Por exemplo:   
 grandes tempos de set-up (tempos de que decorrem  
da necessidade de ajustar ou limpar o equipamento antes iniciar ou 
depois de terminar a prudu\c{c}\~{a}o de um lote de um dado produto),   
ou perda de vendas e mercado por demora no atendimento ao cliente  
favorecem a forma\c{c}\~{a}o de grandes estoques; 
 enquanto altos custos financeiros e de manuten\c{c}\~{a}o  
favorecem o trabalho com pequenos estoques. 

 Para encontrar solu\c{c}\~{o}es \'{o}timas nesta situa\c{c}\~{a}o
paradoxal, que geralmente s\~{a}o solu\c{c}\~{o}es intermedi\'{a}rias
entre as pol\'{\i}ticas extremas, \'{e} necess\'{a}rio equacionar e
otimizar um mo\-delo quantitativo. Discutiremos agora uma classe de
modelos para otimizar pol\'{\i}ticas de planejamento de 
produ\c{c}\~{a}o e estoques em uma empresa industrial ou comercial.
Consi\-deraremos a demanda como uma vari\'{a}vel aleat\'{o}ria, a ser
especificada por uma distribui\c{c}\~{a}o de probabilidade a partir das
previs\~{o}es de demanda.  

A decis\~{a}o (ou controle), em cada per\'{\i}odo $n$, \'{e} a quantidade 
de produto a ser produzida, $y(n)$. Sob certas condi\c{c}\~{o}es \'{e} 
possivel demonstrar que o controle \'{o}timo \'{e} caracterizado pelo 
par ``ponto de recompra -- patamar de recomposi\c{c}\~{a}o'': caso o
estoque seja menor ou igual ao ponto de recompra, $s_n$, devemos
recompor o estoque, comprando ou produzindo  $y(n)$,  at\'{e} o patamar
$S_n$.  Estas s\~{a}o as pol\'{\i}ticas de Scarf. 

A estrutura b\'{a}sica dos modelos $(s,S)$ \'{e} a seguinte:  

\begin{enumerate}

\item Um horizonte finito de planejamento, i.e.  $N$ per\'{\i}odos
  (meses, semanas)  de produ\c{c}\~{a}o, $n\in \{1, 2, \ldots N\}$. 
  Em cada per\'{\i}odo ``herdamos'' um estoque $x(n)$ do per\'{\i}odo 
  $n-1$, produzimos para aumentar o estoque at\'{e} o n\'{\i}vel $y(n)$, 
  e finalmente, ap\'{o}s as vendas do per\'{\i}odo, o estoque baixa 
  para $z(n)$. 
                                 
\item $d(n)$, a demanda no per\'{\i}odo $n$. 

\item  $b(n)$ a fra\c{c}\~{a}o dos clientes recusa espera no 
atendimento. 

\item A equa\c{c}\~{a}o b\'{a}sica de evolu\c{c}\~{a}o do sistema \'{e} a
  equa\c{c}\~{a}o de evolu\c{c}\~{a}o dos estoques, $z(n,y,d)$. 
  Diferentes equa\c{c}\~{o}es de evolu\c{c}\~{a}o de estoques modelam, 
  v\'{a}rios aspectos pertinentes; por exemplo: 

  $$ z(n,y,d) = (y-d) +b(n)*(d-y)^{+}$$ \ \ 

  Para qualquer n\'umero real x definimos $x^{+}$ como sendo a parte 
  positiva de x, isto \'e, $x^{+}= max\{x,0\}$.  
  Estoques negativos representam pedidos n\~{a}o atendidos em espera.

\item $k(n)$, o custo constante de set-up para produ\c{c}\~{a}o. 

\item $w(n)$, o custo unit\'{a}rio de compra ou produ\c{c}\~{a}o.  
  
\item $r(n)$, o pre\c{c}o unit\'{a}rio de venda. 
      O estoque remanecente do \'{u}ltimo per\'{\i}odo 
      \'{e} ``liquidado'' por um pre\c{c}o $r(N+1)$. 

\item $h(z,n)$ o custo de manter o estoque $z(n)$.  Para valores 
  negativos de $z$, este custo representa uma penalidade pela 
  espera do cliente. 

 Assim o custo instant\^{a}neo no per\'{\i}odo $n$ ser\'{a} dado por

 $$c(n) = w(n)*(y-x) +k(n)*(y-x)^> + h(z(n,y,d)) -r(n)*(y-z)$$ \ \    

 Para qualquer n\'umero real x definimos $x^{>}$ como 1 se $x > 0$ 
 e 0 caso contr\'ario.  

\item $a(n)$, o fator de desconto intertemporal computado a partir 
  das taxas de infla\c{c}\~{a}o e juros no per\'{\i}odo, nos define 
  o custo presente de operar o sistema ate o horizonte de 
  planejamento (custo aditivo descontado): 
  $$ f = \sum_{n=1}^{N} \left( c(n) * \prod_{l=1}^n a(l) \right) \ .$$ 

\end{enumerate} 

 A partir desta descri\c{c}\~{a}o do sistema podemos computar uma
pol\'{\i}tica \'{o}tima de estoques na vari\'{a}vel de decis\~{a}o,
$y(n)$, atrav\'{e}s de programa\c{c}\~{a}o din\^{a}mica no contexto  de
cadeias de Markov.

Ilustramos agora a sem\^{a}ntica do modelo de Scarf num contexto
aned\'{o}tico. Consideremos um lojista que vende esquis em Bariloche. 
Ele pode fazer pedidos \`a f\'abrica (em Buenos Aires) semanalmente  
pagando  um custo fixo de k = US\$100,00 ao motorista, mais um 
custo unit\'ario de US\$100,00 por esqui pedido. 
Cada esqui \'e revendido em Bariloche por r = US\$150,00
, mas b = 30\% dos clientes se recusa a esperar caso n\~ao haja 
mercadoria em estoque. A taxa de juros semanal na temporada \'e de 
2\%. A temporada tur\'{\i}stica de inverno \'e de N=10 semanas, 
finda a qual \'e feita uma liquida\c{c}\~ao de estoque ao pre\c{c}o de 
r(N+1)= US\$ 80,00. Assumiremos que n\~ao h\'a custo de manuten\c{c}\~ao 
de estoques e analisaremos duas demandas: a primeira determin\'{\i}stica 
de $\bar{d}$ = 10 unidades por per\'{\i}odo (semana) e a segunda 
estoc\'astica com a seguinte distribui\c{c}\~ao:

$$
  \begin{array}{c|ccccccc} 
 $d$ & 7 & 8 & 9 & 10 & 11 & 12 & 13 \\
\hline 
 $probab$ &  0.05 & 0.10 & 0.20 & 0.30 & 0.20 & 0.10 & 0.05 
 \end{array}  
 , \ \ \bar d = 10 \ .
$$ 

As primeiras duas linhas da tabela seguinte apresentam a ordem de 
compra \'otima  para a primeira semana, sem estoque inicial. VP 
\'e o valor presente esperado do resultado operacional do neg\'ocio. 
$\Delta$ \'e o decr\'escimo esperado deste resultado, caso no primeiro 
per\'{\i}odo seja feita a ordem sub-\'otima de $\bar{d}$  = 10 unidades. 
Os valores n.$\Delta$ nos d\~ao uma estimativa da perda de resultado 
para ordens iguais \`a demanda esperada, $\bar{d}$  = 10, em processo 
cont\'{\i}nuo.  As colunas referentes a s1  e  S1
s\~ao respectivamente o ponto de recompra e o patamar de 
recomposi\c{c}\~ao  no primeiro per\'{\i}odo. 

Nas linhas seguintes da tabela estudamos como estes resultados 
s\~ao afetados pela altera\c{c}\~ao de alguns dos par\^ametros  do 
modelo. Nas \'ultimas duas linhas estudamos a introdu\c{c}\~ao de 
um custo de falta de 10\% do valor de venda por unidade.

Dentro da nota\c{c}\~ao apresentada temos a seguinte situa\c{c}\~ao  
inicial: 

\begin{itemize}

\item[] $N$ = 10

\item[] $b(n)$ = 0.3

\item[] $k(n)$ = US\$ 100.00

\item[] $r(n)$ = US\$ 150.00

\item[] $a(n)$ = 0.98 

\item[] $r(N+1)$ = US\$ 80.00

\item[] $w(n)$ = US\$ 100.00 

\item[] $hn/r$ = 0

\end{itemize}

\begin{table}[ht]
\begin{center}
\begin{tabular}{||l|||c||c|c||c||c||c||}
\hline
\multicolumn{7}{||c||}
{Solu\c{c}\~{a}o \a'{O}tima do Modelo B\a'{a}sico Alterado} \\ 
\hline 
{Altera\c{c}\~{a}o} & 
Demanda 
   &
\multicolumn{2}{|c||}{Pol\a'{\i}tica} & 
Ordem  & 
VP & 
n $\Delta$  \\
\hline 
          &  & $s_1$ & $S_1$ & \a'Otima &  &  \\ 
\hline
\hline 
Dados     & Estoc\a'{a}st.   & 05 & 31&  31 & 3980 & 447   \\ 
\hline  
Originais & Determ.   & 05 & 30 & 30 & 4070 & 389  \\ 
\hline   
\hline  
 $r=200$  & Estoc\a'{a}st.   & 07 & 32 & 32  & 8532 & 494  \\ 
\hline   
          & Determ.   & 07 & 30 & 30  & 8643 & 389  \\ 
\hline  
\hline    
 $a=0.99$ & Estoc\a'{a}st.   & 06 & 52 & 52  & 4299 & 610  \\  
\hline    
          & Determ.   & 07 & 50 & 50  & 4393 & 590  \\  
\hline  
\hline    
 $b=0.7$  & Estoc\a'{a}st.   & 08 & 32 & 32  & 3960 & 522  \\  
\hline    
          & Determ.   & 08 & 30 & 30  & 4070 & 389  \\  
\hline  
\hline
 $k=50$   & Estoc\a'{a}st.   & 07 & 21 & 21  & 4172 & 225  \\  
\hline    
          & Determ.   & 07 & 20 & 20  & 4250 & 195  \\  
\hline  
\hline
 hn/r = 0.1   & Estoc\a'{a}st.   & 07 & 32 & 32  & 3964 & 476  \\  
\hline    
          & Determ.   & 07 & 30 & 30  & 4070 & 389  \\  
\hline  
    
\end{tabular}
\end{center}
\end{table}

 Da compara\c{c}\~{a}o dos resultados no cen\'{a}rios determin\'{\i}stico e 
estocastico, vemos que \'{e} mais facil obter lucros no mundo
determin\'{\i}stico (maior valor presente), estando todavia no mundo com 
incertezas o maior valor de planejamento (maior $\Delta$). 
 Ademais, a an\'{a}lise de sensibilidade aos diversos fatores nos mostra 
que as seguintes altera\c{c}\~{o}es podem induzir a um aumento no 
n\'{\i}vel \'{o}timo de estoques: Aumento do lucro unit\'{a}rio, 
diminui\c{c}\~{a}o da taxa de juros, aumento de liquidez (um mercado com 
clientes cativos \'{e} denominado viscoso, enquanto um mercado com grande 
disponibilidade de fornecedores \'{e} denominado l\'{\i}quido), aumento de 
custo de set-up, e finalmente, aumento do custo de mal atendimento.

%% file: cap6.tex
\chapter{Contole e Estima\c{c}\~ao LQG}

Neste ponto do trabalho, o leitor j\'{a} deve ter indagado quanto 
\`{a} aplicabilidade, sob o ponto de vista computacional, das estruturas 
recursivas que construimos. \'{E} claro que a viabilidade de 
utiliza\c{c}\~ao do m\'{e}todo depende fortemente da estrutura  da 
fun\c{c}\~ao custo e tamb\'{e}m da din\^amica relacionando estados e 
controles. 

 Estudaremos agora sistemas onde o  controle \'{o}timo em cada instante
\'e  trivialmente construido a partir de uma transforma\c{c}\~ao linear
do estado no qual  o sistema se encontra. 
 Trabalharemos com sistemas nos quais os estados e os controles 
s\~{a}o determin\'{\i}sticos e relacionam-se com seu passado segundo  
uma express\~{a}o linear do tipo $ x(t+1) = A(t)*x(t) + B(t)*u(t) $, com 
matrizes $A(t)$ e $B(t)$ conhecidas, e onde os custos s\~ao fun\c{c}\~oes 
quadr\'aticas dos estados. 

As propriedades decorrentes desta estrutura s\~ao muito fortes e acarretam 
um ganho substancial do ponto de vista de efici\^encia computacional.  
Deve-se ainda salientar que um grande n\'umero de problemas reais 
admitem uma modelagem desta natureza, e assim estes modelos conjugam 
efici\^encia e aplicabilidade.

\section{Evolu\c{c}\~{a}o Linear com Custo Quadr\'{a}tico}  

 Num sistema {\bf Linear com custo Quadr\'{a}tico } (LQ) os estados 
evoluem na forma 
 $$ x(t+1) = A(t)*x(t) + B(t)*u(t) $$ sendo o custo instant\^{a}neo 
 uma fun\c{c}\~ao quadr\'atica $c(t,x,u)$. 
 
 Mais precisamente, o custo quadr\'{a}tico instant\^{a}neo \'e  
definido em fun\c{c}\~{a}o de uma matriz $Q(t)$, sim\'{e}trica e positiva 
definida, ou seja: 

 \[ c(t,x(t),u(t)) =
 \left[ \begin{array}{cc} x(t)' & u'(t) \end{array} \right]
 \  
 \left[ \begin{array}{cc} QX(t) & QC(t) \\
                          QC(t)' & QU(t) \end{array} \right]
 \
 \left[ \begin{array}{c} x(t) \\ u(t) \end{array} \right]
 \]

A fun\c{c}\~{a}o de custo adotada \'{e} simplesmente o custo aditivo e 
para cada instante $t$, definimos as fun\c{c}\~{o}es auxiliares 
de custo futuro $$f_{\Pi}(t,x) = \sum_{s=t}^h c(s,x,u) $$ que 
fornecem o custo a partir do instante $t$, quando adotamos a 
pol\'\i tica $\Pi$. 

Para este problema convencionamos que $u(h) = 0$, ou seja, o controle 
exercido no horizonte de planejamento \'{e} nulo.

O seguinte teorema apresenta um resultado muito importante e surpreendente 
que permite 
que para esta particular classe de sistemas, o controle \'otimo em cada 
instante seja obtido trivialmente como uma transforma\c{c}\~ao linear 
do estado em que o sistema se encontra. 

\vspace{1 cm}

\noindent
{\bf Teorema} (Princ\'{\i}pio de Otimalidade)

Em um sistema linear com custo quadr\'atico vale: 
 \begin{itemize}
 \item[{\bf a.}] A pol\'{\i}tica \'{o}tima \'{e} uma transforma\c{c}\~{a}o linear do
      estado presente, i.e.  existe uma matriz $K(t)$ denominada 
      matriz de controle tal que 
             $$u^{*}(t) = K(t) x(t)$$
 \item[{\bf b.}] O custo futuro \'{o}timo \'{e} uma fun\c{c}\~{a}o quadr\'{a}tica do
      estado presente, i.e. existe, para cada instante $t$, uma matriz $FO(t)$, 
      tal que 
             $f^{*}(t) = x(t)'FO(t)x(t)$.
 \end{itemize}

\noindent
 {\bf Prova:}

\noindent
Base de indu\c{c}\~{a}o (condi\c{c}\~{a}o de contorno)

Como por hip\'{o}tese $u(t)=0$, segue trivialmente da defini\c{c}\~{a}o 
do custo que 
$$ f^{*}(x(h)) = x(h)'QX(h)x(h).$$

\noindent
Passo de Indu\c{c}\~{a}o (equa\c{c}\~{a}o  de Bellman):

 \begin{eqnarray*}
 f^{*}(x(t)) & = & \min_{u(t)} c(x(t),u(t)) + x(t+1)'FO(t+1)x(t+1) \ \ = \\
 &  &  \mbox{} \hspace{-25mm} 
  \min_{u(t)} \left[ \begin{array}{cc} x(t)' & u'(t) \end{array}\right]
 \left[ \begin{array}{cc} QX(t) + A(t)'FO(t+1)A(t) & QC(t) + A(t)'FO(t+1)B(t) \\
    QC(t)'+ B(t)'FO(t+1)A(t) & QU(t) + B(t)'FO(t+1)B(t) \end{array}\right]
    \left[ \begin{array}{c} x(t) \\ u(t) \end{array}\right]  \\
  & = & \min_{u(t)}                                                       
    \left[ \begin{array}{cc} x(t)' & u'(t) \end{array}\right]
    \left[ \begin{array}{cc} FX(t) & FC(t) \\ 
                            FC(t)' & FU(t) \end{array}\right]
    \left[ \begin{array}{c} x(t) \\ u(t) \end{array}\right]     \\
  & = & \min_{u(t)} f(x(t),u(t))
 \end{eqnarray*}                                  

A determina\c{c}\~{a}o do controle \'{o}timo \'{e} obtida  atrav\'es de 
$$ {\nabla}_{u}(f(x,u)) = 2FU(t)u(t) + 2FC(t)'x(t) = 0$$

Mas 
 $$ {\nabla}_{u}(f(x,u^*))=0 \Rightarrow  u^*(t)= K(t)x(t) $$
 onde $K(t)=  -FU(t)^{-1}FC(t)'x(t) $
(Note que  $FU(t)$ \'{e} invers\'{\i}vel pois $FU(t)>0$, o que segue da
hip\'otese de positividade de $Q(t)$). 

\noindent
Assim temos: 

 \begin{eqnarray*}
 K(t)x(t) & \equiv & -FU(t)^{-1}FC(t)'x(t) \\
 & = &   -\left[ ( QU(t)+B(t)'FO(t+1)B(t) )^{-1}
            ( QC(t)'+B(t)'FO(t+1)A(t) ) \right]
 \end{eqnarray*}
 Tomando $u(t) = K(t)x(t)$, temos
 $$ f^{*}(x(t)) = x(t)'FO(t)x(t)
                = x(t)'( FX(t) - FC(t)FU(t)^{-1}FC(t)' )x(t)
 $$ 
 
 {\bf Q.E.D.}\\

\noindent
Observe que tomando o controle \'{o}timo  $u(t) = u^{*}(t)$, conforme 
apresentado no teorema, a evolu\c{c}\~{a}o do sistema ser\'{a} dada por
 $$ x(t+1)= A(t)x(t) + B(t)K(t)x(t) = 
    (A(t)+B(t)K(t))x(t) \equiv G(t)x(t) 
 $$
 A matriz $G(t)=A(t)+B(t)K(t)$ \'{e} denominada {\it matriz de ganho 
do sistema}.

A prova do teorema n\~{a}o s\'{o} garante a forma linear do controle
\'{o}timo e a forma quadr\'{a}tica do \'{o}timo da fun\c{c}\~{a}o de custo
futuro, mas fornece explicitamente um sistema de rela\c{c}\~{o}es de
recorr\^{e}ncia para o c\^{o}mputo das matrizes envolvidas. 

Mais precisamente, tanto as matrizes que fornecem os controles e custo 
futuro \'{o}timos ($FO(t)$ e $K(t)$) quanto as matrizes que 
s\~{a}o utilizadas para obt\^{e}-las ( $FX(t)$, $FC(t)$, $FU(t)$, $FO(t)$ )
podem ser escritas em fun\c{c}\~{a}o da matriz $Q(t)$ que define 
o custo quadr\'{a}tico e da matriz $FO(t+1)$. Ou seja, 
$$ [ FX(t), FC(t), FU(t), FO(t), K(t), G(t) ] =
   R [ QX(t), QC(t), QU(t), FO(t+1) ].
 $$ 
 Esta rela\c{c}\~{a}o de recorr\^{e}ncia regredindo no tempo \'{e}
conhecida como a {\it Equa\c{c}\~{a}o de Riccati}.  Note que na
equa\c{c}\~{a}o de Riccati a matriz $FO$ \'{e} o elo da recorr\^{e}ncia, 
sendo a \'{u}nica entidade aparecendo com \'{\i}ndice $t+1$ o que 
se reflete em um ganho computacional substancial. 

\section{Sistemas Homog\^{e}neos no Tempo}

Analisemos o que ocorre quando trabalhamos com sistemas invariantes no 
tempo, ou seja, quando as matrizes de coeficientes s\~ao constantes. 
Chamamos de {\it Sistemas Homog\^{e}neos no Tempo} \`aqueles que 
evoluem segundo uma rela\c{c}\~ao do tipo: $$x(t+1) = A x(t) + B u(t).$$ 

Uma propriedade que seria desej\'avel nestes sistemas \'e denominada 
controlabilidade. Um sistema \'e control\'avel se  e s\'o 
se podemos lev\'a-lo a um estado previamente fixado a partir 
de qualquer estado inicial $x(1)$. Formalmente, 

\begin{itemize}

\item[]{\bf Controlabilidade}

Um sistema linear $x(t+1)=Ax(t)+Bu(t)$ \'{e} {\it k-control\'{a}vel} sse,
para qualquer estado inicial $x(1)$, existe uma sequ\^{e}ncia de $k$
controles, $u(1), u(2), \ldots, u(k)$, que levam $x(1)$ a um estado
arbitrariamente prefixado, $x(k+1)$.  

\'E f\'acil verificar que o sistema \'{e} k-control\'{a}vel 
se e s\'{o} se a equa\c{c}\~{a}o linear
$x(k+1)- A^{k}x(1) = \sum_{i=0}^{k-1}{A^{i}Bu(k-i)}$
for sempre sol\'{u}vel, ou equivalentemente, se a matriz
$M(k) = \left[ B | AB | A^2B | \dots | A^{k-1}B \right]$
 tiver posto pleno.

\end{itemize}

Vimos, ao estudar o Princ\'{\i}pio de Otimalidade, que o custo 
futuro \'otimo se escreve como uma fun\c{c}\~ao quadr\'atica 
do estado presente e que sua determina\c{c}\~ao est\'a diretamente 
associada \`a estrutura recursiva obtida para o c\'alculo das 
matrizes $FX(t), FC(t), FU(t), K(t)$ e $G(t)$. Uma importante quest\~ao 
torna-se ent\~ao a exist\^encia de um ponto fixo na Equa\c{c}\~ao de 
Ricatti, i.e. na express\~ao recursiva. Queremos assegurar, na verdade,  
que para sistemas com horizonte muito grande a fun\c{c}\~ao custo 
futuro convirja.

 Observemos inicialmente que,
fixado um horizonte $h$ e um estado $x$, $f^*(t,x)$ decresce
monotonicamente em $t$. De fato, sendo $x^*(t)$ e $u^*(t)$ a trajet\'{o}ria 
\'{o}tima correspondendo ao custo $f^*(t,x)$, temos que
 \begin{eqnarray*} 
  f^*(t,x) &=& \sum_{k=t}^h c(k,x^*(t),u^*(k)) 
            \geq \sum_{k=t}^{h-1} c(k,x^*(t),u^*(k)) \\  
           &=&  \sum_{k=t+1}^h c(k,x^*(t),u^*(k))
            \geq f^*(t+1,x)  
 \end{eqnarray*} 
 Reciprocamente, fixando $x$ e $t$, $f^*_h(t,x)$ cresce monotonicamente
ao aumentarmos o horizonte, $h$.  Se todavia $(A,B)$ for
control\'{a}vel, $f^*_h(t,x)$ \'{e} limitada
superiormente, e portanto converge para um limite $f^*_{\infty}(t,x)$. 
Logo, em virtude do princ\'{\i}pio de otimalidade, tamb\'{e}m $FO$ 
converge. 
Como todas as outras matrizes na equa\c{c}\~{a}o de Riccati s\~{a}o\
escritas em fun\c{c}\~{a}o de $FO$, temos limites de horizonte infinito
para $FX$, $FU$, $FC$, $K$ e $G$.  No limite de horizonte infinito $FO$
ser\'{a} um ponto fixo da equa\c{c}\~{a}o de Riccati.  Esta \'{e} a
{\it Equa\c{c}\~{a}o de Equil\'{\i}brio de Riccati}:
 $$ FO = R [ QX, QC, QU, FO ].$$ 

O conceito  {\it Estabilidade}  indica se um sistema homog\^eneo 
converge para a origem.

\begin{itemize}

\item[]{\bf Estabilidade}
Um sistema linear $x(t+1)=Ax(t)+Bu(t)$  \'{e} dito
{\it est\'{a}vel} se e somente se existir uma matriz de controle, $K$,
tal que $x(t+1) = (A+BK)^{t} x(0) = G^{t} x(0) \rightarrow 0$

\end{itemize}

Mas como os autovalores de $G^{t}$ s\~ao as t-\'esimas pot\^encias 
dos autovalores de $G$, a \'ultima condi\c{c}\~ao  equivale a termos 
todos os autovalores de G dentro do c\'{\i}rculo unit\'ario. Nestas 
condi\c{c}\~oes, e por abuso de notra\c{c}\~ao a matriz de ganho $G$   
\'e dita est\'avel.

 No caso particular onde a matriz B \'e nula, \'e imediato que a 
seq\"u\^encia $\{x(t)\}$ converge para a origem, a partir de um 
estado inicial arbitr\'{a}rio, $x(1)$, se e somente se 
$\lim_{t\rightarrow \infty}A^t =0$. Mas como os
autovalores de $A^t$ s\~{a}o as $t$-\'{e}simas pot\^{e}ncias dos
autovalores de $A$, a \'{u}ltima condi\c{c}\~{a}o equivale a termos todos
os autovalores de $A$ dentro do c\'{\i}rculo unit\'{a}rio.  Neste caso
$||x(t)|| \leq {|\lambda |}^t ||x(0)||$, onde $\lambda$ \'{e} o autovalor
de $G$ de m\'{a}ximo m\'{o}dulo.  Nestas condi\c{c}\~{o}es o sistema, e por
abuso de nota\c{c}\~{a}o a matriz $A$, \'{e} dito {\bf est\'{a}vel}. 
Estabilidade, assim como controlabilidade \'e uma condi\c{c}\~ao 
que assegura limita\c{c}\~ao de $f^*_h(t,x)$ e consequentemente a 
converg\^encia de $f$.

Para controlar sistemas com horizonte muito grande \'{e} pr\'{a}tico
utilizarmos o controle de horizonte infinito para $t<ha , (h-ha)\ll h$,
o controle (sub-\'{o}timo) ``de cruzeiro'', e o controle \'{o}timo para ``a
aproxima\c{c}\~{a}o final'' em $ha\leq t\leq h$.

\section{Evolu\c{c}\~{a}o Linear com Ru\'{\i}do Gaussiano}  

Estudamos na se\c{c}\~{a}o anterior, modelos onde tanto estados quanto
controles eram determin\'{\i}sticos. A introdu\c{c}\~ao de ru\'{\i}dos 
aleat\'orios em um sistema tr\'as os benef\'{\i}cios de modelos
estoc\'asticos, por\'em exige que empreguemos um ferramental mais 
sofisticado, especialmente quando se pretende realizar estima\c{c}\~oes. 

Admitiremos que s\~ao conhecidas observa\c{c}\~oes de vari\'aveis 
relacionadas com os estados cujos valores queremos prever. Nestas 
observa\c{c}\~{o}es est\~ao  presentes ru\'{\i}dos, e ser\'a 
nosso objetivo filtr\'a-los de forma a conseguir informa\c{c}\~oes 
sobre os estados. Aqui tamb\'em ser\'a considerado essencialmente 
o caso linear, onde tanto as observa\c{c}\~oes $y(t)$ e os estados $x(t)$ num 
dado instante $t$, quanto os estados em instantes distintos ($x(t)$, $x(t-1)$)
est\~ao interligados por rela\c{c}\~oes lineares. 

A literatura comumente apresenta a resolu\c{c}\~ao do problema aqui analisado 
em termos de uma equa\c{c}\~ao recursiva onde as solu\c{c}\~oes, i.e., 
as estimativas dos estados em um dado instante s\~ao obtidas em fun\c{c}\~ao 
da estimativa do instante imediatamente anterior, por\'em com a 
desvantagem de haver a necessidade de se inverter uma matriz a cada passo. 
A formula\c{c}\~ao aqui apresentada, ainda que possa ser colocada dentro 
deste contexto, possibilita que seja explorada a estrutura dos sistemas 
lineares nos quais a resolu\c{c}\~ao do problema recai.

Um sistema Linear com Ru\'{\i}do Gaussiano (LG) evolui na forma
 $$ x(t+1) = A(t)x(t) + B(t)u(t) + v(t)$$
 onde s\~{a}o dados o controle $u(t)$ e as matrizes $A(t)$ e $B(t)$, e
$v(t)$ e um processo estoc\'{a}stico Gaussiano, i.e.  de esperan\c{c}a
zero e distribui\c{c}\~{a}o normal multi-variada. 

No sistema LG a informa\c{c}\~{a}o que temos a cada instante sobre o
estado do sistema, $x(t)$, \'{e} a observa\c{c}\~{a}o :
 $$y(t) = C(t)x(t) + w(t)$$
onde a matriz $C(t)$ \'{e} dada, e $w(t)$ \'{e} um processo Gaussiano.

Para completar a caracteriza\c{c}\~{a}o do sistema LG, \'{e} dada a
covari\^{a}ncia dos ru\'{\i}do, i.e. 

 $$ {Cov}(
 \left[ \begin{array}{c} w(t) \\ -v(t) \end{array}\right] )
 =
 \left[ \begin{array}{cc} VX(t) & VC(t) \\
                            VC(t)' & VY(t) \end{array}\right]
 =  L(t)L(t)'.
 $$
 Os ru\'{\i}dos em tempos distintos s\~{a}o supostos independentes.

Nosso objetivo \'e  encontrar a melhor estimativa para os estados
passados, presente, e o estado do pr\'{o}ximo instante futuro quando se 
disp\~oe de informa\c{c}\~oes at\'e o presente $t$. 
Assim, queremos estimar no instante $t$, os valores, $x(t-k|t)$ ou $x(t|t)$, 
ou $x(t+1|t)$.
Estimar $x(t)$ no instante $t$, $x(t|t)$, \'{e} chamado {\bf
filtra\c{c}\~{a}o }, estimar $x(t+1)$ no instante $t$, $x(t+1|t)$, \'{e}
chamado {\bf predi\c{c}\~{a}o }, enquanto estimar os estados passados,
$x(t-k|t)$, \'{e} chamado {\bf revis\~{a}o}.

 Numa nota\c{c}\~ao compactada, queremos estimar o vetor 
 $x[t+1] = [ x(1)' | \ldots | x(t)'| x(t+1)' ]'$, a partir das
equa\c{c}\~{o}es de evolu\c{c}\~{a}o e das observa\c{c}\~{o}es presente
e passadas,
 $y[t] = [ y(1)' | \ldots | y(t)' ]'$.

Sendo as matrizes $B(t)$ e os controles $u(t)$ conhecidos, est\~ao 
dispon\'{\i}veis as diferen\c{c}as no estado por a\c{c}\~{a}o do controle, 
 \begin{eqnarray*}
 d(t) & = & B(t)u(t) = x(t+1) - A(t)x(t) - v(t) \\
 d(0) & = & x(1) - v(0)
 \end{eqnarray*}
 onde $d(0)$ \'{e} uma estimativa do estado inicial, $x(1)$, sujeita a
um erro Gaussiano $v(0)$ de covari\^{a}ncia $L(0)L(0)'$.

Podemos escrever todas as equa\c{c}\~oes relacionando os estados entre si e 
tamb\'em as observa\c{c}\~oes e estados ao longo do tempo na forma de 
um sistema de equa\c{c}\~oes lineares cuja matriz \'{e} esparsa e 
estruturada:

 $$
 \left[ \begin{array}{c}
   d(0) \\ y(1) \\ d(1) \\ y(2) \\ d(2) \\
   \bullet \\ \bullet \\ y(t) \\ d(t)
 \end{array} \right]
 =
 \left[ \begin{array}{ccccc}
    I & & & & \\
    C(1) & & & & \\
    -A(1) & I & & & \\
      & C(2) & & & \\
      & -A(2) & \bullet & & \\
      & & \bullet & \bullet & \\
      & & & C(t) & \\
      & & & -A(t) & I
 \end{array} \right]
 \left[ \begin{array}{c}
    x(1) \\ x(2) \\ \bullet \\ \bullet \\ x(t) \\ x(t+1)
 \end{array} \right]
 +
 \left[ \begin{array}{c}
      -v(0) \\ w(1) \\ -v(1) \\  w(2) \\ -v(2) \\ \bullet \\
       \bullet \\ w(t) \\ -v(t)
 \end{array} \right]
 $$
 ou, numa nota\c{c}\~{a}o mais compacta,
 $$
  \left\{
  \begin{array}{c}
  d[t] =  A[t]x[t+1] +  v[t]   \\  
  y[t] =  C[t]x[t] +    w[t].
  \end{array}
  \right. 
 $$
  
Duas quest\~{o}es s\~{a}o decisivas ao tratarmos da resolu\c{c}\~{a}o 
do problema: a obten\c{c}\~{a}o de seus par\^{a}metros, 
i.e., das matrizes de coeficientes $A(t)$, $B(t)$ e $C(t)$ que definem 
a din\^{a}mica no 
sistema acima, e a determina\c{c}\~{a}o das estimativas  
para uma particular escolha dos par\^{a}metros. 
Restringiremos nossa discuss\~ao ao caso em que os par\^ametros 
s\~ao conhecidos pois a quest\~ao de estima\c{c}\~ao de par\^Ametros 
foge do  escopo deste trabalho. 

Conhecidos os par\^ametros do modelo, a obten\c{c}\~ao de estimativas 
de estados  \'e 
conseguida atrav\'es da resolu\c{c}\~ao do sistema acima, 
explorando-se esta particular estrutura.

Multiplicando a equa\c{c}\~{a}o acima por 
 $$L[t]^{-1} = {diag}( L(0), L(1), \ldots , L(t) )^{-1}$$
temos
 $$
 \left[ \begin{array}{c}
   \bar d(0) \\ \bar y(1) \\ \bar d(1) \\ \bar y(2) \\ \bar d(2) \\
   \bullet \\ \bullet \\ \bar y(t) \\ \bar d(t)
 \end{array} \right]
 =
 \left[ \begin{array}{ccccc}
    \bar L(0) & & & & \\
    \bar C(1) & & & & \\
    \bar A(1) & \bar L(1) & & & \\
      & \bar C(2) & & & \\
      & \bar A(2) & \bullet & & \\
      & & \bullet & & \\
      & & \bullet & \bullet & \\
      & & & \bar C(t) & \\
      & & & \bar A(t) & \bar L(t)
 \end{array} \right]
 \left[ \begin{array}{c}
    x(1) \\ x(2) \\ \bullet \\ \bullet \\ x(t) \\ x(t+1)
 \end{array} \right]
 +
 \left[ \begin{array}{c}
      \bar v(0) \\ \bar w(1) \\ \bar v(1) \\  \bar w(2) \\ \bar v(2) \\
       \bullet \\ \bullet \\ \bar w(t) \\ \bar v(t)
 \end{array} \right]
 $$
 ou, numa nota\c{c}\~{a}o mais compacta,
 $$
 \left\{
 \begin{array}{c}
 \bar d[t] = \bar A[t]x[t+1] + \bar v[t] \\
 \bar y[t] = \bar C[t]x[t] + \bar w[t]
 \end{array}
 \right.
 $$

Ao colocar o sistema nesta forma obtemos um novo sistema, ainda bem estruturado,
e com a vantagem de possuir ru\'{\i}dos brancos, i.e., com matriz de 
covari\^{a}ncia igual \`a identidade. Isso decorre da propriedade 
da covari\^ancia: ${Cov}(Mx)=M{Cov}(x)M'$, e pode ser verificado facilmente. 
Neste caso, sabemos que
coincidem o estimador de m\'{a}xima verossimilhan\c{c}a , MLE, o
estimador de m\'{\i}nimos quadrados, LSE, e tamb\'{e}m o melhor estimador
linear n\~{a}o tendencioso, BLUE.  Portando a melhor estimativa de
$x[t+1]$, LSE, \'{e} obtida minimizando a norma quatr\'{a}tica de
res\'{\i}duo (ru\'{\i}do) no sistema de equa\c{c}\~{o}es
super-determinado acima.

Vamos ent\~ao resolver o sistema apresentado utilizando transforma\c{c}\~{o}es 
ortogonais nos subsistemas do sistema original. Essas transforma\c{c}\~oes  
deixam inalterada a
norma quadr\'{a}tica de um vetor, e nos permitem colocar o sistema de
equa\c{c}\~{o}es numa forma ainda mais conveniente. 
Consideremos ent\~ao as seguintes fatora\c{c}\~{o}es ortogonais (QR):
 \begin{eqnarray*}
       Q(1)' \left[ \begin{array}{ccc}
             \bar L(0) & 0 & \bar d(0) \\
             \bar C(1) & 0 & \bar y(1) \\
             \bar A(1) & \bar L(1) & \bar d(1)
             \end{array} \right]
 & = &
             \left[ \begin{array}{ccc}
             \tilde L(0) & S(0) & \tilde d(0) \\
             0 & 0 & \tilde y(1) \\
             0 & \hat L(1) & \hat d(1)
             \end{array} \right]
 \\ 
       Q(t)' \left[ \begin{array}{ccc}
             \hat L(t-1) & 0 & \hat d(t-1) \\
             \bar C(t) & 0 & \bar y(t) \\
             \bar A(t) & \bar L(t) & \bar d(t)
             \end{array} \right]
 & = &
             \left[ \begin{array}{ccc}
             \tilde L(t-1) & S(t-1) & \tilde d(t-1) \\
             0 & 0 & \tilde y(t) \\
             0 & \hat L(t) & \hat d(t)
             \end{array} \right]
 \end{eqnarray*}
 As transforma\c{c}\~{o}es acima s\~{a}o facilmente constru\'{\i}das como 
sequ\^{e}ncias de rota\c{c}\~{o}es de Givens que eliminam as colunas 
de $\bar C$ e $\bar A$ da direita para a esquerda, eliminando os elementos 
de cima para baixo. 

 As transforma\c{c}\~{o}es ortogonais do sistema de
equa\c{c}\~{o}es s\~ao definidas pelas composi\c{c}\~{o}es das 
transforma\c{c}\~{o}es ortogonais aplicadas aos blocos,
 $$
 Q[t]' =  Q(t)' \circ Q(t-1)' \circ \ldots \circ Q(1)' \circ Q(0)' \ .  
 $$

Aplicando $Q[t]'$ ao nosso sistema de equa\c{c}\~{o}es temos
 $$
 \left[ \begin{array}{c}
   \tilde d(0) \\ \tilde y(1) \\ \tilde d(1) \\ 
   \bullet \\ \bullet \\  
   \tilde d(t-1) \\ \tilde y(t) \\ \hat d(t)
 \end{array} \right]
 =
 \left[ \begin{array}{ccccc}
    \tilde L(0) & S(0) & & & \\
    0 & 0 & & & \\
      & \tilde L(1) & \bullet & & \\
      & & \bullet & & \\
      & & \bullet & \bullet & \\
      & &  & 0 & \\
      & & & \tilde L(t-1) & S(t-1) \\
      & & & 0 & 0 \\
      & & & & \hat L(t)
 \end{array} \right]
 \left[ \begin{array}{c} 
    x(1) \\ x(2) \\ \bullet \\ \bullet \\ x(t) \\ x(t+1)
 \end{array} \right]
 +
 \left[ \begin{array}{c}
      \tilde v(0) \\ \tilde w(1) \\ \tilde v(1) \\  
      \tilde w(2) \\ \tilde v(2) \\
      \bullet \\ \bullet \\ \tilde w(t) \\ \hat v(t)
 \end{array} \right]
 $$
ou, numa nota\c{c}\~{a}o mais compacta, separando os dois sub-sistemas, 
 $$
 \left\{
 \begin{array}{c}
 \tilde d[t] = \tilde L[t]x[t+1] + \tilde v[t] \\
        \tilde y[t] = 0 + \tilde w[t]
 \end{array}
 \right.
 $$

 O primeiro subsistema, $\tilde d[t]=\tilde L[t] x[t+1]$, \'{e}
triangular e pode ser resolvido exatamente, i.e. com ru\'{\i}do 
 $\tilde v[t]=0$.  Todavia no segundo sub-sistema nada pode ser feito
para acomodar $\tilde y[t]=0$. Portando o estimador de m\'{\i}nimos
quadrados \'{e} simplesmente a  solu\c{c}\~{a}o do sistema triangular. 
 
Para fazer apenas a
filtra\c{c}\~{a}o, isto \'e, estimar $x(t)$ no instante $t$, n\~{a}o 
\'{e} necess\'{a}rio manter toda a matriz
$\tilde L[t]$.  Para a opera\c{c}\~{a}o de predi\c{c}\~{a}o no instante
$t$ basta preservar $\hat L(t)$ e $\hat d(t)$, como fica claro na
rela\c{c}\~{a}o de recorr\^{e}ncia: 
 \begin{eqnarray*}
  x(t+1|t) & = & \hat L(t)^{-1} \hat d(t) \\ 
  x(t|t)   & = & \tilde L(t-1)^{-1} ( \tilde d(t-1) - S(t-1) x(t+1) ) 
 \end{eqnarray*}
 
\section{Princ\'{\i}pio de Equival\^encia} 

 O filtro acima descrito \'{e} uma variante do filtro de Kalman.  Estude
a implementa\c{c}\~{a}o deste filtro em Matlab, e use-o para estimar os
estados presentes e passados, no sistema do exemplo dado, ao longo de
sua evolu\c{c}\~{a}o temporal.  Interprete o filtro que construimos em
termos de programa\c{c}\~{a}o din\^{a}mica.  Interprete cada um dos conceitos
 apresentados no cap\'{\i}tulo 4  no contexto do filtro.

 N\~ao \'e dif\'{\i}cil demonstrar que a teoria de controle LQ \'e 
 compat\'{\i}vel com a teoria de sistemas LG no seguinte sentido: o 
 controle $u^{**}$ que minimiza o custo 
 de controle esperado, em um sistema LQG ( sistema linear com custo 
 quadr\'atico e ru\'{\i}do gaussiano ), \'e igual ao controle $u^{*}$ 
 que minimiza o custo de controle no sistema LQ determin\'{\i}stico, 
 onde o estado do sistema LQ \'e o estado do sistema LQG estimado pelo  
 Filtro de Kalman. Este \'e o Princ\'{\i}pio de Equival\^encia. Neste  
 sentido o problema de controle LQ e o problema de estima\c{c}\~ao LG 
 s\~ao ditos duais.


\section{Generaliza\c{c}\~oes do Filtro de Kalman}

Diversas generaliza\c{c}\~oes do Filtro de Kalman s\~ao comumente 
empregadas. Para os problemas que analisamos em finan\c{c}as, estudamos 
sistemas na forma:

\begin{equation}
\left\{ \begin{array}{l}
    x(t+1) = f_{t}(x(t)) + B(t) u(t)+ \epsilon (t) \\
    y(t) = g_{t}(x(t)) + \xi (t) \hspace{2.5cm}   t \in N 
\end{array} \right.
\end{equation}

O objetivo aqui  \'e encontrar uma estimativa, 
$\hat{x}(t^{*} \mid t )$, para algum $t^{*}$,

\noindent
As fun\c{c}\~{o}es $f_{t}$ e $g_{t}$ podem n\~{a}o ser  lineares, por\'{e}m 
s\~ao deriv\'{a}veis. Tamb\'em aqui os controles $u_{t}$ s\~ao 
determin\'\i sticos e $y(t)$ s\~ao as observa\c{c}\~oes realizadas. 
A abordagem mais simples consiste em  fazer aproxima\c{c}\~{o}es por 
Taylor recaindo no caso linear.

\section{Exerc\'{\i}cios} 

\begin{enumerate} 
 \item Reescreva a teoria de sistemas LQ com custo aditivo descontado.  
 \item Considere o vetor de estados aumentado, 
   $\left[ x_1,\ldots x_n,\, 1\right]'$; e explique como usar esta 
   \'ultima ``coordenada constante'' do vetor de estado para introduzir 
   custos puramente lineares em $x$ ou $u$.   
 \item Implemente a equa\c{c}\~{a}o de Ricatti com custo descontado em Matlab. 
  Monte e controle um exemplo simples. 
 \item  Prove que 
  \begin{enumerate} 
  \item Todo sistema control\'{a}vel \'{e} estabiliz\'{a}vel. 
  \item Todo sistema $k$-control\'{a}vel \'{e} $k+1$-control\'{a}vel. 
  \item Se $x\in R^n$, todo sistema control\'{a}vel \'{e} $n+1$-control\'{a}vel. 
  \end{enumerate} 
 \item  
 De exemplos de 
  \begin{enumerate} 
  \item Um sistema estabiliz\'{a}vel mas n\~{a}o control\'{a}vel. 
  \item Um sistema, em $R^2$, 2-control\'{a}vel mas n\~{a}o 1-control\'{a}vel. 
  \end{enumerate}  

 \item Imagine que a medida que o tempo passa, passemos a descrer 
cada vez mais de nossas observa\c{c}\~{o}es e estimativas passadas: 
Formalmente, estando no instante $t$ tomaremos  
 $$ {Cov}(
 \left[ \begin{array}{c} w(t-k) \\ -v(t-k) \end{array}\right] )
 =
 \delta ^{t-k}
 \left[ \begin{array}{cc} VX(t-k) & VC(t-k) \\
                            VC(t-k)' & VY(t-k) \end{array}\right]\ . 
 $$
 Este modelo \'{e} chamado de {\em mem\'{o}ria evanescente}. 
 Reescreva a teoria de estima\c{c}\~{a}o LG com mem\'{o}ria evanescente, 
e interprete o medelo como um problema de programa\c{c}\~ao 
din\^amica  com custo descontado. 
 
 \item Implemente a Estima\c{c}\~{a}o LG em linguagem C. 
 \begin{itemize} 
  \item[a.] Considere mem\'{o}ria evanescente. 
  \item[b.]  Considere que os par\^{a}metros de evolu\c{c}\~{a}o do sistema,
$A$, $B$ e $L$, bem como o desconto intertemporal $\delta$, s\~{a}o
constantes. Considere que a matriz de observa\c{c}\~{a}o, $C$ \'{e} 
usualmente igual a um padr\~{a}o, $C$, mas que podemos especificar, 
em cada instante, altera\c{c}\~{o}es do padr\~{a}o. 
 \item[c.] Explique como utilizar altera\c{c}\~{o}es em $C$ para modelar 
observa\c{c}\~{o}es perdidas, dias sem preg\~{a}o, etc. 
 \item[d.] A implementa\c{c}\~{a}o deve permitir filtra\c{c}\~{a}o, 
previs\~{a}o para at\'{e} $kp$ passos no futuro (via observa\c{c}\~{o}es 
nulas), e revis\~{a}o para no m\'{a}ximo $kr<<h$ instantes anteriores. 
Explique como esta revis\~{a}o limitada facilita a implementa\c{c}\~{a}o 
do algoritmo. 
 \item[e.] Implemente medidas de erro quadr\'{a}tico m\'edio sobre os 
 res\'{\i}duos
de filtra\c{c}\~{a}o, $k$-revis\~{a}o, e $k$-predi\c{c}\~{a}o.  Explique
como poder\'{\i}amos usar estas medidas de erro para melhor ajustar o modelo
$(A,B,C,L,\delta)$. 
 \end{itemize} 
\end{enumerate}

%% file: cap7.tex
\chapter{ \'{A}rvores de Decis\~{a}o }

Este cap\'{\i}tulo apresenta o algoritmo REAL de aprendizado 
autom\'{a}tico,  para constru\c{c}\~{a}o de \'{a}rvores de 
classifica\c{c}\~{a}o TDIDT (Top Down Induction Decision  Tree) com
atributos a valores reais [1], [7], [8]. 

O Projeto REAL come\c{c}ou como uma aplica\c{c}\~{a}o a ser  utilizada
no mercado de a\c{c}\~{o}es, provendo um bom  algoritmo para pever a
adequa\c{c}\~{a}o de estrat\'{e}gias de opera\c{c}\~{a}o.  Neste
contexto, o sucesso ou fracasso de uma dada opera\c{c}\~{a}o 
corresponde a classe do exemplo, enquanto os atributos s\~{a}o 
``indicadores t\'{e}cnicos'' que assumem valores reais.   As
exigencias dos usu\'{a}rios para a ferramenta de suporte  \`{a}
decis\~{a}o explicam v\'{a}rias caracter\'{\i}sticas \'{u}nicas do
algoritmo. 

O projeto come\c{c}ou testando v\'{a}rios algoritmos de aprendizado 
apresentados no projeto ESPRIT-StatLog [5]; o software CAL5 [6], um 
algoritmo Top-Down para gera\c{c}\~{a}o de \'{a}rvores de
classifica\c{c}\~{a}o,  mostrou-se especialmente adequado para a
aplica\c{c}\~{a}o que tinhamos  em mente. O algoritmo CAL5 teve uma
forte influ\^{e}ncia no projeto,  e foi utilizado como a principal
refer\^{e}ncia para compara\c{c}\~{a}o de  performance. Na nossa
aplica\c{c}\~{a}o o algoritmo REAL apresentou algumas vantagens
importantes: 
 \begin{enumerate} 
 \item Geralmente as \'{a}rvores de classifica\c{c}\~{a}o tem menor taxa
de erro. 
 \item Uma medida \'{u}nica de convic\c{c}\~{a}o mostrou-se mais
conveniente que o tradicional par (probabilidade, confi\^{a}n\c{c}a).  
 \item Os procedimentos de ramifica\c{c}\~{a}o do REAL detem-se 
naturalmente, dispensando um posterior procedimento de poda.  
 \end{enumerate} 

\section{Formula\c{c}\~{a}o do Problema} 

Os problemas de classifica\c{c}\~{a}o s\~{a}o apresentados como uma
matriz  $A$, $n \times (m+1)$. Cada linha, $A(i,:)$,  representa um
examplo,  e cada coluna, $A(:,j)$, um atributo. As primeiras $m$ colunas
s\~{a}o  atributos a valores reais, e a \'{u}ltima coluna, $A(i,m+1)$
\'{e} a classe do  exemplo. Parte destes exemplos, o conjunto de
treinamento, \'{e} usado  pelo algoritmo para gerar a \'{a}rvore de
classifica\c{c}\~{a}o, que \'{e} ent\~{a}o testada  com os exemplos
remanecentes. A taxa de erro de classifica\c{c}\~{a}o no  conjunto de
teste \'{e} uma maneira simples de avaliar a \'{a}rvore de 
classifica\c{c}\~{a}o.

\section{Constru\c{c}\~{a}o da \'{A}rvore}  

Cada itera\c{c}\~{a}o principal do algoritmo REAL corresponde \`{a}
ramifica\c{c}\~{a}o  de um n\'{o} terminal (folha) da \'{a}rvore.  Os
exemplos naquele n\'{o} s\~{a}o classificados de acordo com o valor do 
atributo selecionado, e novos ramos s\~{a}o gerados para intervalos 
especificos. A parti\c{c}\~{a}o do dom\'{\i}nio de um atributo em (sub)
intervalos  adjacentes e n\~{a}o sobrepostos corresponde ao processo de
discretiza\c{c}\~{a}o. 
 Cada itera\c{c}\~{a}o principal do REAL inclui: 
 \begin{enumerate}  
 \item A discretiza\c{c}\~{a}o de cada atributo, e sua avalia\c{c}\~{a}o
por  uma fun\c{c}\~{a}o de perda.  
 \item Sele\c{c}\~{a}o do melhor atributo, e ramifica\c{c}\~{a}o de
acordo com sua  discretiza\c{c}\~{a}o.  
 \item Jun\c{c}\~{a}o de intervalos adjacentes que n\~{a}o
alcan\c{c}aram um limiar  m\'{\i}nimo de convic\c{c}\~{a}o.  
 \end{enumerate}    

\section{Convic\c{c}\~{a}o e Fun\c{c}\~{a}o de Perda}  

Dado um n\'{o} de classe $c$ con $n$ exemplos, $k$ dos quais
incorretamente  classificados e $n-k$ corretamente classificados,
queremos um par\^{a}metro  escalar, $cm$, para medir tanto a
probabilidade de termos uma classifica\c{c}\~{a}o  incorreta como o
n\'{\i}vel de confian\c{c}a desta probabilidade. Uma tal medida 
simplificada de convic\c{c}\~{a}o nos foi colocada como uma necessidade
dos  usu\'{a}rios para operar no mercado de capitais.   
 Seja $q$ a probabilidade de classifica\c{c}\~{a}o incorreta para um
exemplo em  um dado n\'{o}, $p=(1-q)$ a probabilidade de
classifica\c{c}\~{a}o correta, e  assumamos a exist\^{e}ncia para $q$ de
uma distribui\c{c}\~{a}o Bayesiana: 
  $$D(c) = Pr(q \leq c) = Pr(p \geq 1-c)$$  

Definimos a medida de convic\c{c}\~{a}o: $100*(1-cm)\%$, onde 
 $$cm =  \mbox{min}\ c \ \ | \ \  Pr( q \leq c ) \geq 1 -g(c)$$ 

 \noindent 
 e $g(\ )$ \'{e} uma bije\c{c}\~{a}o  monotonicamente crecente de
$[0,1]$ sobre si  mesmo. Nossa experi\^{e}ncia no mercado de capitais
nos ensinou a ser  muito cautelosos com a certeza de afirma\c{c}\~{o}es,
assim tomamos  $g(\ )$ uma fun\c{c}\~{a}o convexa. Neste artigo $D(c)$
\'{e} a distribui\c{c}\~{a}o a  posteriori de uma amostra tomada de uma
distribui\c{c}\~{a}o de  Bernoulli,  com uma distribui\c{c}\~{a}o a
priori uniforme para $q$:  
 \begin{eqnarray*}
  B(n,k,q) &=& comb(n,k) * q^k * p^{n-k} \\  
  D(c,n,k) &=& \int_{q=0}^c B(n,k,q) \ \ \  / \ \ \int_{q=0}^1 B(n,k,q) \\    
           &=& \mbox{betainc}(c,k+1,n-k+1)  
 \end{eqnarray*}  

Tamb\'{e}m neste artigo focalizamos nossa aten\c{c}\~{a}o em:  
 $$g(c)  =  g(c,r)  =  c^r  , \ \  r  \geq 1.0 $$ 
 onde $r$ \'{e} chamado o par\^{a}metro de convexidade. 

Com estas escolhas, a posteriori \'{e} uma fun\c{c}\~{a}o beta
incompleta,  facilmente computavel, e $cm$ \'{e} a raiz da
fun\c{c}\~{a}o monotonicamente  decrecente:  
 \begin{eqnarray*} 
  cm(n,k,r) &=& c \ \ |  \ \ f(c) = 0 \\ 
  f(c)      &=& 1 -g(c) -D(c,n,k) \\   
            &=& 1 -c^r -\mbox{betainc}(c,k+1,n-k+1)  
 \end{eqnarray*}   
 \noindent

Finalmente queremos uma fun\c{c}\~{a}o de perda para a
discretiza\c{c}\~{a}o  baseada na medida de convic\c{c}\~{a}o. Neste
artigo utilizamos a somat\'{o}ria da convic\c{c}\~{a}o da
classifica\c{c}\~{a}o de cada exemplo, i.e. a soma, sobre todos  os
intervalos, da medida de convic\c{c}\~{a}o do intervalo vezes o
n\'{u}mero de  exemplos nele contido. 
 $$loss = \sum_i  n_i*cm_i$$

\section{Procedimento de Discretiza\c{c}\~{a}o} 

Dado um atributo, o primeiro passo do processo de discretiza\c{c}\~{a}o
\'{e}  ordenar os exemplos no n\'{o} pelo valor do atributo, e ent\~{a}o
agrupar  exemplos vizinhos de mesma clase, ou com identico valor  do
atributo. Assim, ao fim deste primeiro passo temos, para o atributo 
escolhido, a melhor discretiza\c{c}\~{a}o ordenada onde cada intervalo
contem  um grupo de exemplos de uma mesma classe, ou com o mesmo valor 
do atributo.    

Nos passos subseq\"{u}entes, procuramos juntar intervalos para reduzir 
o valor da fun\c{c}\~{a}o de perda da discretiza\c{c}\~{a}o. O ganho de
juntar $J$  intervalos adjacentes,  $I_{h+1}$, $I_{h+2}$, \ldots
$I_{h+J}$, \'{e} o decrecimo relativo da fun\c{c}\~{a}o de perda:    
  $$gain(h,j) = \sum_j loss(n_j,k_j,r)\ \  - loss(n,k,r) $$ 
 onde $n= \sum_j n_j$ e $k$ conta os exemplos de classes 
minorit\'{a}rias no novo intervalo. 

A cada passo realizamos a opera\c{c}\~{a}o de agrupamento de m\'{a}ximo
ganho.  O procedimento de discretiza\c{c}\~{a}o termina ao esgorarem-se
as opera\c{c}\~{o}es  de agrupamento com ganho positivo. 
 
Os exemplos seguintes mostram alguns intervalos de classe uniforme  que
seriam agrupados que poderiam ser agrupados durante o  procedimento de
discretiza\c{c}\~{a}o.  A nota\c{c}\~{a}o $(n,k,m,r,\pm )$ significa que
temos dois grupos de mesma  classe, de tamanho $n$ e $m$, separados por
um grupo de tamanho  $k$ com exemplos de outra classe. $r$ \'{e} o
par\^{a}metro de convexidade,  e  $+$ ($-$) indica que poderiamos juntar
(ou n\~{a}o) os tr\^{e}s intervalos.  

\begin{verbatim} 
 ( 2,1, 2,2,+) 
 ( 6,2, 7,2,-) ( 6,2, 8,2,+) ( 6,2,23,2,+) ( 6,2,24,2,-) 
 ( 7,2, 6,2,-) ( 7,2, 7,2,+) ( 7,2,42,2,+) ( 7,2,43,2,-) 
 (23,3,23,2,-) (23,3,43,2,-) (23,3,44,2,+) 
 (11,3,13,3,-) (11,3,14,3,+) (11,3,39,3,+) (11,3,40,3,-) 
 (12,3,12,3,-) (12,3,13,3,+) (12,3,54,3,+) (12,3,55,3,-) 
\end{verbatim} 
 
Nestes exemplos vemos que \'{e} necess\'{a}rio termos nos extremos 
grupos suficientemente grandes e equilibrados para ``absorver''  
o ruido ou impureza no grupo intermedi\'{a}rio. Um par\^{a}merto de 
convexidade alto implica numa perda maior em grupos pequenos, 
ajudanto a absor\c{c}\~{a}o de impurezas esparsas.

\section{Ramifica\c{c}\~{a}o e Reagrupamento}  

Para cada n\'{o} terminal na \'{a}rvore, devemos:  
 \begin{enumerate} 
 \item realizar, para cada atributo dispon\'{\i}vel, o procedimento de 
discretiza\c{c}\~{a}o. 
 \item medir a perda da discretiza\c{c}\~{a}o para cada atributo.   
 \item selecionar o atributo que leva a discretiza\c{c}\~{a}o de
m\'{\i}nima perda, e  
 \item ramificar o n\'{o} de acordo com a discretza\c{c}\~{a}o
correspondente. 
 \end{enumerate} 
 Se nenhum atributo induzir uma discretiza\c{c}\~{a}o que decre\c{c}a a
fun\c{c}\~{a}o de  perda de pelo menos um limite de precis\~{a}o
num\'{e}rica $\epsilon > 0$,  n\~{a}o ocorre ramifica\c{c}\~{a}o.   

 Uma dada discretiza\c{c}\~{a}o, feita em um certo n\'{\i}vel da
\'{a}rvore, pode  impedir o progresso do processo de ramifica\c{c}\~{a}o
nos n\'{\i}veis abaixo.  Por esta raz\~{a}o estabelecemos uma meta
m\'{\i}nima de convic\c{c}\~{a}o, $ct$,  e ap\'{o}s cada
ramifica\c{c}\~{a}o reagupamos os intervalos adjacentes onde  $cm < ct$.
Para evitar um loop infinito, o valor da fun\c{c}\~{a}o de perda  dado
ao intervalo reagrupado \'{e} a soma das perdas nos intervalos  sendo
reagrupados. Nas folhas da \'{a}rvore final esta opera\c{c}\~{a}o de 
reagrupamento \'{e} desfeita. A meta de convic\c{c}\~{a}o termina
naturalmente  o processo de ramifica\c{c}\~{a}o, n\~{a}o havendo
necessidade de um  procedimento dicional para poda da \'{a}rvore, como
na maioria  dos algoritmos TDIDT.

\section{Implementa\c{c}\~{a}o Computacional}

Para os testes num\'{e}ricos, a serem detalhados na se\c{c}\~{a}o 9,
utilizamos  uma implementa\c{c}\~{a}o padr\~{a}o do algoritmo REAL.
Nesta implementa\c{c}\~{a}o  cada problema demora cerca de 2 minutos,
incluindo treinamento e teste,  em um Pentium 200MHz. 

 REAL foi implementado como um c\'{o}digo {\it C++} absolutamente 
portavel, e a aplica\c{c}\~{a}o final final recebeu uma interface
gr\'{a}fica para  o usu\'{a}rio em Microsoft VB-4.0. Esta
implementa\c{c}\~{a}o padr\~{a}o gasta grande  parte do tempo de
processamento computando a fun\c{c}\~{a}o $cm(n,k,r)$.   
 Podemos acelerar substancialmente o algoritmo utilizando tabelas 
pr\'{e}-computadas para valore pequenos do argumento $n$ (digamos 
$n\leq 100$), e tabelas de coeficientes para polinomios interpoladores 
para valores maiores de $n$. Outro expediente de acelera\c{c}\~{a}o
\'{e}  restringir o espa\c{c}o de busca para as opera\c{c}\~{o}es de
agrupamento a uma  vizinhan\c{c}a de apenas $2 \leq J \leq Jmax$
intervalos: Escolhendo um  $Jmax$ conveniente aceleramos o algoritmo sem
nenhuma degrada\c{c}\~{a}o  apreciavel dos resultados. 

\section{Estrat\'{e}gias de Opera\c{c}\~{a}o no Mercado} 

Uma estrat\'{e}gia de opera\c{c}\~{a}o no mercado \'{e} um conjunto
pr\'{e}-definido  de regras determinando as a\c{c}\~{o}es de operador no
mercado. A estrat\'{e}gia  deve conter crit\'{e}rios para classificar
uma opera\c{c}\~{a}o realizada como um  suscesso ou um fracasso. 

 Definamos, como exemplo, a estrat\'{e}gia $buysell(t,d,l,u,c)$: 
 \begin{itemize} 
 \item No instante $t$ compre o ativo {\it A}, a seu pre\c{c}o $p(t)$. 
 \item Venda o ativo {\it A} assim que: 
 \begin{enumerate} 
  \item $t' = t+d$ , ou 
  \item $p(t') = p(t)*(1+u/100)$ , ou  
  \item $p(t') = p(t)*(1-l/100)$ .  
 \end{enumerate} 
 \item A estrat\'{e}gia \'{e} bem suscedida se   
       $c \leq 100*p(t')/(p(t) \leq u$ 
 \end{itemize} 
 Os parametros $u$, $l$, $c$ and $d$ podem ser interpretados como, 
respectivamente, o retorno desejado e o m\'{\i}nimo aceitavel o custo da
opera\c{c}\~{a}o, e o limite de tempo para encerrar a opera\c{c}\~{a}o.
 A figura 1 ilustra poss\'{\i}veis inst\^{a}ncias de aplica\c{c}\~{a}o
da estrat\'{e}gia  $buysell(t,d,l,u,c)$.

\begin{figure}[hbt]
\center{
\tiny
\input{fig01.tex}
\normalsize
{\caption{ Exemplos para a estrategia $buysell(t,d,l,u,c)$.}}}
\end{figure}
\normalsize 

\subsection*{Indicadores T\'{e}cnicos} 

Analistas de Mercado utilizam v\'{a}rias ferramentas para previs\~{a}o
de  pre\c{c}o de ativos, incluindo ferramentas baseadas em dados
atuariais,  an\'{a}lise de insumo-produto ou macro-econ\^{o}nica, etc.,
todas estas  denominadas ferramentas fundamentalistas.  Outra
fam\'{\i}lia de ferramentas s\~{a}o os indicadores t\'{e}cnicos.  Um
indicador t\'{e}cnico \'{e} uma fun\c{c}\~{a}o de uma ou mais
vari\'{a}veis  observ\'{a}veis no mercado. V\'{a}rios distribuidores
comercias difundem  dados diarios e ``intraday'' para todos os mercados
importantes.  Na BOVESPA, a Bolsa de Valores de S\~{a}o Paulo, estes
dados incluem:  $H(t)$, $L(t)$, $O(t)$, $C(t)$, $M(t)$  and $V(t)$,
respectivamente  os pre\c{c}os m\'{a}ximo, m\'{\i}nimo, de abertura,
fechamento e m\'{e}dio, e o  volume negociado no dia $t$. 
 Um exemplo de indicador t\'{e}cnico \'{e} o pre\c{c}o de abertura em
rela\c{c}\~{a}o  ao m\'{a}ximo alcan\c{c}ado nos $r$ dias anteriores: 
 $$ OH_r(t) = O(t) / \max \{ H(t), H(t-1) , \ldots H(t-r) \} $$ 

Indicadores t\'{e}cnicos s\~{a}o ferramentas de an\'{a}lise de mercado 
tradicionais e de grande aceita\c{c}\~{a}o [3]. A grande familiaridade
dos  operadores de mercado estes indicadores explica a motiva\c{c}\~{a}o
conforto e confian\c{c}a dos operadores ao utilizar uma ferramenta de 
suporte a decis\~{a}o que fornecem regras l\'{o}gicas de
opera\c{c}\~{a}o baseadas  no valor de indicadores t\'{e}cnicos
conhecidos (atributos). 
 As regras e os atributos utilizados para uma dada classifica\c{c}\~{a}o
podem  ser facilmente compreendidas (e aceitas ou rejeitadas). Esta
confian\c{c}a  traduz-se em um uso \'{a}gil e eficiente da ferramenta.
De fato, o uso dos  indicadores como atributos da \'{a}rvore TDIDT foi
um pedido do cliente. 

O n\'{u}cleo do sistema de suporte \`{a} decis\~{a}o para
opera\c{c}\~{o}es de mercado  \'{e} um sistema de classifica\c{c}\~{a}o
do tipo TDIDT baseado no algoritmo  REAL. Consideremos alguns problemas
de classifica\c{c}\~{a}o para a  estrat\'{e}gia $buysell(t,d,l,u,c)$.
Nestes problemas, um exemplo no instante  $t$ tem como atributos valores
de v\'{a}rios indicadores t\'{e}cnicos no  instante $t-1$, e como classe
o sucesso ou fracasso da aplica\c{c}\~{a}o da  estrat\'{e}gia no
instante $t$, baseado na informa\c{c}\~{a}o dispon\'{\i}vel entre os 
instantes $t$ e $d+d$. Tomamos os exemplos em segmentos de s\'{e}ries 
temporais sem superposi\c{c}\~{a}o.

\subsection*{Fun\c{c}\~{a}o Objetivo} 

 O procedimento de classifica\c{c}\~{a}o, aplicado a um dado conjunto de
exemplos, gera a matriz de classifica\c{c}\~{a}o, tamb\'{e}m conhecida
como  matriz de confus\~{a}o, como a matriz na tabela 1,  
\begin{table}[ht]
\center{ 
\begin{tabular}{|c||c|c|}
\hline
Verdad./Atribuida & Sucesso & Falha \\ 
\hline 
Sucesso & $n11$ & $n12$ \\ 
\hline
Falha & $n21$ & $n22$ \\ 
\hline 
\end{tabular}
}
{\caption{ Matriz de Classifica\c{c}\~{a}o (confus\~{a}o)}}
\end{table} 
\normalsize 
 %
 onde $n11$ e $n22$ s\~{a}o o n\'{u}mero de exemplos corretamente 
classificados de aplica\c{c}\~{o}es bem sucedidas e fracassadas, $n12$
s\~{a}o  aplica\c{c}\~{o}es bem sucedidas incorretamente classificadas
como falhas,  e $n21$ os erros de classifica\c{c}\~{a}o opostos. O
operador de mercado  espera que o sistema de suporte \`{a} decis\~{a}o o
auxilie a detectar quase  todas as boas oportunidades de
aplica\c{c}\~{a}o da estrat\'{e}gia. Quando  aconselhado a aplicar a
estat\'{e}gia, o operador espera que o sistema  raramente que o sistema
raramente esteja errado. Assim nosso  objetivo \'{e} o de maximizar a
taxa de aproveitamento das oportunidades,  $ry$, e tamb\'{e}m o de
minimizar as falhas de aplica\c{c}\~{a}o, $rf$:    
 $$ ry = n11 / (n11+n12) \ \ \mbox{and} \ \ rf = n21 / (n11+n21) $$   

 Para conciliar estes objetivos m\'{u}ltiplos e antag\^{o}nicos
definimos uma  fun\c{c}\~{a}o de m\'{e}rito que pode ser interpretada
como uma estimativa  conservadora do ganho com as aplica\c{c}\~{o}es da
estrat\'{e}gia  $buysell(t,d,l,u,c)$: 
 $$ merit = c*n11 - l*n21 $$  
  
\section{Testes Num\'{e}ricos} 

  Testamos o algoritmo de classifica\c{c}\~{a}o exposto nas
se\c{c}\~{o}es anteriores  com $buysell(t,d,l,u,c)$, e $d=5\ dias$,
$l=1\%$, $c=1\%$ e $u=3\%$, em segmentos n\~{a}o superpostos de
s\'{e}ries temporais de pre\c{c}os de duas  das a\c{c}\~{o}es mais
liq\"{u}idas (negociadas) na BOVESPA:   Telebras-PN (TEL4), com aprox.
300 exemples 45\% bem sicedidos,  e Petrobras-PN (PET4), com aprox. 400
exemplos 40\%  bem sucedidos.  Dividimos cada conjunto de exemplos em 10
subconjuntos. Em cada rodada do algoritmo geramos a \'{a}rvore usando 9
dos subconjuntos  para tereinamento, e testando a \'{a}rvore no
subconjunto remanecente.  repetimos este procedimento para os
par\^{a}metros do algoritmo em um  grid discreto:   
 \begin{description}
 \item[REAL:]  
   $ (r, ct) \in  \{1.0, 1.5, \ldots, 4.0\} 
                   \times \{0.1, 0.15, 0.2, \ldots, 0.45\} . $   
 \item[Cal5:]  
    $ (S, \alpha) \in \{0.05, 0.10, \ldots, 0.90\} 
                       \times \{0.05, 0.10, \ldots, 0.90\} .$
 \item[NewID:] 
  $ \phi \in \{0\%, 2\%, 4\%, \ldots, 28\%, 30\% \}. $ 
 \end{description} 

 Tamb\'{e}m incluimos como benchmark NewID, uma algoritmo TDITT 
cl\'{a}ssico, primordialmente desenvolvido para classifica\c{c}\~{a}o
categ\'{o}rica,  mas tamb\'{e}m capaz de utilizar atributos reais. Nas
tabelas 2 e 3  mostramos o valor dos par\^{a}metros que otimizam a
m\'{e}dia da fun\c{c}\~{a}o de  m\'{e}rito, e os correspondentes valores
m\'{e}dios de $rf$ e $ry$.  

\begin{table}[!htb]
\center{ 
\begin{tabular}{|c||c|c|c|c|}
\hline
    Algoritmo & $P^*$  & $merit$ &  $rf$   & $ry$ 
    \\ \hline \hline 
    REAL   & $r=3.5,  ct=0.4$  & $3.8$ & $0.22$ & $0.44$   \\ \hline 
    Cal5   & $S=0.3, \alpha=0.1$ & $3.3$ & $0.35$ & $0.63$ \\ \hline
    NewID  & $\phi=6\%$ & $2.9$ & $0.25$ & $0.45$ \\ \hline
\end{tabular}
{\normalsize \caption{Par\^{a}metros \'{o}timos para TEL4}
}}
\end{table}
\normalsize 


\begin{table}[!htb] 
\center{ 
\begin{tabular}{|c||c|c|c|c|}
\hline
    Algoritmo & $P^*$   & $merit$ &  $rf$ & $ry$ 
    \\ \hline \hline 
    REAL  & $r=2, ct=0.30$ & $4.3$ & $0.24$ & $0.39$  \\ \hline 
    Cal5  & $S=0.4, \alpha=0.2$ & $3.8$ & $0.23$ & $0.35$  \\ \hline
    NewID & $\phi=8\%$ & $2.9$ & $0.42$ & $0.28$ \\ \hline
\end{tabular}
\normalsize
{\normalsize \caption{Par\^{a}metros \'{o}timos para PET4} 
}}
\end{table}
\normalsize 

 Para uma an\'{a}lise de sensibilidade mostramos resultados similares em
uma vizinhan\c{c}a dos par\^{a}metros \'{o}timos nas tabelas 4 e 5. 

\begin{table}[!htb] 
\center{ 
\begin{tabular}{|c||c|c|c|} 
\hline 
  $ \begin{array}{lr} 
         & ct \\ 
      r  &  
  \end{array} $ 
& 
0.35  & $ct^*=0.40$ & 0.45 \\ \hline \hline 
3.0   & 
  $ \begin{array}{c} 
     merit = 3.5  \\ 
     rf= 0.18   \\ 
     ry= 0.38   \\ 
  \end{array} $ 
  & 
  $ \begin{array}{c} 
     merit = 3.5   \\ 
     rf= 0.26   \\ 
     ry= 0.47   \\ 
  \end{array} $ 
  & 
  $ \begin{array}{c} 
     merit = 3.7   \\ 
     rf= 0.30   \\ 
     ry= 0.51   \\ 
  \end{array} $ 
\\ \hline 
$r^*=3.5$   & 
  $ \begin{array}{c} 
     merit = 3.5   \\ 
     rf= 0.14   \\ 
     ry= 0.34  \\ 
  \end{array} $ 
  & 
  $ \begin{array}{c} 
     merit = 3.8  \\ 
     rf= 0.22  \\ 
     ry= 0.44  \\ 
  \end{array} $ 
  & 
  $ \begin{array}{c} 
     merit = 3.5  \\ 
     rf= 0.31  \\ 
     ry= 0.51  \\ 
  \end{array} $ 
\\ \hline 
4.0   & 
  $ \begin{array}{c} 
     merit = 3.2  \\ 
     rf= 0.27   \\ 
     ry= 0.38  \\ 
  \end{array} $ 
  & 
  $ \begin{array}{c} 
     merit = 3.5  \\ 
     rf= 0.24  \\ 
     ry= 0.41  \\ 
  \end{array} $ 
  & 
  $ \begin{array}{c} 
     merit = 3.2  \\ 
     rf= 0.27  \\ 
     ry= 0.45  \\ 
  \end{array} $ 
\\ \hline 
\end{tabular}   
{\normalsize \caption{REAL - An\'{a}lise de sensibilidade para TEL4} }
}
\end{table}
\normalsize 

\begin{table}[!htb] 
\center{ 
\begin{tabular}{|c||c|c|c|} 
\hline 
  $ \begin{array}{l r} 
         & ct\\ 
      r  &  
    \end{array} $ 
& 
0.25  & $ct^* = 0.30$ & 0.35 \\ \hline \hline 
1.5   & 
  $ \begin{array}{c} 
     merit = 2.7  \\ 
     rf= 0.30  \\ 
     ry= 0.24  \\ 
  \end{array} $ 
  & 
  $ \begin{array}{c} 
     merit = 2.2  \\ 
     rf= 0.36  \\ 
     ry= 0.35  \\ 
  \end{array} $ 
  & 
  $ \begin{array}{c} 
     merit = 1.8 \\ 
     rf= 0.40  \\ 
     ry= 0.43  \\ 
  \end{array} $ 
\\ \hline 
$r^*=2.0$   & 
  $ \begin{array}{c} 
     merit = 3.0  \\ 
     rf= 0.35  \\ 
     ry= 0.23  \\ 
  \end{array} $ 
  & 
  $ \begin{array}{c} 
     merit = 4.3  \\ 
     rf= 0.24  \\ 
     ry= 0.39  \\ 
  \end{array} $ 
  & 
  $ \begin{array}{c} 
     merit = 3.1   \\ 
     rf= 0.36  \\ 
     ry= 0.45  \\ 
  \end{array} $ 
\\ \hline 
2.5   & 
  $ \begin{array}{c} 
     merit = 0.1  \\ 
     rf= 0.94  \\ 
     ry= 0.02  \\ 
  \end{array} $ 
  & 
  $ \begin{array}{c} 
     merit = 3.2 \\ 
     rf= 0.36  \\ 
     ry= 0.26  \\ 
  \end{array} $ 
  & 
  $ \begin{array}{c} 
     merit = 3.4 \\
     rf= 0.26  \\
     ry= 0.32  \\
  \end{array} $ 
\\ \hline 
\end{tabular}  
{\normalsize \caption{REAL - An\'{a}lise de sensibilidade para PET4} }
}
\end{table}
\normalsize 


\subsection*{Mais Testes Num\'{e}ricos} 

 Tamb\'{e}m testamos os algoritmos REAL e CAL5 com o conjunto de 
exemplos de van Cutsem para detec\c{c}\~{a}o de emerg\^{e}ncias em redes
de  pot\^{e}ncia; para o qual CAL5 teve a melhor performance j\'{a}
publicada [4].   Otimizamos os par\^{a}metros sobre o mesmo grid da
se\c{c}\~{a}o anterior.

 As \'{a}rvores geradas foram testadas de duas maneiras: 
 \begin{enumerate} 
 \item A contagem padr\~{a}o {\it hits} e {\it misses}, i.e.
classifica\c{c}\~{o}es corretas e incorretas.  
 \item Eliminando os exemplos que caem em folhas que n\~{a}o atingem  a
meta m\'{\i}nima de convic\c{c}\~{a}o (para o REAL), ou
probabilidade-confi\^{a}n\c{c}a  (para o CAL5), e tamb\'{e}m eliminando
os exemplos de teste que exibem  um valor de atributos fora do intervalo
do n\'{o} em que \'{e} utilizado na \'{a}rvore  de
classifica\c{c}\~{a}o, e s\'{o} ent\~{a}o contando os erros e acertos
remanecentes,   {\it Hits} e {\it Misses}. 
 \end{enumerate}

\begin{table}[!htb] 
\center{ 
\begin{tabular}{|c||c|c|c|c|c|}
\hline
    {\footnotesize Algoritmo} & $P^*$ & $hit$ & $miss$ & $Hit$ & $Miss$    
    \\ \hline \hline 
    REAL      & $(r,ct)=(1.5,0.20)$ & 
    $241.7$ & $8.3$ & $230.6$ & $4.1$ 
    \\ \hline 
    Cal5      & $(S,\alpha )=(0.65,0.15)$ & 
    $240.6$ & $9.4$ & $236.2$ & $6.9$ 
    \\ \hline 
\end{tabular}
\normalsize
{\normalsize \caption{Par\^{a}metros \'{o}timos para van Cutsem} 
}}
\end{table}
\normalsize 

 Para uma an\'{a}lise de sensibilidade mostramos resultados similares
numa  visinhan\c{c}a dos par\^{a}metros \'{o}timos, nas tabelas 7 e 8.  

\begin{table}[!htb] 
\center{ 
\begin{tabular}{|c||c|c|c|} 
\hline 
  $ \begin{array}{lr} 
         & ct \\ 
      r  &  
  \end{array} $ 
& 
0.15  & $ct^*=0.20$ & 0.25 \\ \hline \hline 
1.0   & 
  $ hits = 240.5 $ 
  & 
  $ hits = 239.0 $ 
  & 
  $ hits = 236.4 $ 
\\ \hline 
$r^*=1.5$   & 
  $ hits = 240.7 $ 
  & 
  $ hits = 241.7 $ 
  & 
  $ hits = 240.3 $ 
\\ \hline 
2.0   & 
  $ hits = 240.1 $ 
  & 
  $ hits = 239.7 $ 
  & 
  $ hits = 239.8 $ 
\\ \hline 
\end{tabular}   
{\normalsize \caption{REAL - An\'{a}lise de sensibilidade para van Cutsem} }
}
\end{table}
\normalsize 

\begin{table}[!htb] 
\center{ 
\begin{tabular}{|c||c|c|c|} 
\hline 
  $ \begin{array}{l r} 
         & \alpha \\ 
      S  &  
    \end{array} $ 
& 
0.10  & ${\alpha}^* = 0.15$ & 0.20 \\ \hline \hline 
0.6   & 
  $ hits = 237.8 $ 
  & 
  $ hits = 238.3 $ 
  & 
  $ hits = 239.2 $ 
\\ \hline 
$S^*=0.65$   & 
  $ hits = 238.5 $ 
  & 
  $ hits = 240.6 $ 
  & 
  $ hits = 239.8 $ 
\\ \hline 
0.7   & 
  $ hits = 239.9 $ 
  & 
  $ hits = 238.8 $ 
  & 
  $ hits = 239.6 $ 
\\ \hline 
\end{tabular}  
{\normalsize \caption{Cal5 - An\'{a}lise de sensibilidade para van Cutsem} }
}
\end{table}
\normalsize 


\section{Conclus\~{o}es e Futuras Pesquisas} 

 Os usu\'{a}rios queriam uma ferramenta de classifica\c{c}\~{a}o que
fornecesse  regras de classifica\c{c}\~{a}o ``compreensiveis'' e
baseadas em atributos  j\'{a} familiares. REAL alcan\c{c}ou estes
objetivos. A medida simplificada  de convic\c{c}\~{a}o para cada
classifica\c{c}\~{a}o foi muito apreciada pelos usu\'{a}rios  quando
tomando decis\~{o}es de mercado em tempo real.  
  REAL mostrou-se mais eficiente que todos os outros algoritmos  TDITD a
que tivemos acesso, e mostrou-se uma ferramenta util para   suporte
\`{a} tomada de decis\~{o}es quanto a aplica\c{c}\~{a}o de
estrat\'{e}gias de  opera\c{c}\~{a}o no mercado de capitais. 

 Resultados preliminares nos levam a crer que as vantagems do REAL 
sobre o CAL5 se acentuam para dados mais ruidosos, mas esta 
afirma\c{c}\~{a}o requer mais testes num\'{e}ricos. No momento estamos
estudando  o comportamento do REAL com variantes do procedimento de 
discretiza\c{c}\~{a}o baseadas em fun\c{c}\~{o}es de custo alternativas,
e tamb\'{e}m  estamos interessados em compara-lo com abordagens
diferentes para  o problema [2].

\section*{Agradecimentos} 

Agradecemos o suporte recebido do DCC-IME-USP -  Departamento de 
Ci\^{e}ncia da Computa\c{c}\~{a}o da Universidade of S\~{a}o Paulo,   do
CNPq -  Conselho Nacional de Desenvolvimento Cient\'{\i}fico  e
Tecnol\'{o}gico,  da FAPESP - Funda\c{c}\~{a}o de Amparo \`{a}  Pesquisa
do Estado  de S\~{a}o Paulo, e da {\it BM\&F} Bolsa de Mercadorias e de
Futuros do  Estado de S\~{a}o Paulo. O software usado para computar os
indicadores  t\'{e}cnicos foi desenvolvido pela Profa. Celma O. Ribeiro,
do Departamento  de Engenharia Industrial da Escola Polit\'{e}cnica da
USP.    

 Somos gratos a Junior Barrera, Alan M. Durham, Fl\'{a}vio S. C. da
Silva, Jacob Zimbarg Sobrinho and Carlos A. B. Pereira, do IME-USP, por
muitos bons coment\'{a}rios, a Wolfgang Mueller, do Fraunhofer Institut,
por um execut\'{a}vel para SUN-Sparc do CAL5, e a Gerd Kock,  do
GMD-FIRST Berlin, por toda a ajuda na Alemanha.

%% file: fig01.tex

\font\thinlinefont=cmr5
\mbox{\beginpicture
\setcoordinatesystem units < 0.700cm, 0.700cm>
\unitlength= 0.700cm
\linethickness=1pt
\setplotsymbol ({\makebox(0,0)[l]{\tencirc\symbol{'160}}})
\setshadesymbol ({\thinlinefont .})
\setlinear
%
%
\linethickness= 0.500pt
\setplotsymbol ({\thinlinefont .})
\setdots < 0.0953cm>
\plot  2.667 21.558  9.874 21.558 /
\linethickness= 0.500pt
\setplotsymbol ({\thinlinefont .})
\setsolid
%
%
\plot 14.002 22.225 13.875 22.733 13.748 22.225 /
\putrule from 13.875 22.733 to 13.875 17.716
\putrule from 13.875 17.716 to 13.875 19.494
\putrule from 13.875 19.494 to 22.193 19.494
%
%
\plot 21.685 19.367 22.193 19.494 21.685 19.622 /
%
%
%
\linethickness= 0.500pt
\setplotsymbol ({\thinlinefont .})
\putrule from 14.446 19.653 to 14.446 19.336
%
%
\linethickness= 0.500pt
\setplotsymbol ({\thinlinefont .})
\putrule from 21.114 19.653 to 21.114 19.336
%
%
\linethickness= 0.500pt
\setplotsymbol ({\thinlinefont .})
%
%
\plot 14.002 22.225 13.875 22.733 13.748 22.225 /
\putrule from 13.875 22.733 to 13.875 17.716
\putrule from 13.875 17.716 to 13.875 19.494
\putrule from 13.875 19.494 to 22.193 19.494
%
%
\plot 21.685 19.367 22.193 19.494 21.685 19.622 /
%
%
%
\linethickness= 0.500pt
\setplotsymbol ({\thinlinefont .})
\putrule from 14.446 19.653 to 14.446 19.336
%
%
\linethickness= 0.500pt
\setplotsymbol ({\thinlinefont .})
\putrule from 21.114 19.653 to 21.114 19.336
%
%
\linethickness= 0.500pt
\setplotsymbol ({\thinlinefont .})
\setdots < 0.0953cm>
\plot 13.875 21.558 21.114 21.558 /
%
%
\linethickness= 0.500pt
\setplotsymbol ({\thinlinefont .})
\plot 21.114 18.605 21.114 21.558 /
%
%
\linethickness= 0.500pt
\setplotsymbol ({\thinlinefont .})
\plot 13.875 18.605 21.114 18.605 /
%
%
\linethickness= 0.500pt
\setplotsymbol ({\thinlinefont .})
\setsolid
%
%
\plot  2.794 22.225  2.667 22.733  2.540 22.225 /
\putrule from  2.667 22.733 to  2.667 17.716
\putrule from  2.667 17.716 to  2.667 19.494
\putrule from  2.667 19.494 to 10.986 19.494
%
%
\plot 10.478 19.367 10.986 19.494 10.478 19.622 /
%
%
%
\linethickness= 0.500pt
\setplotsymbol ({\thinlinefont .})
\putrule from  3.207 19.653 to  3.207 19.336
%
%
\linethickness= 0.500pt
\setplotsymbol ({\thinlinefont .})
\putrule from  9.874 19.653 to  9.874 19.336
%
%
\linethickness= 0.500pt
\setplotsymbol ({\thinlinefont .})
\setdots < 0.0953cm>
\plot  9.874 18.605  9.874 21.558 /
%
%
\linethickness= 0.500pt
\setplotsymbol ({\thinlinefont .})
\plot  2.667 18.605  9.874 18.605 /
%
%
\linethickness= 0.500pt
\setplotsymbol ({\thinlinefont .})
\plot 13.938 20.288 21.145 20.288 /
%
%
\linethickness= 0.500pt
\setplotsymbol ({\thinlinefont .})
\plot  2.667 20.288  9.874 20.288 /
\linethickness= 0.500pt
\setplotsymbol ({\thinlinefont .})
\setsolid
%
%
\plot 14.446 19.494 	14.545 19.513
	14.639 19.530
	14.731 19.546
	14.819 19.562
	14.904 19.577
	14.986 19.591
	15.065 19.605
	15.142 19.617
	15.215 19.629
	15.285 19.640
	15.353 19.650
	15.419 19.660
	15.542 19.677
	15.656 19.692
	15.762 19.703
	15.860 19.712
	15.950 19.719
	16.034 19.723
	16.113 19.725
	16.186 19.724
	16.254 19.722
	16.320 19.717
	16.445 19.698
	16.512 19.682
	16.582 19.663
	16.654 19.641
	16.729 19.617
	16.806 19.589
	16.885 19.560
	16.965 19.528
	17.047 19.495
	17.131 19.460
	17.215 19.425
	17.301 19.388
	17.387 19.351
	17.474 19.314
	17.562 19.276
	17.649 19.239
	17.737 19.203
	17.824 19.168
	17.911 19.133
	17.998 19.101
	18.084 19.070
	18.169 19.041
	18.252 19.015
	18.335 18.991
	18.416 18.970
	18.495 18.952
	18.573 18.938
	18.648 18.928
	18.721 18.922
	18.792 18.920
	18.860 18.923
	18.954 18.933
	19.050 18.950
	19.149 18.972
	19.252 19.002
	19.360 19.039
	19.473 19.083
	19.592 19.135
	19.719 19.196
	19.785 19.230
	19.853 19.266
	19.923 19.304
	19.996 19.345
	20.071 19.388
	20.148 19.433
	20.229 19.481
	20.311 19.532
	20.397 19.585
	20.485 19.641
	20.577 19.699
	20.671 19.760
	20.769 19.824
	20.870 19.891
	20.974 19.961
	21.082 20.034
	/
\linethickness= 0.500pt
\setplotsymbol ({\thinlinefont .})
%
%
\plot  3.207 19.494 	 3.283 19.463
	 3.358 19.433
	 3.429 19.405
	 3.498 19.378
	 3.565 19.352
	 3.630 19.327
	 3.752 19.282
	 3.866 19.242
	 3.973 19.206
	 4.072 19.175
	 4.164 19.149
	 4.251 19.127
	 4.331 19.109
	 4.407 19.095
	 4.478 19.085
	 4.546 19.079
	 4.610 19.077
	 4.731 19.082
	 4.840 19.099
	 4.956 19.127
	 5.079 19.166
	 5.142 19.189
	 5.207 19.215
	 5.272 19.242
	 5.338 19.271
	 5.404 19.302
	 5.471 19.334
	 5.538 19.367
	 5.606 19.402
	 5.673 19.437
	 5.739 19.474
	 5.806 19.511
	 5.872 19.548
	 5.937 19.586
	 6.001 19.624
	 6.125 19.700
	 6.244 19.775
	 6.356 19.847
	 6.459 19.915
	 6.553 19.978
	 6.636 20.034
	 6.701 20.080
	 6.769 20.130
	 6.840 20.185
	 6.916 20.246
	 6.996 20.313
	 7.082 20.386
	 7.174 20.467
	 7.273 20.556
	 7.379 20.654
	 7.493 20.760
	 7.616 20.876
	 7.681 20.938
	 7.748 21.003
	 7.818 21.070
	 7.890 21.140
	 7.965 21.212
	 8.043 21.288
	 8.123 21.367
	 8.206 21.449
	 8.293 21.534
	 8.382 21.622
	/
%
%
\put { \normalsize (a)} [B] at  6.287 17.272
%
%
\put { \normalsize (b)} [B] at 17.621 17.272
%
\put { \normalsize u} [B] at  2.381 21.400
%
%
\put { \normalsize day} [B] at 22.098 18.891
%
%
\put { \normalsize l} [B] at  2.381 18.605
%
%
\put { \normalsize u} [B] at 13.621 21.400
%
%
\put { \normalsize l} [B] at 13.621 18.605
%
%
\put { \small t} [B] at 14.446 19.050
%
%
\put { \normalsize day} [B] at 10.859 18.891
%
%
\put { \small t} [B] at  3.207 19.050
%
%
\put { \small t+d} [B] at  9.874 19.050
%
%
\put { \small t+d} [B] at 21.114 19.050
%
%
\put { \normalsize c} [B] at  2.381 20.130
%
%
\put { \normalsize c} [B] at 13.652 20.130
\linethickness=0pt
\putrectangle corners at  1.619 22.733 and 22.289 17.272
\endpicture}

%% file: cap8.tex
 \chapter{Fundos de Pens\~{a}o}

 A primeira parte deste cap\'{\i}tulo descreve a utiliza\c{c}\~{a}o de
uma ferramenta para a an\'{a}lise  de fluxos de caixa em fundos de
pens\~{a}o no Brasil. Muitos dos fundos de pens\~{a}o  existentes
s\~{a}o do tipo Benef\'{\i}cio Definido (BD), onde o participante 
aposentado ou seus dependentes remanescentes recebem uma renda mensal 
vital\'{\i}cia. 
 O processo estoc\'{a}stico subjacente \'{e} modelado como um processo 
de ramifica\c{c}\~{a}o orientado pelas diversas taxas de falha
dependentes do tempo. 
 Os fluxos de caixa esperados s\~{a}o computados atrav\'{e}s de
fun\c{c}\~{o}es recursivas  que descrevem o processo de
ramifica\c{c}\~{a}o, evitando assim diversas aproxima\c{c}\~{o}es  que
s\~{a}o utilizadas pelos m\'{e}todos atuariais tradicionais. Essas
fun\c{c}\~{o}es  recursivas tamb\'{e}m fornecem um c\'{a}lculo direto da
vari\^{a}ncia do fluxo de  caixa e outras estat\'{\i}sticas.

 As an\'{a}lises e simula\c{c}\~{o}es do passivo atuarial
s\~{a}o usadas como  entradas para a gest\~{a}o de ativos do plano.
 Diversos modelos de otimiza\c{c}\~{a}o, geralmente empregando
programa\c{c}\~{a}o din\^{a}mica e estoc\'{a}stica, s\~{a}o utilizados
com  este objetivo. Estas t\'{e}cnicas s\~{a}o discutidas na segunda 
parte do cap\'{\i}tulo. 

 \section{Passivo Atuarial}

 O principal benef\'{\i}cio para o participante de um plano BD
(benef\'{\i}cio  definido) \'{e} uma renda vital\'{\i}cia durante a
aposentadoria. Antes da  aposentadoria, um membro \'{e} chamado ativo. A
renda na aposentadoria  \'{e} uma fun\c{c}\~{a}o dos sal\'{a}rios ou
contribui\c{c}\~{o}es passadas do participante  enquanto ativo (p.ex.
m\'{e}dia dos \'{u}ltimos per\'{\i}odos). 
 O participante  ativo efetua contribui\c{c}\~{o}es ao plano de
pens\~{a}o, e essas contribui\c{c}\~{o}es  podem ser complementadas via
contribui\c{c}\~{o}es de uma patrocinadora  (p.ex. o empregador ou o
governo). Um participante ativo se tornar\'{a}  inativo quando se
aposentar; a aposentadoria poder\'{a} ser ordin\'{a}ria  (quando o
participante cumpre as car\^{e}ncias de idade e tempo de 
contribui\c{c}\~{a}o) ou por invalidez (p.ex. acidente ou doen\c{c}a).
Um  membro ativo tamb\'{e}m pode desligar-se do plano. 

 O participante tamb\'{e}m pode possuir dependentes (usualmente sua 
fam\'{\i}lia) com direito a uma pens\~{a}o mensal ap\'{o}s a morte do
participante.  Dependentes podem ser vital\'{\i}cios, que receber\~{a}o
uma pens\~{a}o vital\'{\i}cia  (p.ex. esposa/vi\'{u}va, filhos
portadores de defici\^{e}ncia), ou tempor\'{a}rios,  que receber\~{a}o a
pens\~{a}o por um tempo limitado (p.ex. filhos normais at\'{e}  a
maioridade aos 21 anos). A pens\~{a}o de cada dependente \'{e} uma
fra\c{c}\~{a}o da  renda do participante. Um benef\'{\i}cio adicional
pago \`{a} fam\'{\i}lia no momento  do falecimento do participante -
denominado pec\'{u}lio - pode tamb\'{e}m estar  dispon\'{\i}vel. 

 Diversas restri\c{c}\~{o}es e corre\c{c}\~{o}es \cite{Alb 93},
\cite{Bor 92},  \cite{Bow 97}, \cite{Day 94}, \cite{DeF 91}, \cite{Pan
92} aumentam a  complexidade deste modelo b\'{a}sico, como por exemplo: 

 - A aposentadoria e os demais benef\'{\i}cios por ela definidos podem
ser  corrigidos por um \'{\i}ndice de infla\c{c}\~{a}o de longo prazo,
ou ainda ser  reajustados pelo sal\'{a}rio de um participante ativo de
mesmo n\'{\i}vel funcional  do aposentado. 

 - As car\^{e}ncias para aposentadoria ordin\'{a}ria podem ser baseadas
na idade e  tempo de servi\c{c}o do participante, e tamb\'{e}m baseadas
nas condi\c{c}\~{o}es determinadas  pelo estatuto do plano ou pela
legisla\c{c}\~{a}o governamental vigente, ambas  mut\'{a}veis ao longo
do tempo. 

 - Os participantes podem receber uma aposentadoria b\'{a}sica do Estado
(p.ex.  INSS), sendo obriga\c{c}\~{a}o do plano complementar esta
aposentadoria, at\'{e} atingir  os Benef\'{\i}cio Definido (BD) pelo
estatuto.

 - Mudan\c{c}as de h\'{a}bitos sociais e de legisla\c{c}\~{a}o podem
alterar o status dos  dependentes legais (p.ex. concubinas e filhos
gerados fora do matrim\^{o}nio). 

 - Participantes desligados podem reivindicar o resgate de suas
contribui\c{c}\~{o}es  (ou tamb\'{e}m as da patrocinadora) corrigidas
pelos \'{\i}ndices de infla\c{c}\~{a}o ou de  investimentos financeiros.

\section{Grafos e Formula\c{c}\~{a}o Recursiva}

 Um processo de ramifica\c{c}\~{a}o \'{e} descrito por um grafo, onde
cada v\'{e}rtice (ou  n\'{o}) corresponde a um estado, e cada arco (ou
aresta) conectando dois v\'{e}rtices  corresponde a uma poss\'{\i}vel
transi\c{c}\~{a}o de estados. Nos processos atuariais que  estamos
estudando, um estado \'{e} caracterizado pela idade do participante,
tempo  de servi\c{c}o, sal\'{a}rio, fam\'{\i}lia, etc. Uma
transi\c{c}\~{a}o \'{e} caracterizada por sua  probabilidade, bem como
pelos benef\'{\i}cios e contribui\c{c}\~{o}es que a transi\c{c}\~{a}o 
implica. Em geral, \'{e} conveniente tratar os valores dos
benef\'{\i}cios e  contribui\c{c}\~{o}es como fra\c{c}\~{o}es do
benef\'{\i}cio principal (aposentadoria), ou  alguma  outra unidade
adimensional. 

 O valor esperado de uma vari\'{a}vel aleat\'{o}ria (p.ex.
benef\'{\i}cios ou contribui\c{c}\~{o}es)  para um certo participante,
num dado per\'{\i}odo, \'{e} a soma ponderada dos valores  das
vari\'{a}veis aleat\'{o}rias em todas as transi\c{c}\~{o}es
poss\'{\i}veis naquele per\'{\i}odo:  \[E(X(t)) = \sum_{j \in W(t)}
Pr(j)*x(j),\] onde $W$ \'{e} o conjunto de todas as transi\c{c}\~{o}es
poss\'{\i}veis,  $x(j)$ o valor da vari\'{a}vel  aleat\'{o}ria na
transi\c{c}\~{a}o $j$, e $Pr(j)$ a probabilidade da transi\c{c}\~{a}o
$j$. O fluxo  esperado daquela vari\'{a}vel aleat\'{o}ria \'{e} a
s\'{e}rie dos seus valores esperados no  futuro (per\'{\i}odos
subsequentes, usualmente anos). A descri\c{c}\~{a}o dos processos de 
ramifica\c{c}\~{a}o na forma de grafos fornece uma formula\c{c}\~{a}o
algor\'{\i}tmica recursiva  para o c\'{a}lculo de todos esses fluxos de
caixa.

\subsection*{Grafo do Participante Aposentado}

 O estado de um participante aposentado possui sua idade,
benef\'{\i}cios e a lista  de dependentes. Suponhamos que um
participante aposentado possui no m\'{a}ximo um  dependente
vital\'{\i}cio (esposa). Se o participante e sua esposa est\~{a}o ambos
vivos no instante $t$, o participante estar\'{a}, no instante $t+1$, em
um dos quatro estados poss\'{\i}veis, dependendo da sobreviv\^{e}ncia ou
n\~{a}o do mesmo e de sua esposa: sejam  $(x,y)$ as idades do
participante e de sua esposa no instante $t$. No instante $t+1$,  eles
podem alcan\c{c}ar os estados $(x+1,y+1)$, $(x+1,\sim)$, $(\sim,y+1)$ ou
$(\sim,\sim)$, onde o til  $(\sim)$ representa \'{o}bito. A
probabilidade de cada uma das quatro transi\c{c}\~{o}es \'{e} dada  pela
taxa de mortalidade, $h(a)$, nas respectivas idades: 
 \begin{eqnarray}
 & & Pr(t, (x, y), (x+1, y+1)) = (1-h(x))*(1-h(y)); \\
 & & Pr(t, (x, y), (x+1, ~)) = (1-h(x))*h(y); \\
 & & Pr(t, (x, y), (~, y+1)) = h(x)*(1-h(y)); \\
 & & Pr(t, (x, y), (~, ~)) = h(x)*h(y).
 \end{eqnarray}

 Um participante aposentado deixa o sistema (plano previdenci\'{a}rio)
quando cessam  todos os fluxos de caixa por ele gerados, possivelmente
muito tempo depois de  seu pr\'{o}prio falecimento. As folhas da
\'{a}rvore ramificada do participante  correspondem ao estado terminal
$(\sim,\sim)$. Assume-se que os dependentes tempor\'{a}rios  (filhos)
sempre sobrevivem at\'{e} sua maioridade.  

 Como mencionamos na se\c{c}\~{a}o 2, podem ocorrer m\'{u}ltiplos
dependentes vital\'{\i}cios.  Uma possibilidade seria incorporar os
dependentes vital\'{\i}cios m\'{u}ltiplos  diretamente no processo de
ramifica\c{c}\~{a}o, a um alto custo computacional.  Ocorre que os
estatutos dos planos BD levam em considera\c{c}\~{a}o somente o
n\'{u}mero  total de dependentes remanescentes ap\'{o}s o falecimento do
participante. Isto  permite uma simplifica\c{c}\~{a}o significativa.
Modelamos os dependentes vital\'{\i}cios  no processo de
ramifica\c{c}\~{a}o na aposentadoria como um dependente vital\'{\i}cio 
virtual correspondente ao \'{u}ltimo dependente vital\'{\i}cio real
remanescente. No  ap\^{e}ndice \ref{apen5} apresentamos um pequeno
programa Matlab (denominado {\em depvital.m})  para calcular a
distribui\c{c}\~{a}o  conjunta de probabilidades de sobreviv\^{e}ncia de tal
dependente virtual. \'{e} f\'{a}cil generalizar  o procedimento para
tr\^{e}s ou mais dependentes vital\'{\i}cios. Os fluxos de caixa dos 
dependentes vital\'{\i}cios que falecem antes do \'{u}ltimo sobrevivente
podem ent\~{a}o ser  modelados como fluxos de caixa independentes.

\begin{figure}[hbt] 
 \centerline{\pdfximage width 15.0cm height 20.0cm 
 {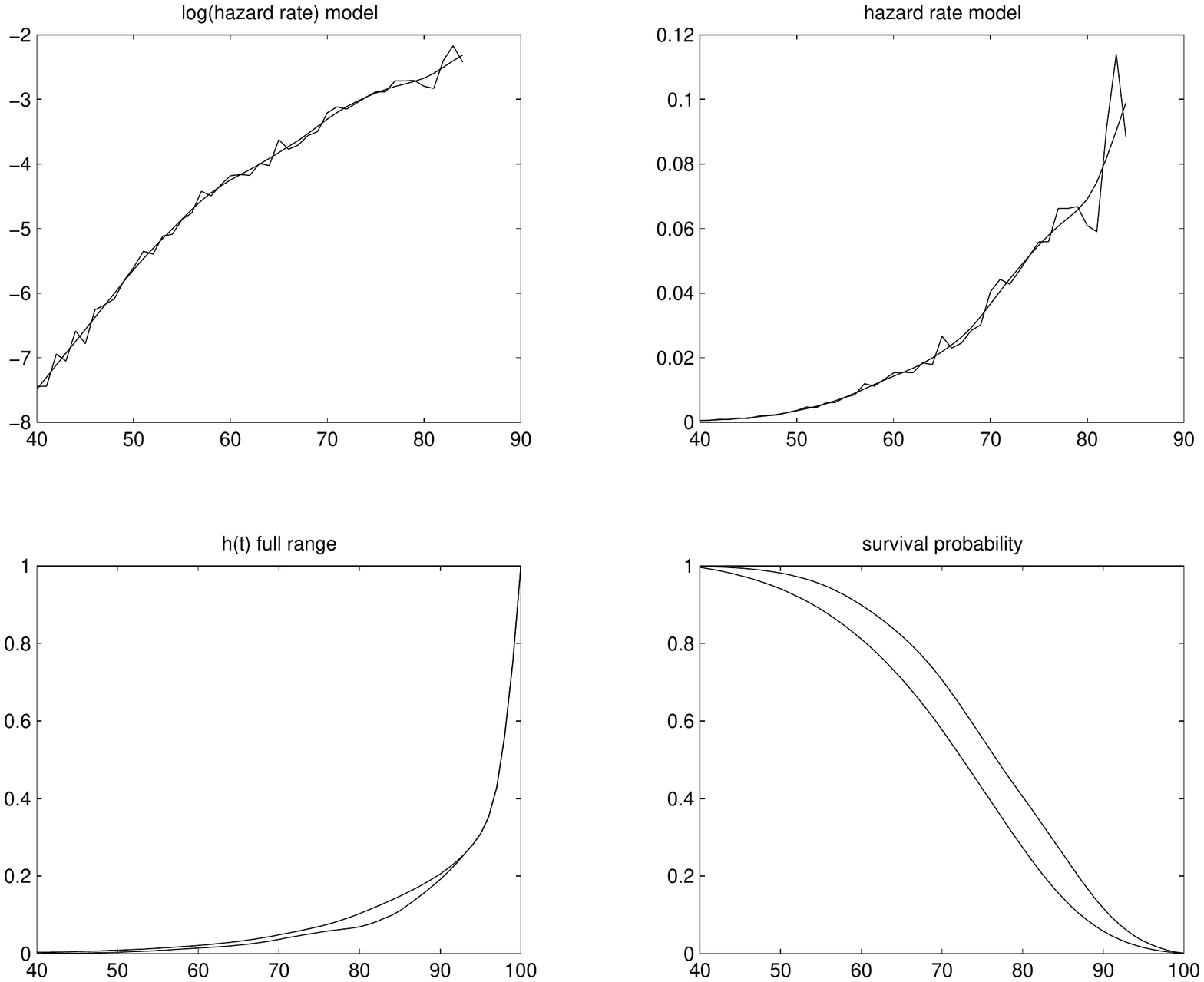}\pdfrefximage\pdflastximage} 
 \caption{ \label{fig8a} 
          Estat\'{\i}stica de ordem para modelagem de 
          dependentes vital\'{\i}cios m\'{u}ltiplos.}
 \end{figure}

 A modelagem precisa dos dependentes vital\'{\i}cios m\'{u}ltiplos
possui um impacto  significativo sobre os fluxos de caixa  de
benef\'{\i}cios daqueles dependentes  (tipicamente, 30{\%}). Uma vez que
essa situa\c{c}\~{a}o est\'{a} se tornando cada vez mais  frequente, tal
an\'{a}lise cuidadosa \'{e} recomend\'{a}vel. A figura \ref{fig8a}
apresenta  distribui\c{c}\~{o}es comparativas de sobreviv\^{e}ncia
calculadas pelo programa do  ap\^{e}ndice \ref{apen5} (as linhas
pontilhadas representam as taxas de mortalidade do dependente mais 
velho e do mais jovem, respectivamente; as linhas s\'{o}lidas
representam as taxas de mortalidade  do primeiro e do \'{u}ltimo
dependente a falecer, respectivamente). Em alguns casos a 
estat\'{\i}stica de ordem \'{e} aproximada pelas taxas  de
sobreviv\^{e}ncia do dependente vital\'{\i}cio mais jovem. A partir dos
gr\'{a}ficos  \ref{fig8b}(a) a \ref{fig8b}(d), pode-se perceber que esta
aproxima\c{c}\~{a}o pode ser  bastante enganosa.

\section*{Grafo do Participante Ativo}

 O estado de um participante ativo possui sua idade, tempo de plano,
tempo  de servi\c{c}o, escolaridade, sal\'{a}rio, etc. Enquanto ativo,
\'{e} dif\'{\i}cil obter  uma rela\c{c}\~{a}o confi\'{a}vel de
dependentes; assim, sup\~{o}e-se que os participantes  ativos possuem
uma fam\'{\i}lia-padr\~{a}o, baseada em dados estat\'{\i}sticos  e  no  
perfil  geral  dos  participantes. Se  um participante est\'{a} ativo no
 instante $t$, com idade $a$ e tempo de servi\c{c}o $e$, ele
alcan\c{c}ar\'{a} no instante  $t+1$ um dos quatro poss\'{\i}veis
estados, dependendo do mesmo ainda estar no  plano, ativo, vivo, e
v\'{a}lido. Falecimento, invalidez e desligamento s\~{a}o  riscos
competitivos com fun\c{c}\~{o}es de probabilidade (condicionais na
n\~{a}o  ocorr\^{e}ncia dos demais riscos precedentes) $hd(a)$, $hb(a)$
e $hw(e)$. Assim,  as probabilidades de transi\c{c}\~{a}o (exceto para a
aposentadoria previs\'{\i}vel,  obtida na maturidade) para falecimento,
invalidez, desligamento e  perman\^{e}ncia em atividade s\~{a}o,
respectivamente: 
 \begin{eqnarray}
 & & hd(a), hb(a), hw(e), and 	 \\
 & & (1-hd(a))*(1-hb(a))*(1-hw(e)).
 \end{eqnarray}

 Se o participante se desliga do plano, recebe um montante baseado em
suas  contribui\c{c}\~{o}es passadas. Se este morre ou se torna
inv\'{a}lido, ele entra em  regime de aposentadoria/pens\~{a}o
prematuramente. O processo de ramifica\c{c}\~{a}o  do membro ativo \'{e}
 portanto limitado ao ramo principal correspondente \`{a}
sobreviv\^{e}ncia sobre  todos os riscos, uma estrutura mais parecida
com um ``bambu'' do que com  uma ``\'{a}rvore''. As folhas do bambu
correspondem ao estado de desligamento,  ou \`{a} raiz de um processo de
ramifica\c{c}\~{a}o para aposentadoria.

 \section{T\'{a}buas Biom\'{e}tricas e Outros Ajustes}

\begin{figure}[hbt] 
 \centerline{\pdfximage width 15.0cm height 20.0cm 
 {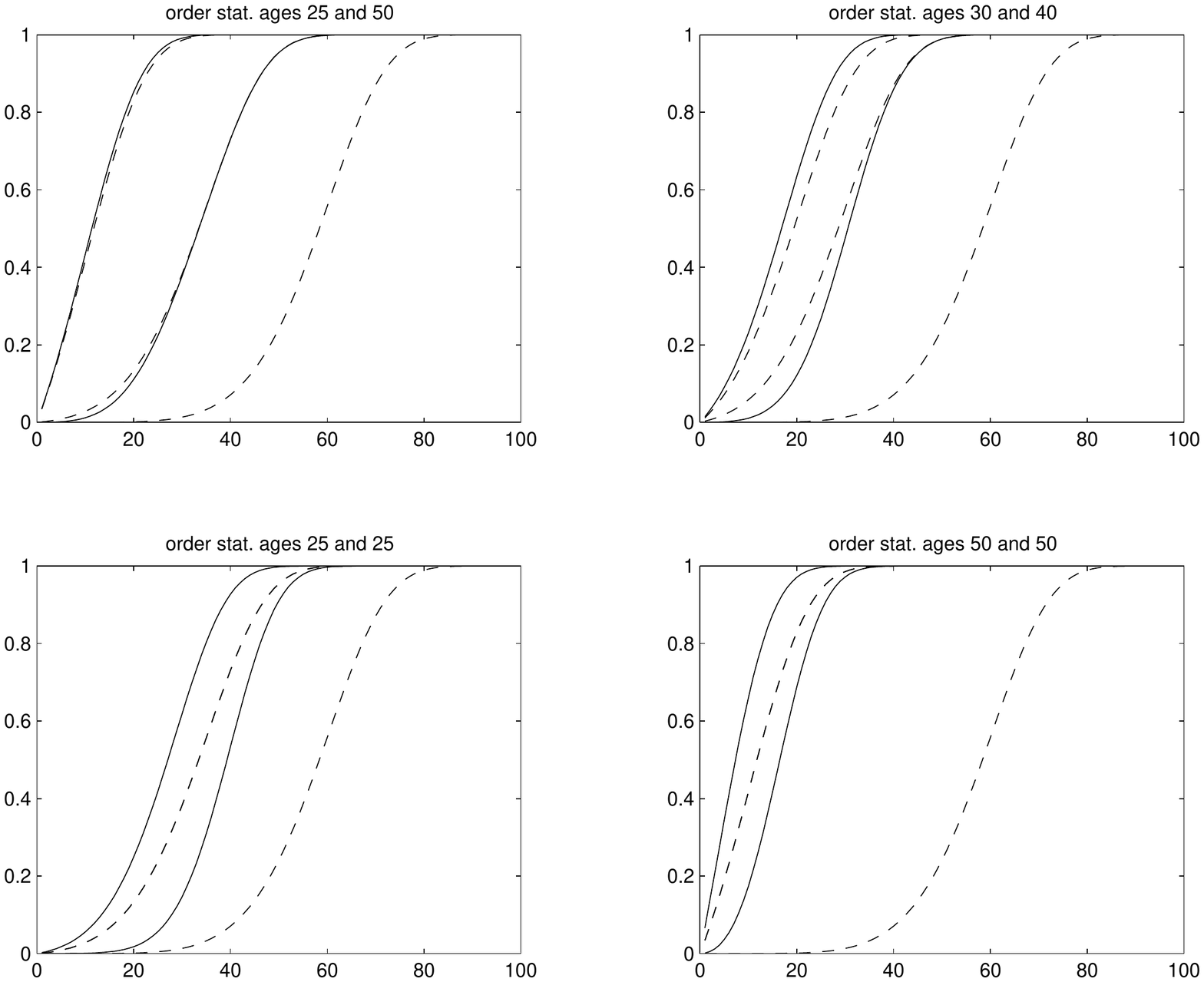}\pdfrefximage\pdflastximage} 
 \caption{ \label{fig8b}
          Compara\c{c}\~{o}es entre t\'{a}buas oficiais 
          e t\'{a}buas obtidas a partir de uma popula\c{c}\~{a}o.}
 \end{figure}

 {\em T\'{a}buas biom\'{e}tricas:} T\'{a}buas de taxas de mortalidade
s\~{a}o dispon\'{\i}veis em  diversos pa\'{\i}ses. Uma das t\'{a}buas
biom\'{e}tricas mais utilizadas no Brasil  \'{e} a EB-7. Todavia, uma
popula\c{c}\~{a}o espec\'{\i}fica, como a massa de  participantes de uma
certa companhia ou plano previdenci\'{a}rio, pode  divergir
significativamente das m\'{e}dias nacionais. Para planos
espec\'{\i}ficos,  alguns com at\'{e} duzentos mil participantes, foi
identificada a necessidade  de ajustar essas tabelas. A figura
\ref{fig8b} apresenta algumas compara\c{c}\~{o}es  dessas
distribui\c{c}\~{o}es de sobreviv\^{e}ncia (as linhas pontilhadas
representam as  frequ\^{e}ncias observadas na popula\c{c}\~{a}o; as
linhas s\'{o}lidas representam as probabilidades ajustadas pelo modelo;
as linhas tracejadas representam as taxas fornecidas  pela t\'{a}bua
biom\'{e}trica EB7-1975). Como \'{e} usual em atu\'{a}ria, estabelecemos
 um corte, limitando a idade individual a um m\'{a}ximo (p.ex. 100
anos). O impacto  desses ajustes sobre o passivo do plano \'{e}
consider\'{a}vel, da ordem de at\'{e} 20{\%}. 

 Utilizamos um modelo GMDH (Group Method Data Handling) polinomial,
usando as  t\'{a}buas dispon\'{\i}veis (informa\c{c}\~{a}o a priori) e o
hist\'{o}rico populacional  (falecimentos observados e censurados)
\cite{Far 84}. Os modelos GMDH polinomiais  possuem complexidade
vari\'{a}vel e diversos par\^{a}metros. O melhor modelo foi  selecionado
automaticamente por uma heur\'{\i}stica de busca controlada pelo 
crit\'{e}rio PSE (Predicted Squared Error) \cite{Bar 84}. O objetivo do
crit\'{e}rio PSE \'{e}  minimizar erros sobre dados ainda n\~{a}o
observados, buscando o equil\'{\i}brio  entre erros sobre dados de
treinamento e uma penalidade de overfit. O  modelo final foi validado
usando m\'{e}todos computacionalmente intensivos  de reamostragem
estat\'{\i}stica \cite{Goo 99} \cite{Urb 94}.

 {\em Corre\c{c}\~{a}o de unidade:} Durante a modelagem de uma
transi\c{c}\~{a}o entre per\'{\i}odos  consecutivos, de $t$ a $t+1$,
(dependendo de como o  modelo \'{e} implementado),  premissas
n\~{a}o-realistas podem ser introduzidas, por exemplo: uma
transi\c{c}\~{a}o  de falecimento pode implicar que o participante
falece no primeiro (ou  \'{u}ltimo) m\^{e}s do ano. Para corrigir tal
dicotomia booleana (0-1), podemos  supor que o falecimento ocorre no
m\^{e}s central, e utilizar um fator de  corre\c{c}\~{a}o 6/12 = 1/2, ou
que o falecimento ocorre no dia central do m\^{e}s  central, e utilizar
um fator de corre\c{c}\~{a}o $(6+1/2)/12 = 13/24$, e assim por  diante.
Esses fatores de corre\c{c}\~{a}o s\~{a}o denominados corre\c{c}\~{o}es
fracionais (ou  corre\c{c}\~{o}es de discretiza\c{c}\~{a}o) \cite{Bow
97}. Seu impacto sobre  os c\'{a}lculos finais \'{e}  normalmente
pequeno, mas eles s\~{a}o importantes para preservar a consist\^{e}ncia 
do modelo. 

{\em Crescimento Salarial:} A renda (ou sal\'{a}rio) de um participante
ativo, que \'{e}  a base para seus benef\'{\i}cos, \'{e} supostamente
crescente ao longo de sua vida  profissional. A renda usualmente aumenta
com o tempo, mas tal crescimento  possui um efeito de satuar\c{c}\~{a}o.
Diversos modelos ajustam-se bem a esta  situa\c{c}\~{a}o \cite{Pas 90},
como os modelos Exponencial Modificado, de Gompertz e  Log\'{\i}stico
(Pearl): 
 \begin{eqnarray}
 M(t) &=& a + b*exp(-c*t) ; 	\\		
 G(t) &=& exp(a + b*exp(-c*t)) ; \\		
 L(t) &=& a / (1 + b*exp(-c*t)) .
 \end{eqnarray}

\section{Programa\c{c}\~{a}o Estoc\'{a}stica}

%% file: cap9.tex
\chapter{Portf\'{o}lios Mistos Contendo Op\c{c}\~{o}es}

 Vamos construir neste cap\'{\i}tulo portf\'olios formados por ativos
fundamentais e op\c c\~oes europ\'eias sobre estes ativos. Para
selecionar portf\'olios eficientes analisando-se m\'edia e vari\^ancia
de retornos, usaremos o modelo de Markowitz do capitulo anterior.
Deduziremos express\~ oes analiticas para a esperan\c ca e covari\^
ancia entre taxas de retornos de ativos e op\c c\~oes, dados necess\'
arios para a constru\c c\~ ao do modelo baseado nos dois primeiros
momentos centrais da carteira.

\section{O mercado de op\c c\~ oes}

Op\c c\~ oes de compra e de venda s\~ ao contratos  referentes
a direitos e deveres de compra e venda adquiridos sobre um determinado ativo,
o objeto do contrato.

Nas op\c c\~ oes de compra (ou ``call'' ) o titular adquire o direito de
comprar um determinado ativo numa data futura, a um determinado pre\c co
estipulado no contrato. E para adquirir tal direito paga-se um pr\^ emio ao
vendedor da op\c c\~ ao. Nas op\c c\~ oes de venda (ou `` put'' ) o titular
tem o direito de vender o ativo-objeto a um pre\c co pr\' e-determinado em
uma data futura. 

Assim as partes integrantes de um contrato de op\c c\~ ao englobam:

\begin{itemize}

\item Titular: aquele que det\' em o direito de comprar ou vender o ativo do
  contrato;

\item Emissor (ou lan\c cador): aquele que oferece o direito de compra ou de venda
  estipulado na op\c c\~ ao;

\item ativo-objeto: ativo que poder\' a ser negociado no contrato;

\item pre\c co de exerc\'{\i}cio: pre\c co pelo qual o ativo-objeto poder\' a ser
  comprado ou vendido pelo titular;

\item pr\^ emio: valor pago ao lan\c cador pelo titular para adquirir o direito de
compra ou venda do ativo-objeto.

\item data de exerc\'{\i}cio: data na qual o titular poder\' a negociar o
  ativo-objeto do contrato.

\end{itemize}

A data de exerc\'{\i}cio em que se pode negociar o ativo d\' a margem a dois
tipos de contratos de op\c c\~ oes: a op\c c\~ ao americana e a europ\'
eia. Na op\c c\~ ao europ\' eia o titular pode exercer o seu direito somente
na data estipulada no contrato. J\' a na op\c c\~ ao americana o titular tem a
liberdade de exercer seu direito a qualquer tempo at\' e a data de
exerc\'{\i}cio do contrato.

O pr\^ emio pago ao emissor do contrato, que \' e o pre\c co da op\c c\~ ao,
leva em considera\c c\~ ao a oferta e a damanda do ativo, o pre\c co de
exerc\'{\i}cio, a cota\c c\~ ao de mercado \` a vista do ativo, a volatilidade
de pre\c cos, entre outros. Os ativos pass\'{\i}veis de negocia\c c\~ ao em
op\c c\~ oes s\~ ao definidos pela institui\c{c}\~ao que as negocia, assim como
o m\^ es de expira\c c\~ ao do contrato e os pre\c cos de exerc\'{\i}cio. 

Em particular, para as op\c c\~ oes sobre a\c c\~ oes, a Bolsa de Valores de
S\~ ao Paulo, por exemplo, fixa a data de vencimento do contrato na
segunda-feira mais pr\' oxima do dia 15 dos meses pares. E os pre\c cos de
exerc\'{\i}cio s\~ ao fixados pr\' oximos ao pre\c co de mercado da a\c c\~ ao
objeto, acima e abaixo, e deixando o pr\^ emio ser um par\^ ametro livre para
negocia\c c\~ ao em preg\~ ao.

As op\c c\~ oes n\~ ao s\~ ao negociadas diretamente pelos seus compradores e
vendedores, e sim por corretoras que atuam em nome de seus clientes  dentro de
normas estabelecidas pela Bolsa.

Vamos fixar a nota\c c\~ ao para os elementos formadores de um contrato de
op\c c\~ ao europ\'eia:

\begin{itemize}

\item $T$: data de exerc\'{\i}cio do contrato;

\item $S(t)$: pre\c co do ativo-objeto na data t;

\item $K_c(t), K_p(t)$: pre\c co de exerc\'{\i}cio de uma op\c c\~ao de compra
  e venda, respectivamente, na data t;

\item $C(t), P(t)$: pre\^emio de  uma op\c c\~ao de compra
  e venda, respectivamente, na data t.

\end{itemize}

Vamos supor que as taxas de corretagens que fazem parte da negocia\c
c\~ ao de op\c c\~ oes j\' a est\~ ao contabilizados no pre\c co de
exerc\'{\i}cio do contrato. Desta forma, uma op\c c\~ ao de compra europ\' eia
pode ser vista como um derivativo que paga ao seu titular, na data de
exerc\'{\i}cio (T), a quantia

\[
X(T) = \max(S(T) - K_c(t_0),0) \quad ( \geq 0 )
\]
onde $t_0$ \'{e} a data de assinatura do contrato.

Se o pre\c co de mercado do ativo-objeto for maior que o pre\c co de
exerc\'{\i}cio do contrato, ou seja, $S(T)>K_c(t_0)$, ent\~ ao o titular pode
comprar o ativo por $K_c(t_0)$, em exerc\'{\i}cio do seu direito, e vend\^
e-lo no mercado por $S(T)$, ganhando a diferen\c ca $S(T) - K_c(t_0)$. Esta
diferen\c ca pode n\~ ao cobrir o dinheiro gasto no pr\^ emio pago ao
emissor.                            

Se, por outro lado, $S(T)  \leq  K_c(t_0)$ perder-se-\' a $S(T) - K_c(t_0) \ \
\ (\leq  0)$ se o direito de compra for exercido. Neste caso, deixa-se o
contrato expirar sem exerc\'{\i}cio do direito de compra.

Analogamente, uma op\c c\~ ao de venda europ\' eia pode ser vista como um
derivativo cujo valor na data de exerc\'{\i}cio \' e dado por

\[
X_{T} = \max(K_p(t_0) - S(T),0) \quad ( \geq 0 )
\]

Vamos especificar as taxas de retorno para o caso de a\c c\~ oes e op\c c\~
oes europ\' eias, com que vamos trabalhar para definir os momentos de primeira
e segunda ordem de carteiras mistas.

Daqui por diante omitires a palavra europ\' eia quando falarmos de op\c c\~
oes, ficando impl\'{\i}cito que \' e sobre este tipo de contrato que estamos
falando.

Para um ativo cujo pre\c co de mercado na data t \' e dado por $S(t)$, a taxa
de retorno $r_s$ no per\'{\i}odo $\tau = T-t$  $ (T > t)$ ser\'{a} dada por

\begin{equation}
r_s = \frac{S(T)-S(t)}{S(t)} \label{eq:fig41}
\end{equation}

Para as op\c{c}\~{o}es de compra e venda, as taxas de retorno ser\~{a}o dadas por

\begin{equation}
r_c = \frac{\max(S(T)-K_c(t_0),0)-C(t_0)}{C(t_0)} \quad \hbox{(``call'' )}
\label{eq:fig42} 
\end{equation}

\begin{equation}
r_p = \frac{\max(K_p(t_0)-S(T),0)-P(t_0)}{P(t_0)} \quad \hbox{(``put'' )}
\label{eq:fig43} 
\end{equation}

onde $t_0$ \'e a data de assinatura dos contratos.

\section{Um Exemplo de Portf\'{o}lio}

Vamos considerar um portf\'olio  formado por dois ativos, chamados de ativo A
e ativo B, e por op\c c\~ oes de compra e venda que t\'em como ativos-objetos os
pr\'oprios ativos A e B. Vamos supor tamb\'em que as datas de assinatura $(t_0)$ e
as datas de exerc\'{\i}cio $(T)$ s\~ao as mesmas para todas as op\c c\~oes. 

Denote por $S_a$ e $S_b$ os pre\c{c}os dos ativos A e B, respectivamente. Os
retornos logar\'{\i}tmicos de A e B, extrapolados para a data $T$, s\~ao dados por

\vspace{5mm}
\begin{equation}
x=ln\left( \frac{S_a(T)}{S_a(t_0)}\right) \cong N(\mu_a,\sigma_a^2)
\label{eq:fig44} 
\end{equation}

e  

\begin{equation}
y=ln\left( \frac{S_b(T)}{S_b(t_0)}\right) \cong N(\mu_b,\sigma_b^2)
\label{eq:fig45} 
\end{equation}

\vspace{5mm}

e as fun\c c\~oes densidade de probabilidade dadas por 

\vspace{5mm}

\begin{equation}
f(x)=\frac{1}{\sigma_a\sqrt{2\pi}}\exp{\left\{-\frac{(x-\mu_a)^2}{2\sigma_a^2}\right\}}
\label{eq:fig46} 
\end{equation}

\begin{equation}
f(y)=\frac{1}{\sigma_b\sqrt{2\pi}}\exp{\left\{-\frac{(y-\mu_b)^2}{2\sigma_b^2}\right\}}
\label{eq:fig47} 
\end{equation}

\vspace{5mm}
Sendo $\rho$ o coeficiente de correla\c c\~ ao entre os ativos A e B, a fun\c c\~ao
distribui\c c\~ao conjunta \' e dada por

$$
\begin{array}{rl}
f(x,y) &=\frac{1}{\sigma_a\sigma_b\sqrt{1-\rho^2}}
\exp{\left\{-\frac{1}{2(1-\rho^2)}
\left[
\left(\frac{x-\mu_a}{\sigma_a}\right)^2
-2\rho\frac{x-\mu_a}{\sigma_a}\frac{y-\mu_b}{\sigma_b}
+\left(\frac{y-\mu_b}{\sigma_b}\right)^2
\right]
\right\}} 
\end{array} 
$$

\[
(-\infty < x,y < +\infty)
\]

Precisamos colocar as taxas de retorno dadas em ~\ref{eq:fig41},
~\ref{eq:fig42} e ~\ref{eq:fig43} em fun\c c\~ao das vari\'{a}veis $x$ e $y$, pois
\' e destas vari\' aveis que temos informa\c c\~oes sobre a densidade de
probabilidade.

De ~\ref{eq:fig44} e ~\ref{eq:fig45} temos 

\begin{equation}
S_a(T) =  S_a(t_0)exp(x) \quad  \hbox{e} \quad 
S_b(T) =  S_b(t_0)exp(y)
\label{eq:fig48}
\end{equation}
 
\vspace{5mm}
Substituindo ~\ref{eq:fig48} em ~\ref{eq:fig41} teremos
\vspace{5mm}

\begin{equation}
r_{s_a} = exp(x)-1 \quad  \hbox{e} \quad 
r_{s_b} = exp(y)-1
\label{eq:fig49}
\end{equation}

onde $r_{s_a}$ e  $r_{s_b}$ s\~ao as taxas de retorno dos ativos A e B,
respectivamente.

Para as op\c c\~oes fixaremos a letra $K$ para denotar pre\c co de
exerc\'{\i}cio do contrato, colocando um \'{\i}ndice superior \`a letra
para diferenciar calls de puts, e um \'{\i}ndice inferior para indicar o
ativo-objeto a que se refere. Por exemplo, $K_a^c(t) $ denota o pre\c co
de exerc\'{\i}cio de uma call sobre o ativo A, na data t, e $K_b^p(t)$
indica o pre\c co de exerc\'{\i}cio de uma put sobre o ativo B, na data
t. 

Os pr\^emios das op\c c\~oes de compra ser\~ao denotados pela letra $C$,
com um \'{\i}ndice inferior indicando o ativo a que se refere. O  mesmo vale
para as op\c coes de venda, trocando a letra $C$ pela letra $P$. Por
exemplo, $C_a(t)$ indica o pr\^{e}mio de uma call sobre o ativo A na data t,
e $P_b(t)$ indica o pr\^{e}mio de uma put sobre o ativo B, na data t.

Portanto, as taxas de retorno das op\c coes de compra (assinadas na data
$t_0$) que t\^em A e B como ativos-objeto ser\~ao dadas por

\vspace{5mm}

\begin{equation}
\begin{array}{rl}
r_{c_a} =& (\max(S_a(t_0)e^x-K_a^c(t_0),0)-C_a(t_0))/C_a(t_0)\\
&\\
r_{c_b} =& (\max(S_b(t_0)e^y-K_b^c(t_0),0)-C_b(t_0))/C_b(t_0)\label{eq:fig410}
\end{array}
\end{equation}

\vspace{5mm}

Do mesmo modo para as puts,

\vspace{5mm}

\begin{equation}
\begin{array}{rl}
r_{p_a} =& (\max(K_a^p(t_0)-S_a(t_0)e^x,0)-P_a(t_0))/P_a(t_0)\\ 
&\\
r_{p_b} =&  (\max(K_b^p(t_0)-S_b(t_0)e^y,0)-P_b(t_0))/P_b(t_0) 
\label{eq:fig411}
\end{array}
\end{equation}

\vspace{5mm}

Quando n\~ ao houver d\'uvidas quanto as datas que estamos usando,
omitiremos a letra $t$ dos pr\^ emios, dos pre\c cos de exerc\'{\i}cio e
das cota\c c\~oes dos ativos para n\~ ao sobrecarregar a nota\c c\~ ao.

Uma outra forma de escrever as express\~oes ~\ref{eq:fig410} e
~\ref{eq:fig411} \'{e} definir os dom\'{\i}nios das v.a. $x$ e $y$ nos quais o
termo $ \max(S_0exp(x)-K,0)$ \'{e} positivo 

\vspace{5mm}
\begin{equation}
\begin{array}{rl}
\max(S\exp(x)-K,0) > 0 &\Longleftrightarrow x >
\ln\left(\frac{K}{S}\right)\\
&\\
\max(K-S\exp(x),0) > 0 &\Longleftrightarrow x <
\ln\left(\frac{K}{S}\right).\\
\end{array}\label{eq:fig412}
\end{equation}
\vspace{5mm}

Considerando  ~\ref{eq:fig412} nas express\~ oes ~\ref{eq:fig410} e
~\ref{eq:fig411}, redefinimos as taxas de retorno das op\c c\~oes por

$$
\begin{array}{rl}
r_{c_a} = &  \left\{
             \begin{array}{ll}
                \frac{1}{C_a(t_0)}\,(S_a(t_0)e^x-K_a^c(t_0)-C_a(t_0))\quad
                    & \hbox{se} \quad
                    \ln{\left(\frac{K_a^c(t_0)}{S_a(t_0)}\right)} < x <
                    +\infty \\ 
                    & \\        
                    0 & \quad \quad \quad \quad \hbox{c.c.}\\
                  \end{array}
            \right.\\
&\\
r_{c_b} = &  \left\{
             \begin{array}{ll}
                \frac{1}{C_b(t_0)}\,(S_b(t_0) e^y-K_b^c(t_0)-C_b(t_0))\quad
                    & \hbox{se} \quad
                    \ln{\left(\frac{K_b^c(t_0)}{S_b(t_0)}\right)} < y <
                    +\infty \\
                    & \\ 
                    0 & \quad \quad \quad \quad \hbox{c.c.}\\
                  \end{array}
            \right.\\
\end{array}
$$

\begin{equation}
\begin{array}{rl}
r_{p_a} = &  \left\{
             \begin{array}{ll}
                \frac{1}{P_a(t_0)}\,(K_a^p(t_0)-S_a(t_0)e^x-P_a(t_0))\quad
                    & \hbox{se} \quad
                    -\infty < x < \ln{\left(\frac{K_a^p(t_0)}{S_a(t_0)}\right)}\\ 
                    & \\
                    0 & \quad \quad \quad \quad \hbox{c.c.}\\
                  \end{array}
            \right.\\
&\\
r_{p_b} = &  \left\{
             \begin{array}{ll}
                \frac{1}{P_b(t_0)} \, (K_b^p(t_0)-S_b(t_0)e^y-P_b(t_0))\quad
                    & \hbox{se} \quad
                    -\infty < y < \ln{\left(\frac{K_b^p(t_0)}{S_b(t_0)}\right)} \\
                   & \\ 
                   0 & \quad \quad \quad \quad \hbox{c.c.}\\
                  \end{array}
            \right.\\
\end{array}\label{eq:fig413}
\end{equation}

\vspace{5mm}

O conhecimento de uma estimativa do retorno logar\'{\i}tmico para
os ativos, no per\'{\i}odo $[t,T]$ considerado, \' e necess\' ario para
podermos definir as taxas de retorno dadas em ~\ref{eq:fig413}. Estes dados
podem ser obtidos atrav\' es de an\' alise de s\' eries hist\' oricas de pre\c
cos.

Observe que, para o processo de pre\c cos do ativo-objeto, que estamos supondo
n\~ao ser um derivativo, a taxa de retorno $r=\frac{S_T}{S_0}$, $S_T > 0$ e
$S_0>0$, n\~ao assume valor zero e nem valor negativo, estando, assim, o
retorno logar\'{\i}tmico $x=ln{r}$ bem definido, independente da oscila\c c\~ao de
pre\c cos dos ativos. Por outro lado, os retornos de op\c c\~oes, definidos pelas
f\'ormulas ~\ref{eq:fig410} e  ~\ref{eq:fig411} podem eventualmente assumir
valor zero ou negativo, dependendo do comportamento dos pre\c cos do ativo, do
pre\c{c}o de exerc\'{\i}cio e do pr\^emio. Por isso os retornos
logar\'{\i}tmicos $\ln{r_c}$  e  $\ln{r_p}$ para op\c{c}\~oes n\~ao est\~ao bem
definidos, podendo haver indetermina\c c\~oes quando $r_c \rightarrow 0$ e
$r_p \rightarrow 0$. Para evitar inconsist\^encias e padronizar o tipo de
retorno utilizado para op\c{c}\~oes e ativo-objeto adotamos os retornos dados
pelas express\~oes ~\ref{eq:fig41}, ~\ref{eq:fig42} e ~\ref{eq:fig43}.

A seguir derivaremos as express\~ oes anal\'{\i}ticas dos {\it valores esperados} das taxas dadas em ~\ref{eq:fig49} e ~\ref{eq:fig413} e das covari\^ ancias entre  elas, para o caso deste portf\' olio exemplo. Elas ser\~ ao usadas para  calcular o retorno esperado $E(r)$ e a matriz de covari\^ ancia $Cov(r)$ de  portf\' olios.

\section{Esperan\c{c}a e Covari\^{a}ncia do Retorno
         de Op\c{c}\~{o}es}

Vamos voltar ao portf\'{o}lio formado pelos ativos A e B e por op\c{c}\~{o}es sobre estes
ativos. Manteremos a nota\c{c}\~{a}o introduzida na se\c{c}\~{a}o anterior, com 
a exce\c{c}\~{a}o de que omitiremos o termo $t_0$ (data de assinatura das calls e das
puts). Portanto os pre\c{c}os das a\c{c}\~{o}es ser\~{a}o dados por $S_a$ e $S_b$, os pr\^{e}mios
das op\c{c}\~{o}es por $C_a$, $C_b$, $P_a$ e $P_b$ e os pre\c{c}os de exerc\'{\i}cio por $K_a^c$,
$K_a^p$, $K_b^c$ e $K_b^p$.  

Vale lembrar ainda que dada uma v.a cont\'{\i}nua $x$ em $\Re$ e sua respectiva fun\c{c}\~{a}o
densidade de probabilidade $p(x)$, a esperan\c{c}a de uma fun\c{c}\~{a}o $g(x)$ \'{e} dada por

\[
E[g(x)]=\int_{-\infty}^{+\infty} g(x)p(x)\,dx
\]

Em virtude das repeti\c{c}\~{o}es dos c\'{a}lculos alg\'{e}bricos que envolvem a dedu\c{c}\~{a}o das
express\~{o}es das covari\^{a}ncias, detalhamos esses c\'{a}lculos em algumas delas. As
demais usam racioc\'{\i}nio an\'{a}logo.
\vspace{10mm}

{\bf Esperan\c{c}a das taxas de retorno }

\vspace{5mm}
Nota\c{c}\~{a}o: $ \hbox{erf(a)}=\int_{a}^{+\infty} \exp(-\psi^2)\,d\psi$

\vspace{5mm}

\begin{eqnarray*}
\mu_{sa} &=& E[r_{s_a}] = \int_{-\infty}^{+\infty} \exp(x) f(x) \,dx - 1\\
& &\\
&=& \frac{1}{\sigma_{a} \sqrt{2\pi}} \int_{-\infty}^{+\infty} \exp\left\{x -
\frac{(x-\mu_{a})^{2}}{2\sigma_{a}^{2}}\right\} \,dx -1 \\
& &\\
&\stackrel{\mathrm{1''}}{=}& \frac{\exp\left( 
\frac{1}{2}\sigma_{a}^{2} + \mu_{a}\right)}{\sqrt{\pi}}
\int_{-\infty}^{+\infty} \exp\left(-t^{2}\right) \,dt -1 \\
&=& \exp{\left(\mu_a+\frac{1}{2}\sigma_a^2\right)}-1
\end{eqnarray*}

\begin{eqnarray*}
\mu_{ca} &=& E[r_{c_a}] = \frac{1}{C_a}\int_{\ln\left(\frac{K_a^c}{S_a}\right)}^{+\infty} 
(S_a\exp(x)-K_a^c) f(x) \,dx -1 \\
& &\\
&=& \frac{S_a}{C_a\sigma_{a} \sqrt{2\pi}}
\int_{\ln\left(\frac{K_a^c}{S_a}\right)}^{+\infty}
 \exp\left\{x -
\frac{(x-\mu_{a})^{2}}{2\sigma_{a}^{2}}\right\} \,dx\\
& & - \frac{K_a^c}{C_a\sigma_{a} \sqrt{2\pi}}\int_{\ln\left(\frac{K_a^c}{S_a}\right)}^{+\infty}
 \exp\left\{\frac{(x-\mu_{a})^{2}}{2\sigma_{a}^{2}}\right\}\,dx  -1 \\
& &\\
&\stackrel{\mathrm{1''}}{=}&
\frac{S_a}{2C_a}
\exp{\left(\mu_a+\frac{1}{2}\sigma_a^2\right)}
\left\{1-erf\left[\frac{1}{\sigma_a\sqrt{2}}\left(\ln\left(\frac{K_a^c}{S_a}\right)-\mu_a-\sigma_a^2\right)
\right] 
\right\}\\
& &\\
& &-\frac{1}{2C_a}(K_a^c) \left\{1-erf\left[\frac{1}{\sigma_a\sqrt{2}}\left(\ln\left(\frac{K_a^c}{S_a}\right)-\sigma_a^2\right)
\right] 
\right\}-1\\
\end{eqnarray*}

\vspace{5mm}
Mudan\c{c}a de vari\'{a}vel usada:

\[
1'': \left\{
        \begin{array}{ll}
             t &= \frac{1}{\sigma_a\sqrt{2}}(x-\mu_a-\sigma_a^2)\\
             & \\ 
             w &= \frac{1}{\sigma_a \sqrt{2}}(x-\sigma_a)\\
        \end{array}
\right.
\]

\vspace{5mm}

\begin{eqnarray*}
\mu_{pa} &=& E[r_{p_a}] = \frac{1}{P_a}\int_{-\infty}^{\ln\left(\frac{K_a^p}{S_a}\right)} 
(K_a^p-S_a\exp(x)) f(x) \,dx -1\\
& &\\
&\stackrel{\mathrm{1''}}{=}&
\frac{1}{2P_a}(K_a^p)
\exp{\left(\mu_a+\frac{1}{2}\sigma_a^2\right)}
\left\{1+erf\left[\frac{1}{\sigma_a\sqrt{2}}\left(\ln\left(\frac{K_a^p}{S_a}\right)-\mu_a-\sigma_a^2\right)
\right] 
\right\}\\
& &\\
& &
-\frac{S_a}{2P_a}\left\{1+erf\left[\frac{1}{\sigma_a\sqrt{2}}\left(\ln\left(\frac{K_a^p}{S_a}\right)-\sigma_a^2\right)
\right] 
\right\} -1 \\
\end{eqnarray*}

\vspace{10mm}

{\bf Covari\^{a}ncias}

No lugar de calcularmos as express\~{o}es de covari\^{a}ncias para as taxas de retorno
simples dadas em ~\ref{eq:fig49},~\ref{eq:fig410} e ~\ref{eq:fig411}
deduziremos express\~{o}es para o retorno simples (= $ \frac{\hbox{pre\c{c}o
    final}}{\hbox{pre\c{c}o inicial}} $) de a\c{c}\~{o}es e op\c{c}\~{o}es,
sendo a diferen\c{c}a entre as duas grandezas apenas uma constante, isto \'{e},

$$
\hbox{Taxa de retorno simples =  Retorno simples - 1.}
$$

O resultado num\'{e}rico de covari\^{a}ncias n\~{a}o muda, visto que para duas v.a. X e Y e
constantes $k_1$ e $k_2$ vale a igualdade
$$
cov(X+k_1,Y+k_2)=cov(X,Y).
$$

\begin{itemize}
\item $Cov(r_{s_a},r_{s_b})$ \\

\begin{eqnarray*}
cov(r_{s_a},r_{s_a})& =& var(r_{s_a}) \\
&=& E_{2x}-E_{x}^2\\
\end{eqnarray*}

\begin{eqnarray*}
E_{2x}&=& \int_{-\infty}^{+\infty} exp(2x)f(x)\,dx \\
& &\\
& =&\frac{1}{\sigma_a\sqrt{2\pi}}\int_{-\infty}^{+\infty}  \exp\left\{2x -
\frac{(x-\mu_{a})^{2}}{2\sigma_{a}^{2}}\right\} \,dx \\
& &\\
&\stackrel{\mathrm{2''}}{=}&
\frac{1}{\sqrt{\pi}}\exp{\left(2\mu_a+2\sigma_a^2\right)}
\int_{-\infty}^{+\infty} \exp(-\psi^2)\,d\psi\\
& &\\
&=& \exp{\left(2\mu_a+2\sigma_a^2\right)}\\ 
\end{eqnarray*}
\begin{eqnarray*}
E_x& =& \exp{\left(\mu_a+0.5\sigma_a^2\right)}\\
\end{eqnarray*}

onde 2'' indica a mudan\c{c}a de vari\'{a}vel

\[
\psi = \frac{x-\mu_a-2\sigma_a^2}{\sigma_a\sqrt{2}}
\]

Logo,

\[
var(r_{s_a})=\exp\left(2\mu_a+\sigma_a^2 \right)\left(\exp(\sigma_a^2)-1\right)
\]

\item $Cov(r_{s_a},r_{s_b})$

\begin{eqnarray*}
cov(r_{s_a},r_{s_b})& = &E[r_{s_a}r_{s_b}]- E[r_{s_a}] E[r_{s_b}]\\
& = & E_{xy}-E_{x} E_{y}\\
\end{eqnarray*}

sendo

\begin{eqnarray*}
E_{xy}&=& \int_{-\infty}^{+\infty}\int_{-\infty}^{+\infty}
f(x,y)\exp(x+y)\,dy \,dx\\
& &\\
&=& \frac{1}{2\pi\sigma_a\sigma_b\sqrt{1-\rho^2}}
\int_{-\infty}^{+\infty}\int_{-\infty}^{+\infty}
\exp(x+y)\\
& & \cdot \exp\left\{-\frac{1}{2(1-\rho^2)} 
\left[(\frac{x-\mu_a}{\sigma_a})^2
-2\rho\frac{x-\mu_a}{\sigma_a}\frac{y-\mu_b}{\sigma_b} +
(\frac{y-\mu_b}{\sigma_b})^2\right]\right\} \,dy \,dx\\
& &\\
&\stackrel{\mathrm{3''}}{=}& 
\frac{\exp(\mu_{a}+\mu_{b})}{2\pi\sqrt{1-\rho^{2}}}\int_{-\infty}^{+\infty}\int_{-\infty}^{+\infty}
\exp \left\{\sigma_{a}u+\sigma_{b}v
-\frac{1}{2(1-\rho^{2})}(u^{2} - 2\rho uv + v^{2})\right\} \,dv \,du\\
& &\\
&\stackrel{\mathrm{4''}}{=}&
\frac{1}{\pi\,\sqrt{2}}\exp\left(\mu_a+\mu_b+\frac{1}{2}\sigma_b^2(1-\rho^2)
\right)\\
& &\cdot \int_{-\infty}^{+\infty}\left(
\int_{-\infty}^{+\infty}\exp(-\psi^2)\,d\psi \right)
\exp(-\frac{1}{2}(u^2-2u(\sigma_a +\sigma_b\rho)))\,du \\
& & \\
&\stackrel{\mathrm{5''}}{=}&
\frac{1}{\sqrt{\pi}} 
\exp\left(\mu_a+\mu_b+\frac{1}{2}\sigma_b^2(1-\rho^2)+ 
\frac{1}{2}(\sigma_a+\sigma_b\rho)^2
\right) \int_{-\infty}^{+\infty}\exp(-\psi^2) \,d\psi \\
& &\\
&=&\exp\left(\mu_a+\mu_b+\frac{1}{2}(\sigma_b^2+\sigma_a^2)+\sigma_a\sigma_b\rho
\right) 
\end{eqnarray*}

\begin{eqnarray*}
E_{x} &=& \int_{-\infty}^{+\infty}\int_{-\infty}^{+\infty}
f(x,y)\exp(x)\,dy \,dx\\
& & \\
&\stackrel{\mathrm{3''}}{=}& 
\frac{\exp(\mu_{a})}{2\pi\sqrt{1-\rho^{2}}}\int_{-\infty}^{+\infty}\int_{-\infty}^{+\infty}
\exp \left\{\sigma_{a}u -\frac{1}{2(1-\rho^{2})}(u^{2} - \rho uv +
v^{2})\right\} \,dv \,du\\ 
& &\\
&\stackrel{\mathrm{6''}}{=}&
\frac{1}{\pi\,\sqrt{2}}exp(\mu_a)\int_{-\infty}^{+\infty}\left(
\int_{-\infty}^{+\infty}\exp(-\psi^2)\,d\psi \right)
\exp(-\frac{1}{2}(u^2-2u\sigma_a))\,du \\
& & \\
&\stackrel{\mathrm{7''}}{=}&
\frac{1}{\sqrt{\pi}} 
\exp\left(\mu_a+\frac{1}{2}\sigma_a^2 
\right) \int_{-\infty}^{+\infty}\exp(-\psi^2) \,d\psi \\
& &\\
&=&\exp\left(\mu_a+\frac{1}{2}(\sigma_a^2)
\right) \\
\end{eqnarray*}
\begin{eqnarray*}
E_y&=& \exp(\mu_b+\frac{1}{2}\sigma_b^2)
\end{eqnarray*}

Mudan\c{c}as de vari\'{a}vel:

\[
3'': \left\{
        \begin{array}{ll}
             u &= (x-\mu_a)/\sigma_a\\ 
             & \\ 
             v &= (y-\mu_b)/\sigma_b
        \end{array}
\right.
\]

\[
4'':\quad \psi=\frac{1}{\sqrt{2(1-\rho^2)}}(v-u\rho-\sigma_b(1-\rho^2))
\]

\[
5'': \quad w=\frac{1}{\sqrt{2}}(u-\sigma_a-\sigma_b\rho)
\]

\[
6'':\quad \psi=\frac{1}{\sqrt{2(1-\rho^2)}}(v-u\rho)
\]

\[
7'': \quad w=\frac{1}{\sqrt{2}}(u-\sigma_a)
\]

Portanto,

\[
cov(r_{s_a},r_{s_b})=\exp(\mu_a+\mu_b+\frac{1}{2}(\sigma_a^2+\sigma_v^2))
\left(\exp(\sigma_a\sigma_b\rho)-1)\right)
\]

\item $Cov(r_{s_a},r_{c_a})$

\begin{eqnarray*}
cov(r_{s_a},r_{c_a})& = &
\frac{1}{C_a}\left\{S_aE_{2x}-K_a^cE_x\right\}-\mu_{sa}\mu_{ca} \\ 
\end{eqnarray*}

onde

\begin{eqnarray*}
E_{2x} &= & \frac{1}{2}\exp(2\mu_a+2\sigma_a^2)\left\{1-erf\left[\frac{1}{\sigma_a\sqrt{2}}\left(\ln\left(\frac{K_a^c}{S_a}\right)-\mu_a-2\sigma_a^2\right)
\right] 
\right\}\\
\end{eqnarray*}
\begin{eqnarray*}
E_{x} &= & \frac{1}{2}\exp(\mu_a+\frac{1}{2}\sigma_a^2)\left\{1-erf\left[\frac{1}{\sigma_a\sqrt{2}}\left(\ln\left(\frac{K_a^c}{S_a}\right)-\mu_a-\sigma_a^2\right)
\right] 
\right\}\\
\end{eqnarray*}

\item $Cov(r_{s_a},r_{p_a})$

\begin{eqnarray*}
cov(r_{s_a},r_{p_a})& =
& \frac{1}{P_a}\left\{-S_aE_{2x}+K_a^pE_x\right\}-\mu_{sa}\mu_{pa} \\
\end{eqnarray*}

onde

\begin{eqnarray*}
E_{2x} &= & \frac{1}{2}\exp(2\mu_a+2\sigma_a^2)\left\{1+erf\left[\frac{1}{\sigma_a\sqrt{2}}\left(\ln\left(\frac{K_a^p}{S_a}\right)-\mu_a-2\sigma_a^2\right)
\right] 
\right\}\\
\end{eqnarray*}
\begin{eqnarray*}
E_{x} &= & \frac{1}{2}\exp(\mu_a+\frac{1}{2}\sigma_a^2)\left\{1+erf\left[\frac{1}{\sigma_a\sqrt{2}}\left(\ln\left(\frac{K_a^p}{S_a}\right)-\mu_a-\sigma_a^2\right)
\right] 
\right\}\\
\end{eqnarray*}

\item $Cov(r_{s_a},r_{c_b})$

\begin{eqnarray*}
cov(r_{s_a},r_{c_b})& =&
\frac{1}{C_b}\left\{S_bE_{xy}-E_{x}K_b^c\right\}-\mu_{cb}
\bar{E}_x-\frac{\mu_{sa}}{C_b}\left\{S_bE_y-K_b^cE_1 \right\}+\mu_{sa}\mu_{cb}
\end{eqnarray*}

onde

\begin{eqnarray*}
R_b^c &=& \frac{1}{\sigma_b}(\ln\left(\frac{K_b^c}{S_b}\right)-\mu_b)\\
\end{eqnarray*}
\begin{eqnarray*}
E_{xy} &= &
\frac{1}{2}\exp(\mu_a+\mu_b+\frac{1}{2}(\sigma_a^2+\sigma_b^2)+\rho\sigma_a\sigma_b)\left\{1-erf\left[\frac{1}{\sqrt{2}}(R_b^c-\sigma_b-\sigma_a\rho)
\right]\right\}\\
\end{eqnarray*}
\begin{eqnarray*}
E_{x} &= & \frac{1}{2}\exp(\mu_a+\frac{1}{2}\sigma_a^2)\left\{1-erf\left[\frac{1}{\sqrt{2}}\left(R_b^c-\rho\sigma_a^2\right)
\right] 
\right\}\\
\end{eqnarray*}
\begin{eqnarray*}
E_{y} &= & \frac{1}{2}\exp(\mu_b+\frac{1}{2}\sigma_b^2)\left\{1-erf\left[\frac{1}{\sqrt{2}}\left(R_b^c-\sigma_b\right)
\right] 
\right\}\\
\end{eqnarray*}
\begin{eqnarray*}
\bar{E}_{x} &= & \frac{1}{2}\exp(\mu_a+\frac{1}{2}\sigma_a^2)\\
\end{eqnarray*}
\begin{eqnarray*}
E_1 &= & \frac{1}{2}\left\{1-erf\left[\frac{1}{\sqrt{2}}\left(R_b^c\right)
\right] 
\right\}\\
\end{eqnarray*}

\item $Cov(r_{s_a},r_{p_b})$

\begin{eqnarray*}
cov(r_{s_a},r_{p_b})& =&
\frac{1}{P_b}\left\{-S_bE_{xy}+E_{x}K_b^p\right\}-\mu_{pb}
\bar{E}_x-\frac{\mu_{sa}}{P_b}\left\{-S_bE_y+K_b^cE_1 \right\}+\mu_{sa}\mu_{pb}
\end{eqnarray*}

onde

\begin{eqnarray*}
R_b^p &=& \frac{1}{\sigma_b}(\ln\left(\frac{K_b^p}{S_b}\right)-\mu_b)\\
\end{eqnarray*}
\begin{eqnarray*}
E_{xy}  &= &
\frac{1}{2}\exp(\mu_a+\mu_b+\frac{1}{2}(\sigma_a^2+\sigma_b^2)+\rho\sigma_a\sigma_b)\left\{1+erf\left[\frac{1}{\sqrt{2}}(R_b^c-\sigma_b-\sigma_a\rho)
\right]\right\}\\
\end{eqnarray*}
\begin{eqnarray*}
E_{x} &= & \frac{1}{2}\exp(\mu_a+\frac{1}{2}\sigma_a^2)\left\{1+erf\left[\frac{1}{\sqrt{2}}\left(R_b^p-\rho\sigma_a\right)
\right] 
\right\}\\
\end{eqnarray*}
\begin{eqnarray*}
E_{y} &= & \frac{1}{2}\exp(\mu_b+\frac{1}{2}\sigma_b^2)\left\{1+erf\left[\frac{1}{\sqrt{2}}\left(R_b^p-\sigma_b\right)
\right] 
\right\}\\
\end{eqnarray*}
\begin{eqnarray*}
\bar{E}_{x} &= & \frac{1}{2}\exp(\mu_a+\frac{1}{2}\sigma_a^2)\\
\end{eqnarray*}
\begin{eqnarray*}
E_1 &= & \frac{1}{2}\left\{1+erf\left[\frac{1}{\sqrt{2}}\left(R_b^c\right)
\right] 
\right\}\\
\end{eqnarray*}

\item $Cov(r_{c_a},r_{p_a})$

\[
cov(r_{c_a},r_{p_a}) = \left\{
        \begin{array}{ll}
          \frac{1}{C_aP_a}\left\{-S_a^2E_{2x}+S_aE_x(K_a^c+K_a^p)-K_a^cK_a^pE_1\right\}-\mu_{ca}\mu_{pa},  & \hbox{se} \quad K_a^c > K_a^p\\ 
             & \\ 
            -\mu_{ca}\mu_{pa}, &  \hbox{se} \quad K_a^c \leq K_a^p
        \end{array}
\right.
\]

onde

\begin{eqnarray*}
E_{2x} &= & \frac{1}{2}\exp(2\mu_a+2\sigma_a^2)\\
& &\cdot \left\{
erf\left[\frac{1}{\sigma_a\sqrt{2}}\left(\ln\left(\frac{K_a^p}{S_a}\right)- 
\mu_a-2\sigma_a^2\right)\right]-erf\left[\frac{1}{\sigma_a\sqrt{2}}\left(\ln\left(\frac{K_a^c}{S_a}\right)- 
\mu_a-2\sigma_a^2\right)\right]
\right\}\\
\end{eqnarray*}
\begin{eqnarray*}
E_{x} &= & \frac{1}{2}\exp(\mu_a+\frac{1}{2}\sigma_a^2)\\
& & \cdot \left\{
erf\left[\frac{1}{\sigma_a\sqrt{2}}\left(\ln\left(\frac{K_a^p}{S_a}\right)-\mu_a-\sigma_a^2\right)
\right]-erf\left[\frac{1}{\sigma_a\sqrt{2}}\left(\ln\left(\frac{K_a^c}{S_a}\right)-\mu_a-\sigma_a^2\right)
\right] 
\right\}\\
\end{eqnarray*}
\begin{eqnarray*}
E_1 &= & \frac{1}{2}\left\{
erf\left[\frac{1}{\sigma_a\sqrt{2}}\left(\ln\left(\frac{K_a^p}{S_a}\right)-\mu_a\right)
\right] -erf\left[\frac{1}{\sigma_a\sqrt{2}}\left(\ln\left(\frac{K_a^c}{S_a}\right)-\mu_a\right)
\right] 
\right\}\\
\end{eqnarray*}

\item $Cov(r_{c_a},r_{c_b})$

\begin{eqnarray*}
cov(r_{c_a},r_{c_b})& =&\frac{1}{C_aC_b}\left\{S_aS_bE_{xy}-S_aE_xK_b^c-S_bE_yK_a^c\right\}\\
& &-\frac{\mu_{cb}}{C_a}(S_a\bar{E}_x-K_a^c\bar{E}_1)-\frac{\mu_{ca}}{C_b}(S_b\bar{\bar{E}}_y-K_b^c\bar{\bar{E}}_1)+\mu_{ca}\mu_{cb}\\
\end{eqnarray*}

onde

\begin{eqnarray*}
R_a^c &=& \frac{1}{\sigma_b}(\ln\left(\frac{K_a^c}{S_b}\right)-\mu_a)\\
\end{eqnarray*}
\begin{eqnarray*}
R_b^c &=& \frac{1}{\sigma_b}(\ln\left(\frac{K_b^c}{S_b}\right)-\mu_b)\\
\end{eqnarray*}
\begin{eqnarray*}
E_{xy} &= &
\frac{1}{2\sqrt{2\pi}}\exp(\mu_a+\mu_b+\frac{1}{2}(\sigma_a^2+\sigma_b^2)+\rho\sigma_a\sigma_b)\\
& & \cdot \int_{R_a^c}^{+\infty}
\left\{1-erf\left[\frac{1}{\sqrt{2(1-\rho^2)}}(R_b^c-\sigma_b(1-\rho^2)-\rho u)
\right]\right\}\\
& & \cdot exp(-\frac{1}{2}(u-\sigma_a-\rho\sigma_b)^2)\,du\\
\end{eqnarray*}
\begin{eqnarray*}
E_{x} &= & \frac{1}{2\sqrt{2\pi}}\exp(\mu_a+\frac{1}{2}\sigma_a^2)\\
& & \cdot\int_{R_a^c}^{+\infty} 
\left\{1-erf\left[\frac{1}{\sqrt{2(1-\rho^2)}}\left(R_b^c-\rho u\right)
\right] 
\right\}\\
& & \cdot \exp(-\frac{1}{2}(u-\sigma_a)^2)\,du\\
\end{eqnarray*}
\begin{eqnarray*}
E_{y} &= & \frac{1}{2\sqrt{2\pi}}\exp(\mu_b+\frac{1}{2}\sigma_b^2)\\
& & \cdot \int_{R_a^c}^{+\infty} 
 \left\{1-erf\left[\frac{1}{\sqrt{2(1-\rho^2)}}\left(R_b^c-\rho u- \sigma_b(1-\rho^2)\right)
\right]
\right\}\\
& & \cdot \exp(-\frac{1}{2}(u-\sigma_b\rho)^2)\,du\\ 
\end{eqnarray*} 
\begin{eqnarray*}
E_1 &= & \frac{1}{2\sqrt{2\pi}}\int_{R_a^c}^{+\infty} 
\left\{1-erf\left[\frac{1}{\sqrt{2(1-\rho^2)}}\left(R_b^c-\rho u \right)\right]\right\} 
\exp(-\frac{1}{2}u^2)
\,du\\
\end{eqnarray*}
\begin{eqnarray*}
\bar{E}_{x} &= & \frac{1}{2}\exp(\mu_a+\frac{1}{2}\sigma_a^2)\left\{1-erf\left[\frac{1}{\sqrt{2}}\left(R_a^c-\sigma_a\right)
\right] 
\right\}\\
\end{eqnarray*}
\begin{eqnarray*}
\bar{E}_{1} &= &
\frac{1}{2}\left\{1-erf\left[\frac{R_a^c}{\sqrt{2}}\right]\right\}\\ 
\end{eqnarray*}
\begin{eqnarray*}
\bar{\bar{E}}_{y} &= & \frac{1}{2}\exp(\mu_b+\frac{1}{2}\sigma_b^2)\left\{1-erf\left[\frac{1}{\sqrt{2}}\left(R_b^c-\sigma_b\right)
\right] 
\right\}\\ 
\end{eqnarray*}
\begin{eqnarray*}
\bar{\bar{E}}_{1} &= & \frac{1}{2}\left\{1-erf\left[\frac{R_b^c}{\sqrt{2}}\right] 
\right\}\\
\end{eqnarray*}

\item $Cov(r_{c_a},r_{p_b})$

\begin{eqnarray*}
cov(r_{c_a},r_{p_b})& =&
\frac{1}{C_aP_b}\left\{-S_aS_bE_{xy}+S_aE_xK_b^p+S_bE_yK_a^c-K_a^cK_b^pE_1\right\}\\
&
&-\frac{\mu_{pb}}{C_a}(S_a\bar{E}_{x}-K_a^c\bar{E}_1)-\frac{\mu_{ca}}{P_b}(-S_b\bar{\bar{E}}_{y}+K_b^p\bar{\bar{E}}_1)+\mu_{ca}\mu_{pb}\\
\end{eqnarray*}

onde

\begin{eqnarray*}
R_a^c &=& \frac{1}{\sigma_b}(\ln\left(\frac{K_a^c}{S_b}\right)-\mu_a)\\
& &\\
& &\\
R_b^p &=& \frac{1}{\sigma_b}(\ln\left(\frac{K_b^p}{S_b}\right)-\mu_b)\\
\end{eqnarray*}

Os valores de $E_{xy}$, $E_{x}$, $E_{y}$ , $E_1 $, $ \bar{E}_x$ ,$\bar{E}_1$
,$\bar{\bar{E}}_x$ e $\bar{\bar{E}}_1$ s\~{a}o iguais \`{a}s
express\~{o}es de $cov(r_{c_a},r_{c_b})$, trocando o sinal da fun\c{c}\~{a}o erf e
trocando tamb\'{e}m $R_b^c$ por $R_b^p$.

\item $Cov(r_{p_a},r_{p_b})$

\begin{eqnarray*}
cov(r_{p_a},r_{p_b})& =&
\frac{1}{P_aP_b}\left\{S_aS_bE_{xy}-S_aE_xK_b^p-S_bE_yK_a^p+K_a^pK_b^pE_1\right\}\\
&
&+\frac{\mu_{pb}}{P_a}(S_a\bar{E}_{x}-K_a^p\bar{E}_1)+\frac{\mu_{pa}}{P_b}(S_b\bar{\bar{E}}_{y}-K_b^p\bar{\bar{E}}_1)+\mu_{pa}\mu_{pb}\\
\end{eqnarray*}

onde

\begin{eqnarray*}
R_a^p &=& \frac{1}{\sigma_b}(\ln\left(\frac{K_a^p}{S_a}\right)-\mu_a)\\
\end{eqnarray*}
\begin{eqnarray*}
R_b^p &=& \frac{1}{\sigma_b}(\ln\left(\frac{K_b^p}{S_b}\right)-\mu_b)\\ 
\end{eqnarray*}
\begin{eqnarray*}
E_{xy} &= &
\frac{1}{2\sqrt{2\pi}}\exp(\mu_a+\mu_b+\frac{1}{2}(\sigma_a^2+\sigma_b^2)+\rho\sigma_a\sigma_b)\\
& & \cdot \int_{-\infty}^{R_a^p}
\left\{1+erf\left[\frac{1}{\sqrt{2(1-\rho^2)}}(R_b^p-\sigma_b(1-\rho^2)-\rho u)
\right]\right\}\\
& & \cdot exp(-\frac{1}{2}(u-\sigma_a-\rho\sigma_b)^2)\,du\\ 
\end{eqnarray*}
\begin{eqnarray*}
E_{x} &= & \frac{1}{2\sqrt{2\pi}}\exp(\mu_a+\frac{1}{2}\sigma_a^2)\\
& & \cdot\int_{-\infty}^{R_a^p}
\left\{1+erf\left[\frac{1}{\sqrt{2(1-\rho^2)}}\left(R_b^p-\rho u\right)
\right] 
\right\}\\
& & \cdot \exp(-\frac{1}{2}(u-\sigma_a)^2)\,du\\
\end{eqnarray*}
\begin{eqnarray*}
E_{y} &= & \frac{1}{2\sqrt{2\pi}}\exp(\mu_b+\frac{1}{2}\sigma_b^2)\\
& & \cdot \int_{-\infty}^{R_a^p} 
 \left\{1+erf\left[\frac{1}{\sqrt{2(1-\rho^2)}}\left(R_b^p-\rho u- \sigma_b(1-\rho^2)\right)
\right]
\right\}\\
& & \cdot \exp(-\frac{1}{2}(u-\sigma_b\rho)^2)\,du\\ 
\end{eqnarray*}
\begin{eqnarray*}
E_1 &= & \frac{1}{2\sqrt{2\pi}}\int_{-\infty}^{R_a^p} 
\left\{1+erf\left[\frac{1}{\sqrt{2(1-\rho^2)}}\left(R_b^p-\rho u \right)
\exp(-\frac{1}{2}u^2)
\right] 
\right\}\,du\\
\end{eqnarray*}
\begin{eqnarray*}
\bar{E}_{x} &= & \frac{1}{2}\exp(\mu_a+\frac{1}{2}\sigma_a^2)\left\{1+erf\left[\frac{1}{\sqrt{2}}\left(R_a^p-\sigma_a\right)
\right] 
\right\}\\
\end{eqnarray*}
 
\begin{eqnarray*}
\bar{E}_{1} &= &
\frac{1}{2}\left\{1+erf\left[\frac{R_a^p}{\sqrt{2}}\right]\right\}\\ 
\bar{\bar{E}}_{y} &= & \frac{1}{2}\exp(\mu_b+\frac{1}{2}\sigma_b^2)\left\{1+erf\left[\frac{1}{\sqrt{2}}\left(R_b^p-\sigma_b\right)
\right] 
\right\}\\
\end{eqnarray*}
\begin{eqnarray*}
\bar{\bar{E}}_{1} &= & \frac{1}{2}\left\{1+erf\left[\frac{R_b^p}{\sqrt{2}}\right] 
\right\}\\
\end{eqnarray*}

\item $Cov(r_{c_{a1}},r_{c_{a2}})$

Neste caso, vamos calcular a covari\^{a}ncia entre as taxas de retorno de duas
op\c{c}\~{o}es de compra sobre o mesmo ativo A, mesma data de expira\c{c}\~{a}o, mas com
pre\c{c}os de exerc\'{\i}cio diferentes e, portanto, pr\^{e}mios diferentes.

Sejam $K_1^c$ e $K_2^c$ os pre\c{c}os de exerc\'{\i}cio e $C_1$ e $C_2$ os pr\^{e}mios das
calls.

\begin{eqnarray*}
cov(r_{c_{a1}},r_{c_{a2}})& = & \frac{1}{C_1C_2}\left\{S_a^2E_{2x}-S_aE_x(K_1^c+K_2^c)+K_1^cK_2^cE_1\right\}-\mu_{c1}\mu_{c2}\\
\end{eqnarray*}

\[
K=\max\{K_1^c,K_2^c\}
\]

onde

\begin{eqnarray*}
E_{2x} &= & \frac{1}{2}\exp(2\mu_a+2\sigma_a^2)\left\{1-erf\left[\frac{1}{\sigma_a\sqrt{2}}\left(\ln\left(\frac{K}{S_a}\right)-\mu_a-2\sigma_a^2\right)
\right] 
\right\}\\  
\end{eqnarray*}
\begin{eqnarray*}
E_{x} &= & \frac{1}{2}\exp(\mu_a+\frac{1}{2}\sigma_a^2)\left\{1-erf\left[\frac{1}{\sigma_a\sqrt{2}}\left(\ln\left(\frac{K}{S_a}\right)-\mu_a-\sigma_a^2\right)
\right] 
\right\}\\
\end{eqnarray*}
\begin{eqnarray*}
E_1 &= & \frac{1}{2}\left\{1-erf\left[\frac{1}{\sigma_a\sqrt{2}}\left(\ln\left(\frac{K}{S_a}\right)-\mu_a\right)
\right] 
\right\}\\
\end{eqnarray*}

\item $Cov(r_{p_{a1}},r_{p_{a2}})$

Aqui calculamos a covari\^{a}ncia entre as taxas de retorno de duas
op\c{c}\~{o}es de venda sobre o mesmo ativo A, mesma data de expira\c{c}\~{a}o, mas com
pre\c{c}os de exerc\'{\i}cio e pr\^{e}mios diferentes.

Sejam $K_1^p$ e $K_2^p$ os pre\c{c}os de exerc\'{\i}cio e $P_1$ e $P_2$ os pr\^{e}mios das
calls.

\begin{eqnarray*}
cov(r_{p_{a1}},r_{p_{a2}})& = & \frac{1}{P_1P_2}\left\{S_a^2E_{2x}-S_aE_x(K_1^p+K_2^p)+K_1^pK_2^pE_1\right\}-\mu_{p1}\mu_{p2}\\
\end{eqnarray*}

\[
K=\min\{K_1^p,K_2^p\}
\]

onde

\begin{eqnarray*}
E_{2x}&= & \frac{1}{2}\exp(2\mu_a+2\sigma_a^2)\left\{1+erf\left[\frac{1}{\sigma_a\sqrt{2}}\left(\ln\left(\frac{K}{S_a}\right)-\mu_a-2\sigma_a^2\right)
\right] 
\right\}\\
& &\\
& &\\
E_{x} &= & \frac{1}{2}\exp(\mu_a+\frac{1}{2}\sigma_a^2)\left\{1+erf\left[\frac{1}{\sigma_a\sqrt{2}}\left(\ln\left(\frac{K}{S_a}\right)-\mu_a-\sigma_a^2\right)
\right] 
\right\}\\
& &\\
& &\\
E_1 &= & \frac{1}{2}\left\{1+erf\left[\frac{1}{\sigma_a\sqrt{2}}\left(\ln\left(\frac{K}{S_a}\right)-\mu_a\right)
\right] 
\right\}\\
\end{eqnarray*}

\item $Cov(r_{c_{aT}},r_{c_{a\tau}})$

Neste caso calculamos a covari\^{a}ncia entre duas op\c{c}\~{o}es de compra sobre o mesmo
ativo, mas com datas de expira\c{c}\~{a}o diferentes (dadas por $T$ e $\tau$, $\tau <
T$).

Suponha que temos a informa\c{c}\~{a}o

\[
x_T = \ln\frac{S_a(T)}{S_a(t_0)} \cong  N(\mu,\sigma^2)
\]

Denote por $x_\tau$ o retorno logar\'{\i}timico do ativo no per\'{\i}odo $t_0$ a $\tau$
e  $x_{T-\tau}$ o retorno logar\'{\i}timico do ativo no per\'{\i}odo $\tau$ a $T$.
Ent\~{a}o

\[
x_\tau = \ln\frac{S_a(\tau)}{S_a(t_0)} \cong
N(\mu_\tau,\sigma_{\tau}^2) 
\]

\[
x_{T-\tau} = \ln\frac{S_a(T-\tau)}{S_a(\tau)} \cong
N(\mu_{T-\tau},\sigma_{T-\tau}^2) 
\]
 
onde

\begin{eqnarray*}
\mu_\tau & = & \mu_T \frac{\tau-t_0}{T-t_0}\\
\sigma_\tau^2 & = & \sigma_T^2\frac{\tau-t_0}{T-t_0}\\
& &\\
\mu_{T-\tau} & = & \mu_T\frac{T-\tau}{T-t_0}\\
\sigma_{T-\tau}^2 & = & \sigma_T^2\frac{T-\tau}{T-t_0}
\end{eqnarray*}

E a correla\c{c}\~{a}o entre $x_\tau$ e  $x_T$ \'{e} dada por

\[
corr(x_T,x_\tau) =
\frac{\sigma_T^2+\sigma_\tau^2-\sigma_{T-\tau}^2}{2\sigma_T\sigma_\tau}
\]

Desta forma, podemos calcular a covari\^{a}ncia entre as duas op\c{c}\~{o}es usando a
mesma express\~{a}o da $Cov(r_{c_{a}},r_{c_{b}})$, considerando $x=x_T$,
$y=x_\tau$ e $\rho=corr(x_T,x_\tau)$.

\item $Cov(r_{p_{aT}},r_{p_{a\tau}})$

Idem ao caso anterior, onde a express\~{a}o usada para o c\'{a}lculo da covari\^{a}ncia \'{e}
a dada por $Cov(r_{p_{a}},r_{p_{b}})$.

\item $Cov(r_{c_{aT}},r_{p_{a\tau}})$

Idem aos casos anteriores, usando a express\~{a}o dada em
$Cov(r_{c_{a}},r_{p_{b}})$ para o c\'{a}lculo da covari\^{a}ncia.

\end{itemize}

%% file: cap12.tex
\chapter{Assimetria}

O modelo de forma\c{c}\~ ao de carteiras visto no cap\'{\i}tulo anterior
pode ser estendido fazendo-se uso do terceiro momento central
(assimetria) do retorno de um portf\' olio. Vamos definir express\~ oes
anal\'{\i}ticas para assimetria entre retornos de ativos e op\c{c}\~ oes
europ\' eias, do mesmo modo feito para as covari\^ ancias. Vamos
tamb\'{e}m definir um nova fun\c{c}\~ ao utilidade a ser maximizada pelo
investidor. 

\section{Assimetria de retorno de Portf\'{o}lios}

Para um portf\' olio formado por $n$ ativos de risco, os tr\^ es primeiros
momentos do retorno $R_p$ da carteira s\~ ao dados por

\[
E(R_p)=\sum_{i=1}^n x_i E(r_i) = x'E(r)
\]

\[
\sigma_p^2=\sum_{i=1}^n\sum_{j=1}^n x_ix_j\sigma_{ij} = x' Cov(r) x
\]

\[
m_p^3=\sum_{i=1}^n\sum_{j=1}^n \sum_{k=1}^n x_ix_jx_k m_{ijk}
\]

onde 

\begin{description}

\item  $E(r_i)$ = retorno esperado do i-\' esimo ativo

\item $ \sigma_{ij}$ = covari\^ ancia entre o retorno do i-\' esimo e do
  j-\'esimo ativo 

\item $m_{ijk} $ = co-assimetria\protect\footnote[2]{Aqui assimetria ({\it skewness})
\' e definido como o terceiro momento central n\~ ao normalizado} entre o retorno do
i-\' esimo, j-\' esimo e o k-\' esimo ativo
 
\item $ x_i$ = quantidade investida no i-\' esimo ativo.
 
\end{description}

Dada uma certa fun\c{c}\~ ao utilidade $U$ do retorno do portf\' olio
($U(R_p)$), a sua expans\~ ao em s\' erie de pot\^ encias em torno do retorno
esperado $ E(R_p)=\bar{ R_p}$ nos gera a seguinte aproxima\c{c}\~ ao de
terceira ordem:

\vspace{5mm}

\begin{eqnarray}
U(R_p)& \cong & U({\bar R}_p )+U^{'}({\bar R}_p )(R_p-{\bar R}_p )+
\frac{U^{''}({\bar R}_p)}{2!}(R_p-{\bar R}_p)^2 +\frac{U^{'''}({\bar
R}_p)}{3!}(R_p-{\bar R}_p)^3 \nonumber \\ \label{fig:eq51} 
\end{eqnarray}

\vspace{5mm}

Aplicando o operador esperan\c{c}a na express\~ ao ~\ref{fig:eq51}, definimos a
fun\c{c}\~ ao 

\vspace{5mm}

\begin{eqnarray}
E[U(R_p)]& \cong & E[U({\bar R}_p )]+
\frac{U^{''}({\bar R}_p)}{2!}E[(R_p-{\bar R}_p)^2] +\frac{U^{'''}({\bar
R}_p)}{3!}E[(R_p-{\bar R}_p)^3] \nonumber \\
& = & E[U({\bar R}_p )]+ \frac{U^{''}({\bar R}_p)}{2!}\sigma^2_p + \frac{U^{'''}({\bar R}_p)}{3!}m_p^3. \nonumber \\ \label{fig:eq52}
\end{eqnarray}

\vspace{5mm}

O modelo de Markowitz usa uma fun\c{c}\~ ao utilidade semelhante a
aproxima\c{c}\~{a}o de segunda ordem de $ E(U(R_p))$:

\vspace{5mm}

\begin{eqnarray}
U(R_p)& \cong & U({\bar R}_p )+\frac{U^{''}({\bar R}_p)}{2!}(R_p-{\bar R}_p)^2
\\
& = &  E[U({\bar R}_p )]+ \frac{U^{''}({\bar R}_p)}{2!}\sigma^2_p  \nonumber
\\ \label{fig:eq53}   
\end{eqnarray}

\vspace{5mm}

Supondo que um investidor seja avesso ao risco, o comportamento qualitativo da
fun\c{c}\~ ao utilidade para este perfil satisfaz $U^{'}(\bar{R}_p) > 0$,$U^{''}(\bar{R}_p) < 0$, $U^{'''}(\bar{R}_p) > 0$. 

Baseado neste comportamento, definimos uma fun\c{c}\~ ao utilidade em rela\c
c\~ ao a m\' edia, vari\^ ancia e assimetria. Para quantificar as
prefer\^encias de um investidor quanto \`a propens\~{a}o ao risco,
nossa fun\c{c}\~{a}o objetivo a ser maximizada ser\'{a} dada por 

\begin{eqnarray}
U(x) & = &  \eta {\bar R}_p - \sigma_p^2 + \alpha m_p^3 \nonumber \\ \label{fig:eq54}
\end{eqnarray}

onde $\eta$ e $\alpha $ ponderam, respectivamente, a taxa de retorno esperada
e a assimetria do retorno do portf\' olio.

Supondo que as prefer\^ encias do investidor est\~ ao representadas pela fun\c
c\~ ao $U(\bar{R}_p, \sigma_p^2, m_p^3)$, o conjunto de porft\' olios \'
otimos dever\' a:

\begin{itemize}
\item maximizar $ m_p^3$ para ${\bar R}_p$ e $\sigma_p^2$ fixos
\item maximizar ${\bar R}_p$ para $m_p^3$ e $\sigma_p^2$ fixos 
\item  minimizar $\sigma_p^2$ para ${\bar R}_p$ e $m_p^3 $ fixos. 
\end{itemize}

Nosso portf\' olio \' otimo ser\' a definido como a carteira $x = (x_1, x_2,
\ldots, x_n)$ que maximiza a fun\c{c}\~ ao utilidade U dada em ~\ref{fig:eq54},
estando sujeita a restri\c{c}\~ oes lineares de investimentos nos ativos e
derivativos.

O problema de aloca\c{c}\~ ao de carteiras mistas, baseado em m\' edia-vari\^
ancia-assimetria, cuja solu\c{c}\~{a}o define portf\' olios \' otimos, \' e dado
por

\begin{equation}
(PMVA) \left\{ 
       \begin{array}{ll}
        \max & \eta \sum_i x_iE(r_i) - \sum_i\sum_j x_i x_j \sigma_{ij} +
        \alpha \sum_i\sum_j\sum_k x_i x_j x_k m_{ijk}^3 \\
        s.a  & T_e * x = t_e \\
             & T_l * x \leq t_l \\ 
        & x_i \geq 0 \quad \quad (i=1..n)
       \end{array}
\right. 
\end{equation}

\vspace{5mm}

Os dados de entrada necess\' arios para a constru\c{c}\~{a}o da fun\c{c}\~{a}o
objetivo do problema (PMVA) s\~ ao: vetor dos retornos esperados de ativos e
op\c{c}\~ oes, matriz de vari\^ ancia - covari\^ ancia, assimetrias e
co-assimetrias entre os retornos dos \'{\i}tens formadores da carteira. Do mesmo
modo feito no caso das covari\^ ancias, vamos deduzir express\~ oes
anal\'{\i}ticas para o c\' alculo das assimetrias e co-assimetrias.

Nosso objetivo agora \' e resolver um problema de programa\c{c}\~ ao n\~ ao
linear param\' etrico. A cada solu\c{c}\~ ao \' otima encontrada de (PMVA)
temos associados portf\' olios cujo retorno esperado, vari\^ ancia e
assimetria s\~ ao dados por

\vspace{5mm}
  
\begin{eqnarray*}
{\bar R}_p & = & x(\eta, \alpha)' E(r) \\
\sigma_p^2 & = & x(\eta, \alpha)' Cov(r) x(\eta, \alpha) \\
 m_p^3 & = & \sum_i \sum_j \sum_k x_i(\eta, \alpha)x_j(\eta, \alpha)x_k(\eta,
\alpha) m_{ijk}\\ 
\end{eqnarray*}

\vspace{5mm}

Fixando $\eta$, podemos obter portf\' olios \' otimos pelo (PMVA) variando o
par\^ ametro $\alpha$, gerando assim uma superf\'{\i}cie eficiente no espa\c
co $ retorno \times risco \times assimetria$, extens\~{a}o do conceito de
fronteira eficiente obtida no modelo baseado em m\'edia-vari\^ ancia.

Podemos defender o uso do terceiro momento na otimiza\c{c}\~ ao de portf\'
olios pelo fato de que os ativos e derivativos que comp\~ oem estas
carteiras t\^ em retornos assim\' etricos. Para os pre\c{c}os de a\c{c}\~ oes
(ativos-objeto usados nos exemplos), supomos um comportamento lognormal. Os
retornos de op\c{c}\~ oes t\^ em um comportamento intrinsicamente assim\'
etrico. 


\section{Express\~ oes de Co-assimetrias}

Vamos agora voltar ao exemplo de portf\'{o}lio da se\c{c}\~{a}o 2 do cap\'{\i}tulo anterior e ampli\'{a}-lo
para conter tr\^{e}s tipos de ativos-objeto, que chamaremos de ativos $ A$, $B$ e
$C$, e op\c{c}\~{o}es europ\'{e}ias sobre estes ativos.  Os respectivos retornos
logar\'{\i}tmicos  s\~{a}o dados por

\vspace{5mm}

\begin{eqnarray*}
x&=&ln\left( \frac{S_a(T)}{S_a(t_0)}\right) \cong N(\mu_a,\sigma_a^2)\\
& &\\
y&=&ln\left( \frac{S_b(T)}{S_b(t_0)}\right) \cong N(\mu_b,\sigma_b^2)\\ 
& &\\
z&=&ln\left( \frac{S_c(T)}{S_c(t_0)}\right) \cong N(\mu_c,\sigma_c^2) 
\end{eqnarray*}

\vspace{5mm}

e fun\c{c}\~{a}o densidade de probabilidade conjunta dada por

\[
\begin{array}{ll}
f(x,y,z) & = \frac{1}{\sigma_{a}\sigma_{b}\sigma_{c} \sqrt{8\pi^3\gamma}}
\exp{\left[
\left(\frac{x-\mu_a}{\sigma_a}\right)^2(1-\rho_{bc}^2)+
\left(\frac{x-\mu_b}{\sigma_b}\right)^2(1-\rho_{ac}^2)+ 
\left(\frac{x-\mu_c}{\sigma_c}\right)^2(1-\rho_{ab}^2)\right]}\cdot\\
& \exp{\left[ 
2\frac{x-\mu_a}{\sigma_a}\frac{y-\mu_b}{\sigma_b}(\rho_{ac}\rho_{bc}-\rho_{ab})
+2\frac{x-\mu_a}{\sigma_a}\frac{z-\mu_c}{\sigma_c}(\rho_{ab}\rho_{bc}-\rho_{ac})
+2\frac{y-\mu_b}{\sigma_a}\frac{z-\mu_c}{\sigma_b}(\rho_{ac}\rho_{ab}-\rho_{bc})
\right]}\\
\end{array}
\]

\[
(-\infty < x,y,z < +\infty)
\]

onde 

\vspace{5mm}

\begin{eqnarray*}
\gamma &=&
1+2\rho_{ab}\rho_{bc}\rho_{ac}-\rho_{ab}^2-\rho_{bc}^2-\rho_{ac}^2\\
& &\\
\rho_{ab},\rho_{ac},\rho_{bc}&=& \hbox{coeficiente de correla\c{c}\~{a}o entre $x$
e $y$,  entre  $x$ e $z$ e  entre $y$ e $z$, respectivamente.} \\
\end{eqnarray*}

\vspace{5mm}

Vamos denotar assimetria pela letra {\it m}. Por exemplo, 
\[
m(r_{s_a},r_{s_b},r_{s_c})=E[(r_{s_a}-E(r_{s_a}))(r_{s_b}-E(r_{s_b}))(r_{s_c}-E(r_{s_c}))]
\] 

\'{e} a co-assimetria entre os retornos $r_{s_a}, r_{s_b}$ e  $r_{s_c}$ dos ativos
$A$, $B$ e $C$.

Aqui estamos supondo que os ativos $ A, B$ e $C$ s\~{a}o distintos entre si. 
Abaixo est\~{a}o as express\~{o}es de alguns casos de co-assimetria. Pela extensiva
repeti\c{c}\~{a}o de c\'{a}lculos simb\'{o}licos na deriva\c{c}\~{a}o das express\~{o}es abaixo, os
detalhes das mudan\c{c}as de vari\'{a}veis usadas foram explicitadas em alguns
casos. Os outros seguem racioc\'{\i}nio an\'{a}logo.

\begin{itemize}

\item {$m(r_{s_a},r_{s_a},r_{s_a})$}

\begin{eqnarray*}
{m(r_{s_a},r_{s_a},r_{s_a})} &=& m^3_{s_a} = E(e^{3x})-3E(e^{2x})E(e^x)+2E^3(e^x)
\end{eqnarray*}

\[
(-\infty < x < +\infty)
\]

onde

\begin{eqnarray*}
E(e^{3x})&=&\exp\left( 3\mu_a+\frac{9}{2}\sigma_a^2 \right)\\
E(e^{2x})&=&\exp\left( 2\mu_a + 2\sigma_a^2\right)\\
E(e^{x})&=&\exp\left( \mu_a + 0.5\sigma_a^2\right)\\
\end{eqnarray*}

\item {$m(r_{s_a},r_{s_a},r_{c_a})$}

\begin{eqnarray*}
& &{m(r_{s_a},r_{s_a},r_{c_a})}=\\
& &\frac{1}{C_a}[S_aE(e^{3x})-2E(e^{2x})(K_a^c+C_a+2S_a)+E(e^x)(S_a+2k_a^c+2C_a)\\
&
&-E(1)(K_a^c+C_a)-2(E(e^x)-E(1))(S_aE(e^{2x})-E(e^x)(K_a^c+S_a+C_a)\\
& &+E(1)(K_a^c+C_a))-(S_aE(e^x)-E(1)(K_a^c+S_a+C_a)(E(e^{2x})-2E(e^x)+1))\\
& & (E(e^x)-E(1))^2(S_aE(e^{x})-E(1)(K_a^c+C_a)(3-E(1)))]
\end{eqnarray*}

\[
(\ln\left(\frac{K_a^c}{S_a}\right) < x < +\infty)
\]

onde

\begin{eqnarray*}
g &=& \frac{1}{\sigma_a\sqrt{2}}(\ln\frac{K_a^c}{S_a}-\mu_a)\\
& &\\
E(e^{3x})&=&\frac{1}{2}[1-erf(g-\frac{3\sigma_a}{\sqrt{2}})]\\
& &\\
E(e^{2x})&=&\frac{1}{2}[1-erf(g-\frac{2\sigma_a}{\sqrt{2}})]\\
& &\\
E(e^{x})&=&\frac{1}{2}[1-erf(g-\frac{\sigma_a}{\sqrt{2}})]\\
& &\\
E(1)&=&\frac{1}{2}[1-erf(g)]\\
\end{eqnarray*}

\item {\bf $m(r_{s_a},r_{s_a},r_{p_a})$}

\begin{eqnarray*}
& &{m(r_{s_a},r_{s_a},r_{p_a})} = \\
& &\frac{1}{P_a}[-S_aE(e^{3x})-2E(e^{2x})(-K_a^p+P_a-2S_a)+E(e^x)(-S_a-2k_a^p+2P_a)\\
&
&-E(1)(-K_a^p+P_a)-2(E(e^x)-E(1))(-S_aE(e^{2x})-E(e^x)(-K_a^p-S_a+P_a)\\
& &+E(1)(-K_a^p+P_a))-(-S_aE(e^x)-E(1)(-K_a^p+P_a)(E(e^{2x})-2E(e^x)+1))\\
& & (E(e^x)-E(1))^2(-S_aE(e^{x})-E(1)(-K_a^p+P_a)(3-E(1)))]
\end{eqnarray*}

\[
(-\infty < x < \ln\left(\frac{K_a^p}{S_a}\right))
\]

onde

\begin{description}

\item
$
g = \frac{1}{\sigma_a\sqrt{2}}(\ln\frac{K_a^p}{S_a}-\mu_a)
$
\end{description}

e as express\~{o}es de $E(e^{3x})$, $E(e^{2x})$, $E(e^x)$ e $E(1)$ s\~{a}o iguais as de
$m(r_{s_a},r_{s_a},r_{c_a})$, trocando o sinal da fun\c{c}\~{a}o $erf$.

\item {\bf $m(r_{s_a},r_{c_{a1}},r_{c_{a2}})$}

\begin{eqnarray*}
& &{m(r_{s_a},r_{s_a},r_{p_a})} = \\
& &\frac{1}{C_{1}C_{2}}[S_a^2E(e^{3x})-E(e^{2x})(K_1^c+K_2^c+C_1+C_2+S_a)\\
& &+E(e^x)((K_1^c+C_1)(K_2^c+C_2)+S_a(K_1^c+C_1+C_1+K_2^c+C_2))\\
& &-E(1)(K_1^c+C_2)(K_2^c+C_2)(S_aE(e^x)-E(1)(K_1^c+C_1))\\
& &\cdot (S_aE(e^{2x})-E(e^x)(K_2^c+S_a+C_2)+E(1)(K_2^c+C_2))\\
& &(S_aE(e^x)-E(1)(K_2^c+C_2))(S_aE(e^{2x})-E(e^x)(K_1^c+S_a+C_1)\\
& &+E(1)(K_1^c+C_1))(E(e^x)-E(1))\\
& &\cdot(S_aE(e^x)-E(1)(K_1^c+C_1))(S_aE(e^x)-E(1)(K_2^c+C_2))(3-E(1))]
\end{eqnarray*}

\[
(\ln\left(\frac{K}{S_a}\right) < x < +\infty)
\]

onde

\begin{description}

\item
$
K=\max\{K_1^c,K_2^c\}
$

\item
$
g = \frac{1}{\sigma_a\sqrt{2}}(\ln\frac{K}{S_a}-\mu_a)
$

\end{description}

e as express\~{o}es de $E(e^{3x})$, $E(e^{2x})$, $E(e^x)$ e $E(1)$ s\~{a}o iguais as de
$m(r_{s_a},r_{s_a},r_{c_a})$.

\item {\bf $m(r_{s_a},r_{p_{a1}},r_{p_{a2}})$}

\begin{eqnarray*}
& &{m(r_{s_a},r_{p_{a1}},r_{p_{a2}})} = \\
& &\frac{1}{C_{1}C_{2}}[S_a^2E(e^{3x})+S_aE(e^{2x})(-K_1^p-K_2^p+P_1+P_2-S_a)\\
& &+E(e^x)((-K_1^p+P_1)(-K_2^p+P_2)-S_a(-K_1^p+P_1+P_1-K_2^p+P_2))\\
& &-E(1)(-K_1^p+P_1)(-K_2^p+P_2)-(-S_aE(e^x)-E(1)(-K_1^p+P_1))\\
& &\cdot(-S_aE(e^{2x})-E(e^x)(-K_2^p-S_a+P_2)+E(1)(-K_2^p+P_2))\\
& &-(-S_aE(e^x)-E(1)(-K_2^p+P_2))(-S_aE(e^{2x})-E(e^x)(-K_1^p-S_a+P_1)\\
& &+E(1)(-K_1^p+P_1))(E(e^x)-E(1))(-S_aE(e^x)-E(1)(-K_1^p+P_1))\\
& &\cdot(-S_aE(e^x)-E(1)(-K_2^p+P_2))(3-E(1))]
\end{eqnarray*}

\[
(-\infty < x < \ln\left(\frac{K}{S_a}\right))
\]

onde

\begin{description}

\item
$
K=\min\{K_1^p,K_2^p\}
$

\item
$
g = \frac{1}{\sigma_a\sqrt{2}}(\ln\frac{K}{S_a}-\mu_a)
$

\end{description}

e as express\~{o}es de $E(e^{3x})$, $E(e^{2x})$, $E(e^x)$ e $E(1)$ s\~{a}o iguais as de
$m(r_{s_a},r_{s_a},r_{p_a})$.

\item {\bf $m(r_{s_a},r_{c_a},r_{p_a})$}

\begin{eqnarray*}
& &{m(r_{s_a},r_{p_a},r_{p_a})} = \\
& &\frac{1}{C_{a}P_{a}}[-S_a^2E(e^{3x})+S_aE(e^{2x})(K_a^p+K_a^c-P_a+C_a+S_a)\\
& &+E(e^x)((K_a^c+C_a)(-K_a^p+P_a)+S_a(-K_a^p+P_a-C_a-K_a^c))\\
& &-E(1)(-K_a^p+P_a)(K_a^c+C_a)-(E(e^x)-E(1))\\
& &\cdot(-S_a^2E(e^{2x})+S_aE(e^x)(K_a^p-P_a+C_a+K_a^c))\\
& &-E(1)(K_a^c+C_a)(K_a^p-P_a)\\
& &-(-S_aE(e^x)+E(1)(K_a^p-P_a))(S_aE(e^{2x})-E(e^x)(K_a^c+S_a+C_a)\\
& &+E(1)(K_a^c+C_a))-(S_aE(e^x)-E(1)(K_2^c+C_a))\\
& &\cdot (-S_aE(e^{2x})-E(e^x)(-K_a^p-S_a+P_a)\\
& &+E(1)(-K_a^p+P_a))(E(e^x)-E(1))\\
& &\cdot (S_aE(e^x)-E(1)(K_a^c+C_a))(-S_aE(e^x)+E(1)(K_a^p-P_a))(3-E(1))]
\end{eqnarray*}

se $ K_a^c < K_a^p$

sen\~{a}o $m(r_{s_a},r_{p_a},r_{c_a}) = 0$

\[
(\ln\left(\frac{K_a^c}{S_a}\right) < x < \ln\left(\frac{K_a^c}{S_a}\right))
\]

onde

\begin{eqnarray*}
g &=& \frac{1}{\sigma_a\sqrt{2}}(\ln\frac{K_a^c}{S_a}-\mu_a)\\
& &\\
E(e^{3x})&=&\frac{1}{2}[erf(h-\frac{3\sigma_a}{\sqrt{2}})-erf(g-\frac{3\sigma_a}{\sqrt{2}})]\\
& &\\
E(e^{2x})&=&\frac{1}{2}[erf(h-\frac{2\sigma_a}{\sqrt{2}})-erf(g-\frac{2\sigma_a}{\sqrt{2}})]\\
& &\\
E(e^{x})&=&\frac{1}{2}[erf(h-\frac{\sigma_a}{\sqrt{2}})-erf(g-\frac{\sigma_a}{\sqrt{2}})]\\
& &\\
E(1)&=&\frac{1}{2}[erf(g)-erf(h)]\\
\end{eqnarray*}

\item{$m(r_{c_{a1}},r_{c_{a2}},r_{c_{a3}})$}

Co-assimetria entre calls com o mesmo ativo b\'{a}sico, mas com pre\c{c}os de
exerc\'{\i}cio e pr\^{e}mios diferentes. 

Nota\c{c}\~{a}o:
\begin{description}
\item$r_{c_{ai}}=$ retorno da op\c{c}\~{a}o de compra sobre o ativo b\'{a}sico `` $ A$''   com
pr\^{e}mio dado por $C_i$ e pre\c{c}o de exerc\'{\i}cio dado por $K_i^c$ (i=1,2 e 3).
\item $K=max{\{K_1^c,K_2^c,K_3^c\}}$
\end{description}

\begin{eqnarray*}
& & {m(r_{c_{a1}},r_{c_{a2}},r_{c_{a3}})} =\\
&
&\frac{1}{C_1C_2C_3}[S_a^3E_{3x}-S_a^2(K_1^c+K_2^c+K_3^c+C_1+C_2+C_3)E_{2x}\\
& &+S_aE_x((K_3^c-C_3)(K_1^c+K_2^c+C_1+C_2)+(K_1+C_1)(K_2+C_2))\\
& &-E_1((K_1^c+C_1)(K_2^c+C_2)(K_3^c+C_3))
-(S_aE_x-(K_1^c+C_1)E(1))\\
& &\cdot (S_a^2E(e^{2x})-S_aE(e^x)(K_2+K_3+C^2+C^3)+E_1(K_2+C_2)(K_3+C_3))\\
& &-(S_aE_x-(K_2^c+C_2)E_1)
(S_a^2E(e^{2x})-S_aE(e^x)(K_1^2+K_3^2+C_2+C_3)\\ 
& &+E(1)(K_2+C_2)(K_3+C_3))
-(S_aE(e^x)-E(1)(K_3^c+C_3))(S_a^2E(e_{2x})\\
& &-S_aE(e^x)(K_1+C_1+K_2+C_2)+E(1)(K_1^c+C_1)(K_2+C_2))\\
&
&+(S_aE(e^x)-(K_1^c+C_1)E(1))(S_aE(e^x)-(K_2^c+C_2)E(1))(S_aE(e^x)\\
& &-(K_3^c+C_3)E(1))(3-E(1))]
\end{eqnarray*}

\begin{eqnarray*}
R &=& \frac{\ln{\frac{K}{S_a}}-\mu_a}{\sigma_a}\\
& &\\
d &=& \frac{R-\sigma_a}{\sqrt{2}}\\
& &\\
E(e^{3x})&=&0.5\exp{(3\mu_a+4.5\sigma_a^2)}(1-erf{[\frac{1}{\sqrt{2}}(R-3\sigma_a)]})\\
& &\\
E(e^{2x})&=&0.5\exp{(2\mu_a+2\sigma_a^2)}(1-erf{[\frac{1}{\sqrt{2}}(d-\sigma_a)]})\\
& &\\
E(e^{x})&=&0.5\exp{(\mu_a+0.5\sigma_a^2)}(1-erf{[d]})\\
& &\\
E(1)&=&0.5(1-erf{[\frac{1}{\sqrt{2}}R]})\\
\end{eqnarray*}

\item{$m(r_{p_{a1}},r_{p_{a2}},r_{p_{a3}})$}

Co-assimetria entre puts com o mesmo ativo b\'{a}sico, mas com pre\c{c}os de
exerc\'{\i}cio e pr\^{e}mios diferentes. 

Nota\c{c}\~{a}o:
\begin{description}
\item$r_{p_{ai}}=$ retorno da op\c{c}\~{a}o de compra sobre o ativo b\'{a}sico `` $ A$''   com
pr\^{e}mio dado por $P_i$ e pre\c{c}o de exerc\'{\i}cio dado por $K_i^p$ (i=1,2 e 3).
\item $K=min{\{K_1^p,K_2^p,K_3^p\}}$ 
\end{description}

\begin{eqnarray*}
& & {m(r_{p_{a1}},r_{p_{a2}},r_{p_{a3}})} =\\
&
&\frac{1}{P_1P_2P_3}[-S_a^3E_{3x}-S_a^2(-K_1^p-K_2^p-K_3^p+P_1+P_2+P_3)E(e^{2x})\\
& &-S_aE(e^x)((-K_3^p+P_3)(-K_1^p-K_2^p+P_1+P_2)+(-K_1^p+P_1)(-K_2^p+P_2))\\
& &-E_1((-K_1^p+P_1)(-K_2^p+P_2)(-K_3^p+P_3))
-(-S_aE(e^x)-(-K_1^p+P_1)E(1))\\
& &\cdot (S_a^2E(e^{2x})+S_aE(e^x)(-K_2^p-K_3^p+P^2+P^3)+E(1)(-K_2^p+P_2)(-K_3^p+P_3))\\
& &-(-S_aE(e^x)-(-K_2^p+P_2)E(1))
(S_a^2E(e^{2x})+S_aE(e^x)(-K_1^p-K_3^p+P_2+P_3)\\ 
& &+E(1)(-K_2^p+P_2)(-K_3^p+P_3))
-(-S_aE(e^x)-E(1)(-K_3^p+P_3))(S_a^2E(e_{2x})\\
& &+S_aE(e^x)(-K_1^p+P_1-K_2^p+P_2)+E(1)(-K_1^p+P_1)(-K_2^p+P_2))\\
&
&+(-S_aE(e^x)-(-K_1^p+P_1)E(1))(-S_aE(e^x)-(-K_2^p+P_2)E(1))(-S_aE(e^x)\\
& &-(-K_3^p+P_3)E(1))(3-E(1))]
\end{eqnarray*}

e as express\~{o}es de $E(e^{3x})$, $E(e^{2x})$, $E(e^x)$ e $E(1)$ s\~{a}o iguais as de
$m(r_{s_a},r_{p_{a1}},r_{p_{a2}})$.

\item{$m(r_{c_{a1}},r_{c_{a2}},r_{p_a})$}

Co-assimetria entre calls com o mesmo ativo b\'{a}sico, mas com pre\c{c}os de
exerc\'{\i}cio e pr\^{e}mios diferentes, e uma put sobre o mesmo ativo. 

Nota\c{c}\~{a}o:
\begin{description}
\item$r_{c_{ai}}=$ retorno da op\c{c}\~{a}o de compra sobre o ativo b\'{a}sico `` $ A$''   com
pr\^{e}mio dado por $C_i$ e pre\c{c}o de exerc\'{\i}cio dado por $K_i^c$ (i=1,2).
\item $P_a$ \'{e} o pr\^{e}mio da put e $K_a^p$ \'{e} o seu pre\c{c}o de exerc\'{\i}cio.
\item $K=max{\{K_1^c,K_2^c\}}$ 
\end{description}

\begin{eqnarray*}
& & {m(r_{c_{a1}},r_{c_{a2}},r_{p_{a}})} =\\
&
&\frac{1}{C_1C_2P_a}[-S_a^3E_{3x}+S_a^2(K_1^c+K_2^c+K_a^p+C_1+C_2-P_a)E(e^{2x})\\
& &-S_aE(e^x)((K_a^p-P_a)(K_1^c+K_2^c+C_1+C2)+(K_1^c+C_1)(K_2^c+C_2))\\
& &+E(1)((K_1^c+C_1)(K_2^c+C_2)(K_a^p+P_a))
-(S_aE(e^x)-(K_1^c+C_1)E(1))\\
& &\cdot (-S_a^2E(e^{2x})+S_aE(e^x)(K_a^p+K_2^c-P^a+C^2)-E(1)(K_2^c+C_2)(K_a^p-P_a))\\
& &-(S_aE(e^x)-(K_2^c+C_2)E(1))
(-S_a^2E(e^{2x})+S_aE(e^x)(K_a^p+K_1^c-P_a+C_1)\\ 
& &-E(1)(K_1^c+C_1)(K_a^p-P_a))
-(-S_aE(e^x)+E(1)(K_a^p-P_a))(S_a^2E(e_{2x})\\
& &-S_aE(e^x)(K_1^c+C_1+K_2^c+C_2)+E(1)(K_1^c+C_1)(K_2^c+C_2))\\
&
&+(S_aE(e^x)-(K_1^c+C_1)E(1))(S_aE(e^x)-(K_2^c+C_2)E(1))(-S_aE(e^x)\\
& &+(K_a^p-P_a)E(1))(3-E(1))]
\end{eqnarray*}

se $ K < K_a^p$  e com isso $(\ln\frac{K}{S_a} < x <\ln\frac{K_a^p}{S_a} )$

sen\~{a}o $m(r_{c_{a1}},r_{c_{a2}},r_{p_{a}}) = 0$.

\begin{description}

\item
$
g = \frac{1}{\sigma_a\sqrt{2}}(\ln\frac{K}{S_a}-\mu_a)
$

\item
$
h = \frac{1}{\sigma_a\sqrt{2}}(\ln\frac{K_a^p}{S_a}-\mu_a)
$

\end{description}

As express\~{o}es de $E(e^{3x})$, $E(e^{2x})$, $E(e^x)$ e $E(1)$ s\~{a}o iguais as de
$m(r_{s_a},r_{c_{a}},r_{p_{a}})$.

\item{$m(r_{c_{a}},r_{p_{a1}},r_{p_{a2}})$}

Co-assimetria entre puts com o mesmo ativo b\'{a}sico, mas com pre\c{c}os de
exerc\'{\i}cio e pr\^{e}mios diferentes, e uma call sobre o mesmo ativo. 

Nota\c{c}\~{a}o:
\begin{description}
\item$r_{p_{ai}}=$ retorno da op\c{c}\~{a}o de venda sobre o ativo b\'{a}sico `` $ A$''   com
pr\^{e}mio dado por $P_i$ e pre\c{c}o de exerc\'{\i}cio dado por $K_i^p$ (i=1,2).
\item $C_a$ \'{e} o pr\^{e}mio da put e $K_a^c$ \'{e} o seu pre\c{c}o de exerc\'{\i}cio.
\item $K=min{\{K_1^p,K_2^p\}}$ 
\end{description}

\begin{eqnarray*}
& & {m(r_{c_{a}},r_{p_{a1}},r_{p_{a2}})} =\\
&
&\frac{1}{P_1P_2C_a}[S_a^3E_{3x}-S_a^2(K_1^p+K_2^[+K_a^c-P_1-P_2+C_a)E(e^{2x})\\
& &+S_aE(e^x)((K_2^p-P_2)(K_1^p+K_a^c-P_1+Ca)+(K_a^c+C_a)(K_1^p-P_1))\\
& &-E(1)((K_1^p-P_1)(K_2^p-P_2)(K_a^c+C_a))
-(S_aE(e^x)-(K_a^c+C_a)E(1))\\
& &\cdot (S_a^2E(e^{2x})-S_aE(e^x)(K_1^p+K_2^p-P_1-P_2)+E(1)(K_2^p-P_2)(K_1^p-P_1))\\
& &-(S_aE(e^x)+(K_1^p-P_1)E(1))
(-S_a^2E(e^{2x})+S_aE(e^x)(K_a^c+K_2^p-P_2+C_a)\\ 
& &-E(1)(K_2^p+P_2)(K_a^c+C_a))
-(-S_aE(e^x)+E(1)(K_2^p-P_2))(-S_a^2E(e_{2x})\\
& &+S_aE(e^x)(K_a^c+C_a+K_1^p-P_1)-E(1)(K_1^p-P_1)(K_a^c+C_a))\\
&
&+(S_aE(e^x)-(K_a^c+C_a)E(1))(-S_aE(e^x)+(K_1^p-P_1)E(1))(-S_aE(e^x)\\
& &+(K_2^p-P_2)E(1))(3-E(1))]
\end{eqnarray*}

se $ K_a^c < K $  e com isso $(\ln\frac{K_a^c}{S_a} < x <\ln\frac{K}{S_a} )$

sen\~{a}o $m(r_{p_{a1}},r_{p_{a2}},r_{c_{a}}) = 0$.

\begin{description}

\item
$
g = \frac{1}{\sigma_a\sqrt{2}}(\ln\frac{K}{S_a}-\mu_a)
$

\item
$
h = \frac{1}{\sigma_a\sqrt{2}}(\ln\frac{K_a^c}{S_a}-\mu_a)
$

\end{description}

As express\~{o}es de $E(e^{3x})$, $E(e^{2x})$, $E(e^x)$ e $E(1)$ s\~{a}o iguais as de
$m(r_{c_{a1}},r_{c_{a2}},r_{p_{a}})$.

\item {\bf $m(r_{s_a},r_{s_a},r_{s_b})$}

\begin{eqnarray*}
{m(r_{s_a},r_{s_a},r_{s_b})} &=&
E(e^{2x+y})-E(e^{2x})E(e^y)-2E(e^x)E(e^{x+y})+2E(e^x)^2E(e^y) 
\end{eqnarray*}

\[
(-\infty < x,y < +\infty)
\]

onde

\begin{eqnarray*}
E(e^{2x+y})&=&\exp\left( 2\mu_a+ \mu_b +2\sigma_a^2 + \frac{1}{2}\sigma_b^2 +2\sigma_a\sigma_b\rho_{ab}\right)\\
& &\\
E(e^{x+y})&=&\exp\left( \mu_a+ \mu_b +\frac{1}{2}(\sigma_a^2 +\sigma_b^2) +\sigma_a\sigma_b\rho_{ab}\right)\\
& &\\
E(e^{2x})&=&\exp\left( 2\mu_a+ \mu_b +2\sigma_a^2 \right)\\
& &\\
E(e^{x})&=&\exp\left( \mu_a + \sigma_a^2 \right)\\
& &\\
E(e^{y})&=&\exp\left( \mu_b + \sigma_b^2 \right)\\
& &\\
\end{eqnarray*}

\item {\bf $m(r_{s_a},r_{s_a},r_{c_b})$}

\begin{eqnarray*}
{m(r_{s_a},r_{s_a},r_{c_b})} &=&
\frac{1}{C_b}[S_bE(e^{2x+y})-E(e^{2x})(K_b^c+C_b)-2S_bE(e^{x+y})+2E(e^x)(K_b^c+C_b)
\\
& &+S_bE(e^y) -
E(1)(K_b^c+C_b)-2(E(e^x)-E(1))(S_bE(e^{x+y})\\
& &
-E(e^x)(K_b^c+C_b)-S_bE(e^y)+E(1)(K_b^c+C_b))\\
& &-(S_bE(e^y)-E(1)(K_b^c+C_b))(E(e^{2x})-2E(e^x)\\
& &+E(1))+(E(e^x)-E(1))^2(S_bE(e^y)-(K_b^c+C_b)E(1))(3-E(1))]
\end{eqnarray*}

\[
(-\infty < y < +\infty)
\]

\[
(\ln\frac{K_b^c}{S_b} < y < +\infty)
\]

onde

\begin{eqnarray*}
R_b^c&=&\frac{1}{\sigma_b}(ln\frac{K_b^c}{S_b}-\mu_b)\\
& &\\
E(e^{2x+y})&=&0.5\exp\left( 2\mu_a+ \mu_b
+2\sigma_a^2+\frac{1}{2}\sigma_b^2+2\sigma_a\sigma_b\rho_{ab}\right)\\
& &
\quad \quad \quad \quad \quad \cdot[1-erf(\frac{1}{\sqrt{2}}(R_b^c-\sigma_b-2\sigma_a\rho_{ab}))]\\
& &\\
E(e^{x+y})&=&0.5\exp\left( \mu_a+ \mu_b +\frac{1}{2}(\sigma_a^2 +\sigma_b^2)
+\sigma_a\sigma_b\rho_{ab}\right)[1-erf(\frac{1}{\sqrt{2}}(R_b^c-\sigma_b-\sigma_a\rho_{ab}))]\\
& &\\
E(e^{2x})&=&0.5\exp\left( 2\mu_a+2\sigma_a^2 \right)[1-erf(\frac{1}{\sqrt{2}}(R_b^c-2\sigma_a\rho_{ab}))]\\
& &\\
E(e^{x})&=&0.5\exp\left( \mu_a + \sigma_a^2 \right)[1-erf(\frac{1}{\sqrt{2}}(R_b^c-\sigma_a\rho_{ab}))]\\
& &\\
E(e^{y})&=&0.5\exp\left( \mu_b + \sigma_b^2 \right)[1-erf(\frac{1}{\sqrt{2}}(R_b^c-\sigma_b))]\\
& &\\
E(1)&=&0.5[1-erf(\frac{1}{\sqrt{2}}(R_b^c))]\\
\end{eqnarray*}

\item {$m(r_{s_a},r_{s_a},r_{p_b})$}

\begin{eqnarray*}
{m(r_{s_a},r_{s_a},r_{p_b})} &=&
\frac{1}{P_b}[-S_bE(e^{2x+y})-E(e^{2x})(-K_b^c+P_b)+2S_bE(e^{x+y})+2E(e^x)(-K_b^p+P_b)
\\
& &-S_bE(e^y) +
E(1)(K_b^p-P_b)-2(E(e^x)-E(1))(-S_bE(e^{x+y})\\
& &
+E(e^x)(K_b^p-P_b)+S_bE(e^y)+E(1)(K_b^p-P_b))\\
& &-(-S_bE(e^y)+E(1)(K_b^p-P_b))(E(e^{2x})-2E(e^x)\\
& &+E(1))+(E(e^x)-E(1))^2(-S_bE(e^y)+(K_b^p-P_b)E(1))(3-E(1))]
\end{eqnarray*}

\[
(-\infty < x < +\infty) \ \ \   
(-\infty < y < ln\frac{K_b^c}{S_b})
\]

As express\~{o}es de $E(e^{3x})$, $E(e^{2x})$, $E(e^x)$ e $E(1)$ s\~{a}o iguais as de
$m(r_{s_{a}},r_{s_{a}},r_{c{b}})$, trocando o sinal da fun\c{c}\~{a}o erf e trocando
tamb\'{e}m $R_b^c$ por $R_b^p=\frac{1}{\sigma_b}(ln\frac{K_b^p}{S_b}-\mu_b) $.

\item $m(r_{s_a},r_{c_a},r_{c_b})$

\begin{eqnarray*}
m(r_{s_a},r_{c_a},r_{c_b}) &=& \frac{1}{C_aC_b}[S_aS_bE(e^{2x+y})-S_aE(e^{2x})(K_b^c+C_b)-S_bE(e^{x+y})(K_a^c+C_a+S_a)\\
& &
+E(e^{x})(K_a^c+C_a+S_a)(K_b^c+C_b)+E(e^{y})S_b(K_a^c+C_a)\\
& &
-E(1)(K_a^c+C_a)(K_b^c+C_b)-(E(e^x)-E(1))(S_aS_bE(e^{x+y})\\
& &-S_a(K_b^c+C_b)E(e^{x})-S_b(K_a^c+C_a)E(e^{y})+(K_a^c+C_a)(K_b^c+C_b)E(1))\\
& &
-(S_aE(e^{x})-E(1)(K_a^c+C_a))(S_bE(e^{x+y})-(K_b^c+C_b)E(e^{x})\\
& &
-S_bE(e^{y})+E(1)(K_b^c+C_b))-(S_bE(e^{y})-E(1)(K_b^c+C_b))\\
& &(S_aE(e^{2x})-(K_a^c+C_a+S_a)E(e^{x})+E(1)(K_a^c+C_a))\\
&
&+(E(e^{x})-E(1))(S_aE(e^{x})-E(1)(K_a^c+C_a))(S_bE(e^{y})\\
& &-E(1)(K_b^c+C_b))(3-E(1))]
\end{eqnarray*}

\[
(\ln\frac{K_a^c}{S_a}< x < +\infty) \ \ \  
(\ln\frac{K_b^c}{S_b}< y < +\infty)
\]

onde

\begin{eqnarray*}
R_a^c &=& \frac{1}{\sigma_b}(ln\frac{K_a^c}{S_a}-\mu_a)\\
& &\\
& &\\
R_b^c &=& \frac{1}{\sigma_b}(ln\frac{K_b^c}{S_b}-\mu_b)\\
& &\\
& &\\
E(e^{2x+y})&=&\exp\left(2\mu_a+ \mu_b
+2\sigma_a^2+\frac{1}{2}\sigma_b^2+2\sigma_a\sigma_b\rho_{ab}\right)\int_{R_a^c}^{+\infty}\exp[-0.5(u-2\sigma_a-\sigma_b\rho_{ab})^2]\\
& &\cdot \left[1-erf\left(\frac{1}{\sqrt{2(1-\rho_{ab})}}(R_a^b-u\rho_{ab}-\sigma_b(1-\rho_{ab}^2))\right)\right]\,du
\\
& &\\
& &\\
E(e^{x+y})& =&\exp\left( \mu_a+ \mu_b +\frac{1}{2}(\sigma_a^2 +\sigma_b^2) +\sigma_a\sigma_b\rho_{ab}\right)\int_{R_a^c}^{+\infty}\exp[-0.5(u-\sigma_a-\sigma_b\rho_{ab})^2]\\
& &\cdot \left[1-erf\left(\frac{1}{\sqrt{2(1-\rho_{ab})}}(R_a^b-u\rho_{ab}-\sigma_b(1-\rho_{ab}^2))\right)\right]\,du
\\
& &\\
& &\\
E(e^{2x})&=&\exp\left( 2\mu_a+ \mu_b +2\sigma_a^2 \right)\int_{R_a^c}^{+\infty}\exp[-0.5(u-2\sigma_a)^2]\\
& &\cdot \left[1-erf\left(\frac{1}{\sqrt{2(1-\rho_{ab})}}(R_a^b-u\rho_{ab}))\right)\right]\,du
\\
& &\\
& &\\
E(e^{x})&=&\exp\left( \mu_a + \sigma_a^2 \right)\int_{R_a^c}^{+\infty}\exp[-0.5(u-\sigma_a)^2]\\
& &\cdot \left[1-erf\left(\frac{1}{\sqrt{2(1-\rho_{ab})}}(R_a^b-u\rho_{ab}))\right)\right]\,du\\
& &\\
& &\\
E(e^{y})&=&\exp\left( \mu_b + \sigma_b^2 \right))\int_{R_a^c}^{+\infty}\exp[-0.5(u-\sigma_b\rho_{ab})^2]\\
& &\cdot \left[1-erf\left(\frac{1}{\sqrt{2(1-\rho_{ab})}}(R_a^b-u\rho_{ab}-\sigma_b(1-\rho_{ab}^2))\right)\right]\,du\\
& &\\
& &\\
E(1)&=&\frac{1}{2\sqrt{2\pi}}\int_{R_a^c}^{+\infty}
\exp[-0.5u^2]\left[1-erf\left(\frac{1}{\sqrt{2(1-\rho_{ab})}}(R_a^b-u\rho_{ab})\right)\right]\,du\\
\end{eqnarray*}

\item $m(r_{c_a},r_{c_b},r_{c_c})$

\begin{eqnarray*}
m(r_{c_a},r_{c_b},r_{c_c}) &=&
\frac{1}{C_aC_bC_c}[S_aS_bS_bE(e^{x+y+z})-S_aS_bE(e^{x+y})(K_c^c+C_c)\\
& &-S_aS_cE(e^{x+z})(K_b^c+C_b)-S_bS_cE(e^{y+z})(K_a^c+C_a)\\
& &
+S_aE(e^{x})(K_b^c+C_b)(K_c^c+C_c)+E(e^{y})S_b(K_a^c+C_a)(K_c^c+C_c)\\
& &
+S_cE(e^{z})(K_a^c+C_a)(K_b^c+C_b)
-E(1)(K_a^c+C_a)(K_b^c+C_b)(K_c^c+C_c)\\
& &-(S_aE(e^x)-E(1)(K_a^c+C_a))(S_bS_cE(e^{y+z})\\
& &-S_c(K_b^c+C_b)E(e^{z})-S_b(K_c^c+C_c)E(e^{y})+(K_c^c+C_c)(K_b^c+C_b)E(1))\\
& &
-(S_bE(e^{y})-E(1)(K_b^c+C_b))(S_aS_cE(e^{x+z})-S_a(K_c^c+C_c)E(e^{x})\\
& &
-S_cE(e^{z})(K_a^c+C_a)+E(1)(K_a^c+C_a)(K_c^c+C_c))\\
& &-(S_cE(e^{z})-E(1)(K_c^c+C_c))(S_aS_bE(e^{x+z})\\
& &-S_a(K_b^c+C_b)E(e^{x})-S_b(K_a^c+C_a)E(e^{y})+E(1)(K_a^c+C_a)(K_b^c+C_b))\\
&
&+(S_aE(e^{x})-E(1)(K_a^c+C_a))(S_bE(e^{y})\\
& &-E(1)(K_b^c+C_b))(S_cE(e^{z})-E(1)(K_c^c+C_c))(3-E(1))]
\end{eqnarray*}

\[
(\ln\frac{K_a^c}{S_a}< x < +\infty) \ \ \  
(\ln\frac{K_b^c}{S_b}< y < +\infty) \ \ \ 
(\ln\frac{K_c^c}{S_c}< z < +\infty)
\] 

Nota\c{c}\~{a}o:

\begin{eqnarray*}
R_a^c &=& \frac{1}{\sigma_a}(ln\frac{K_a^c}{S_a}-\mu_a)\\
& &\\
R_b^c &=& \frac{1}{\sigma_b}(ln\frac{K_b^c}{S_b}-\mu_b)\\
& &\\
R_c^c &=& \frac{1}{\sigma_c}(ln\frac{K_c^c}{S_c}-\mu_c)\\
& &\\
a&=&1-\rho_{ab}^2 \\
b&=&1-\rho_{ac}^2 \\
c&=&1-\rho_{bc}^2 \\
& &\\
A&=& \rho_{ac}\rho_{bc}-\rho_{ab}\\
B&=& \rho_{ab}\rho_{bc}-\rho_{ac}\\
C&=& \rho_{ab}\rho_{ac}-\rho_{bc}\\
& &\\
T&=&\frac{1}{c}(\sigma_cB-\sigma_ac+\rho_{ab}(\sigma_cC-sigma_bc))\\
& &\\
L&=&\frac{1}{\sqrt{2}}(R_a^c+T)\\
\end{eqnarray*}

\begin{eqnarray*}
E(e^{x+y+z})&=&\int_{ln\frac{K_a^c}{S_a}}^{+\infty}\int_{ln\frac{K_b^c}{S_b}}^{+\infty}\int_{ln\frac{K_c^c}{S_c}}^{+\infty}f(x,y,z)exp(x+y+z)\,dz\,dy\,dx\\
& &\\
&\stackrel{\mathrm{1''}}{=}&\frac{exp{(\mu_a+\mu_b+\mu_c)}}{\sqrt{8R\pi^3}}\int_{R_a^c}^{+\infty}\int_{R_b^c}^{+\infty}\int_{R_c^c}^{+\infty}exp\{-\frac{1}{2R}[au^2+bv^2+cw^2\\
& &  \quad \quad \quad \quad +2uvA+2uwB+2vwC-2R\sigma_au-2R\sigma_bv-2R\sigma_cw]\}\,dw\,dv\,du\\
& &\\
&\stackrel{\mathrm{2''}}{=}&\frac{1}{2\pi\sqrt{c\pi}}exp(\mu_a+\mu_b+\mu_c+\frac{1}{2c}\sigma_c^2R)\int_{R_a^c}^{+\infty}\int_{R_b^c}^{+\infty}\left(\int_{F(u,v)}^{+\infty}e^{-\zeta^2}\,d\zeta
\right)\\
& & \cdot
exp\{-\frac{1}{2R}[u^2(a-\frac{B^2}{c})+v^2(b-\frac{C^2}{c})+2v(u(A-\frac{BC}{c})-R\sigma_b+\frac{RC\sigma_c}{c})\\
& & \quad \quad \quad \quad +2uR(\sigma\frac{B}{c}-\sigma_a)]\}\,dv\,du\\
& &\\
&=&\frac{1}{4\pi\sqrt{c}}exp(\mu_a+\mu_b+\mu_c+\frac{1}{2c}\sigma_c^2R)\int_{R_a^c}^{+\infty}\int_{R_b^c}^{+\infty}[1-erf(F(u,v))]\\
& & \cdot
exp\{-\frac{1}{2c}[u^2+v^2+2v(-u\rho_{ab}-c\sigma_b+C\sigma_c)+2u(\sigma_c
B-\sigma_a c)]\}\,dv\,du\\
& &\\
&\stackrel{\mathrm{3''}}{=}&\frac{1}{2\pi\sqrt{2}}exp(\mu_a+\mu_b+\mu_c+\frac{1}{2c}\sigma_c^2R+\frac{1}{2c}
(\sigma_cC-c\sigma_b)^2)\\
& &\cdot \int_{R_a^c}^{+\infty}\int_{G(u)}^{+\infty}[1-erf(F(u,\xi))]\\
& & \cdot
exp\{-\frac{1}{2c}[u^2c+v^2+2u(\sigma_cB-\sigma_a
c+\rho_{ab}(\sigma_cC-\sigma_b c))]-\xi^2\}\,d\xi \,du\\
& &\\
&=&\frac{1}{2\pi\sqrt{2}}exp(\mu_a+\mu_b+\mu_c+\frac{1}{2c}\sigma_c^2R+\frac{1}{2c}
(\sigma_cC-c\sigma_b)^2)\\
& &\cdot \int_{R_a^c}^{+\infty}\int_{G(u)}^{+\infty}[1-erf(F(u,\xi))]exp\{-\frac{1}{2}(u^2+2uT)-\xi^2\}\,d\xi \,du\\
& &\\
&\stackrel{\mathrm{4''}}{=}&\frac{1}{2\pi\sqrt{2}}exp(\mu_a+\mu_b+\mu_c+\frac{1}{2c}\sigma_c^2R+\frac{1}{2c}
(\sigma_cC-c\sigma_b)^2+0.5T^2)\\
& &\cdot
\int_{L}^{+\infty}\int_{G(\psi)}^{+\infty}[1-erf(F(\psi,\xi))]exp\{-\psi^2-\xi^2\}\,d\xi
\,d\psi\\
& &\\
&\stackrel{\mathrm{5''}}{=}&\frac{1}{2\pi}exp(\mu_a+\mu_b+\mu_c+0.5(\sigma_a^2+\sigma_b^2+\sigma_c^2)+\sigma_a\sigma_b\rho_{ab}+\sigma_a\sigma_c\rho_{ac}+\sigma_b\sigma_c\rho_{bc}-L^2)\\
& & \cdot \int_{0}^{+\infty}H_1(\varphi)exp(-\varphi^2)\,d\varphi\\
& &\\
& &\\
\end{eqnarray*}

onde 

\begin{eqnarray*}
H_1(\varphi)&=&exp(-2L\varphi)\int_{G_1(\varphi)}^{+\infty}[1-erf(F_1(\varphi,\xi))]exp(-\xi^2)\,d\xi\\
& &\\
G_1(\varphi)&=&\frac{1}{\sqrt{2c}}(R_b^c-c\sigma_b-C\sigma_c-\rho_{ab}(\sqrt{2}\varphi+L)+\rho_{ab}T)\\
& &\\
F_1(\varphi,\xi)&=&\sqrt{\frac{c}{2R}}[R_c^c+\frac{1}{c}(\varphi\sqrt{2}(B+\rho_{ac}C)
+(\sqrt{2}L-T)(B+C\rho_{ab})\\
& & \quad \quad +\xi C\sqrt{2c}+Cc\sigma_b-C^2\sigma_c)-\sigma_cR]\\
\end{eqnarray*}
As mudan\c{c}as de vari\'{a}veis indicadas acima s\~{a}o as seguintes:

\begin{eqnarray*}
\begin{array}{rl}
1^{''} &  \left\{
             \begin{array}{ll}
                u = \frac{1}{\sigma_a}(x-\mu_a)\\
                v = \frac{1}{\sigma_b}(y-\mu_b)\\
                w = \frac{1}{\sigma_c}(z-\mu_c)\\
                  \end{array}
            \right.\\
& \\
2^{''} &  \left\{
             \begin{array}{ll}
                \zeta = \sqrt{\frac{c}{2R}}(w+\frac{1}{c}(uB+vC-\sigma_cR))\\
             \end{array}
           \right.\\
& \\
3^{''} &  \left\{
             \begin{array}{ll}
                \xi = \frac{1}{\sqrt{2c}}(v-c\sigma_b+C\sigma_c-u\rho_{ab})\\
\end{array}          
\right.\\
& \\
4^{''} &  \left\{
             \begin{array}{ll}
                \psi = \frac{1}{\sqrt{2}}(u+T)\\
\end{array}          
\right.\\
& \\
5^{''} &  \left\{
             \begin{array}{ll}
                \varphi = \psi -L
\end{array}          
\right.\\
\end{array}
\end{eqnarray*}

\begin{eqnarray*}
E(e^{x+y})& =&\frac{1}{2\pi}\exp\left( \mu_a+ \mu_b +\frac{1}{2}(\sigma_a^2
+\sigma_b^2)
+\sigma_a\sigma_b\rho_{ab}\right)\int_{0}^{+\infty}H_2(\varphi)exp(-\varphi^2)\,d\varphi\\
& &\\
\hbox{onde}& & \\
& &\\
H_2(\varphi)&=&exp(-2L_2\varphi)\int_{G_2(\varphi)}^{+\infty}[1-erf(F_2(\varphi,\xi))]exp(-\xi^2)\,d\xi\\
& &\\
G_2(\varphi)&=&G_1(\varphi)-\frac{1}{\sqrt{2c}}\sigma_cC\\
& &\\
F_2(\varphi,\xi)&=&F_1(\varphi,\xi)+\sqrt{\frac{c}{2R}}\sigma_c(R+\frac{C^2}{c})\\
& &\\
T_2&=&-\sigma_a-\sigma_b\rho_{ab}\\
& &\\
L_2&=&\frac{1}{\sqrt{2}}(R_a^c+T_2)
\end{eqnarray*}

\begin{eqnarray*}
E(e^{x+z})& =&\frac{1}{2\pi}\exp\left( \mu_a+ \mu_c +\frac{1}{2}(\sigma_a^2
+\sigma_c^2)
+\sigma_a\sigma_c\rho_{ac}\right)\int_{0}^{+\infty}H_3(\varphi)exp(-\varphi^2)\,d\varphi\\
& &\\
\hbox{onde}& & \\
& &\\
H_3(\varphi)&=&exp(-2L_2\varphi)\int_{G_3(\varphi)}^{+\infty}[1-erf(F_3(\varphi,\xi))]exp(-\xi^2)\,d\xi\\
& &\\
G_3(\varphi)&=&G_1(\varphi)+\frac{1}{\sqrt{2c}}\sigma_bC\\
& &\\
F_3(\varphi,\xi)&=&F_1(\varphi,\xi)+\sqrt{\frac{c}{2R}}\sigma_bC\\
& &\\
T_3&=&T_1+\sigma_b\rho_{ab}\\
& &\\
L_3&=&\frac{1}{\sqrt{2}}(R_a^c+T_3)
\end{eqnarray*}

\begin{eqnarray*}
E(e^{y+z})& =&\frac{1}{2\pi}\exp\left( \mu_b+ \mu_c +\frac{1}{2}(\sigma_b^2
+\sigma_c^2)
+\sigma_b\sigma_c\rho_{bc}\right)\int_{0}^{+\infty}H_4(\varphi)exp(-\varphi^2)\,d\varphi\\
& &\\
\hbox{onde}& & \\
& &\\
H_4(\varphi)&=&exp(-2L_4\varphi)\int_{G_4(\varphi)}^{+\infty}[1-erf(F_4(\varphi,\xi))]exp(-\xi^2)\,d\xi\\
& &\\
G_4(\varphi)&=&G_1(\varphi)\\
& &\\
F_4(\varphi,\xi)&=&F_1(\varphi,\xi)\\
& &\\
T_4&=&T_1+\sigma_a\\
& &\\
L_4&=&\frac{1}{\sqrt{2}}(R_a^c+T_4)
\end{eqnarray*}

\begin{eqnarray*}
E(e^{x})& =&\frac{1}{2\pi}\exp(\mu_a+\frac{1}{2}\sigma_a^2-L_5^2)\int_{0}^{+\infty}H_5(\varphi)exp(-\varphi^2)\,d\varphi\\
& &\\
\hbox{onde}& & \\
& &\\
H_5(\varphi)&=&exp(-2L_5\varphi)\int_{G_5(\varphi)}^{+\infty}[1-erf(F_5(\varphi,\xi))]exp(-\xi^2)\,d\xi\\
& &\\
G_5(\varphi)&=&G_1(\varphi)-\frac{1}{\sqrt{2c}}(\sigma_cC-\sigma_bc)\\
& &\\
F_5(\varphi,\xi)&=&F_1(\varphi,\xi)-\sqrt{\frac{c}{2R}}[\frac{1}{c}(\sigma_bCc-C^2\sigma_c)-R\sigma_c]\\
& &\\
T_5&=&-\sigma_a\\
& &\\
L_5&=&\frac{1}{\sqrt{2}}(R_a^c+T_5)
\end{eqnarray*}

\begin{eqnarray*}
E(e^{y})& =&\frac{1}{2\pi}\exp(\mu_b+\frac{1}{2}\sigma_b^2-L_6^2)\int_{0}^{+\infty}H_6(\varphi)exp(-\varphi^2)\,d\varphi\\
& &\\
\hbox{onde}& & \\
& &\\
H_6(\varphi)&=&exp(-2L_6\varphi)\int_{G_6(\varphi)}^{+\infty}[1-erf(F_6(\varphi,\xi))]exp(-\xi^2)\,d\xi\\
& &\\
G_6(\varphi)&=&G_1(\varphi)-\frac{1}{\sqrt{2c}}\sigma_cC\\
& &\\
F_6(\varphi,\xi)&=&F_1(\varphi,\xi)+\sqrt{\frac{c}{2R}}\sigma_c(\frac{C^2}{c}-R)\\
& &\\
T_6&=&-\sigma_b\rho_{ab}\\
& &\\
L_6&=&\frac{1}{\sqrt{2}}(R_a^c+T_6)
\end{eqnarray*}

\begin{eqnarray*}
E(e^{z})& =&\frac{1}{2\pi}\exp(\mu_c+\frac{1}{2}\sigma_c^2-L_7^2)\int_{0}^{+\infty}H_7(\varphi)exp(-\varphi^2)\,d\varphi\\
& &\\
\hbox{onde}& & \\
& &\\
H_7(\varphi)&=&exp(-2L_7\varphi)\int_{G_7(\varphi)}^{+\infty}[1-erf(F_7(\varphi,\xi))]exp(-\xi^2)\,d\xi\\
& &\\
G_7(\varphi)&=&G_1(\varphi)-\frac{1}{\sqrt{2c}}\sigma_bC\\
& &\\
F_7(\varphi,\xi)&=&F_1(\varphi,\xi)+\sqrt{\frac{c}{2R}}\sigma_bC\\
& &\\
T_7&=&\frac{1}{c}(\sigma_cB+C\sigma_c\rho_{ab}\\
& &\\
L_7&=&\frac{1}{\sqrt{2}}(R_a^c+T_7)
\end{eqnarray*}

\begin{eqnarray*}
E(1)& =&\frac{1}{2\pi}\exp(-L_8^2)\int_{0}^{+\infty}H_8(\varphi)exp(-\varphi^2)\,d\varphi\\
& &\\
\hbox{onde}& & \\
& &\\
H_8(\varphi)&=&exp(-2L_8\varphi)\int_{G_8(\varphi)}^{+\infty}[1-erf(F_8(\varphi,\xi))]exp(-\xi^2)\,d\xi\\
& &\\
G_8(\varphi)&=&G_1(\varphi)-\frac{1}{\sqrt{2c}}(\sigma_cC-\sigma_bc)\\
& &\\
F_8(\varphi,\xi)&=&F_1(\varphi,\xi)-\sqrt{\frac{c}{2R}}(C\sigma_b-\sigma_c(\frac{C^2}{c}-R))\\
& &\\
T_8&=&0\\
& &\\
L_8&=&\frac{1}{\sqrt{2}}(R_a^c)
\end{eqnarray*}

\end{itemize}

%% file: apen1.tex
\appendix

\chapter{Matlab}  

\section{Hist\'{o}rico} 

Matlab, de Matrix Laboratory, \'{e} um ambiente interativo para
computa\c{c}\~{a}o envolvendo matrizes.  Matlab foi desenvolvido no
in\'{\i}cio da d\'{e}cada de 80 por Cleve Moler, no Departamento de
Ci\^{e}ncia da Computa\c{c}\~{a}o da Universidade do Novo M\'{e}xico,
EUA. 
 \index{Matlab} 
 
Matlab coloca \`{a} disposi\c{c}\~{a}o do usu\'{a}rio, num ambiente
interativo, as bibliotecas desenvolvidas nos projetos LINPACK e EISPACK. 
Estes projetos elaboraram bibliotecas de dom\'{\i}nio p\'{u}blico para
\'{A}lgebra Linear. LINPACK tem rotinas para solu\c{c}\~{a}o de
sistemas de equa\c{c}\~{o}es lineares, e EISPACK tem rotinas para
c\'{a}lculo de autovalores.  Os manuais destes projetos s\~{a}o portanto
documenta\c{c}\~{a}o complementar \`{a} documenta\c{c}\~{a}o do Matlab. 

Vers\~{o}es posteriores de Matlab foram desenvolvidas na firma comercial
MathWorks Inc., atualmente na vers\~{a}o 5.1, que det\^{e}m os direitos
autorais destas implementa\c{c}\~{o}es.  As vers\~{o}es recentes do
produto MATLAB melhoram significativamente o ambiente interativo,
incluindo facilidades gr\'{a}ficas de visualiza\c{c}\~{a}o e
impress\~{a}o; todavia a ``Linguagem Matlab'' manteve-se quase
inalterada.  Existem v\'{a}rios interpretadores da linguagem Matlab em
dom\'{\i}nio publico, como Matlab 1.0, Octave, rlab e Scilab.  Existem
tamb\'{e}m outros interpretadores comerciais de Matlab, como CLAM. 
Existem ainda v\'{a}rias Tool Boxes, bibliotecas vendidas pela MathWorks
e por terceiros, com rotinas em Matlab para \'{a}reas espec\'{\i}ficas. 

Usaremos a grafia de nome pr\'{o}prio, {\bf Matlab}, como refer\^{e}ncia
a linguagem, o nome em mai\'{u}sculas, {\bf MATLAB}, como refer\^{e}ncia
ao produto comercial da MathWorks, e a grafia em min\'{u}sculas, 
{\bf matlab}, como refer\^{e}ncia a um interpretador gen\'{e}rico da
linguagem Matlab.

\section{O Ambiente} 

Para entrar no ambiente matlab, simplesmente digite ``matlab''.  O
prompt do matlab \'{e} \verb#>># , que espera por comandos.  Para sair
use o comando \verb#quit#.  Dentro do ambiente matlab, um comando
precedido do bang, \verb#!#, \'{e} executado pelo sistema operacional,
assim: usando \verb#!dir# ou \verb#!ls# ficaremos sabendo os arquivos no
diret\'{o}rio corrente, e usando \verb#!edit# , \verb#!vi# ou
\verb#!emacs# , seremos capazes de editar um arquivo.  Normalmente
Matlab distingue mai\'{u}sculas de min\'{u}sculas. 

O comando \verb#help# exibe todos os comandos e s\'{\i}mbolos
sint\'{a}ticos dispon\'{\i}veis.  O comando \verb#help nomecom# fornece
informa\c{c}\~{o}es sobre o comando de nome {\em nomecom}.  O comando
\verb#diary nomearq# abre um arquivo de nome {\em nomearq}, e ecoa tudo
que aparece na tela para este arquivo.  Repetindo o comando \verb#diary#
fechamos este arquivo. O formato dos n\'{u}meros na tela pode ser 
alterado com o comando \verb#format#. 

Os comandos \verb#who# e \verb#whos# listam as vari\'{a}veis em
exist\^{e}ncia no espa\c{c}o de trabalho.  O comando \verb#clear# limpa o
espa\c{c}o de trabalho, extinguindo todas as vari\'{a}veis.  O comando
\verb#save nomearq# guarda o espa\c{c}o de trabalho no arquivo 
{\em nomearq}, e o comando \verb#load nomearq# carrega um espa\c{c}o de
trabalho previamente guardado com o comando \verb#save#. 

Em Matlab h\'{a} dois terminadores de comando: a v\'{\i}rgula, \verb#,#
, e o ponto-e-v\'{\i}rgula, \verb#;# .  O resultado de um comando
terminado por v\'{\i}rgula \'{e} ecoado para a tela.  O terminador
ponto-e-v\'{\i}rgula n\~{a}o causa eco.  Resultados ecoados na tela
s\~{a}o atribu\'{\i}dos a vari\'{a}vel do sistema \verb#ans# (de answer,
ou resposta).  O terminador v\'{\i}rgula pode ser suprimido no
\'{u}ltimo comando da linha.  Para continuar um comando na linha
seguinte devemos terminar a linha corrente com tr\^{e}s pontos,
\verb#...# , o s\'{\i}mbolo de continua\c{c}\~{a}o.  O sinal de
porcento, \verb#%# , indica que o resto da linha \'{e} coment\'{a}rio.

\section{Matrizes} 

O tipo b\'{a}sico do Matlab \'{e} uma matriz de n\'{u}meros complexos. 
Uma matriz real \'{e} uma matriz que tem a parte imagin\'{a}ria de todos
os seus elementos nula.  Um vetor linha \'{e} uma matriz $1\times n$,
um vetor coluna uma matriz $n\times 1$, e um escalar uma matriz
$1\times 1$. 

As atribui\c{c}\~{o}es \\  
 \verb#A = [1, 2, 3; 4, 5, 6; 7, 8, 9];#  ou \\  
 \verb#A = [1 2 3; 4 5 6; 7 8 9];# , \\  
s\~{a}o equivalentes, e atribuem \`{a} vari\'{a}vel $A$ o valor
 $$ A = \left[ \begin{array}{ccc} 
    1 & 2 & 3 \\ 4 & 5 & 6 \\ 7 & 8 & 9  \end{array} \right] 
 $$

Matrizes s\~{a}o delimitadas por colchetes, elementos de uma mesma linha
s\~{a}o separados por v\'{\i}rgulas (ou apenas por espa\c{c}os em
branco), e linhas s\~{a}o separadas por ponto-e-v\'{\i}rgula (ou pelo
caracter nova-linha).  O ap\'{o}strofe, \verb#'# , transp\~{o}em uma
matriz.  \'{E} f\'{a}cil em Matlab compor matrizes blocadas, desde que
os blocos tenham dimens\~{o}es compat\'{\i}veis! Por exemplo:
 
 $$
 A = \left[ \begin{array}{cc} 1 & 2 \\ 3 & 4 \end{array} \right] 
 \ , \ \ 
 B = \left[ \begin{array}{c} 5 \\ 6 \end{array} \right] 
 \ , \ \ 
 C = \left[ \begin{array}{c} 7 \\ 8 \\ 9  \end{array} \right] 
 \ , \ \ 
 \verb#D=[A,B;C'];# 
 \ \ D = \left[ \begin{array}{ccc} 1 & 2 & 5 \\ 3 & 4 & 6 \\ 
                 7 & 8 & 9 \end{array} \right]
 $$ 
Cuidado ao concatenar matrizes com os espacos em branco, pois estes 
s\~{a}o equivalentes a v\'{\i}rgulas, separando elementos. Assim: 
\verb#[1,2+3]==[1 5]# mas \verb#[1,2 +3]==[1,2,3]#. 
 \index{Matriz!blocada} 

H\'{a} v\'{a}rias fun\c{c}\~{o}es para gerar matrizes e vetores
especiais: \verb#zeros(m,n)#, \verb#ones(m,n)# e \verb#rand(m,n)#
s\~{a}o matrizes de dimens\~{a}o $m\times n$ com, respectivamente,
zeros, uns, e n\'{u}meros aleat\'{o}rios em [0,1].  O vetor \verb#i:k:j#
\'{e} o vetor linha $\left[i,i+k,i+2k,\ldots i+nk\right]$, onde 
 $n=\max m \mid i+mk\leq j$.  Podemos suprimir o ``passo'' $k=1$,
escrevendo o vetor \verb#i:1:j# simplesmente como \verb#i:j#.  Se $v$
\'{e} um vetor, \verb#diag(v)# \'{e} a matriz diagonal com diagonal $v$,
se $A$ \'{e} uma matriz quadrada, \verb#diag(A)# \'{e} o vetor com os
elementos da diagonal principal de $A$.  A matriz identidade de ordem
$n$ \'{e} \verb#eye(n)#, o mesmo que \verb#diag(ones(1,n))#. 
  \index{Matriz!\'{\i}ndices}

\verb#A(i,j)# \'{e} o elemento na $i$-\'{e}sima linha, $j$-\'{e}sima
coluna de $A$, $m\times n$.  Se $vi$ e $vj$ s\~{a}o vetores de
\'{\i}ndices em $A$, i.e.  vetores linha com elementos inteiros
positivos, em $vi$ n\~{a}o excedendo $m$, e em $vj$ n\~{a}o excedendo
$n$, \verb#A(vi,vj)# \'{e} a sub-matriz do elementos de $A$ com
\'{\i}ndice de linha em $vi$ e \'{\i}ndice de coluna em $vj$.  Em
particular \verb#A(1:m,j)#, ou \verb#A(:,j)# \'{e} a $j$-\'{e}sima coluna
de de $A$, e \verb#A(i,1:j)#, ou \verb#A(i,:)#, \'{e} a $i$-\'{e}sima
linha de $A$. Exemplo: 
 $$ 
 A = \left[ \begin{array}{ccc} 
     11 & 12 & 13 \\ 21 & 22 & 23 \\ 31 & 32 & 33  
     \end{array} \right] \ \ 
 vi = \left[ \begin{array}{c} 3 \\ 1 \end{array} \right] \ \  
 vj = \left[ \begin{array}{c} 2 \\ 3 \end{array} \right] \ \  
 A(vi,vj) = 
 \left[ \begin{array}{cc} 32 & 33 \\ 12 & 13 \end{array} \right]
 $$ 

As opera\c{c}\~{o}es de adi\c{c}\~{a}o, subtra\c{c}\~{a}o, produto,
divis\~{a}o e pot\^{e}ncia, \verb#+ - * / ^#, s\~{a}o as usuais da
\'{a}lgebra linear.  
 O operador de divis\~{a}o \`{a} esquerda, \verb#\#, fornece em 
 \verb#x = A\b;# uma solu\c{c}\~{a}o $x \mid A*x=b$.  
 O operador de divis\~{a}o a direita, \verb#/# , fornece em 
 \verb#a = b/A;# uma solu\c{c}\~{a}o $x \mid x*A=b$. Quando o sistema
\'{e} bem determinado isto \'{e} o mesmo que, respectivamente, 
 \verb#inv(A)*b# ou \verb#b*inv(A)#.  
 (Quando o sistema \'{e} super-determinado x \'{e} uma solu\c{c}\~{a}o no
sentido dos m\'{\i}nimos quadrados.)  
 Os operadores aritm\'{e}ticos de produto, pot\^{e}ncia e divis\~{a}o
tem a vers\~{a}o pontual, \verb#.* .^ ./ .\#  , que s\~{a}o executados
elemento a elemento.  Exemplo: 
 $$ x=\left[ \begin{array}{ccc} 5 & 5 & 5 \end{array} \right] \ , \ \ 
    y=\left[ \begin{array}{ccc} -1 & 0 & 1 \end{array} \right] \ , \ \ 
 x\, .*\, y =\left[ \begin{array}{ccc} -5 & 0 & 5 \end{array} \right]  
 $$ 

Matlab \'{e} uma linguagem declarativa (em oposi\c{c}\~{a}o a
imperativa) onde as vari\'{a}veis s\~{a}o declaradas, dimensionadas e
redimensionadas dinamicamente pelo interpretador.  Assim, se $A$
presentemente n\~{a}o existe, \verb#A=11;# declara e inicializa uma
matriz real $1\times 1$.  Em seguida, o comando \verb#A(2,3)=23;# 
redimensionaria $A$ como a matriz $2\times 3$ \newline 
\verb#[11, 0, 0; 0, 0, 23]#.  
A nome da matriz nula \'{e} \verb#[ ]#.  A atribui\c{c}\~{a}o 
 \verb#A(:,2)=[];# anularia a segunda coluna de $A$, tornado-a a matriz 
 $2\times 2$ \verb#[11, 0; 0, 23]#.

\section{Controle de Fluxo}

Os operadores relacionais da linguagem s\~{a}o \verb#< <= > >= == ~=#, 
que retornam valor $0$ ou $1$ conforme a condi\c{c}\~{a}o seja 
verdadeira ou falsa. Os operadores l\'{o}gicos, {\em n\~{a}o e ou}, 
s\~{a}o, respectivamente, \verb#~ & | #.  

Matlab possui os comandos de fluxo {\em for -- end}, {\em while -- end}
e {\em if -- elseif -- else -- end}, que tem a mesma sintaxe do Fortran,
exemplificada a seguir.  Lembre-se de {\bf n\~{a}o} escrever a palavra
\verb#elseif# como duas palavras separadas. 

\begin{verbatim}                               
for i=1:n                 if(x<0)          fatn=1;  
  for j=1:m                 signx=-1;      while(n>1) 
    H(i,j)=1/(i+j-1);     elseif(x>0)        fatn=fatn*n; 
  end                       signx=1;         n=n-1; 
end                       else             end 
                            signx=0; 
                          end                        
\end{verbatim} 

Uma considera\c{c}\~{a}o sobre efici\^{e}ncia: Matlab \'{e} uma
linguagem interpretada, e tempo de interpreta\c{c}\~{a}o de um comando
simples pode ser bem maior que seu tempo de execu\c{c}\~{a}o.  Para
tornar o programa r\'{a}pido, tente operar sobre matrizes ou blocos,
evitando loops expl\'{\i}citos que operem numa matriz elemento por
elemento.  Em outras palavras, tente evitar loops que repetem um comando
que atribui valores a elementos, por atribui\c{c}\~{o}es a vetores ou
matrizes.  As facilidades no uso de vetores de \'{\i}ndices, os
operadores aritm\'{e}ticos pontuais, e fun\c{c}\~{o}es como \verb#max# ,
\verb#min# , \verb#sort# , etc., tornam esta tarefa f\'{a}cil, e os
programas bem mais curtos e leg\'{\i}veis. 

\section{Scripts e Fun\c{c}\~{o}es}

O fluxo do programa tamb\'{e}m \'{e} desviado pela invoca\c{c}\~{a}o de
subrotinas.  A subrotina de nome {\em nomsubr} deve estar guardada no
arquivo \verb#nomsubr.m#  ; por isto subrotinas s\~{a}o tamb\'{e}m
denominadas M-arquivos (M-files).  H\'{a} dois tipos de subrotinas:
Scripts e Fun\c{c}\~{o}es. 

Um script \'{e} simplesmente uma seq\"{u}\^{e}ncia de comandos, que
ser\~{a}o executados como se fossem digitados ao prompt do matlab. 
Subrotinas podem invocar outras subrotinas, inclusive recursivamente. 

Um M-arquivo de fun\c{c}\~{a}o em Matlab come\c{c}a com a
declara\c{c}\~{a}o da forma \\ 
 \verb#[ps1, ps2,... psm] = nomefunc( pe1, pe2,... pen )#  \\  
 A lista entre par\^{e}nteses \'{e} a lista de par\^{a}metros de entrada
da fun\c{c}\~{a}o, e a lista entre colchetes s\~{a}o os par\^{a}metros
de sa\'{\i}da.  Par\^{a}metros s\~{a}o vari\'{a}veis locais, assim como
s\~{a}o vari\'{a}veis locais todas as vari\'{a}veis no corpo da
fun\c{c}\~{a}o. 

Ao invocarmos a fun\c{c}\~{a}o {\em nomefunc} com o comando \\ 
  \verb#[as1,... asm] = nomefunc(ae1,... aen);# \\  
 Os argumentos de entrada, $ae1,\ldots aen$, s\~{a}o passados por valor
aos (i.e.  copiados nos) par\^{a}metros de entrada, e ao fim do
M-arquivo, ou ao encontrar o comando \verb#return# , os par\^{a}metros
de sa\'{\i}da s\~{a}o passados aos argumentos de sa\'{\i}da.  Exemplos:

\begin{verbatim} 
 function [mabel, mabin] = vmax(v) 
 % procura o maior elemento, em valor absoluto, 
 % dentro de um vetor, linha ou coluna.  
 %Obs: Esta funcao NAO segue os conselhos 
 %para operar sobre blocos, e nao elementos! 
                                             
 [m,n]=size(v);                             
 if( m ~= 1  &  n ~= 1 ) 
   erro; 
 else 
   K=max([m,n]); 
   mabel= abs(v(1)); mabin= 1;  
   for k=2:K 
     if( abs(v(k)) > mabel ) 
       mabel= abs(v(k)); mabin= k; 
     end%if 
   end%for  
 end%else 
\end{verbatim} 

Para nos referir, dentro de uma fun\c{c}\~{a}o, a uma vari\'{a}vel
externa, esta deve ter sido declarada uma vari\'{a}vel global com o
comando \verb#global#.  A forma de especificar uma vari\'{a}vel como global
muda um pouco de interpretador para interpretador, e mesmo de vers\~{a}o
para vers\~{a}o: Diga \verb#help global# para saber os detalhes de como
funciona a sua implementa\c{c}\~{a}o.

%% file: apen2.tex
\chapter{Manual do Critical-Point for Windows}                          
			  
\begin{center}
   V. 3.00 \ 01-12-1998
\end{center}

 Critical-Point for Windows \'e um programa integrado para resolver e analisar 
modelos de otimiza\c{c}\~ao de portf\'olios de tipo m\'edia-vari\^ancia 
(modelo de Markowitz). 
 O programa \'{e} distribuido nas vers\~{o}es estudantil e profissional.
 As vers\~{o}es s\~{a}o praticamente id\^{e}nticas, tanto na forma como
na apresenta\c{c}\~{a}o da an\'{a}lise.  A vers\~ao profissional pode
resolver modelos de grande porte,  sendo orientada para uso comercial.
J\'a a  vers\~ao estudantil permite a an\'{a}lise de modelos com at\'{e}
12 ativos, 4 op\c{c}\~{o}es e 20 restri\c{c}\~{o}es, mais ativos sem
risco. A vers\~ao estudantil deve ser utilizada exclusivamente para
pesquisa  acad\^emica e atividades sem fim lucrativo, estando
dispon\'{\i}vel para  estes prop\'ositos. Gra\c{c}as ao apoio da  BM\&F,
Bolsa de Mercadorias   e de Futuros de S\~ao Paulo, e  ao NOPEF-USP  -
N\'ucleo de Otimiza\c{c}\~ao  e Processos Estoc\'asticos Aplicados a
Economia e Finan\c{c}as da Universidade  de S\~ao Paulo. Os direitos
autorais das duas vers\~oes pertencem ao  autor do programa,  Julio M.
Stern, que pode ser contactado em {\it jstern@ime.usp.br} ou 
{\it www.ime.usp.br/~jstern}

 \section{Instala\c{c}\~{a}o}

	A configura\c{c}\~{a}o m\'{\i}nima exigida pela vers\~{a}o
estudantil \'{e} um micro IBM-PC Pentium com 16MB de RAM, 3MB livres no
disco r\'{\i}gido, e MS-Windows 95. 

Para iniciar a instala\c{c}\~ao siga os seguintes passos:
\begin{enumerate}
\item     Coloque o disco de instala\c{c}\~{a}o no drive A ;
\item     Execute (do Windows) A:$\backslash$SETUP  .
\end{enumerate}

O Critical-Point vers\~{a}o estudantil ser\'{a} instalado no
sub-diret\'{o}rio C:$\backslash$SCRIPO, e a vers\~{a}o profissional no
diret\'{o}rio C:$\backslash$CRIPO. 
N\~{a}o altere o nome e a localiza\c{c}\~{a}o
destes diret\'{o}rios no processo de instala\c{c}\~{a}o. 

Observa\c{c}\~ao:   Se o programa de instala\c{c}\~{a}o abortar com  uma
mensagem de erro a res\-peito de um arquivo *.DLL ou *.OCX, saia do
Windows,  apague este arquivo do diret\'{o}rio \\
C:$\backslash$WINDOWS$\backslash$SYSTEM  e recomece a
instala\c{c}\~{a}o.

\subsection*{Abertura}

Para abrir o aplicativo Critical-Point: 

\begin{enumerate}
\item    Abra o Windows e duplo-clique o \'{\i}cone do
Critical-Point;
\item    Digite o Password: ``student''. Use apenas letras
min\'{u}sculas e sem as aspas;  
\item   A barra de menu do Critical-Point aparecer\'{a} no canto superior 
esquerdo da tela.
\end{enumerate}

 \section{Gera\c{c}\~{a}o da Fronteira Eficiente}
 
Para tra\c{c}ar a Fronteira Eficiente de Markowitz, \'{e}
necess\'{a}rio clicar, em sequ\^{e}ncia, os bot\~{o}es dos
programas:  
\begin{enumerate}
\item    Filtro (F), 
\item    incluir Derivativos (D), 
\item    Compilador MDL (M),  
\item    Programa\c{c}\~{a}o Quadr\'{a}tica Param\'etrica (Q) 
\end{enumerate}
 na barra de bot\~{o}es.  
 Alternativamente, pode-se executar estes programas atrav\'es do menu 
Run. Esta sequ\^encia resolve o modelo especificado nos 
arquivos MODEL.CP e MODPS.CP como ser\'a apresentado adiante.

 \subsection*{Visualiza\c{c}\~{a}o da Fronteira Eficiente}
 
 Clique o bot\~ao Graph (G). O gr\'{a}fico contendo
a fronteira eficiente aparecer\'{a}.  Os pontos cr\'{\i}ticos,
utilizados para o c\'{a}lculo da fronteira s\~{a}o identificados pelo
s\'{\i}mbolo ``+''.  Cada Ponto Cr\'{\i}tico \'{e} caracterizado por
alguma mudan\c{c}a qualitativa no portf\'{o}lio \'{o}timo como  
um ativo que entrou ou saiu do portf\'olio, ou uma restri\c{c}\~{a}o 
que passou de ativa (justa) a inativa(folgaada), ou vice-versa. 

 Os t\'{\i}tulos, cores, formato num\'{e}rico,
visualiza\c{c}\~{a}o da reta tangentes e muitas outras op\c{c}\~oes 
poder\~{a}o ser modificadas atrav\'{e}s do menu Options Graph.  As
novas op\c{c}\~{o}es do usu\'{a}rio ser\~{a}o salvas automaticamente.
Utilize o comando Options Reset Graph   para  retornar aos valores default. 

 \subsection*{Numeric Template}
 
	Esta \'{e} a janela que fornece as informa\c{c}\~{o}es
num\'{e}ricas sobre a Fronteira Eficiente.  Para cham\'{a}-la, acione o
bot\~ao Numeric Template (N). A janela do Numeric
Template \'{e} composta de seis linhas e oito colunas 
mais os bot\~{o}es Recalculate, Plot, Report e Save.  As linhas s\~ao 
denominadas Sele\c{c}\~ao A at\'e Sele\c{c}\~ao F e podem ser utilizadas para 
analisar at\'e seis portf\'olios eficientes simultaneamente.

 A primeira coluna apresenta  valor de $\eta$, o par\^ametro de 
pondera\c{c}\~ao  entre o retorno esperado e o risco do portf\'olio. 
Trata-se de um par\^ametro 
de propens\~ao ao risco (em oposi\c{c}\~ao \`a avers\~ao ao risco). 
Quanto maior $\eta$, maior a import\^ancia do retorno esperado do 
portf\'olio em rela\c{c}\~ao ao risco do portf\'olio. 

 A segunda coluna apresenta o retorno esperado do portf\'olio (e). 

 A terceira exibe a vari\^ancia do portf\'olio (v). 

A quarta coluna  mostra o desvio-padr\~{a}o (s) do portf\'{o}lio.

 A quinta  coluna mostra o valor da taxa de retorno sem risco, $r$. 
 Esta \'e a taxa de juros para o qual um determinado portf\'{o}lio \'{e}
tangente. Alternativamente, dada uma taxa de juros, o Critical-Point
calcula qual o portf\'{o}lio tangente correspondente. 

 A sexta e s\'etimas  colunas mostram os par\^{a}metros $k$ e $l$. 
 O par\^ametro $k$ indica que  o portf\'olio situa-se entre os pontos
cr\'{\i}ticos $k$ e $k+1$. O  par\^ametro $l$ fornece a posi\c{c}\~ao
relativa entre estes dois  extremos, de tal forma que $l=0$ significa
que o portf\'olio \'e o  ponto cr\'{\i}tico  $k$, $l=0.5$ significa que
o o portf\'olio est\'a  a meio  caminho entre $k$ e $k+1$, e $l=1$
significa que o portf\'olio \'e o  ponto cr\'{\i}tico $k+1$. 

 A oitava coluna indica o status da sele\c{c}\~{a}o.  
 O s\'{\i}mbolo:
 
 \begin{itemize}

 \item[]  $\emptyset$ \ \ \ mostra que a sele\c{c}\~{a}o ainda n\~{a}o foi
recalculada; 

 \item[]  + \ \ \      significa que est\'{a} em cima de um ponto
cr\'{\i}tico; 

 \item[]  $\sqrt{}$ \ \  significa que a sele\c{c}\~{a}o  n\~ao 
\'{e} um ponto cr\'{\i}tico; 

 \item[]  $\uparrow  \downarrow$   \ \ indicam que o par\^{a}metro de entrada  
est\'{a} muito alto ou muito baixo.
\end{itemize}

 Podemos selecionar um portf\'olio digitando o valor de qualquer um destes 
par\^ametros: acione o mouse sobre o campo de entrada e digite o 
valor do par\^ametro (este campo aparecer\'a em azul e sublinhado). Em 
seguida acione o bot\~ao Recalculate e todos os demais par\^ametros 
(que aparecer\~ao em preto) ser\~ao recalculados consistentemente 
com  o par\^ametro digitado. Sempre que uma sele\c{c}\~ao \'e recalculada, 
uma janela auxiliar apresentar\'a a composi\c{c}\~ao do portf\'olio 
selecionado.

  Os par\^ametro e e s podem ser selecionados diretamente na janela  
Graph: use o cursor do mouse para apontar para uma posi\c{c}\~ao 
selecionada na fronteira eficiente; em seguida clique o bot\~ao da 
esquerda para copiar o par\^ametro do retorno esperado no campo 
apropriado da linha selecionada (utilize o bot\~ao da direita  
para copiar a coordenada s).

 O bot\~{a}o Plot tra\c{c}a a Fronteira Eficiente al\'em de exibir os 
pontos selecionados (indicados por pequenos
c\'{\i}rculos sobre a curva), a reta tangente \`a fronteira eficiente,  etc. 
Quando um portf\'olio \'e 
selecionado por sua taxa de juros (r), tamb\'em a linha tangente \'e 
desenhada.

 O bot\~{a}o Report gera uma descri\c{c}\~{a}o de todos os portf\'olios 
selecionados  num arquivo ASCII chamado REPORT.CP. Este arquivo pode ser
lido e impresso atrav\'{e}s do MS-DOS Editor que \'{e} acionado
automaticamente atrav\'{e}s do menu Open File na barra de  bot\~oes. 
 O editor default pode ser modificado nas op\c{c}\~oes do menu.  
 O bot\~{a}o Save grava os resultados antes da sa\'{\i}da do programa. 

 O formato num\'{e}rico do Numeric Template pode ser alterado
atrav\'{e}s do menu Options Numeric. Para voltar ao formato original, 
use o comando Options Reset Numeric. 

 \subsection*{Impress\~{a}o}
 
 Acione File Print Graph no menu ou clique o \'{u}ltimo bot\~{a}o da
barra de bot\~{o}es para a impress\~{a}o da Fronteira Eficiente.  
 A impressora e os defaults de impress\~{a}o usados ser\~{a}o aqueles
definidos para o ambiente Windows.  O comando File Print Setup do menu
permite a reconfigura\c{c}\~{a}o da impressora.   Atrav\'{e}s do comando
Options To Clipboard, pode-se enviar o gr\'{a}fico para o Clipboard do
Windows no formato Bitmap ou Metafile.

 \section{Descri\c{c}\~ao do modelo e linguagem MDL}

 Os arquivos DERIV.CP e MODEL.CP cont\'{e}m  as informa\c{c}\~{o}es que
o usu\'{a}rio deve fornecer ao Critical-Point.  Estes arquivos ASCII
podem ser editado pelo editor MS-DOS Editor ou qualquer outro  editor
sem formata\c{c}\~ao.
 O arquivo MODEL.CP cont\'{e}m os  nomes e informa\c{c}\~oes 
a respeito de cada ativo que podem compor o portf\'{o}lio, bem como
restri\c{c}\~{o}es que o portf\'{o}lio deve obedecer.  Um exemplo \'{e} o
seguinte:

\begin{verbatim}
 all  = {TEL4,ELE6,PET4,BB4,BBD4,SCO4,CEV4,BRH4}; 

 #sectors
 energy  = {ELE6,PET4};
 bank = {BB4,BBD4};
 food = {SCO4,CEV4,BRH4};
 state = {TEL4,ELE6,PET4,BB4};
 private = ~state;     

 # constraints 
 normal: sum[all] $ ==1;
 statelim: sum[state] $ <= 0.5;
 statebanks= state & banks;
 statebklim:  sum[statebanks] $ <= 0.1; 
 foodsbound:  sum[food] $ >= 0.2; 

 liquindex[all] = {1.0@TEL4,0.6@ELE6,0.5@PET4,0.4@BB4,
                   0.4@BBD4,0.2@SCO4,0.2@CEV4,0.3@BRH4}; 
 liqconstr: for[all] $ <= 0.5*liquindex;  
\end{verbatim}

 A primeira linha define o conjunto universo, isto \'e, quais s\~{a}o 
os ativos que podem compor o portf\'olio. 

 A quarta linha define o subconjunto ou setor  de energia e as 
linhas seguintes definem outros setores (banc\'ario,alimenta\c{c}\~ao, 
estatais, privadas). 

 A primeira restri\c{c}\~ao (normal) estabelece que a soma das 
participa\c{c}\~{o}es de todos os ativos no portf\'{o}lio \'{e} igual 
a 1. Esta \'{e} a restri\c{c}\~{a}o normal que tem de estar sempre
presente, indicando o total a ser investido no portf\'{o}lio.  Este
total pode ser um n\'{u}mero de unidades monet\'arias, mas geralmente
\'{e} tomado igual a 1 ou  100  de modo que, as quantidades a serem
investidas em cada ativo  representem fra\c{c}\~{o}es da unidade ou 
percentuais do investimento total.  O s\'{\i}mbolo ``=='' \'e o
s\'{\i}mbolo de igualdade, n\~ao devendo  ser confundido com ``='', o
sinal de atribui\c{c}\~ao. 

 A segunda restri\c{c}\~ao determina que a soma dos investimentos nos 
ativos estatais ( definido como  {\tt \{TEL4,ELE6,PET4,BB4\} } )
deve ser menor ou igual a 0.5

 A linha que segue a segunda restri\c{c}\~ao define o conjunto de 
bancos estatais, {\tt statebanks}, contendo os ativos que pertencem aos 
conjuntos de  bancos e  de estatais. 
 A terceira  restri\c{c}\~ao estabelece que a soma dos investimentos 
nos bancos estatais deve ser menor ou igual a 0.1. 

 A restri\c{c}\~ao em alimenta\c{c}\~ao, {\tt foodsbound},  determina que 
a soma dos investimentos neste setor deve exceder 0.2. 
 
 A \'ultima restri\c{c}\~ao estabelece que o investimento em cada 
ativo do universo n\~{a} deve exceder ao \'{\i}ndice de liquides definido 
no vetor {\tt liquindex}.

 \subsection*{Retornos Esperados e Risco}

O arquivo MODPS.CP cont\'{e}m informa\c{c}\~{o}es sobre o
retorno esperado de cada ativo, ao fim do per\'{\i}odo em
considera\c{c}\~{a}o no modelo, bem como o desvio padr\~{a}o (ou a
volaatilidade, uma  medida de risco) associado a este progn\'{o}stico.

\begin{verbatim}
all={
TEL4,ELE6,PET4,BB4,BBD4,SCO4,CEV4,BRH4};

er[all]={
4.521883e-003@TEL4, 9.349340e-002@ELE6, 1.414101e-001@PET4, 
4.441184e-002@BB4, 4.125617e-002@BBD4, 2.153917e-002@SCO4, 
-1.467467e-001@CEV4, 7.254108e-002@BRH4 };

std[all]={
2.105123e-001@TEL4, 3.214724e-001@ELE6, 2.988641e-001@PET4, 
2.952717e-001@BB4, 2.019181e-001@BBD4, 1.709987e-001@SCO4, 
2.471665e-001@CEV4, 1.866201e-001@BRH4 };

Hurst=0.50;
extrap=30;
sample=90;

\end{verbatim}
	
    O vetor er cont\'em os retornos esperados, e o vetor std o
desvio padr\~{a}o destes retornos.  Dizemos que esperamos do ativo
asseta um retorno de er[asseta],  ``mais ou menos'' std[asseta].  
	
   Ao escrever um n\'{u}mero decimal, n\~{a}o ESQUE\c{C}A O ZERO ANTES
DO PONTO DECIMAL.

 \section*{A Linguagem de Descri\c{c}\~ao do Modelo}

	Vejamos em maior detalhe a linguagem MDL ou Model 
Description Language. Esta linguagem 
tem por objetivo descrever, f\'{a}cil e intuitivamente, as 
restri\c{c}\~{o}es que o portf\'{o}lio deve obedecer. 
   
 \subsection*{Nomes}

Um nome v\'{a}lido \'{e} uma sequ\^encia de letras,
d\'{\i}gitos, sublinhas, e pontos, come\c{c}ando com uma letra, e tendo
no m\'{a}ximo 15 caracteres. N\~ao \'e permitido utilizar o mesmo nome 
para dois ou mais objetos em um programa na linguagem MDL. Tamb\'em n\~ao 
\'e permitido utilizar como nomes palavras reservadas da linguagem. As 
palavras reservadas s\~ao: abs, exp, for, ln, max, min, maxe, mine, print, 
sign, sum. 

 \subsection*{\'Indices}

Os modelos tipo Markowitz constroem portf\'{o}lios
\'{o}timos a partir de ativos dentro de um dado universo.  Este
conjunto de todos os ativos em considera\c{c}\~{a}o, tamb\'em denominado 
conjunto universo,  \'{e} o conjunto {\it all}, ou o conjunto de 
\'{\i}ndices do modelo. A defini\c{c}\~ao do conjunto universo 
deve ser a primeira linha em um programa MDL.   

 \subsection*{Conjuntos}

Um conjunto \'e definido por uma atribui\c{c}\~ao constituida por:         

\begin{enumerate}
\item   O nome do conjunto
\item   O s\'{\i}mbolo de atribui\c{c}\~ao ``=''. 
\item   Uma lista de nomes de ativos
\item   O terminador de senten\c{c}a ``;'' 
\end{enumerate}
\noindent        
Uma lista de nomes \'{e} constituida de: 
\begin{enumerate}
\item    O s\'{\i}mbolo de come\c{c}o de lista,  ``\{'', 
\item    Uma sequ\^{e}ncia de nomes de ativos, separados por
v\'{\i}rgulas.
\item    O  s\'{\i}mbolo de fim de lista, ``\}''. 
\end{enumerate}
Al\'em de definir conjuntos  pela lista de seus elementos,
podemos ter conjuntos definidos por opera\c{c}\~{o}es entre 
outros conjuntos. Estas opera\c{c}\~{o}es s\~{a}o: 

\begin{itemize}
\item[\~\ ]
O complemento de um dado conjunto, \'{e} o conjunto de todos
os elementos pertencentes a all que n\~ao est\~ao presentes 
no conjunto.
	
\item[$\mid$]
A uni\~{a}o
de dois conjuntos \'{e} o conjunto dos elementos no
primeiro,  no segundo, ou em ambos os conjuntos. 
	 
\item[$\&$]
A intersec\c{c}\~{a}o de dois conjuntos \'{e} o conjunto dos 
elementos que pertencem tanto ao primeiro quanto ao
segundo conjunto. 
	
\item[$\backslash$]
A diferen\c{c}a entre dois conjuntos \'{e} constitu\'{\i}da pelos
elementos que pertencem ao primeiro mas n\~{a}o ao segundo
conjunto.

\end{itemize}

Nos exemplos seguintes definimos alguns conjuntos novos 
atrav\'es de conjuntos pr\'e-existentes, por exemplo: 

\begin{verbatim}
private = ~state; 
statebanks = state & banks; 
foodandenergy = food | energy;
\end{verbatim}

 A preced\^{e}ncia destes operadores \'{e} a mesma da aritm\'{e}tica
comum: em uma dada express\~ao i.e.  primeiro  fazemos todas as
opera\c{c}\~{o}es de complemento, depois as  opera\c{c}\~{o}es de
intersec\c{c}\~ao, e  finalmente as uni\~oes e diferen\c{c}as. 
 A preced\^encia de  c\'alculo pode ser modificada utilizando  
par\^{e}nteses como sinal de pontua\c{c}\~{a}o. 

 \subsection*{Escalares}
 
 Um escalar \'{e} simplesmente um n\'{u}mero inteiro ou real. N\'{u}meros 
inteiros s\~{a}o formados por uma sequ\^encia de
um ou mais d\'{\i}gitos decimais: 1,2,...,9,0. 
Um n\'{u}mero inteiro pode ainda ser precedido de um sinal, + ou
-, sem espa\c{c}o entre o sinal e a sequ\^{e}ncia de d\'{\i}gitos. 
Assim s\~{a}o exemplos de n\'{u}meros inteiros: 0; -998; 754.   

\noindent
Um n\'{u}mero real pode ser representado em dois  formatos:
\begin{itemize}
\item[]
Nota\c{c}\~ao de ponto flutuante
\item[]
Nota\c{c}\~ao de ponto fixo
\end{itemize}

\begin{enumerate}

\item   Nota\c{c}\~{a}o de ponto fixo: 

Um n\'{u}mero escrito
nesta forma \'{e} uma sequ\^{e}ncia de:  um inteiro, o ponto decimal
e um inteiro sem sinal. 

Note que, em ponto fixo, \'{e} necess\'{a}rio escrever ao menos
um d\'{\i}gito, possivelmente o zero, tanto antes como depois do ponto.
Note tamb\'em que o separador decimal \'{e} o ponto, e n\~{a}o a
v\'{\i}rgula ou qualquer outro s\'{\i}mbolo.  Assim, s\~{a}o exemplos
corretamente grafados 0.0, 37.65; -8.83, mas s\~{a}o {\bf incorretos} os
exemplos 10. ; .15 e -.88. 

\item 
Nota\c{c}\~{a}o  de ponto flutuante (cient\'{\i}fica): 
\begin{enumerate}
\item    Um n\'{u}mero inteiro ou um real em ponto fixo.  Esta
\'{e} a mantissa do n\'{u}mero em nota\c{c}\~{a}o cient\'{\i}fica. 
\item    A letra E, mai\'{u}scula ou min\'{u}scula. 
\item    Um n\'{u}mero inteiro.  Este \'{e} o expoente do
n\'{u}mero em nota\c{c}\~{a}o cient\'{\i}fica. 
\end{enumerate}

	O expoente do n\'{u}mero em nota\c{c}\~{a}o cient\'{\i}fica
indica de quantos d\'{\i}gitos, para a direita ou para a esquerda,
devemos deslocar o ponto na mantissa.  Esta nota\c{c}\~{a}o \'{e}
\'{u}til para escrever n\'{u}meros muito grandes ou muito pequenos, pois
evita longas seq\"u\^{e}ncias de zeros antes ou depois do ponto. 
Assim, podemos representar 1000000 por 1E6, -1000000 por -1E6, 0.000001
por -1E-6 e -0.000001 por -1E-6. 
\end{enumerate}

 \subsection*{Vetores Associativos}

Vetores associam um valor num\'{e}rico a cada ativo. 
 Atribui\c{c}\~oes a vetores s\~ao sempre 
restritas ao conjunto de dom\'{\i}nio da defini\c{c}\~ao 
que \'e indicado no lado 
esquerdo da atribui\c{c}\~ao. Por exemplo: usando o vetor j\'a 
definido anteriormente, temos: 

\begin{verbatim}
std[all]={
2.105123e-001@TEL4, 3.214724e-001@ELE6, 2.988641e-001@PET4, 
2.952717e-001@BB4, 2.019181e-001@BBD4, 1.709987e-001@SCO4, 
2.471665e-001@CEV4, 1.866201e-001@BRH4 };
mincapt[private] = { 0.05@CEV4, 0.03@SCO4, 0.02@BBD4 }; 
maxmin[private] = max( minpart , mincapt ); 
\end{verbatim}

 A express\~ao  $<$valor$>$@$<$ativo$>$ associa o valor do vetor ao ativo.

 As seguintes regras se aplicam a atribui\c{c}\~oes: 
 \begin{enumerate}
 \item[1-]  Uma atribui\c{c}\~ao n\~ao muda os valores 
de um vetor fora dos limites do dom\'{\i}nio da atribui\c{c}\~ao. 
 \item[2-] 
 Quando o dom\'{\i}nio \'e omitido, \'e tomado 
como o universo, all.
 \item[3-]
 O valor inicial de um vetor em todos os ativos \'e zero.  
 \end{enumerate}

 Podemos tamb\'{e}m atribuir valores a vetores atrav\'es 
opera\c{c}\~{o}es aritm\'{e}ticas (adi\c{c}\~ao, subtra\c{c}\~ao, 
produto, divis\~ao e exponencia\c{c}\~ao entre
dois vetores.  As opera\c{c}\~oes s\~ao calculadas para cada 
posi\c{c}\~ao, isto \'e, elemento a elemento.  
Por exemplo, ap\'os a atribui\c{c}\~ao 
 \begin{verbatim}
 example[private] = std*er;
 \end{verbatim}
 os valores dos elementos do vetor example nas posi\c{c}\~oes  
correspondentes \`as empresas  privadas \'e 
 {\tt std*er;} e os valores 
 em outros elementos permanecem inalterados ou nulos (caso 
seja a primeira atribui\c{c}\~ao ao vetor). 

 Podemos tamb\'em realizar opera\c{c}\~oes aritm\'eticas entre 
escalares e vetores. Neste caso o escalar \'e interpretado como 
um vetor constante.

Exemplo:

\begin{verbatim} 
 er3[a]=er*3    # cada elemento de p que pertence ao subconjunto a
	     # eh multiplicado por tres e armazenado em p3
\end{verbatim} 

 Finalmente, podemos acessar e modificar o valor de um vetor em um 
particular ativo. Por exemplo: {\tt std[CEV4]=0.55;}

 \subsection*{Fun\c{c}\~oes Escalares}

 As fun\c{c}\~oes  logar\'{\i}tmo,  ln(),  exponencial, exp(),  valor
absoluto, abs(),  sinal, sign(), maior elemento maxe(), menor elemento
mine() e  soma interna sum() podem ser aplicadas a escalares e vetores.
 Assim  como express\~oes aritm\'eticas, estas fun\c{c}\~oes s\~ao
aplicadas a  cada elemento de um vetor. Por exemplo: 
 \begin{verbatim}lner[a] = ln(er);\end{verbatim}

As fun\c{c}\~oes m\'aximo e m\'{\i}nimo podem ser calculadas em pares 
de escalares ou vetores. Estas fun\c{c}\~oes tamb\'em s\~ao calculadas 
pontualmente  quando assumem argumentos que s\~ao vetores. Por exemplo: 

\begin{verbatim}
 a1={1@AA, 2@BB, 3@DD};    # zero@CC is implicit
 b1={5@BB, 1@DD};
 c1=max(a1, b1); 

 # c1  is  {1@AA, 5@BB, 0@CC, 3@DD};
\end{verbatim}

 \subsection*{Restri\c{c}\~oes}

 As restri\c{c}\~{o}es do problema definem as condi\c{c}\~{o}es a
serem satisfeitas pelo portf\'{o}lio. A primeira restri\c{c}\~{a}o,
imprescind\'{\i}vel, \'{e} a condi\c{c}\~{a}o de normaliza\c{c}\~{a}o:  
 \begin{verbatim}
 normal: sum[all] $ == 1;
 \end{verbatim}

 Ela indica a quantidade total a ser investida no portf\'olio, que em 
geral \'e tomada como 1 ou 100, de forma que a composi\c{c}\~ao de 
portf\'olio selecionada aparece como fra\c{c}\~oes da unidade ou 
por fra\c{c}\~ao percentual. 
 S\~{a}o as vari\'aveis de decis\~{a}o do
modelo, representando o valor investido no i-\'{e}simo
ativo do portf\'{o}lio. Equa\c{c}\~{o}es de igualdade s\~{a}o indicadas
pelo sinal ``=='', e inequa\c{c}\~{o}es de desigualdade s\~{a}o indicadas
pelos sinais ``$<=$'' e ``$>=$''. 
       
\noindent
 Observa\c{c}\~ao importante: 
 \begin{enumerate}
 \item N\~ao confunda a atribui\c{c}\~ao ``='' com o sinal de 
igualdade  ``==''.
 \item As restri\c{c}\~oes de igualdade   devem ser definidas antes 
das desigualdades. 
 \end{enumerate}
 Podemos definir restri\c{c}\~{o}es em termos de somas ponderadas dos
investimentos feitos em um conjunto de ativos, como em
{\tt foodsbound : sum[food] \$ $<=$ 0.2;}
 estabelecendo que a soma 
dos investimentos no setor de alimenta\c{c}\~ao n\~ao deve ultrapassar 
20 por cento.  

 Podemos, da mesma forma, estabelecer conjuntos  de 
restri\c{c}\~oes como: 
 \begin{verbatim}
 ineqmin: $[TEL4]>=0.3;
 upperbound: for[all] $ <=0.4;
 ineqmax: for[private] $ <= 0.1*x;  
 \end{verbatim}
 estabelecendo que o investimento em {\tt TEL4} deve ser de pelo menos 
30\%, que cada ativo individualmente n\~ao deve exceder 40\%   
do portf\'olio e que o investimento em cada ativo do setor privado n\~ao deve 
ultrapassar 10\% do valor que o vetor "x" associa \`aquele banco.

 \section{Modelo de Markowitz}

 As vari\'aveis de decis\~ao em um modelo do tipo Markowitz s\~ao 
interpretadas como as participa\c{c}\~{o}es que os diversos 
ativos t\^{e}m no portf\'{o}lio. O objetivo \'e otimizar 
a fun\c{c}\~ao utilidade sobre o retorno esperado do portf\'olio, 
comparado \`a vari\^ancia deste retorno. As principais entradas 
deste modelo s\~ao os retornos esperados, em uma matriz p, e a 
correspondente matriz de covari\^ancia S. O total investido em cada 
ativo, digamos, x, deve ser sempre positivo. Estas s\~ao as 
restri\c{c}\~oes de sinal. Outras restri\c{c}\~oes de igualdade e 
de desigualdade podem ser adicionadas na forma $$ Te*x == te$$ e  
 $$ Tl * x < = tl.$$ A forma geral do problema de otimiza\c{c}\~ao 
\'e a seguinte: 
 $$ max \ \ U(x) = \eta x'*p -x'*S*x$$
 onde 
 \begin{itemize}
 \item[]
x \'{e} o vetor de participa\c{c}\~{o}es dos ativos no
portf\'{o}lio (a ser encontrado),

 \item[] er=E(retorno) \'{e} o vetor de taxas de retorno esperadas dos ativos

 \item[] S=Cov(retorno) \'{e} a matriz de covari\^{a}ncia dos
retornos dos ativos

 \item[] Te \'{e} a matriz de coeficientes das restri\c{c}\~{o}es de
igualdade

 \item[] te \'{e} o vetor de termos independentes das restri\c{c}\~{o}es
de igualdade

 \item[]  Tl \'{e} a matriz de coeficientes das restri\c{c}\~{o}es de
desigualdade

 \item[] tl \'{e} o vetor de termos independentes das restri\c{c}\~{o}es
de desigualdade

 \item[] $\eta$ (l\^e-se  eta) \'{e} o par\^{a}metro de
pondera\c{c}\~{a}o  entre a utilidade da esperan\c{c}a do retorno do
portf\'{o}lio versus a avers\~{a}o ao risco no investimento: quanto
maior $\eta$, maior a propens\~ao do  investidor ao risco para valores
de retorno esperado elevados. Valores  reduzidos de $\eta$ quantificam
uma alta avers\~ao ao risco por parte do  investidor. 
 \end{itemize}

 Trata-se portanto de um problema de Otimiza\c{c}\~ao 
Quadr\'{a}tica Param\'{e}trica. A solu\c{c}\~ao, para um dado valor de 
$\eta$ \'e o vetor vi\'avel x (i.e. um vetor x satisfazendo todas as 
restri\c{c}\~oes do modelo)  que otimiza a fun\c{c}\~ao utilidade. 
 Logo, x@asseta \'e o total investido em asseta no portf\'olio x. 
 Uma vez que para cada valor de eta temos uma solu\c{c}\~ao \'otima 
diferente, devemos escrev\^e-la como x($\eta$). 

 Cada uma destas solu\c{c}\~{o}es \'{o}timas, ou eficientes,
representa um portf\'{o}lio cuja esperan\c{c}a de retorno \'{e}
 e($\eta$)=x($\eta$)' * p, com vari\^{a}ncia 
 v($\eta$)=x($\eta$)' * S * x($\eta$), 
e desvio padr\~{a}o
 s($\eta$)= $\sqrt{v(\eta)}$.  A curva e($\eta$) versus s($\eta$) 
\'{e} a Fronteira Eficiente do modelo.

 A Fronteira Eficiente \'{e} formada por segmentos de
par\'{a}bolas que se unem formando uma curva c\^{o}ncava.  Ela
representa a fronteira superior do conjunto de portf\'{o}lios
vi\'{a}veis. Para
calcul\'{a}-la \'{e} necess\'{a}rio primeiro determinar os pontos
cr\'{\i}ticos.  Cada ponto cr\'{\i}tico \'{e} um ponto na curva
correspondente a um $\eta$  cr\'{\i}tico.  

	Os etas cr\'{\i}ticos s\~{a}o aqueles em que ocorre alguma
mudan\c{c}a qualitativa no portf\'{o}lio x($\eta$), i.e.,  
	
 \begin{enumerate}
 \item    um ativo entra ou sai do portf\'{o}lio, ou 

 \item  uma restri\c{c}\~{a}o de desigualdade passa de uma
desigualdade estrita (folgada ou inativa) a uma igualdade (desigualdade
justa ou ativa), ou vice-versa. 
 \end{enumerate} 

 Para referir-se \`{a} posi\c{c}\~{a}o de um dado portf\'{o}lio,   \'{e}
usual a  frase ``o portf\'{o}lio de retorno {\it e} mais ou menos {\it
s}'' para  indicar o portf\'{o}lio de retorno esperado {\it e} e desvio
padr\~{a}o  {\it s}.  

 \section*{Exemplo de Aplica\c{c}\~ao}

 Tomemos um exemplo do modelo da  se\c{c}\~ao que explica o arquivo MODEL.CP, 
com retorno esperado e desvio padr\~ao obtidos atrav\'es dos par\^ametros 
padr\~ao do filtro.

 Utilizando o Numerical Template para selecionar alguns pontos 
da fronteira eficiente podemos obter:

\begin{verbatim}
Selected Portfolios: Parameters, Assets and Composition

Parameters of portfolio A
eta        esp      var         std         rate      k     l 
4,22E-01  9,39E-02  4,28E-02  2,07E-01  -7,51E-03    1     0,00E+00

Assets and Composition of portfolio A
5,00E-01@PET4  2,00E-02@BBD4  3,00E-02@SCO4   5,00E-02@CEV4  
4,00E-01@BRH4                                                 

Parameters of portfolio B
eta      esp       var       std      rate      k        l     
3,21E-01   8,72E-02    3,78E-02   1,95E-01   -3,05E-02   2    0,00E+00 

Assets and Composition of portfolio B
4,03E-01@PET4  2,00E-02@BBD4  3,00E-02@SCO4  5,00E-02@CEV4  4,97E-01@BRH4   
	       
\end{verbatim}
 e assim por diante.

 \section{Filtro}
     
 O modelo exige, como insumo fornecido pelo usu\'{a}rio, a matriz de
correla\c{c}\~{a}o dos retornos esperados dos ativos, o vetor de
retornos esperados e o vetor de desvios padr\~{a}o.  As estimativas
destes par\^{a}metros podem ser feitas com base nos dados hist\'{o}ricos
de cada ativo, ou por qualquer outro meio que o usu\'{a}rio julgar
conveniente.  Uma ferramenta simples e rubusta para este prop\'{o}sito,
i.e. um filtro, \'{e} distribuido junto com o programa. 

 O filtro opera sobre longas s\'{e}ries hist\'{o}ricas de
pre\c{c}os dos ativos sendo considerados.  O disquete de
instala\c{c}\~{a}o cont\'{e}m alguns exemplos de s\'eries 
hist\'{o}ricas de alguns dos ativos do mercado brasileiro.  
Hist\'{o}ricos atualizados diariamente podem ser comprados de firmas
divulgadoras de dados financeiros. 

 Os par\^{a}metros do filtros s\~{a}o ajustados no menu Options
Filter. S\~{a}o eles:

\[
\begin{array}{ll}
${\bf Final date}$  &    $data da \'{u}ltima cota\c{c}\~{a}o a
ser considerada$ \\
${\bf Interval}$    &    $intervalo entre cota\c{c}\~{o}es
a serem consideradas$ \\
${\bf Samples}$     &   $N\'{u}mero de cota\c{c}\~{o}es a serem
consideradas$ \\
${\bf Extrap}$  &  $N\'{u}mero de intervalos, adiante da 
data final, para que$ \\ 
	    &   $se projetam retorno esperado, desvio
padr\~{a}o e covari\^{a}ncias$ \\
${\bf Deflator}$    &     $Nome do arquivo contendo as datas de
cota\c{c}\~{a}o $\\
${\bf Extension}$   &    $Extens\~{a}o dos arquivos de
cota\c{c}\~{o}es $ \\
${\bf  Hurst }$ &    $Par\^ametro caracter\'{\i}stico do movimento Browniano
(entre 0.5 e 1)$ \\
\end{array}
\]

 O filtro (F) calcula o retorno m\'edio e o desvio padr\~ao 
para cada ativo na s\'erie hist\'orica. As amostras 
dos pre\c{c}os s\~ao tomadas no dom\'{\i}nio do tempo T, 
especificado pelos par\^ametros do filtro: 

 t in T = [data-final, datafinal-intervalo, datafinal -2intervalos, ..., 
datafinal-nro-de-amostras*intervalo]

 Os filtro estima retornos logar\'{\i}tmicos (ou cont\'{\i}nuos).
Denotando  por $price(i,t)$ e $r(i,t)$ o pre\c{c}o e o retorno do ativo
i no  per\'{\i}odo t, o filtro calcula:  
 $$ r(i,t)= ln \left(\frac{price(i,t+1)}{price(i,t)} \right).$$ 
 A seguir \'e calculado o retorno m\'edio de cada ativo 
 meanl(i) = $\frac{1}{samples} \sum_{t} r(i,t)$
e a matriz de covari\^ancia, desvios padr\~ao e matriz de correla\c{c}\~ao 
das amostras: 
\begin{itemize}
\item[]
$ Sl(i,j) = \frac{1}{samples}
\sum_t (rl(i,t)-meanl(i))*(rl(j,t)-meanl(j))$
\item[]
 $ sl(i) =  \sqrt{Sl(i,i)}$  
\item[]
$corrl(i,j) = \frac{Sl(i,j)}{sl(i)*sl(j)}$
\end{itemize}
 Em seguida o filtro extrapola o retorno esperado e o desvio padr\~ao 
por extrap intervalos adiante: 

 $$erl(i) = extrap*meanl \ ; \ \ std(i) = extrap^{Hurst} *sl(i);$$

 Finalmente o filtro convertetodas as medidas logar\'{\i}tmicas de 
retorno e risco para medidas simples (de retorno e risco). 
 Para pre\c{c}os deterministicos em $t=0$ e $t=T$, o retorno 
 logar\'{\i}tmico \'{e} $rl=ln(S(T)/S(0))$, e o retorno simples \'{e} 
 $r=(S(T)-S(0))/S(0)$, portanto $r=exp(rl)-1$. 
 O filtro assume que os pre\c{c}os tem uma distribui\c{c}\~{a}o log-normal, 
 de modo que as regras de transforma\c{c}\~{a}o fical algo mais complicadas: 
 $$r = exp(erl +0.5stdl^2) -1 \ ; \ \ s^2 = exp(2r +sl^2)(exp(sl^2) -1).$$ 

 Finalmente o filtro salva o retorno m\'edio e o desvio padr\~ao
extrapolados  como vetores associativos no arquivo FILTER.CP a matriz de
correla\c{c}\~ao no arquivo matricial CORRELF.M . 

 O conjunto universo, i.e. o conjunto  all, utilizado pelo filtro \'e  o
conjunto original  que tamb\'em aparece no arquivo FILTER.CP. 
 No arquivo MODEL.CP podemos  restringir o universo original eliminando
alguns de seus elementos.  Quando  restringimos o universo original, o
compilador MDL  extrai automaticamente de er, std e correl, apenas as
submatrizes e  subvetores de que necessita. Os usu\'arios da  vers\~ao
profissional  do Critical Point podem receber atrav\'es de modem
atualiza\c{c}\~oes  destas matrizes e vetores, que s\~ao geradas para
diversos ativos  brasileiros. Os filtros utilizados para calcul\'a-los
s\~ao mais  sofisticados  que os filtros utilizados pelo Critical Point.
 Estes filtros incorporam tecnologias avan\c{c}adas como an\'alise
fractal, filtros adaptativos, tratamentos especiais para valores
faltantes etc.

 \section{Derivativos} 

 A ferramenta para derivativos expande os vetores de retorno esperado e 
desvio padr\~{a}o, bem como a matriz de correla\c{c}\~{a}o, para incluir 
os ativos derivativos sobre ativos do conjunto fundamental. 

\begin{verbatim}  
put={OTP19@TEL4, OTP24@TEL4 };
call={OTC16@TEL4, OTC17@TEL4 };

exdays={40@OTP19, 30@OTP24, 40@OTC16, 40@OTC17};

O={6.7@OTP19, 10.9@OTP24,10.9@OTC16, 6.7@OTC17};
S={63.2@OTP19, 63.2@OTP24, 63.2@OTC16, 63.2@OTC17};
K={64@OTP19, 72@OTP24, 56@OTC16, 68@OTC17};
\end{verbatim} 

No arquivo exemplo DERIV.CP acima, temos o conjunto put declarando 
as op\c{c}\~{o}es de venda (europeias) OTP19 e OTP24 sobre o ativo 
fundamental TEL4; da mesma forma, declaramos op\c{c}\~{o}es de compra 
OTC16 e OTC17. Para definir op\c{c}\~{o}es temos quew especificar o 
pre\c{c}o de contrato de op\c{c}\~{a}o (O), o pre\c{c}o do ativo 
fundamental (S), o pre\c{c}o de exerc\'{\i}cio (K), e o n\'{u}mero 
de dias at\'{e} o exerc\'{\i}cio (exdays).

 \subsection*{Covari\^{a}ncia de Derivativos} 

 A ferramenta para derivativos escreve o conjunto universo extendido (all), 
com o nome de todos os ativos (fundamentais e derivativos), bem como 
seus retornos esperados e desvios padr\~{a}o, no arquivo ERSX.CP, 
e a matriz de correla\c{c}\~{a}o extendida em CORRELX.M.

Como exemplo simples, a express\~{a}o para retorno de em call \'{e}: 

$$rc= ( max( S(t1) -K(t0) , 0 ) -O(t0) ) / O(t0)$$   

Os retornos experados, desvios padr\~{a}o e correla\c{c}\~{o}es para 
derivativos s\~{a}o calculados asssumindo distribui\c{c}\~{a}o log-normal, 
usando as express\~{o}es desenvolvidas no cap\'{\i}tulo 4 e t\'{e}cnicas 
avan\c{c}adas de integra\c{c}\~{a}o num\'{e}rica.

 \section{Arquivos de Dados e Matriciais}

 O compilador MDL  compila, i.e. traduz,  o modelo descrito  na
linguagem MDL para a forma matricial. O problema descrito desta forma, 
no arquivo MARKIN.M \'e entrada para o solver do Programa de 
Programa\c{c}\~ao Param\'etrica (Q). Para construir modelos mais 
sofisticados devemos entender a estrutura dos arquivos  matriciais. 

 A ordem na qual os ativos aparecem no vetor ou nas matrizes \'e aquela  
na qual eles aparecem no universo original, segundo o arquivo  
MODPS.CP. O filtro tem um proce\-dimento para computar a matriz de 
correla\c{c}\~ao. Cada elemento da matriz de correla\c{c}\~ao 
$ Cor(i,j)$  pode ser interpretado como uma medida da 
interfer\^encia m\'utua 
entre o retorno do i-\'esimo e do j-\'esimo ativos. Por exemplo, se 
os dois ativos t\^em retornos proporcionais $  p_{i} = a * p_{j},$ 
 $a >  0$ ent\~ao $Cor(i,j)=1$. Se, por outro lado, a constante 
de proporcionalidade for negativa,  $ a < 0 $ (isto \'e o pre\c{c}o do 
ativo i crescer implica que o pre\c{c}o j decresce) ent\~ao 
$ Cor(i,j) = -1$. 
Finalmente se ambos os retornos s\~ao mutuamente independentes, 
$ Cor(i,j)=0$.

 Um usu\'ario mais sofisticado pode construir sua pr\'opria matriz 
de correla\c{c}\~ao. O Critical Point permite manipula\c{c}\~ao de arquivos 
matriciais, entretanto, a utiliza\c{c}\~ao de  uma matriz que n\~ao 
satisfa\c{c}a todas as propriedades matem\'aticas da matriz de 
correla\c{c}\~ao pode causar problemas num\'ericos no 
procedimento de programa\c{c}\~ao quadr\'atica param\'etrica (Q). 
 Dentre as propriedades da matriz de correla\c{c}\~ao temos: 
 \begin{tabbing}
 1- Domin\^ancia diagonal \hspace{0.5cm} 
 \=  cor(i,i) ==1; \hspace{0.3cm} abs(cor(i,j)) $<$ 1, i $\neq$ j;  \\
 2- Simetria: \>  cor(i,j) == cor(j,i); \\
 3- Positividade \>  min(eig(cor)) $\geq$ 0 
 \end{tabbing}

 \subsection*{Retornos Hist\'oricos versus An\'alise de Mercado}

 Ap\'os a leitura dos arquivos MODPS.CP e MODEL.CP, o 
compilador MDL l\^e o arquivo CORREL.M a fim de reconstruir a 
matriz de covari\^ancia.  
 $$S(i,j) = s(i)^{'} * Corr(i,j)* s(j)$$;

 Finalmente o compilador MDL salva as restri\c{c}\~oes estabelecidas 
no MODEL.CP, bem como o vetor de retornos esperados e a matriz de 
covari\^ancia,  no arquivo MARKIN.M, que ser\'a entrada do 
procedimento de programa\c{c}\~ao quadr\'atica  param\'etrica (Q). 

 O compilador MDL trata os vetores de retornos esperados  e desvios 
padr\~ao como qualquer vetor associativo de forma que ele tamb\'em 
possa ser modificado utilizando os recursos da linguagem MDL. 
 Para evitar inconsist\^{e}ncias no c\'{a}lculo dos retornos esperados e
desvios padr\~{a}o, estas modifica\c{c}\~{o}es devem ser feitas no 
arquivo DERIV.CP. 
 Cabe   mencionar que analistas  de mercado profissionais
raramente  utilizam   m\'edias extrapoladas como previs\~ao de retorno.
Os retornos m\'edios  e tamb\'em os desvios padr\~ao extrapolados pelo
filtro, s\~ao  utilizados somente para ``calibrar'' o modelo mas estes
valores podem ser  ajustados pelos analistas. Entre as mais populares
t\'ecnicas de  ajuste est\~ao: 
 \begin{enumerate} 
 \item An\'alise fundamentalista
 \item Predi\c{c}\~ao atrav\'es de an\'alise gr\'afica
 \item Previs\~ao atrav\'es de an\'alise t\'ecnica
 \item Modelos matem\'aticos de redes neurais, filtros adaptativos 
       e \'arvores de classifica\c{c}\~ao. 
 \end{enumerate} 
 Estas t\'ecnicas fogem do escopo deste manual e s\~ao objeto de 
estudo de disciplinas espec\'{\i}ficas dentro do NOPEF-USP. 

 \subsection*{Arquivos de Dados}

O filtro  l\^e s\'eries hist\'oricas de pre\c{c}os do diret\'orio DATA. 
O arquivo  dolof.ofc cont\'em a taxa de convers\~ao de Reais para
D\'olares americanos. Este  arquivo  tamb\'em define o calend\'ario
oficial , isto \'e, os  dias de negocia\c{c}\~ao. Outros arquivos *.ofc
t\^em os arquivos com  s\'eries de pre\c{c}os deflacionados por d\'olar
para alguns ativos   brasileiros. As tr\^es primeiras linhas de cada
arquivo definem: 
 \begin{enumerate} 
 \item O nome do ativo
 \item O deflator de pre\c{c}os 
 \item O n\'umero de a\c{c}\~oes 
 \end{enumerate} 

 O terminador de  s\'erie \'e o s\'{\i}mbolo asterisco, *.  O pr\'oximo 
exemplo mostra o arquivo TEL3.ofc, com os pre\c{c}os, em d\'olares, 
por 10000 a\c{c}\~oes da TELebr\'as, na BOVESPA, nos \'ultimos 4 dias de  
 1994. As datas devem estar no formato dd/mm/yy e os 
pre\c{c}os em nota\c{c}\~ao de ponto flutuante. 
\pagebreak
\begin{verbatim}
Asset: TEL3      
Deflator: DOLOF.OFC 
Shares: 1E+4        
Date        Price      
30/12/94    4.314E+02     
29/12/94    4.304E+02       
28/12/94    4.030E+02     
27/12/94    4.025E+02     
*        (T.S. terminator)     

(End of File) 
\end{verbatim}  

 \section{Restri\c{c}\~oes l\'ogicas sobre conjuntos}

A linguagem MDL permite restringir conjuntos atrav\'es de express\~oes 
l\'ogicas. Exemplo:
{\tt exampleset[energy|bank]=(arrbank$>$0.5)|(arrenergy$<$0.2);}
ap\'os este comando, somente os ativos do setor banc\'ario 
que tamb\'em satisfazem a condi\c{c}\~ao 
{\tt arrbank>0.5} ou os ativos  do setor de energia 
que satisfazem 
{\tt arrenergy $<$ 0.2} 
permanecem no conjunto  {\tt exampleset}. 

Condi\c{c}\~oes l\'ogicas podem ser constru\'{\i}das utilizando os operadores
de compara\c{c}\~ao: 

\begin{itemize}        
\item[] $> \ \  $ \hspace{0.5cm}  maior que
\item[] $>=$ \hspace{0.5cm} maior ou igual    
\item[] $< \ \ $ \hspace{0.5cm} menor que  
\item[] $<=$  \hspace{0.5cm} menor ou igual    
\item[] $==$  \hspace{0.5cm} igual   
\item[] \~\  $=$  \hspace{0.5cm} diferente
\end{itemize}        
Express\~oes mais complexas podem ser realizadas entre conjuntos e 
vetores.  A representa\c{c}\~ao de um conjunto nada mais \'e que um 
vetor  associativo booleano (0 ou 1). Logo em atribui\c{c}\~oes a 
conjuntos devemos interpretar   o nome de um ativo, asseta, como sendo 
equivalente \`a express\~ao 1@asseta. 
Tamb\'em interpretamos opera\c{c}\~oes l\'ogicas entre  conjuntos 
utilizando os operadores l\'ogicos ``n\~ao'' (\~\ ),  ``e'', ($\&$) e 
``ou'' ($\mid$), como assumindo 
o valor 0 quando falso e qualquer valor n\~ao nulo  quando verdadeiro. 

Quando restringimos o conjunto universo, all, proibimos qualquer ativo 
eliminado de entrar no portf\'olio, e somente os subvetores e submatrizes 
necess\'arios de p, std, e cov ser\~ao escritos no arquivo MARKIN.M pelo  
compilador MDL. Para manter este arquivo consistente, devemos primeiramente 
restringir o universo em MODEL.CP, e s\'o ent\~ao estabelecer as 
restri\c{c}\~oes do modelo. 

 \section{Depura\c{c}\~ao}

 Mesmo os mais experientes programadores comentem erros e, apesar de 
que muito poucos modelos requeiram todos os recursos da linguagem 
MDL, alguns deles tornam-se muito complexos. O compilador MDL possui 
algumas ferramentas simples para auxiliar a detec\c{c}\~ao de erros. 
 Quando o compilador n\~ao consegue analisar uma frase, exibe uma mensagem 
de erro, contendo: 
 \begin{enumerate} 
 \item o nome do arquivo onde ocorreu o erro. 
 \item o n\'umero da linha onde este erro aconteceu.
 \item a \'ultima palavra  v\'alida, a \'ultima lida e o c\'odigo ASCII. 
 \end{enumerate} 
 O comando print x; for\c{c}a o compilador MDL a escrever o objeto x 
(escalar, conjunto ou vetor) no arquivo PRINT.LOG, exatamente como 
este objeto se encontra no momento que o comando \'e acionado 
pelo programa MDL. 

 \subsection*{A calculadora de m\'edia vari\^ancia}

 A  calculadora de m\'edia vari\^ancia ( bot\~ao C) fornece uma maneira 
simples de calcular o desvio padr\~ao e e retorno esperado de 
portf\'olios pr\'e-definidos. Os portf\'olios s\~ao definidos no 
arquivo FOLIOS.CP, por exemplo:

 {\tt papas=\{5.0E-1@PET4,2.0E-2@BBD4,3.0EE-2@SCO4,5.0E-2@CEV4,4.0E-0@BRH4\}}

 A calculadora imprime o valor calculado no arquivo RFOLIOS.CP, por 
exemplo:

 {\tt papas: \hspace{1cm} e = 9.39E-02 \hspace{1cm}  s = 2.07E-01}

 \section{Short Selling e Tracking} 

 Na descri\c{c}\~{a}o do modelo, antes da declara\c{c}\~{a}o de 
restri\c{c}\~{o}es, o subconjunto short pode ser declarado. 
 O efeito de ter um ativo em short \'{e} trocar o sinal de seu retorno 
esperado, bem como o de sua correla\c{c}\~{a}o com outros ativos. 
 
 A interpreta\c{c}\~{a}o financeira de ter um ativo em short \'{e} 
a de vender este ativo a descoberto no sentido de Linter: Uma venda 
a descoberto implica em um dep\'{o}sito colateral, igual ao valor 
vendido. Assim a venda a descoberto n\~{a}o pode ser usada como uma 
fonte de recursos, mas \'{e} sim mais um uso de recurso, a ser incluido 
no portf\'{o}lio.  

 Caso o investidor receba um juro pelo colateral, o vetor de retornos 
esperados deve ser ajustado (uma simples atribui\c{c}\~{a}o usando o 
conjunto short como dom\'{\i}nio). No pr\'{o}ximo exemplo uma taxa 
fixa de 1\% \'{e} paga sobre o colateral.

\begin{verbatim} 
all={ TEL4, ELE6, PET4, BB4, BBD4, SCO4, CEV4, BRH4, IBOVESPA }; 
short={TEL4,ELE6};
er[short]= er -0.01;           #adding 1% to  -er
normal: sum[all] $==1; 
\end{verbatim}

 Em muitas aplica\c{c}\~{o}es queremos formar um portf\'{o}lio para
perseguir (track) um investimento  de refer\^{e}ncia.  A diferen\c{c}a de
retorno entre o portf\'{o}lio e sua refer\^{e}ncia \'{e} denominad erro 
de persegui\c{c}\~{a}o (tracking error). 
 No pr\'{o}ximo exemplo procuramos um portf\'{o}lio para perseguir o 
\'{\i}ndice IBOVESPA. O gr\'{a}fico de m\'{e}dia-vari\^{a}ncia obtido 
para este modelo n\~{a}o mais se refere a retornos absolutos, mas a 
retornos relativos ao investimento de refer\^{e}ncia (IBOVESPA).

\begin{verbatim} 
all={ TEL4, ELE6, PET4, BB4, BBD4, SCO4, CEV4, BRH4, IBOVESPA }; 
short={IBOVESPA};
normal: sum[all] $==2;
track: for[short] $==1;
\end{verbatim}

 \section*{Autores do programa}

 Julio Michael Stern escreveu a vers\~ao original do solver (em Matlab) e 
especificou a linguagem MDL e a interface com o usu\'ario. Jacob 
Zimbarg Sobrinho traduziu o solver para a linguagem C (Watcom 10.0) e 
escreveu o filtro. Fabio Nakano escreveu o compilador MDL 
(utilizando GNU lex e yacc) e a vers\~ao atual da interface com o 
usu\'ario (em Visual Basic). Cibele Dunder escreveu a ferramenta para 
ativos derivativos. 
 Por favor envie sugest\~oes e comunique eventuais erros a 
 jstern@ime.usp.br (internet).

%% file: apen3.tex
\chapter{\'Algebra Linear Computacional}

\section{Nota\c{c}\~{a}o e Opera\c{c}\~{o}es B\'{a}sicas} 

 Este par\'{a}grafo define algumas nota\c{c}\~{o}es matriciais. 
 Indicamos por $(1 \ate n)$ a lista $[1,\ldots n]$, e $j\in (1 \ate n)$
 indica que o \'{\i}ndice $j$ est\'{a} neste dom\'{\i}nio. 
 Uma lista de matrizes tem um (ou mais) \'{\i}ndices superscritos, 
 $S^1\ldots S^m$. 
 Assim $S^k_{h,i}$ \'{e} o elemento na linha $h$ e coluna $i$ 
 da matriz $S^k$. 
 A matriz 
 \[ A = \left[ \begin{array}{ccc} 
    A^{1,1} & \cdots & A_{1,s} \\ 
    \vdots  & \ddots & \vdots  \\ 
    A^{r,1} & \cdots & A_{r,s}  
    \end{array} \right] 
  \] 
 \'{e} uma matriz blocada, onde $A^{p,q}$ \'{e} o $p-q$-\'{e}simo 
 bloco, ou sub-matriz.  

 Quando estamos falando de apenas uma matriz, $X$, costumamos esceve-la 
 com o \'{\i}ndice de linha subscrito, e o \'{\i}ndice de coluna superscrito
 Assim $x_i$, $x^j$, e $x_i^j$ s\~{a}o a linha $i$, a coluna $j$,   
 e o elemento $(i,j)$ da matriz $X$. 
 ESta nota\c{c}\~{a}o \'{e} mais compacta, e resalta o fato de vermos 
 a matriz $X$ como uma matriz blocada por vetores coluna. 
 A matriz $X_{h \ate i}^{j \ate k}$ \'{e} um bloco extraido da matriz 
 $X$, fazendo os \'{\i}ndices de linha e coluna percorrer os 
 dom\'{\i}nios indicados.  
 ${\bf 0}$ e ${\bf 1}$ s\~{a}o matrizes de zeros e uns, 
 geralmente vetores coluna , $n\times 1$. 
 Quando a dimens\~{a}o n\~{a}o est\'{a} 
 indicada, ela pode ser deduzida do contexto.    
 $V>0$ \'{e} uma matriz positiva definida.     
 Definimos a p-norma de um vetor $x$ por 
 $||x||_p = \left( \sum |x_i|^p \right)^{-p}$. 
 Assim, se $x$ para um vetor n\~{a}o negativo, podemos escrever 
 sua 1-norma como $||x||_1 = \uno' x$. 
 
 O produto de Kroneker de duas matrizes \'{e} uma matriz blocada onde 
 o bloco $(i,j)$ \'{e} a segunda matriz multiblicada pelo elemento 
 $(1,j)$ da primeira matriz:  
 \[ 
    A \otimes B \ = \ 
  \left[ \begin{array}{ccc} 
  A^1_1 B & A^2_1 B & \cdots \\ 
  A^1_2 B & A^2_2 B & \cdots \\ 
  \vdots & \vdots & \ddots 
  \end{array} \right] 
 \]   
 As seguintes propriedades podem ser facilmente verificadas: 
 \begin{itemize}  
 \item $(A\kron B)(C\kron D) = (AC)\kron(BD)$  
 \item $(A\kron B)' = A'\kron B'$  
 \item $(A\kron B)^{-1} = A^{-1}\kron B^{-1}$  
 \end{itemize}   
  
 O operador $\Vec$ ``empilha'' as colunas de uma matriz 
 em um \'{u}nico vetor coluna, i.e., se $A$ \'{e} $m\times n$, 
 \[ \Vec(A) = \left[ \begin{array}{c} 
    A^1 \\ \vdots \\ A^n \end{array} \right] 
 \]   
 As seguintes propriedades podem ser facilmente verificadas: 
 \begin{itemize} 
 \item $\Vec(A+B) = \Vec(A) +\Vec(B)$ 
 \item $\Vec(AB) 
 =  \left[ \begin{array}{c} AB^1 \\ \vdots \\ AB^n \end{array} \right] 
 = (I\kron A) \Vec(B)$  
 \end{itemize}

 \section{Espa\c{c}os Vetoriais com Produto Interno}

 Dados dois vetores $x,y \in {\Re }^n$, o seu {\bf produto escalar}
 \'{e} definido como
 \[  < x \mid y > \equiv x'y = \sum_{i=1}^{n} x_{i}y^{i} \ . \] 
Com esta defini\c{c}\~{a}o v\^{e}-se  que o produto escalar \'{e} um operador que
satisfaz as propriedades fundamentais de {\bf produto interno}, a saber: 
\begin{enumerate}
\item $<x\mid y> = <y\mid x>$, simetria. 
\item $<\alpha x+\beta y\mid z> = 
      \alpha <x\mid z> + \beta <y\mid z>$, linearidade. 
\item $<x\mid x> \geq 0$ , semi-positividade.
\item $<x\mid x>=0 \Leftrightarrow x=0$ , positividade.
\end{enumerate}
 
Atrav\'{e}s do produto interno, definimos a norma:
 \[ \| x \| \equiv <x\mid x>^{1/2} \ ; \]  
e definimos tamb\'{e}m o \^angulo entre dois vetores:
 \[  \Theta(x,y) \equiv \arccos ( <x\mid y>  /  \| x\| \| y\| ) \ . \]

\section{Projetores}

 Consideremos o subespa\c{c}o linear gerado pelas colunas de uma matriz
 $A$, $m$ por $n$, $m\geq n$: 
 \[ C(A) = \{ y=Ax, x\in {\Re }^n \} \ .  \] 
 Denominamos $C(A)$ de imagem de $A$, e o complemento de $C(A)$, $N(A)$,
de espa\c{c}o nulo de $A$, 
 \[  N(A) = \{ y \mid  A'y=0 \} \ .  \] 

 Definimos a proje\c{c}\~{a}o de de um vetor $b\in {\Re }^m$ no espa\c{c}o
das colunas de $A$, pelas rela\c{c}\~{o}es: 
 \[ y = P_{C(A)}b \leftrightarrow y\in C(A) \wedge (b-y)\perp C(A) \]  
ou, equivalentemente, 
 \[ y= P_{C(A)}b \leftrightarrow y=Ax \wedge A'(b-y)=0 \ . \] 

No que se segue suporemos que $A$ tem posto pleno, i.e.  que suas
colunas s\~{a}o li\-nearmente independentes.  Provemos que o projetor de
$b$ em $C(A)$ \'{e} dado pela aplica\c{c}\~{a}o linear 
 \[ P_A = A(A'A)^{-1}A' \ .  \]  
Se $y = A((A'A)^{-1}A'b)$, ent\~{a}o obviamente $y\in C(A)$.  
Por outro lado, 
 \[ A'(b-y) = A'(I-A(A'A)^{-1}A')b = (A' - IA')b = 0 \ . \]

\section{Matrizes Ortogonais}

Dizemos que uma matriz quadrada e real \'{e} {\bf ortogonal} sse sua
inversa \'{e} igual a sua transposta.Dada $Q$ uma matriz ortogonal, suas
colunas formam uma base ortonormal de ${\Re }^n$, como pode ser visto da
identidade $Q'Q=I$.  A norma quadr\'{a}tica de um vetor $v$, ou seu
quadrado 
 \[ {\| v \|}^{2} \equiv \sum (v_{i})^2 = v'v \]  
 permanece inalterada por uma transforma\c{c}\~{a}o ortogonal, pois  
 \[ (Qv)'(Qv) = v'Q'Qv = v'Iv = v'v \ . \]

Dado um vetor em ${\Re }^2$, 
$\left[ \begin{array}{c} x_1 \\ x_2 \end{array} \right] $, 
a rota\c{c}\~{a}o deste vetor por um \^{a}ngulo $\theta$ \'{e} dada pela
transforma\c{c}\~{a}o linear 
 \[  
 G(\theta )x = \left[ \begin{array}{cc} 
 \cos (\theta ) & \sin (\theta ) \\ -\sin (\theta ) & \cos (\theta ) 
 \end{array} \right] 
 \left[ \begin{array}{c} x_1 \\ x_2 \end{array} \right] .
 \] 

 Notemos que uma rota\c{c}\~{a}o \'{e} uma transforma\c{c}\~{a}o 
ortogonal, pois
 \[  
 G(\theta )' G(\theta )
 =
 \left[ \begin{array}{cc} {\cos (\theta )}^2 + {\sin (\theta 
 )}^2 & 0 \\ 
       0 & {\cos (\theta )}^2 + {\sin (\theta )}^2 
 \end{array} \right] 
 =
 \left[ \begin{array}{cc} 1 & 0 \\  0 & 1 \end{array} 
 \right] \ .
 \] 
 
 A rota\c{c}\~{a}o de Givens \'{e} um operador linear cuja matriz
coincide com a identidade, exceto num par de linhas onde imergimos uma
matriz de rota\c{c}\~{a}o bidimensional: 
 \[ 
 G(i,j,\theta ) =
 \left[ \begin{array}{cccccccc} 
 1 & & & & & & & \\
   & \ddots & & & & & & \\
   & & \cos (\theta ) & & \sin (\theta ) & & & \\
   & & & \ddots & & & & \\
   & & -\sin (\theta ) & & \cos (\theta ) & & & \\
   &  & &  & &  & \ddots &  \\
   &  & &  & &  &  & 1 \\
 \end{array} \right] \ .
 \] 

 Dizemos que a aplica\c{c}\~{a}o deste operador numa matriz $A$, $G'A$,
roda as linhas $i$ e $j$ de $A$ de um \^{a}ngulo $\theta$ (no sentido 
ant-hor\'{a}rio). 

Como o produto de transforma\c{c}\~{o}es ortogonais continua ortogonal
(prove), podemos usar uma seq\"{u}\^{e}ncia de rota\c{c}\~{o}es para
construir transforma\c{c}\~{o}es ortogonais. 

 Listamos agora algumas rota\c{c}\~{o}es em $\Re^2$, construidas a
partir de  simples rela\c{c}\~{o}es trigonom\'{e}tricas, que ser\~{a}o  
utilizadas como blocos de constru\c{c}\~{a}o para diversos agoritmos. 

 Consideremos, em $\Re^2$, $v$ um vetor, $S$ uma matriz sim\'{e}trica, 
e $A$ uma matriz assim\'{e}trica, 
 \[ 
 v= \left[ \begin{array}{c} x \\ y \end{array} \right]  
 \  , \ \  
 S= \left[ \begin{array}{cc} p & q \\ q & r \end{array} \right] 
 \ , \ \ 
 A= \left[ \begin{array}{cc} a & b \\ c & d \end{array} \right] 
 \] 

 Para anular a segunda componente do vetor $v$ por uma rota\c{c}\~{a}o 
 \'{a} esquerda, $G(\theta_v)'\,v$, basta  tomar um \^{a}ngulo  
 \[ \theta_v=\arctan\left(\frac{y}{x}\right) \ . \]  
 
 Para diagonalizar a matriz sim\'{e}trica por uma rota\c{c}\~{a}o 
sim\'{e}trica, 
 $G( \theta_{diag} )'\,S\,G( \theta_{diag} )$, 
 basta  tomar um \^{a}ngulo  
 \[ \theta_{diag}=\frac{1}{2}\arctan\left(\frac{2q}{r-p}\right) \ . \]  
 
 Para simetrizar a matriz assim\'{e}trica por uma por uma 
rota\c{c}\~{a}o \`{a} esquerda, 
 $G(\theta_{sym})'\,A$, 
 basta  tomar um \^{a}ngulo  
 \[ \theta_{sym}=\arctan\left(\frac{b-c}{a+d}\right) \ . \]  

 Assim, para diagonalizar a matriz assim\'{e}trica basta 
simetriza-la, e em seguida diagonaliza-la. 
 Alternativamente, podemos diagonalizar a matriz assim\'{e}trica 
por um par de ``rota\c{c}\~{o}es de Jacobi'', 
\`{a} esquerda e \`{a} direita, 
 $J(\theta_{r})'\,A\,J(\theta_{l})$;  
 bastando  tomar os \^{a}ngulos  
 \[ 
    \theta_{sum} = \theta_{r} +\theta_{l}  
                      = \arctan\left(\frac{c+b}{d-a}\right)  \ , \ \  
    \theta_{dif} = \theta_{r} -\theta_{l}  
                      = \arctan\left(\frac{c-b}{d+a}\right)  \ \Or        
 \]  
 \[ 
   J(\theta_r)' =  G(\theta_{sum}/2)'\,G(-\theta_{dif}/2)' \ , \ \  
   J(\theta_l)  =  G(\theta_{dif}/2) \,G(\theta_{dif}/2) \ .  
 \] 

 No c\'{a}lculo das rota\c{c}\~{o}es, as fun\c{c}\~{o}es
trigonom\'{e}tricas, Seno, Coseno e Arco-Tangente  n\~{a}o s\~{a}o
realmente utilizadas, j\'{a} que nunca utilizamos os \^{a}ngulos
propriamente ditos, mas apenas 
 $c=\sin(\theta)$ \ e \ $s=\sin{\theta}$, que podemos computar 
 diretamente como 
 \[ 
    c = \frac{x}{\sqrt{x^2 +y^2}} \ , \ \ 
    s= \frac{-y}{\sqrt{x^2 +y^2}} \ . 
 \]   
 Para prevenir overflow podemos utilizar o c\'{a}lculo: 
  \begin{itemize} 
  \item Se $y==0$\ , \ \Then $c=1 \ , \ s=0$ \ . 
  \item Se $y\geq x$ \ , \ \Then 
        $t=-x/y \ , \ s=1/\sqrt{1+t^2} \ , \ c=st$ \ . 
  \item Se $y<x$ \ , \ \Then 
        $t=-y/x \ , \ c=1/\sqrt{1+t^2} \ , \ s=ct$ \ . 
  \end{itemize}

\section{Fatora\c{c}\~{a}o QR}

Dada $A$ uma matriz real de posto pleno $m\times n$, $m\geq n$, podemos
sempre encontrar uma matriz ortogonal $Q$ tal que $A=Q \left[
\begin{array}{c} R \\ 0 \end{array} \right] $, onde $R$ \'{e} uma matriz
quadrada e triangular superior.  Esta decomposi\c{c}\~{a}o \'{e} dita
uma fatora\c{c}\~{a}o ``QR'', ou fatora\c{c}\~{a}o ortogonal, da matriz
$A$.  O fator ortogonal $Q=[C\mid N]$ nos da uma base ortonormal de
${\Re }^m$ onde as $n$ primeiras colunas s\~{a}o uma base ortonormal de
$C(A)$, e as $m-n$ \'ultimas colunas s\~{a}o uma base de $N(A)$, como pode
ser visto diretamente da identidade 
$Q'A=\left[ \begin{array}{c} R\\ 0 \end{array} \right]$.  
Construiremos a seguir um m\'{e}todo para fatora\c{c}\~{a}o ortogonal.

Abaixo ilustramos uma seq\"{u}\^{e}ncia de rota\c{c}\~{o}es de linhas
necess\'{a}rias que leva uma matriz $5\times 3$ \`{a} forma triangular
superior.  Cada par de \'{\i}ndices, $(i,j)$, indica que rodamos estas
linhas do \^{a}ngulo apropriado para zerar a posi\c{c}\~{a}o na linha
$i$, coluna $j$.  Supomos que inicialmente a matriz \'{e} densa, i.e. 
todos os seus elementos s\~{a}o diferentes de zero, e ilustramos o
padr\~{a}o de esparsidade da matriz nos est\'{a}gios assinalados com um
asterisco na seq\"{u}\^{e}ncia de rota\c{c}\~{o}es. 
 \[ 
(1,5) * (1,4) (1,3) (1,2) * (2,5) (2,4) (2,3) * (3,5) (3,4) *
 \] 
 \[ 
\left[ \begin{array}{ccc} 
  x & x & x \\ x & x & x \\ x & x & x \\ x & x & x \\ 0 & x & x
\end{array} \right] \ \  
\left[ \begin{array}{ccc} 
  x & x & x \\ 0 & x & x \\ 0 & x & x \\ 0 & x & x \\ 0 & x & x
\end{array} \right] \ \  
\left[ \begin{array}{ccc} 
  x & x & x \\ 0 & x & x \\ 0 & 0 & x \\ 0 & 0 & x \\ 0 & 0 & x
\end{array} \right] \ \  
\left[ \begin{array}{ccc} 
  x & x & x \\ 0 & x & x \\ 0 & 0 & x \\ 0 & 0 & 0 \\ 0 & 0 & 0
\end{array} \right]  
 \] 

\subsection{M\'{\i}nimos Quadrados}

 Dado um sistema superdeterminado, $Ax=b$ onde a matriz $A$ $m\times n$
tem $m>n$, dizemos que $x^*$ ``resolve'' o sistema no sentido dos
m\'{\i}nimos quadrados, ou que $x^*$ \'{e} a ``solu\c{c}\~{a}o'' de
m\'{\i}nimos quadrados, sse $x^*$ minimiza a norma quadr\'{a}tica do
res\'{\i}duo,
 \[ x^* = Arg \min_{x\in {\Re }^n} \| Ax - b {\|} \ , \] 
Dizemos tamb\'{e}m que $y=Ax^*$ \'{e} a melhor aproxima\c{c}\~{a}o, no
sentido dos m\'{\i}nimos quadrados de $b$ em $C(A)$. 

Como a multiplica\c{c}\~{a}o por uma matriz ortogonal deixa inalterada a
norma quadr\'{a}tica de um vetor, podemos procurar a solu\c{c}\~{a}o
deste sistema (no sentido dos m\'{\i}nimos quadrados) minimizando a
transforma\c{c}\~{a}o ortogonal do res\'{\i}duo usada na
fatora\c{c}\~{a}o QR de $A$,
 \[ 
 \| Q'(Ax-b) {\|}^2 = 
 \| \left[ \begin{array}{c} R \\ 0 \end{array} \right] x - 
    \left[ \begin{array}{c} c \\ d \end{array} \right] 
 {\|}^2 = 
 \| Rx-c {\|}^2  +  \| 0x-d {\|}^2 .
 \] 

 Da \'{u}ltima express\~{a}o v\^e-se que a solu\c{c}\~{a}o, a
aproxima\c{c}\~{a}o e o res\'{\i}duo do problema original s\~{a}o dados
por, respectivamente
 \[ 
 x^* = R^{-1}c \ , \ \ 
 y = Ax^* \mbox{\ \ e \ \ }
 z = Q \left[ \begin{array}{c} 0 \\ d \end{array} \right] .
 \] 

Como j\'{a} hav\'{\i}amos observado, as $m-n$ \'{u}ltimas colunas de $Q$
formam uma base ortonormal de $N(A)$, logo $z \perp C(A)$, de modo que
conclu\'{\i}mos que $y=P_{A}b$!

\section{Fatora\c{c}\~{o}es LU e Cholesky}

Dada uma Matriz $A$, a {\bf opera\c{c}\~{a}o elementar} determinada pelo
{\bf multiplicador} $m^i_j$, \'{e} subtrair da linha $j$ a linha
$i$ multiplicada por $m^i_j$.
A opera\c{c}\~{a}o elementar aplicada a matriz identidade gera a correspondente 
{\bf matriz elementar},
 \[ 
 M(i,j)=
 \left[ \begin{array}{ccccccc} 
   1 & & & & & &  \\
   & \ddots & & & & &  \\
   & & 1 & & & &  \\
   & & \vdots & \ddots & & &  \\
   & & -m_j^i & & 1 & &  \\
   & & \vdots & & & \ddots &  \\
   & & & & & & 1 \\
 \end{array} \right] \ 
 \begin{array}{c} \\ \\ i \\ \\ j \\ \\ \\ \end{array} \ . 
 \]  
Aplicar uma opera\c{c}\~{a}o elementar a matriz $A$ equivale a 
multiplic\'a-la a esquerda pela correspondente matriz elementar. 

Na {\bf Fatora\c{c}\~{a}o de Gauss}, ou fatora\c{c}\~{a}o LU, usamos uma 
seq\"{u}\^{e}ncia de opera\c{c}\~{o}es elementares para levar $A$ a forma triangular superior,
 \[ 
    MA=M(n-1,n)\cdots M(2,3)\cdots M(2,n)M(1,1)
    \cdots M(1,n-1)M(1,n)A=U \ . 
 \]  
 A inversa da produt\'oria desta seq\"{u}\^{e}ncia de matrizes
elementares tem  a forma triangular inferior (verifique)
 \[ 
 L= M^{-1}=
 \left[ \begin{array}{lllll}
 1 & & & & \\
 m^1_2 & 1 & & \\
 \vdots & \vdots & \ddots & & \\
 m^1_{n-1} & m^2_{n-1} &  & 1 & \\  
 m^1_n & m^2_n & \cdots & m^{n-1}_n & 1 \\
 \end{array} \right] \ .
 \] 

No caso de fatorarmos uma matriz sim\'{e}trica, $S=LU$, podemos por 
em evidencia os elementos diagonais de $U$ obtendo $S=LDL'$. 
Se $S$ for positiva definida assim ser\'a $D$, de modo que podemos 
escrever $D=D^{1/2}D^{1/2}$, $D^{1/2}$ a matriz diagonal contendo a 
ra\'{\i}z dos elementos em $D$.
Definindo $C=LD^{1/2}$, temos $S=CC'$, a {\bf fatora\c{c}\~{a}o de 
Choleski} de $S$.

\subsection{Programa\c{c}\~{a}o Quadr\'{a}tica}

O problema de {\bf programa\c{c}\~{a}o quadr\'{a}tica} consiste em 
minimizar a fun\c{c}\~{a}o
 \index{Programa\c{c}\~{a}o Quadr\'{a}tica}
 \[  f(y) \equiv (1/2)y'Wy + c'y \ , \ \ W=W' \] 
sujeitos \`{a}s {\bf restri\c{c}\~{o}es} 
 \[  g_{i}(y) \equiv N_{i}'y = d_i . \] 
Os gradientes de $f$ e $g_i$ s\~{a}o dados, respectivamente, por 
 \[  {\nabla}_{y}f = y'W + c' \ , \ \mbox{e} \  
     {\nabla}_{y}g_i = N_{i}' \ .
 \] 
 As {\bf condi\c{c}\~{o}es de otimalidade} de primeira ordem
(condi\c{c}\~{o}es de Lagrange) estabelecem que as restri\c{c}\~{o}es
sejam obedecidas, e que o gradiente da fun\c{c}\~{a}o sendo minimizada
seja uma combina\c{c}\~{a}o linear dos gradientes das
restri\c{c}\~{o}es.  Assim a solu\c{c}\~{a}o pode ser obtida em
fun\c{c}\~{a}o do {\bf multiplicador de Lagrange}, i.e. do  
vetor $l$ de coeficientes desta combina\c{c}\~{a}o linear, como
 \index{Condi\c{c}\~{a}o!de otimalidade} 
 \index{Condi\c{c}\~{a}o!de Lagrange} 
 \index{Multiplicadores!de Lagrange} 
 \[ 
   N'y = d \ \wedge \ 
   y'W + c' = l' N' \ , 
 \]  
 ou em forma matricial, 
 \[ 
 \left[ \begin{array}{cc} N' & 0 \\ W & N \end{array} \right] 
 \left[ \begin{array}{c} y \\ l \end{array} \right] = 
 \left[ \begin{array}{c} d \\ c \end{array} \right] \ .  
 \] 
 Este sistema de equa\c{c}\~{o}es \'{e} conhecido como o {\bf sistema
normal}.  O sistema normal tem por matriz de coeficientes uma matriz
sim\'{e}trica.  Se a forma quadr\'{a}tica $W$ for {\bf positiva definida}, 
i.e.se $\forall x\ x'Wx \geq 0 \ \wedge \ x'Wx=0 \Leftrightarrow x=0$, e
as restri\c{c}\~{o}es $N$ forem lineramente independentes, a matriz de
coeficientes do sistema normal ser\'{a} tamb\'{e}m positiva definida. 

\section{Fatora\c{c}\~{a}o SVD} 

 A fatora\c{c}\~{a}o SVD decomp\~{o}em uma matriz real $A$, 
 $m\times n,\, m\geq n$, em um produto $D=U'AV$, onde $D$ \'{e} 
 diagonal, e $U,\, V$ s\~{a}o matrizes ortogonais. 
 Consideremos primeiramente o caso $m=n$, i.e. uma matriz 
 quadrada. 

 O algoritmo de Jacobi \'{e} um algoritmo iterativo que, 
 a cada itera\c{c}\~{a}o, ``concentra a matriz na diagonal'', 
 atrav\'{e}s de rota\c{c}\~{o}es de Jacobi.  
 \[ 
    J(i,j,\theta_r)'\,A^k\,J(i,j,\theta_l) = A^{k+1} = 
    \left[ \begin{array}{ccccccc} 
    A^{k+1}_{1,1} & \cdots & A^{k+1}_{1,i} & \cdots & 
    A^{k+1}_{1,j} & \cdots & A^{k+1}_{1,n} \\  
    \vdots        & \ddots & \vdots        & \ddots & 
    \vdots        & \ddots  & \vdots \\ 
    A^{k+1}_{i,1} & \cdots & A^{k+1}_{i,i} & \cdots & 
    0             & \cdots & A^{k+1}_{i,n} \\  
    \vdots        & \ddots & \vdots        & \ddots & 
    \vdots        & \ddots  & \vdots \\ 
    A^{k+1}_{j,1} & \cdots & 0             & \cdots & 
    A^{k+1}_{j,j} & \cdots & A^{k+1}_{j,n} \\  
    \vdots        & \ddots & \vdots        & \ddots & 
    \vdots        & \ddots  & \vdots \\ 
    A^{k+1}_{n,1} & \cdots & A^{k+1}_{n,i} & \cdots & 
    A^{k+1}_{n,j} & \cdots & A^{k+1}_{n,n}   
    \end{array} \right] 
  \]   

  Consideremos a soma dos quadrados dos elementos fora da diagonal 
  na matriz $A$\ $\mbox{Off}_2(A)$.  
  Vemos que 
  \[ 
     \mbox{Off}_2(A^{k+1}) = \mbox{Off}_2(A^{k})     
                         -(A^k_{i,j})^2 -(A^k_{j,i})^2 
  \] 
   Assim, escolhendo a cada itera\c{c}\~{a}o o par de \'{\i}ndices
   que maximiza a soma dos quadrados do par fora da diagonal a 
   ser anulado, temos um algoritmo que converge linearmente 
   para uma matriz diagonal. 

   O algoritmo de Jacobi nos d\'{a} uma prova construtiva da 
   exist\^{e}ncia da fatora\c{c}\~{a}o SVD, e \'{e} a b\'{a}se 
   para v\'{a}rios algoritmos mais eficientes de 
   fatora\c{c}\~{a}o SVD. 

   Se $A$ \'{e} uma matriz retangular, basta inicialmente 
   fatorar $A=QR$, e aplicar o algoritmo de Jacobi ao bloco 
   quadrado superior de $R$.     
   Se $A$ \'{e} quadrada e sim\'{e}trica, a fatora\c{c}\~{a}o 
   obtida \'{e} denominada decomposi\c{c}\~{a}o de autovalores 
   de $A$.     

   As matrizes $U$ e $V$ podem ser interpretadas como bases ortogonais 
   dos respectivos espa\c{c}os de dimens\~{a}o $m$ e $n$. 
   Os valores na diagonal de $S$ s\~{a}o denominados valores singulares 
   da matriz $A$, e podem ser interpretados geom\'{e}tricamente como 
   fatores multiplicadores do mapa $A=UDV'$, que leva cada versor 
   da base $V$ para um m\'{u}ltiplo de um versor da b\'{a}se $U$.

 \section{Matrizes Complexas}

 A maioria das t\'{e}cnicas aqui desenvolvidas para matrizes reais, 
podem ser extendidas para matrizes complexas. Uma forma pr\'{a}tica e 
elegante para tal s\~{a}o as transform\c{c}\~{o}es Q (TQ) de Hemkumar,  
 aplicadas a uma dada matriz $M$, complexa $2\times 2$, na forma de 
 um par ``interno'' de transforma\c{c}\~{o}es de fase unit\'{a}rias, e 
 um par ``externo'' de rota\c{c}\~{o}es, 
 \[ 
 \left[ \begin{array}{cc} 
  c(\phi) & -s(\phi) \\ s(\phi) & c(\phi) 
 \end{array} \right]  
 \left[ \begin{array}{cc} 
  e(i\alpha) & 0 \\ 0 & e(i\beta)  
 \end{array} \right]  
 \left[ \begin{array}{cc} 
  Ae(ia) & Be(ib) \\ Ce(ic) & De(id)   
 \end{array} \right]  
 \left[ \begin{array}{cc} 
  e(i\gamma) & 0 \\ 0 & e(i\delta)  
 \end{array} \right]  
 \left[ \begin{array}{cc} 
  c(\psi) & -s(\psi) \\ s(\psi) & c(\psi)  
 \end{array} \right]   
 \]

 A eleg\^{a}ncia da TQ vem da seguinte observa\c{c}\~{a}o: 
 Enquanto a transforma\c{c}\~{a}o interna afeta apenas os 
 expoentes imagin\'{a}rios dos elementos da matriz, 
 a transforma\c{c}\~{a}o externa pode ser aplicada independentemente 
 \`{a} parte real e \`{a} parte imagin\'{a}ria da matriz, i.e. 
 \[ 
 \left[ \begin{array}{cc} 
  e(i\alpha) & 0 \\ 0 & e(i\beta)  
 \end{array} \right]  
 \left[ \begin{array}{cc} 
  Ae(ia) & Be(ib) \\ Ce(ic) & De(id)   
 \end{array} \right]  
 \left[ \begin{array}{cc} 
  e(i\gamma) & 0 \\ 0 & e(i\delta)  
 \end{array} \right]  = 
 \]  
 \[ 
 \left[ \begin{array}{cc} 
  Ae(ia') & Be(ib') \\ Ce(ic') & De(id')   
 \end{array} \right] =  
 \left[ \begin{array}{cc} 
  Ae(i(a+\alpha+\gamma)) & Be(i(b+\alpha+\delta)) \\ 
  Ce(i(c+\beta+\gamma)) & De(i(d+\beta+\gamma))   
 \end{array} \right] 
 \] 
 \[ 
 \left[ \begin{array}{cc} 
  c(\phi) & -s(\phi) \\ s(\phi) & c(\phi) 
 \end{array} \right]  
 \left[ \begin{array}{cc} 
  A'_r +iA'_i & B'_r +iB'_i \\ C'_r +iC'_i & D'_r +iD'_i   
 \end{array} \right]  
 \left[ \begin{array}{cc} 
  c(\psi) & -s(\psi) \\ s(\psi) & c(\psi)  
 \end{array} \right] =  
 \] 
 \[ 
 \left[ \begin{array}{cc} 
  c(\phi) & -s(\phi) \\ s(\phi) & c(\phi) 
 \end{array} \right]  
 \left[ \begin{array}{cc} 
  A'_r & B'_r \\ C'_r & D'_r    
 \end{array} \right]  
 \left[ \begin{array}{cc} 
  c(\psi) & -s(\psi) \\ s(\psi) & c(\psi)  
 \end{array} \right] 
 \] 
 \[ 
 + i \left(   
 \left[ \begin{array}{cc} 
  c(\phi) & -s(\phi) \\ s(\phi) & c(\phi)  
 \end{array} \right]  
 \left[ \begin{array}{cc} 
  A'_i & B'_i \\ C'_i & D'_i     
 \end{array} \right]  
 \left[ \begin{array}{cc} 
  c(\psi) & -s(\psi) \\ s(\psi) & c(\psi)  
 \end{array} \right] \right)   
 \] 
 
 A tabela seguinte define algumas transforma\c{c}\~{o}es internas
\'{u}teis. Transforma\c{c}\~{o}es do tipo $I$ trocam  os  expoentes
imagin\'{a}rios dos elementos da matriz em uma das  diagonais.
Transforma\c{c}\~{o}es do tipo $R$, $C$ e $D$ tornam  reais os elementos
em uma linha, coluna ou diagonal.

 \begin{center} 
 \begin{tabular}{|c|c|} 
 \hline 
 Tipo & Valores \\ 
 \hline $I_{main}$ & 
 $\alpha = -\beta = \gamma = -\delta = (d-a)/2$ \\ 
 \hline $I_{off}$ & 
 $\alpha = -\beta = -\gamma = \delta = (c-b)/2$ \\ 
 \hline $R_{up}$ & 
 $\alpha = \beta = -(b+a)/2 \ ; \  \gamma = -\delta = (b-a)/2$ \\ 
 \hline $R_{low}$ & 
 $\alpha = \beta = -(d+c)/2 \ ; \  \gamma = -\delta = (d-c)/2$ \\ 
 \hline $C_{left}$ & 
 $\alpha = -\beta = (c-a)/2 \ ; \  \gamma = \delta = -(c+a)/2$ \\ 
 \hline $C_{right}$ & 
 $\alpha = -\beta = (d-b)/2 \ ; \  \gamma = \delta = -(d+b)/2$ \\ 
 \hline $D_{main}$ & 
 $\alpha = \beta = -(d+a)/2 \ ; \  \gamma = -\delta = (d-a)/2$ \\ 
 \hline $D_{off}$ & 
 $\alpha = \beta = -(b+c)/2 \ ; \  \gamma = -\delta = (b-c)/2$ \\ 
 \hline 
 \end{tabular} 
 \end{center}

 \'{E} f\'{a}cil ver que aplicar uma seq\"{u}\^{e}ncia de 
transforma\c{c}\~{o}es internas corresponde \`{a} uma \'{u}nica 
transforma\c{c}\~{a}o interna, cujos par\'{a}metros s\~{a}o a soma dos 
respectivos par\'{a}metros das transforma\c{c}\~{o}es na seq\"{u}\^{e}ncia. 

 Combinando transforma\c{c}\~{o}es internas e externas, podemos  criar
TQs para v\'{a}rios algoritmos interessantes. 
 Por exemplo, as QTs de tipo I, II e III na tabela seguinte, 
podem ser usadas para obter a fatora\c{c}\~{a}o SVD de matrizes complexas, 
 \`{a} semelhan\c{c}a do algoritmo de Jacobi.  
 A TQ-I aplica $R_{low}$ seguida de uma rota\c{c}\~{a}o, 
tendo o efeito de tormar a matriz triangular superior. 
 A TQ-II aplica $D_{main}$, $I_{off}$ seguida uma diagonaliza\c{c}\~{a}o. 
 Para matrizes Hermitianas (auto-adjuntas), a diagonaliza\c{c}\~{a}o  
 \'{e} obtida com apenas uma TQ do tipo III.

 \begin{center} 
 \begin{tabular}{|c|c|c|} 
 \hline 
 Tipo & Interna & Externa \\ 
 \hline I & 
 $\alpha = \beta = -(d+c)/2 \ ; \  \gamma = -\delta = (d-c)/2$ &  
 $\phi=0 \ ; \ \psi = \arctan(C/D)$ \\ 
 \hline II & 
 $\alpha = -(a+b)/2 \ ; \ \beta = \gamma = -\delta = (b-a)/2$ &  
 $\phi \pm \psi = \arctan(B/(D \mp A))$ \\ 
 \hline III & 
 $\alpha = -\beta = -\gamma = \delta = -b/2$ &  
 $\phi = \psi = \arctan(2B/(D-A))/2$ \\ 
 \hline 
 \end{tabular} 
 \end{center}

 \section{Probabilidades em Sub-Espa\c{c}os Lineares}

\section{Exerc\'{\i}cios}

 \begin{enumerate}
 \item Use as propriedades fundamentais do produto interno para provar: 
  \begin{enumerate}
  \item A desigualdade de Cauchy-Scwartz:
   $|<x\mid y>| \leq \| x\| \| y\|$.
   Sugest\~{a}o: Calcule ${\| x-\alpha y \|}^2$ para 
   $\alpha =<x\mid y>^2 / \| y\|$.
  \item A Desigualdade Triagular:
   $\| x+y\| \leq \| x\| +\| y\| $.
  \item Em que caso temos igualdade na desigualdade de Cauchy-Schwartz?
   Relacione sua resposta com a defini\c{c}\~{a}o de \^angulo entre vetores.
  \end{enumerate}
\item  Use a defini\c{c}\~{a}o do produto interno em ${\Re }^n$ para 
 provar a Lei do Paralelogramo: 
 ${\| x+y\|}^2 + {\| x-y\|}^2 = 2{\| x\|}^2 + 2{\| y\|}^2 $.
 \item Uma matriz \'e idempotente,ou um projetor n\~ao 
ortogonal sse $P^{2} = P$. Prove que: 
\begin{enumerate}
 \item $R = (I-P)$ \'{e} idempotente.
 \item ${\Re }^n = C(P) + C(R)$.
 \item Todos os autovalores de $P$ s\~{a}o $0$ ou $+1$. Sugest\~{a}o:
  Mostre que se $0$ \'{e} uma raiz do polin\^{o}mio caracter\'{\i}stico 
  de $P$, ${\varphi}_{P}(\lambda )\equiv \det (P-\lambda I)$, ent\~{a}o
  $(1-\lambda)=1$ \'{e} raiz de ${\varphi}_{R}(\lambda )$.
\end{enumerate}

\item Prove que $\forall P$ idempotente e sim\'{e}trico, 
  $P = P_{C(P)}$. Sugest\~{a}o: Mostre que $P'(I-P)=0$.

\item Prove que o operador de proje\c{c}\~{a}o num dado sub-espa\c{c}o 
  vetorial $V$, $P_{V}$, \'{e} \'unico e sim\'{e}trico.

\item Prove o theorema de Pit\'agoras: 
  $\forall b \in {\Re }^{m} , u \in V$ temos que
  ${\| b-u \| }^2 = {\| b-P_{V}b \| }^2 + {\| P_{V}b - u \| }^2$.

\item Suponha termos a fatora\c{c}\~{a}o QR de uma matriz $A$.
  Considere uma nova matriz $\tilde A$ obtida de $A$ pela substitui\c{c}\~{a}o 
  de uma \'unica coluna. Como podemos atualizar nossa fatora\c{c}\~{a}o ortogonal
  usando apenas $3n$ rota\c{c}\~{o}es de linha? Sugest\~{a}o: $(a)$ Remova a coluna
  alterada de $A$ e atualize a fatora\c{c}\~{a}o usando no m\'aximo $n$ 
rota\c{c}\~oes. 
  $(b)$ Compute a nova coluna alterada pelo fator ortogonal corrente,
  $\tilde a = Q'a = R^{-t}A'a$. $(c)$ Adicione $\tilde a$ como a ultima 
  coluna de $\tilde A$, e torne a atualizar a fatora\c{c}\~{a}o com $2n$ 
  rota\c{c}\~{o}es.

\item Compute as fatora\c{c}\~{o}es $LDL$ e Cholesky da matriz
      \[ \left[ \begin{array}{rrrr}
         4 & 12 & 8 & 12 \\ 12 & 37 & 29 & 38 \\
         8 & 29 & 45 & 50 \\ 12 & 38 & 50 & 113 \\
         \end{array} \right] \ . 
      \] 
\item Prove que
  \begin{enumerate}
  \item $(AB)'=B'A'$.
  \item $(AB)^{-1}=B^{-1}A^{-1}$.
  \item $A^{-t}\equiv (A^{-1})' = (A')^{-1}$.
  \end{enumerate}
\item Descreva quatro algoritmos, para computar $L^{-1}x$ e $L^{-t}x$,
   acessando a matriz $L$, triangular inferior de diagonal unit\'aria,
   por linha ou por coluna.
\end{enumerate}

%% file: apen4.tex
\chapter{Probabilidade}

\section{Interpreta\c{c}\~{a}o Freq\"{u}entista} 

Em muitas circunst\^{a}ncias estamos interessados em estudar
situa\c{c}\~{o}es que produzem resultados imprevis\'{\i}veis.  Chamamos
estas situa\c{c}\~{o}es de {\em experimentos}.  Por exemplo, ao
lan\c{c}armos um dado, consideramos seis resultados poss\'{\i}veis, cada
resultado correspondendo a uma determinada face do dado voltada para
cima.  Chamamos de {\em Espa\c{c}o Amostral}, $A$, ao conjunto de todos
os poss\'{\i}veis resultados.  
 Espa\c{c}os amostrais {\em discretos} cont\^{e}m um n\'{u}mero finito
(ou enumer\'{a}vel) de resultados. Sempre que poss\'{i}vel
apresentaremos a teoria em espa\c{c}os discretos, onde ela \'{e} muito
mais simples que em espa\c{c}os cont\'{i}nuos. 
 Chamamos {\em evento} um subconjunto do espa\c{c}o amostral,
$E\subseteq A$.  Assim, no exemplo do dado, uma representa\c{c}\~{a}o do
espa\c{c}o amostral \'{e} o conjunto $\{ F1, F2, F3, F4, F5, F6 \}$;
nesta representa\c{c}\~{a}o, o evento ``obter uma face \'{\i}mpar''
corresponde ao subconjunto \newline $\{ F1, F3, F5 \}$. 

 Nos experimentos que estudaremos \'{e} \'{u}til atribuirmos a cada um
dos eventos um valor num\'{e}rico.  Esta atribui\c{c}\~{a}o, ou
fun\c{c}\~{a}o, chama-se {\em vari\'{a}vel aleat\'{o}ria}.  No exemplo
do dado \'{e} usual atribuir valores, entre 1 e 6, a cada um dos
eventos, i.e. atribuir $X(Fk)=k$. 

 Consideremos outro exemplo: lan\c{c}amos dois dados, um verde e outro
amarelo, cada qual tendo suas faces numeradas de 1 a 6.  No exemplo dos
dados verde e amarelo, entre muitas outras, poder\'{\i}amos considerar
as seguintes vari\'{a}veis aleat\'{o}rias:
 \begin{enumerate} 
 \item O n\'{u}mero decimal de dois d\'{\i}gitos cujo primeiro
d\'{\i}gito corresponde a face de cima do dado verde, e o segundo
d\'{\i}gito corresponde a face de cima do dado amarelo. 
 \item A soma dos n\'{u}meros na face de cima dos dois dados. 
 \end{enumerate}

 Note que constru\'{\i}mos duas vari\'{a}veis aleat\'{o}rias distintas,
sobre um mesmo experimento: o lan\c{c}amento dos dados verde e amarelo. 
Note tamb\'{e}m que pelo valor da primeira vari\'{a}vel aleat\'{o}ria
podemos saber exatamente o resultado obtido no experimento; o mesmo
j\'{a} n\~{a}o \'{e} verdade para a segunda vari\'{a}vel aleat\'{o}ria. 
Em geral, quando lidamos com um experimento, temos uma dada vari\'{a}vel
aleat\'{o}ria em mente e, n\~{a}o havendo ambig\"{u}idade, quando
mencionamos o experimento j\'{a} subentendemos a vari\'{a}vel
aleat\'{o}ria apropriada. 

 Definimos a imagem de um evento $E\subseteq A$ por $X$, $X(E)$, como o 
conjunto de todos os valores assumidos por $X$ dentro de $E$, i.e., 
 $X(E)\equiv \{ x=X(e), e\in E \}$. 
 Analogamente definimos a pr\'{e}-imagem de $C$ por $X$, $X^{-1}(C)$,
como o evento formado pelos resultados do experimento onde a
vari\'{a}vel $X$ assume valores em $C$, i.e.,
 $X^{-1}(C) = \{ a\in A \mid X(a)\in C \}$ .  

 Dado um experimento que possa ser repetido, interpretamos a {\em
probabilidade} de um evento como a freq\"{u}\^{e}ncia com que o obtemos
como resultado do experimento.    
 A probabilidade de que uma vari\'{a}vel
aleat\'{o}ria $X$ assuma um valor dentro de um conjunto $C$, \'{e}
simplesmente a probabilidade da pr\'{e}-imagem de $C$ por $X$. 
 A {\em distribui\c{c}\~{a}o de probabilidade} de uma
vari\'{a}vel aleat\'{o}ria (discreta) \'{e} uma tabela especificando a
probabilidade de que a vari\'{a}vel assuma cada um dos seus 
poss\'{i}veis valores. 

 Exemplos: 
 \begin{itemize}
 \item Um dado \'{e} dito honesto se a freq\"{u}\^{e}ncia com que gera
cada um dos valores $\left\{ 1,\ldots ,6 \right\}$ \'{e} a mesma.  A
probabilidade de obtermos um dado valor com um dado honesto \'{e}
portanto de $1/6$. 
 \item No experimento descrito no exemplo dos dados verde e amarelo,
assumindo que ambos os dados s\~{a}o honestos, a probabilidade de cada
vari\'{a}vel aleat\'{o}ria assumir cada um dos valores poss\'{\i}veis
\'{e} dada pelas tabelas seguintes: 
 \begin{enumerate} 
  \item $Pr(X=x)=1/36$, $\forall x \in 
   \{ 11, 12,\ldots ,16, 21,\ldots ,26,\ldots ,61,\ldots ,66 \}$. 
  \item $X^{-1}(2)=\{(1,1)\}\Rightarrow Pr(X=2)=1/36$,  
	$X^{-1}(3)=\{(1,2),(2,1)\}\Rightarrow Pr(X=2)=2/36$, 
	\ldots ,  
	$X^{-1}(12)=\{(6,6)\}\Rightarrow Pr(X=12)=1/36$.  
 \end{enumerate} 
 \end{itemize}

\begin{itemize}

\item[]{\bf Propriedades da Probabilidade} 

 Para assegurar a possibilidade de interpretar probabilidade de um evento
como a de sua freq\"{u}\^{e}ncia relativa, exigiremos que uma (medida
de) probabilidade sobre o espa\c{c}o amostral de um experimento
aleat\'{o}rio, satisfa\c{c}a as tr\^{e}s propriedades (axiomas)
seguintes.  Notaremos por $\emptyset$ o conjunto vazio, e por $\bar B =
A-B$ o complementar de um evento $B$ no espa\c{c}o amostral $A$ do
experimento. 
  \begin{enumerate} 
 \item $\forall B \subseteq A, \ \ Pr(B)\geq 0$,  
 \item $Pr(A)=1$,  
 \item Se $B\cap C = \emptyset$, ent\~{a}o 
	  $Pr(B\cup C)=Pr(B)+Pr(C)$.
 \end{enumerate}

 Exerc\'{\i}cio: Prove, a partir dos axiomas, as seguintes propriedades
da probabilidade:
 \begin{enumerate} \setcounter{enumi}{3} 
 \item $Pr(\emptyset )=0$, 
 \item $Pr(\bar B) = 1- Pr(B)$, 
 \item $Pr( B\cup C) = Pr(B)+Pr(C)-Pr(B\cap C)$
 \end{enumerate} 

\end{itemize}

 \section{Infer\^{e}ncia} 

 Em muitas situa\c{c}\~{o}es temos que lidar com experimentos cujas
especifica\c{c}\~{o}es n\~{a}o conhecemos, ou conhecemos
incompletamente.  Por exemplo: Suponha sabermos que o dado verde foi
constru\'{\i}do tomando todos os cuidados para garantir que,
exceto por pequenas marcas que distinguem suas faces, ele \'{e}
absolutamente regular.  Suponha tamb\'{e}m que o dado amarelo foi
trazido por um desconhecido.  Sabemos de antem\~{a}o que o dado verde
\'{e} honesto, mas n\~{a}o sabemos a distribui\c{c}\~{a}o de
probabilidade para o segundo dado. 

 Para obter alguma informa\c{c}\~{a}o sobre o dado amarelo podemos
lan\c{c}\'{a}-lo repetidas vezes, anotando a seq\"{u}\^{e}ncia de
valores.  Este hist\'{o}rico de valores \'{e} o que se chama em
estat\'{\i}stica de uma {\em amostra}. 

 A partir da amostra poder\'{\i}amos induzir alguma informa\c{c}\~{a}o
sobre a distribui\c{c}\~{a}o de probabilidade do dado amarelo:
poder\'{\i}amos por exemplo calcular a freq\"{u}\^{e}ncia relativa de
cada um dos valores na amostra; e ent\~{a}o usar estas
freq\"{u}\^{e}ncias relativas como {\em estimativa} da
distribui\c{c}\~{a}o de probabilidade do experimento.  Este \'{e} um
exemplo do que chamamos {\em infer\^{e}ncia estat\'{\i}stica}. 

 Note que a probabilidade da vari\'{a}vel aleat\'{o}ria $X$ assumir o
valor $x$, $\mbox{Pr}(X=x)$, e a estimativa utilizada,
freq\"{u}\^{e}ncia relativa na amostra, s\~{a}o conceitos distintos: A
distribui\c{c}\~{a}o de probabilidade \'{e} um atributo do experimento
em si, enquanto a estimativa baseia-se numa particular amostra de
resultados do experimento.  No uso pr\'{a}tico, geralmente, a amostra
resume nossas observa\c{c}\~{o}es passadas, queremos basear alguma
decis\~{a}o futura na distribui\c{c}\~{a}o de probabilidade do
experimento (que todavia desconhecemos), e usamos a estimativa como elo
de liga\c{c}\~{a}o para tomar esta decis\~{a}o.

\begin{itemize}

\item[] {\bf M\'{a}xima Verossimilhan\c{c}a} 

Como fazer e justificar estimativas e infer\^{e}ncias que sejam
\'{o}timas, segundo crit\'{e}rios pr\'e-estabelecidos, \'{e} o objeto de
estudo da estat\'{\i}stica matem\'{a}tica.  No caso do dado amarelo, o
crit\'{e}rio usado para fazer a estimativa foi o da {\em m\'{a}xima
verossimilhan\c{c}a}: A melhor estimativa da distribui\c{c}\~{a}o,
segundo o crit\'{e}rio da m\'{a}xima verossimilhan\c{c}a, \'{e} a que
maximiza a probabilidade de obter, num experimento com a
distribui\c{c}\~{a}o estimada, uma amostra id\^{e}ntica a amostra
original. 

Se a amostra original foi uma seq\"{u}\^{e}ncia de $n$ resultados,  
com $n_1,\ n_2,\ldots ,n_6$ ocor\-r\^{e}n\-cias dos valores $1,\ 2,\ldots ,6$, 
a probabilidade de obter uma amostra com as mesmas feq\"{u}\^{e}ncias, 
independentemente da ordem, \'{e} 

 $$Pr(X=[n_1,\, n_2,\ldots n_6]) = 
  \frac{n!}{n_1!\, n_2!\ldots n_6!} 
  p_1^{n_1}p_2^{n_2}\ldots p_6^{n_6} 
 $$ 
\'{E} um exerc\'{\i}cio de c\'{a}lculo mostrar que a
distribui\c{c}\~{a}o que maximiza esta probabilidade \'{e} a das
freq\"{u}\^{e}ncias relativas,
 $p_i=f_i$, onde $f_i=n_i/n$. 
Prove este resultado no caso de uma moeda.
\end{itemize}

\section{Esperan\c{c}a}  

Dado um experimento, o valor esperado de uma vari\'{a}vel aleat\'{o}ria
$X$, ou sua {\em esperan\c{c}a}, \'{e} a m\'{e}dia aritm\'{e}tica sobre
o conjunto dos valores poss\'{\i}veis, $x\in A$, ponderados pela
distribui\c{c}\~{a}o de probabilidade: 
 $$ E(X) = \sum_{x\in X(A)} x*\mbox{Pr}(X=x)\ \ .$$

Dada uma amostra, i.e.  uma seq\"{u}\^{e}ncia de valores, 
 $$ x = \left[ \begin{array}{c} x_1 \\ x_2 \\ 
      \vdots \\ x_n  \end{array} \right] \ \ ,$$
definimos a sua {\bf m\'{e}dia} aritm\'{e}tica como sendo 
 $$ \mu_x = \frac{1}{n} \sum_{i=1}^{n} x_i \ \ .$$

Dada uma amostra, usaremos a sua m\'{e}dia como estimativa da
esperan\c{c}a.  Esta estimativa \'{e} consistente com a estimativa da
distribui\c{c}\~{a}o de probabilidade id\^{e}ntica \`{a}s
freq\"{u}\^{e}ncias relativas na amostra, que por sua vez tinha
justificativa no crit\'{e}rio da m\'{a}xima verossimilhan\c{c}a. 

\begin{itemize}

\item[]{\bf Propriedades de Transforma\c{c}\~{a}o} 

Estaremos sempre interessados em estudar transforma\c{c}\~{o}es lineares
de vari\'{a}veis alea\-t\'{o}\-rias, $Z = \alpha X + \beta Y + \gamma$ e
como estas transforma\c{c}\~{o}es se refletem nas quantidades
estat\'{\i}sticas a serem definidas, por exemplo:
 $$ E( \alpha X + \beta Y + \gamma ) = 
    \alpha E(X) + \beta E(Y) +\gamma $$ 

A prova pode ser dividida em duas partes: 
 \begin{eqnarray*} 
 E(\alpha X + \gamma ) 
 &=& \sum_x ( \alpha x +\gamma) \mbox{Pr}(X=x) \\ 
 &=& \gamma + \alpha \sum_x x \mbox{Pr}(X=x) \\ 
 &=& \gamma + \alpha E(X) \\  
 & & \mbox{} \\ E(X+Y) 
 &=& \sum_{x,y} (x+y) \mbox{Pr}(X=x \wedge Y=y) \\ 
 &=& \sum_{x,y} x \mbox{Pr}(X=x \wedge Y=y) \ \  
     + \sum_{x,y} y \mbox{Pr}(X=x \wedge Y=y) \\ 
 &=& \sum_x x \mbox{Pr}(X=x \wedge Y\in Y(A) ) 
    + \sum_y y \mbox{Pr}(Y=y \wedge X \in X(A) ) \\ 
 &=&  E(X) + E(Y) 
 \end{eqnarray*} 

\end{itemize}

\section{Vari\^{a}ncia} 

A vari\^{a}ncia de uma vari\'{a}vel aleat\'{o}ria \'{e} uma medida de erro ou
dispers\~{a}o desta vari\'{a}vel em torno da sua esperan\c{c}a:
 $$Var(X) = E(( X - E(X) )^2) \ \ . $$ 

\'{E} f\'{a}cil ver que tamb\'{e}m podemos calcular a vari\^{a}ncia como
 \begin{eqnarray*}
 Var(X) & = & E(( X - E(X) )^2) \\
 & = & E( X^2 - 2 X E(X) + E(X)^2 ) \\
 & = & E(X^2) - E(X)^2
 \end{eqnarray*}

\begin{itemize}

\item[]{\bf Desvio Padr\~{a}o}

Desvio padr\~{a}o, $ \sigma_x = \sqrt{var(x)}$ tem a mesma dimens\~{a}o,
ou unidade de medida, que $E(x)$ ou $x$, i.e., \'{e} uma medida de
desvio comensur\'{a}vel com a m\'{e}dia ou os valores assumidos pela
vari\'{a}vel aleat\'{o}ria, sendo portanto de interpreta\c{c}\~{a}o mais
natural. 

\item[]{\bf Justificativa da Norma Quadr\'{a}tica} 

Existem outras alternativas para medir o tamanho do erro, $x$, al\'{e}m
da norma $L_2 = (x'x)^{1/2}$; Por exemplo as normas
 $L_1 = {\bf 1}'\mbox{abs}(x)$, ou 
 $L_\infty = \mbox{max}(\, \mbox{abs}(x)\, )$.  
 Tanto $L_1$ como $L_\infty$ s\~{a}o usadas na estat\'{\i}stica em
certas situa\c{c}\~{o}es especiais, todavia, $L_2$ \'{e} usada na
maioria das situa\c{c}\~{o}es.  Al\'{e}m de ser computacionalmente mais
simples, \'{e} a \'{u}nica que tem a propriedade de ser invariante
por uma (pelo grupo de) rota\c{c}\~{a}o.

\end{itemize}

\subsection*{Covari\^{a}ncia}  

A covari\^{a}ncia entre duas vari\'{a}veis aleat\'{o}rias, $X$ e $Y$, 
\'{e} definida como 
 $$ Cov(X,Y) = E( ( X -E(X) )*( Y -E(Y) ) ) \ \ .$$ 
 A covari\^{a}ncia \'{e} uma medida da ``interdepend\^{e}ncia'' dos 
desvios de ambas as vari\'{a}veis em rela\c{c}\~{a}o a m\'{e}dia. 
Adiaremos uma interpreta\c{c}\~{a}o intuitiva mais detalhada para 
o conceito de correla\c{c}\~{a}o, discutido adiante.  
Por hora, podemos verificar as seguintes propriedades:
 \begin{eqnarray*} 
 Cov(X,Y) 
 &=& E( XY -YE(X) -XE(Y) +E(X)E(Y) ) \\
 &=& E(XY) -E(Y)E(X)   
 \end{eqnarray*} 
donde 
  $$ Cov(X,X)  = E(X^2) - E(X)^2   = Var(X) $$

\begin{itemize}
\item[] {\bf Propriedades de Transforma\c{c}\~{a}o} 

Lema : 
 \begin{eqnarray*}  
 \lefteqn{ Cov(\alpha X +\beta Y +\gamma , Z ) } \\  
 &=& E( (\alpha X +\beta Y +\gamma) Z )  
     -E(\alpha X +\beta Y +\gamma) E(Z) \\ 
 &=& \alpha E(XZ) +\beta E(YZ) +\gamma E(Z) 
     -\alpha E(X)E(Z) -\beta E(Y)E(Z) -\gamma E(Z) \\  
 &=& \alpha Cov(X,Z) +\beta Cov(Y,Z) 
 \end{eqnarray*} 

Do lema segue que:  
 $$ Var( \alpha X +\beta Y +\gamma ) = 
   \alpha ^2 Var(X) +\beta ^2 Var(y) +2\alpha \beta Cov(X,Y) $$ 


\item[] {\bf Nota\c{c}\~{a}o Matricial} 
 
 $$Cov(X)_{i,j} \equiv Cov(X_i, X_j)$$    

Dado $X$ um vetor de $n$ vari\'{a}veis aleat\'{o}rias, $A$ uma matriz
real $m\times n$, e $b$ um vetor real $m\times 1$, os resultados
precedentes implicam na forma gen\'{e}rica da esperan\c{c}a e
vari\^{a}ncia de uma transforma\c{c}\~{a}o linear, que \'{e} dada por:
 $$ E( AX +b ) = AE(X) +b $$ 
 $$ Cov( AX +b ) = A\, Cov(X)\, A' $$  

Em estat\'{\i}stica \'{e} usual a nota\c{c}\~{a}o 
 $Cov(X)_{i,j}=\sigma_{i,j}$.  
Nesta nota\c{c}\~{a}o o desvio padr\~{a}o \'{e} 
 $\sigma_i=\sqrt{\sigma_{i,i}}$. 

Consideremos realizar $n$ vezes e independentemente, 
um experimento cujos resultados s\~{a}o medidos pelo vetor de
vari\'{a}veis aleat\'{o}rias $X= [ X_1, \ldots , X_n ]$, a vari\^{a}ncia
da m\'{e}dia \'{e} dada por

 $$ E(\bar X) = E( \ (1/n)\ {\bf 1}' X \ ) 
	      = (1/n)\ {\bf 1}' E(X) = E(X_1)$$  
 na ultima equa\c{c}\~{a}o denotamos a somat\'{o}ria 
 $\sum X$ por ${\bf 1}'X$, onde ${\bf 1}'=[1,1,\ldots 1]$.

 \begin{eqnarray*}  
 Var( \bar X ) 
 &\equiv & Var ( \ (1/n)\ {\bf 1}' X \ ) \\ 
 &=& (1/n^2)\ {\bf 1}' \ 
     \mbox{diag}([\sigma_{1,1},\ldots \sigma_{n,n}]) \ {\bf 1} \\ 
 &=& (1/n^2) ( \ \sigma_{1,1} +\ldots +\sigma_{n,n} \ ) \\  
 &=& (1/n) \sigma_{1,1}   
 \end{eqnarray*} 

\item[]  {\bf Correla\c{c}\~{a}o}

A correla\c{c}\~{a}o entre duas vari\'{a}veis aleat\'{o}rias, $Cor(X,Y)$ ou 
$\rho_{i,j}\equiv \ Cor(X_i,X_j)$, \'{e} a covari\^{a}ncia 
``normalizada'' pelo desvio padr\~{a}o:   

 $$ Cor(X,Y) = \frac{Cov(X,Y)}{\sigma_X \sigma_Y} \ \
 \mbox{ou}\ \ \rho_{i,j} = \frac{\sigma_{i,j}}{\sigma_i\sigma_j} 
 $$
A correla\c{c}\~{a}o \'{e} uma medida incompleta de independ\^{e}ncia 
entre vari\'{a}veis aleat\'{o}rias, como mostraremos a seguir:  

Dois eventos, $B$ e $C$, s\~{a}o {\em independentes} sse 
 $Pr( B\cap C) = Pr(B)*Pr(C)$.  
Duas vari\'{a}veis aleat\'{o}rias, $X$ e $Y$, s\~{a}o independentes
sse, para qualquer conjunto de valores, $C$ e $D$, 
 $Pr( X\in C \mbox{e} Y\in D ) = Pr( X\in C )*Pr( Y\in D )$. 
Temos ent\~{a}o as seguintes propriedades: 
 \begin{itemize}
 \item Se duas vari\'{a}veis aleat\'{o}rias (de vari\^{a}ncia limitada),
$X$ e $Y$, s\~{a}o independentes, ent\~{a}o $Cov(X,Y) = 0$, pois
$E(XY)=E(X)E(Y)$. 
 \item Todavia, correla\c{c}\~{a}o nula n\~{a}o garante
independ\^{e}ncia! Considere duas vari\'{a}veis aleat\'{o}rias definidas
sobre os resultados de um dado honesto: $X$ assumindo valor $-1$ em $F1$,
$1$ em $F6$, e valor $0$ em todas as outras faces; $Y$ assumindo valor
$1$ em $\{ F1, F6 \}$, e valor $0$ em todas as outras faces.  As
vari\'{a}veis aleat\'{o}rias $X$ e $Y$ n\~{a}o s\~{a}o independentes,
embora tenham correla\c{c}\~{a}o nula (verifique). 
 \item No caso de uma depend\^{e}ncia linear,  
 $Y= \alpha X +\gamma$,
temos que:
 $$Cor(Y,X) 
 = \frac{Cov(\alpha X +\gamma , X)}{\sigma (\alpha X +\gamma)\sigma (X)} 
 = \frac{\alpha Var(X)}{\sqrt{ \alpha ^2} \sigma_X \sigma_X }
 = sign(\alpha)$$
 onde $sign(x) \equiv \ \ 1\ \mbox{se}\ x>0,\ \  
      -1\ \mbox{se}\ x<0,\ \mbox{e}\ \ 0\ \mbox{se}\ x=0.$      
 \end{itemize} 
 
Provemos finalmente que $-1\leq \rho_{i,j}\leq 1$: \\  
Tomemos $X=[X_1,X_2]'$, e $Y=[a_1,a_2]X$.  
 \begin{eqnarray*} 
  Var(Y) &=&  
  \left[ \begin{array}{cc} a_1 & a_2 \end{array} \right]  
  \left[ \begin{array}{cc} 
  \sigma_{1,1} & \sigma_{1,2} \\ \sigma_{2,1} & \sigma_{2,2} 
  \end{array} \right]   
  \left[ \begin{array}{c} a_1 \\ a_2 \end{array} \right] \\ 
  &=& 
  \left[ \begin{array}{cc} a_1 & a_2 \end{array} \right]  
  \left[ \begin{array}{cc} 
  \sigma_{1} & 0 \\ 0 & \sigma_{2}  
  \end{array} \right]   
  \left[ \begin{array}{cc} 
  1 & \rho_{1,2} \\ \rho_{2,1} & 1  
  \end{array} \right]   
  \left[ \begin{array}{cc} 
  \sigma_{1} & 0 \\ 0 & \sigma_{2}  
  \end{array} \right]   
  \left[ \begin{array}{c} a_1 \\ a_2 \end{array} \right] \\ 
  &=& 
  \left[ \begin{array}{cc} b_1 & b_2 \end{array} \right]  
  \left[ \begin{array}{cc} 
  1 & \rho \\ \rho & 1  
  \end{array} \right]   
  \left[ \begin{array}{c} b_1 \\ b_2 \end{array} \right] \\   
  &=& 
  b_1^2 +2\rho b_1 b_2 +b_2^2  
\end{eqnarray*} 
Observemos agora que, pela defini\c{c}\~{a}o de vari\^{a}ncia, $Var(Y)\geq 0$. 
Observemos ainda que se $-1\leq \rho \leq 1$, podemos sempre 
``completar os quadrados'' de modo a reescrever 
 $Var(Y)= (b_1 \pm b_2)^2 +\mbox{abs} ( (1\pm \rho)b_1b_2 )$, 
quantidade obviamente positiva. 
 Todavia se $\mbox{abs} (\rho) >1$, podemos sempre escolher valores 
para $b_1$ e $b_2$ em $\{-1,1\}$ que tornam 
 $ b_1^2 +2\rho b_1 b_2 +b_2^2 < 0$, 
uma contradi\c{c}\~{a}o. 


\end{itemize}

\section{Espa\c{c}os de probabilidade} 
A exposi\c{c}\~ao das se\c{c}\~oes precedentes \'e v\'alida para 
espa\c{c}os amostrais finitos. Para espa\c{c}os n\~ao 
finitos, especialmente os n\~ao enumer\'aveis, uma estrutura mais 
mais complexa \'e necess\'aria (vide [Billingsley]).
Um espa\c{c}o de probabilidade, $(\Omega , {\cal A}, P)$, 
\'{e} um espa\c{c}o amostral, $\Omega$, uma $\sigma$-algebra, ${\cal A}$ e uma
medida de probabilidade, $P:{\cal A}\mapsto [0,1]$. 
Uma vari\'avel aleat\'oria \'{e} uma fun\c{c}\~{a}o $x: \Omega \mapsto {\Re}$,
tal que $x^{-1}(t)\in {\cal A} \ , \ \forall t \in \Re$.

A distribui\c{c}\~{a}o (cumulativa) de uma vari\'avel aleat\'oria $x$, 
$F: {\Re} \mapsto [0,1]$, \'{e} definida por 
$F(t) \equiv P(\{\omega \mid x(\omega ) \leq t \} )$.

A esperan\c{c}a de uma vari\'avel aleat\'oria, $E(x)$, \'{e} definida por
$$E(x) \equiv \int_{-\infty }^{\infty } t dF(t).$$
No caso de uma distribui\c{c}\~{a}o diferenci\'avel, ou discreta, 
temos, respectivamente
$$E(x) = \int t f(t) dt \ \ \mbox{ou} \ \  E(x)=\sum t f(t) \ .$$ 

 O $k$-\'{e}simo {\bf momento} de uma vari\'avel aleat\'oria \'{e} a
esperan\c{c}a  de sua $k$-\'{e}sima pot\^encia (omitiremos o \'{\i}ndice
$k$ para $k=1$), ${\mu}_k(x) \equiv E(x^k) .$
 O $k$-\'{e}simo {\bf momento central} de uma vari\'avel aleat\'oria \'{e} a
esperan\c{c}a da $k$-\'{e}sima pot\^encia do desvio em rela\c{c}\~{a}o 
a sua esperan\c{c}a, ${\mu}^c_k(x) \equiv E((x-\mu_1)^k) .$ 
 A vari\^{a}ncia corresponde ao segundo momento central.

Podemos agora considerar um espa\c{c}o vetorial sobre as vari\'aveis 
aleat\'orias (neste Espa\c{c}o de probabilidade) com segundo momento limitado,
$L^2(\Omega, {\cal A}, P)$, de elementos
$\{ x \mid E(x^2) < \infty \}$.
 A origem de $L^2$ \'{e} a vari\'avel aleat\'oria identicamente nula,
$x\equiv 0$, e o oposto de uma vari\'avel aleat\'oria $x$ \'{e}
$-x=(-1)x$ (explique).
 
As opera\c{c}\~{o}es usuais de soma e produto por escalar
de vari\'aveis aleat\'orias est\'a bem definida neste espa\c{c}o, pois
\begin{enumerate}
\item $E((\alpha x)^2) = {\alpha}^2 E(x^2) < \infty$.
\item $E((x+y)^2) \leq E(2x^2 +2y^2) \leq 
       2E(x^2)+2E(y^2) < \infty$.
\end{enumerate}

Em $L^2$ adotamos a seguinte defini\c{c}\~{a}o de produto interno:
$$ <X\mid Y> \equiv E(XY) \ ,$$
que satisfaz trivialmente as propriedades de simetria, linearidade,
e semi-positividade (prove). 
Algumas tecnicalidades s\~{a}o necess\'arias para assegurar a 
positividade deste produto interno.

\section{Utilidade e Decis\~{o}es} 

Consideremos a seguinte aposta: Um banqueiro lan\c{c}a um dado honesto e
paga o resultado, $X$ o n\'{u}mero na face de cima, em gramas de ouro.  
Um apostador, para apostar, deve pagar 4 gramas de ouro.  Consideraremos
nesta se\c{c}\~{a}o um modelo matem\'{a}tico, a {\em teoria
estat\'{\i}stica da decis\~{a}o}, que nos d\'{a} crit\'{e}rios para 
tomar a decis\~{a}o de participar ou n\~{a}o da aposta. 

Al\'{e}m da descri\c{c}\~{a}o da aposta em si, precisamos de uma
descri\c{c}\~{a}o das prefer\^{e}ncias do apostador.  Esta
descri\c{c}\~{a}o \'{e} dada por uma {\em fun\c{c}\~{a}o de ganho},
$G(x)$.  $G(x)$ nos d\'{a} o ganho atribu\'{\i}do pelo apostador a cada
um dos poss\'{\i}veis valores assumidos pela vari\'{a}vel aleat\'{o}ria
que caracteriza a aposta. 
 
Consideremos um apostador, A, que se importe apenas com a quantidade de
ouro ganha ou perdida no final da aposta.  As prefer\^{e}ncias de A
s\~{a}o adequadamente descritas pela fun\c{c}\~{a}o de ganho 
$G_A(x)=x-4$. 

Considere um segundo apostador, B, que possui 4 gramas de ouro, mas que
s\'{o} poder\'{a} participar de um neg\'{o}cio que com certeza lhe
triplicar\'{a} o capital se dispuser de 6 gramas.  A fun\c{c}\~{a}o
de ganho de B \'{e}:
 $G_B(x)=x-4$ se $x\leq 5$, e $G(6)=3*6-4=14$. 

Um crit\'{e}rio para medir a utilidade de um apostador, com fun\c{c}\~{a}o
de ganho $G()$, participar de uma aposta caracterizada pela
vari\'{a}vel aleat\'{o}ria $X$, \'{e} o {\em ganho esperado}:
 $$U(X) = E(G(X)) = \sum_x G(x)* \mbox{Pr}(X=x)$$ 
Nos dois exemplos anteriores ter\'{\i}amos
 \begin{eqnarray*} 
 U_A(X) &=& E(G_A(X)) = \sum_{x=1}^6 (x-4)*(1/6)= -0.5 \\ 
 U_B(X) &=& E(G_B(X)) = \sum_{x=1}^5 (x-4)*(1/6) \ +14*(1/6) = 1.5 
 \end{eqnarray*}
 ou seja, como resultado final de seu envolvimento na aposta, $A$ espera
perder $0.5$ gramas, enquanto $B$ espera ganhar $1.5$ gramas de ouro. 
Usando o crit\'{e}rio da utilidade esperada $B$ deveria fazer a aposta,
enquanto $A$ todavia n\~{a}o. 
 
\subsection*{Crit\'{e}rio de M\'{e}dia-e-Vari\^{a}ncia}

 Em finan\c{c}as e economia \'{e} usual considerar investidores (ou
agentes econ\^{o}micos, produtores, consumidores, etc.) que desejam
maximizar seus ganhos (ou minimizar seus custos).  Quando os ganhos do
investidor dependem do valor de vari\'{a}veis aleat\'{o}rias, \'{e}
razo\'avel assumir que este deseje maximizar o valor esperado do ganho.
 Todavia, apenas a esperan\c{c}a do ganho, pode ser uma
caracteriza\c{c}\~{a}o incompleta das prefer\^encias do investidor. 
Muitos investidores s\~{a}o ``avessos ao risco'', preferindo usar um
processo cujo ganho tem um dado valor esperado e vari\^{a}ncia, a um
segundo processo cujo ganho tem valor esperado ligeiramente superior,
mas mas vari\^{a}ncia bem maior.  Usaremos a seguir a teoria de
utilidade para estudar situa\c{c}\~{o}es deste tipo. 

A fun\c{c}\~{a}o utilidade (ou crit\'{e}rio) da
m\'{e}dia-e-vari\^{a}ncia \'{e} definida como: 
 $$ U(G,X,\alpha,\beta) = \alpha E(G(X)) -\beta Var(G(X)) $$
 Esta fun\c{c}\~{a}o utilidade (ou crit\'{e}rio) \'{e} muito empregado
em Economia e Finan\c{c}as, onde o par\^{a}metro $\beta / \alpha$ \'{e}
interpretado como uma medida da {\em avers\~{a}o ao risco} do
investidor.  Tivemos oportunidade de desenvolver modelos deste tipo no
cap\'{\i}tulo 3. 

Um crit\'{e}rio quadr\'{a}tico pode tamb\'{e}m ser visto 
como a aproxima\c{c}\~{a}o truncada da s\'{e}rie de Taylor de uma 
fun\c{c}\~{a}o gen\'{e}rica. Consideremos a vari\'{a}vel aleat\'{o}ria 
auxiliar $y=G(x)$, e a s\'{e}rie  de taylor:   
 $$U(y) = U[E(Y)]  +\sum_{k=1}^{\infty} U^{(k)}[E(Y)]*(y-E(Y))^k / k! $$
Truncando a s\'{e}rie ap\'{o}s o termo quadr\'{a}tico, temos 
 $$U(y) \approx U[E(Y)] +U'[E(Y)]*(y-E(Y)) +
		 U''[E(Y)]*(y-E(Y))^2/2  $$ 
Do valor esperado desta aproxima\c{c}\~{a}o obtemos 
 $$E(U(Y)) \approx U[E(Y)] +(U''[E(Y)]/2)Var(Y) $$ 

A \'{u}ltima aproxima\c{c}\~{a}o nos d\'{a} uma nova
interpreta\c{c}\~{a}o para o par\^{a}metro $\beta$ no crit\'{e}rio da
m\'{e}dia-e-vari\^{a}ncia.  Em particular, obtemos uma leg\'{\i}tima
avers\~{a}o ao risco, i.e.  $\alpha>0$ e $\beta>0$, na situa\c{c}\~{a}o
convencional de uma utilidade c\^{o}ncava, onde $U()$ \'{e}
crescente, $U'()>0$, com incrementos decrescentes, $U''()<0$.
 Poder\'{\i}amos aprofundar o estudo de como obter os par\^{a}metros
da fun\c{c}\~{a}o quadr\'{a}tica a partir de pressupostos sobrea a
``real'' fun\c{c}\~{a}o utilidade do investidor; todavia, para
aplica\c{c}\~{o}es a finan\c{c}as, temos dispon\'{\i}veis m\'{e}todos
bem mais diretos, conforme vimos no cap\'{\i}tulo 3.

\section{Modelos Lineares}

Um modelo linear tem a forma $$ \tilde y = Ap + \tilde v $$
onde $A$, $m\times n$, $A$ \'{e} conhecida e de posto pleno, $\rho (A)= n$.

A {\it observa\c{c}\~{a}o} $y$ \'{e} uma realiza\c{c}\~{a}o do vetor
aleat\'orio $\tilde y$, $p$ s\~{a}o os {\it par\^ametros} do modelo a
serem estimados, e $A$ a {\it matriz de coeficientes}.  O {\it ru\'{\i}do}
$\tilde v$ \'{e} um vetor de vari\'aveis aleat\'orias, n\~{a}o observadas, e
de primeiro e segundo momentos $E(\tilde v)=0$ e 
$Cov(\tilde v)={\sigma}^2I$.  O ru\'{\i}do \'{e} interpretado como erros de medida,
flutua\c{c}\~{o}es no processo etc.  Um ru\'{\i}do cuja matriz de covari\^ancia
\'{e} a identidade \'{e} dito {\bf branco}.  Estudaremos agora modelos
lineares simples, cujo ru\'{\i}do \'{e} branco. 

Como $E(\tilde y) = E(Ap+\tilde v) = Ap \in C(A)$, \'{e} geometricamente
intuitivo considerarmos $\hat y = P_A y$ ou equivalentemente, 
$\hat y = Arg \min_{z\in C(A)} \| y-z\|$, ou ainda, \newline 
$\hat y = A\hat p,  \  \hat p = Arg \min_p \| y-Ap\|$, como
estimativas, respectivamente, do valor m\'edio de $\tilde y$ e dos
par\^ametros do modelo. Estes s\~{a}o os {\bf estimadores de m\'{\i}nimos
quadrados} (LSE).  

Um estimador de um escalar $\pi=l'p$ em fun\c{c}\~{a}o de $y$ \'{e} 
{\bf linear} sse tem a forma $\Pi (y)=\alpha +a'y$.
Um estimador $\Pi (y)$, do escalar $\pi=l'p$, \'{e} 
{\bf n\~{a}o tendencioso} sse $E(\Pi (y))=\pi$.
$\beta +b'y$ \'{e} o melhor estimador linear n\~{a}o tendencioso, 
{\bf BLUE}, de $\pi=l'p$ sse dentre todos os estimadores lineares e n\~{a}o 
tendenciosos de $\pi$, este for de m\'{\i}nima vari\^ancia. 
Mostremos agora o seguinte resultado (Gauss-Markov):

{\it Num dado modelo linear, o BLUE de uma combina\c{c}\~{a}o linear dos
par\^ametros, $\pi = l'p$, \'{e} dado por $\hat \pi$, definido pela mesma 
combina\c{c}\~{a}o linear do LSE de $p$, $\hat \pi = \Pi (y) = l'\hat p$.}

Em primeiro lugar notemos que $\hat y = P_A y$ \'{e} uma transforma\c{c}\~{a}o
linear de $y$; Ademais, como assumimos que $A$ tem posto pleno, \'{e}
sempre poss\'{\i}vel encontrar um vetor $a' \mid l'=a'A$. 
Assim $\hat \pi$, $l'\hat p= a'Ap= a'\hat y= a'P_A y$ 
\'{e} realmente um estimador linear em $y$.
$\hat \pi$ \'{e} tamb\'{e}m n\~{a}o tendencioso, pois 
$$E(\hat \pi )= E(l'\hat p)= E(a'P_A\hat y)= a'P_A E(\hat y)= 
  a'P_A Ap= a'Ap = l'p \equiv \pi . 
$$
Finalmente, da formula do projetor, podemos explicitar  
$$\hat \pi= a'P_Ay= a'A(A'A)^{-1}A'y= l'(A'A)^{-1}A'y.$$

Consideremos agora o BLUE de  $\pi=l'p$, $\pi ^*= \beta +b'y$.
Como o BLUE \'{e} por defini\c{c}\~{a}o n\~{a}o tendencioso, devemos ter
$l'p= E(\beta +b'\tilde y)= \beta + b'Ap$, qualquer que seja o verdadeiro
par\^ametro $p$, portanto $b'A=l'$.
A vari\^ancia deste estimador \'{e}
$$ Var(\pi ^*)= b'Ib = b'b.$$
Portanto o BLUE de $\pi=l'p$ ser\'a dado por $\pi ^*=b'y$, onde 
$b'$ \'{e} especificado pelo problema de programa\c{c}\~{a}o quadr\'atica
$$ b= Arg\min_{c\mid A'c=l} c'c .$$
\'{e} f\'acil verificar que $b'=l'(A'A)^{-1}A'$ resolve o sistema
normal deste problema (verifique), o que demonstra que 
$\pi ^*= \hat \pi$, (teorema de Gauss-Markov).

Q.E.D. 

Usaremos agora a fatora\c{c}\~{a}o de Cholesky para transformar um modelo
linear geral, cujo ru\'{\i}do n\~{a}o \'{e} branco, num modelo linear simples
equivalente. Consideremos
$$ y = Ap + v \ , \ E(v)=0 \ , \ Cov(v)= V= L'L \ ,$$
e o modelo simplificado
$$ \bar y \equiv L^{-1}y= L^{-1}Ap+ L^{-1}v = \bar A p+ \bar v \ .$$
Note que 
$$ Cov(\bar v)= L^{-1}(LL')L^{-t}= (L^{-1}L)(L'L^{-t})= I \ ;$$
Assim, transformando os dados e a matriz de coeficientes do modelo geral
pela inversa do fator de Cholesky da matriz de covari\^{a}ncia do ru\'{\i}do,
podemos estimar os par\^{a}metros num modelo simples equivalente.
No modelo simplificado estamos pois minimizando
$$\| \bar y -\bar Ap\| ^2 = (y-Ap)'L^{-t}L^{-1}(y-Ap)
			  = (y-Ap)'W(y-Ap)\ ,$$
onde $W= V^{-1}$ \'{e} denominada a {\bf matriz de informa\c{c}\~{a}o}
do ru\'{\i}do original.

\section{Interpreta\c{c}\~{a}o Bayesiana}

 Consideremos uma parti\c{c}\~{a}o do espa\c{c}o amostral, 
 $$A= \bigcup_1^n A_i\ , \ \ Pr(A_i)>0  \ , \ 
  A_i \cap A_j = \emptyset, i\neq j.$$ 
 A probabilidade condicional de $A_i$ dado o evento $B$ \'{e}, 
 por defini\c{c}\~{a}o, $Pr(A_i \mid B) = Pr(A_i \cap B) / Pr(B)$. 
 Usando $Pr(B) = \sum_1^n Pr(A_i)Pr(B \mid A_i)$, temos que 
 $$ Pr(A_i \mid B) = 
    Pr(A_i)Pr(B \mid A_i) / \sum_{j=1}^n Pr(A_j)Pr(B \mid A_j)$$    
 Este \'{e} o chamado teorema ou regra de Bayes, que nos permite 
 calcular $Pr(A_i \mid B)$ a partir da informa\c{c}\~{a}o 
 ``inversa'', $Pr(B \mid A_j),\ j=1\ldots n$. 

 Consideremos agora uma distribui\c{c}\~{a}o de probabilidades que 
 \'{e} fun\c{c}\~{a}o de um par\^{a}metro dentro de um dom\'{\i}nio 
 bem especificado, o espa\c{c}o param\'{e}trico, 
 $$f(x \mid \theta)\ , \ \ \theta \in \Theta.$$  
 Quando estimamos o par\^{a}metro da distribui\c{c}\~{a}o, 
podemos expressar nossa incerteza assumindo que o pr\'{o}prio 
para\^{a}metro $\theta$ tem uma distribui\c{c}\~{a}o $p(\theta)$ no 
espa\c{c}o param\'{e}trico. 
 Assim, a express\~{a}o $f(x \mid \theta)$ deve ser interpretada como
a probabilidade condicional de $x$ para um dado valor de $\theta$.     

 Dada uma amostra de $n$ experimentos independentes, $x=[x_1 \ldots x_n]$, 
sua distribui\c{c}\~{a}o conjunta \'{e} 
 $$f(x \mid \theta) = \prod_{i=1}^n f(x_i \mid \theta),$$ 
 que podemos integrar sobre $\theta$ obtendo  
 $$ g(x) = \int_{\Theta} f(x \mid \theta) p(\theta) d\theta.$$    
 Usando a regra de Bayes, podemos tamb\'{e}m obter a probabilidade 
 condicional
 $$ f(\theta \mid x) = f(x \mid \theta) p(\theta) / g(x).$$ 

 Neste contexto $p(\theta)$ \'{e} denominada a distribui\c{c}\~{a}o 
 ``a priori'' (inicial) de $\theta$, $f(\theta \mid x)$ 
 \'{e} a probabilidade ``a posteriori'' (depois de observar $x$), 
 e o c\'{a}lculo de $f(\theta \mid x)$ \'{e} denominado a 
 ``opera\c{c}\~{a}o Bayesiana''.    
 
 A opera\c{c}\~{a}o Bayesiana \'{e} ``recursiva'', i.e. se um novo vetor
de observa\c{c}\~{o}es $y$ estiver disponivel, podemos calcular a nova
posteriori $f(\theta \mid x,y)$ usando como priori na opera\c{c}\~{a}o 
Bayesiana $f(\theta \mid x)$, que j\'{a} incorporou toda a
informa\c{c}\~{a}o contida em $x$. 
 Esta recursividade implica em grande efici\^{e}ncia  computacional
pois, ao adquirirmos uma nova observa\c{c}\~{a}o, podemos ``atualizar''
a posteriori  realizando uma opera\c{c}\~{a}o Bayesiana sobre esta
\'{u}nica observa\c{c}\~{a}o, sem  ter que refazer o c\'{a}lculo usando
todas as observa\c{c}\~{o}es anteriores.
 Podemos tamb\'{e}m postergar o c\'{a}lculo da constante de
normaliza\c{c}\~{a}o $g(x)$, que envolve uma integra\c{c}\~{a}o que pode
ser trabalhosa, usando  nos passos intermediarios de aquisi\c{c}\~{a}o
de dados a fun\c{c}\~{a}o de verossimilhan\c{c}a, 
 $$l(\theta \mid x) = f(x \mid \theta) p(\theta) 
   \propto f(\theta \mid x).$$

 A no\c{c}\~{a}o de priori \'{e} algo controversa. Alguns argumentam que
nossa incertesa  sobre o verdadeiro valor do par\^{a}metro n\~{a}o
implica que este possa ser tratado como uma vari\'{a}vel aleat\'{o}ria,
tendo uma distribui\c{c}\~{a}o de probabilidade. 
 Normalmente utiliza-se uma priori ``neutra'' como uma
distribui\c{c}\~{a}o uniforme sobre todo o espa\c{c}o  param\'{e}trico. 
 Outros argumentam que nada h\'{a} de neutro nesta escolha, sendo a
priori uniforme t\~{a}o subjetiva como qualquer outra. 
  Um contra-argumento \'{e} a coveni\^{e}ncia de incorporar a
subjetividade da priori na modelagem, incorporando na priori
``conhecimento subjetivo'' i.e. previamente adquirido fora de um
contexto estat\'{\i}stico formal, mas que nem por isto deveria ser
menosprezado. 

 A linha subjetivista postula que o pr\'{o}prio conceito de
probabilidade \'{e} essencialmente subjetivo. Esta linha encara  o
conceito de probabilidade como retratando nossa ignorancia (ou
informa\c{c}\~{a}o  parcial) a respeito do mundo, bem \`{a} maneira da
f\'{\i}sica estat\'{\i}stica no contexto das leis determin\'{\i}sticas
da f\'{\i}sica  cl\'{a}ssica.
 O debate sobre a validade de uma distribui\c{c}\~{a}o no espa\c{c}o 
param\'{e}trico lembra as cr\'{\i}ticas hist\'{o}ricas ao conceito de uma
fun\c{c}\~{a}o de onda n\~{a}o observavel em  mec\^{a}nica qu\^{a}ntica. 
 A postura subjetivista novamente lembra  a historia da mec\^{a}nica
qu\^{a}ntica, os c\'{e}lebres debates Einstein vs Heisemberg sobre o
princ\'{\i}pio de incerteza, s\'{o} que agora com posi\c{c}\~{o}es
invertidas, i.e. a  posi\c{c}\~{a}o Bayesiana parece inovadora ao propor
uma distribui\c{c}\~{a}o no espa\c{c}o param\'{e}trico, sendo todavia
conservadora ao procurar ref\'{u}gio na interpreta\c{c}\~{a}o
subjetivista para o conceito de probabilidade.

\section{Exerc\'{\i}cios} 
\begin{enumerate}
 
\item Formule o problema de m\'{\i}nimos quadrados como um problema 
de programa\c{c}\~ao quadr\'atica. 
 \begin{enumerate}
 \item Assuma dada uma base $N$ de $N(A)$.
 \item Calcule diretamente o res\'{\i}duo $z = b - y$ em fun\c{c}\~ao de $A$. 
  \end{enumerate}
\item Prove que a distribui\c{c}\~{a}o $F$ \'{e}
  \begin{enumerate}
  \item N\~{a}o decrescente, com $F(-\infty )=0$ e $F(\infty )=1$.
  \item Sempre cont\'{\i}nua \`a esquerda e continua \`a direita exceto num
   numero enumer\'avel de pontos.
  \end{enumerate}
\item Se as vari\'aveis aleat\'orias $x$ e $y$ t\^em, respectivamente, 
   distribui\c{c}\~{o}es $F$ e $G$, determine a distribui\c{c}\~{a}o de   
   \begin{enumerate}
   \item $\alpha x$. 
   \item $x+y$.
   \item $xy$. 
   \end{enumerate}
\item Dadas $x$ e $y$, vari\'aveis aleat\'orias, mostre que:
  \begin{enumerate}
  \item $E(\alpha x +\beta y) = \alpha E(x) +\beta E(y)$.
  \item $Var(\alpha x +\beta y) = {\alpha}^2 Var(x) +
	 {\beta}^2 Var(y) + 2\alpha \beta Cov(x,y) $.
  \end{enumerate}
\item Dado $x$, um vetor de vari\'aveis aleat\'orias, mostre que: 
  \begin{enumerate}
  \item $E(Ax) = A E(x)$.
  \item $Cov(Ax) = A Cov(x)A' $. 
  \end{enumerate}
\item O {\bf tra\c{c}o} de uma matriz $A$ \'{e} definido por
      $ tr(A) \equiv \sum A_i^i$. Mostre  que
  \begin{enumerate}
  \item $tr(A+B) = tr(A) + tr(B)$.
  \item $tr(AB) = tr(BA)$.
  \item $x'Ay = tr(Ayx') = tr(yx'A)$.
  \item Se $A$, $m\times n$, tem posto pleno, $\rho (A) = n$, ent\~{a}o
	$tr(P_A)=n$.
  \item Nas condi\c{c}\~{o}es do \'{\i}tem anterior, definindo $R_A=(I-P_A)$,
	temos que $tr(R_A)=m-n$.
  \end{enumerate}
\item Dado $x$ um vetor de vari\'aveis aleat\'orias, $E (x)=\mu$,
      $Cov(x)=V$, e $S$ uma matriz sim\'{e}trica, temos que
      $$E(x'Sx)= tr(SV) + {\mu}'S\mu .$$
      Sugest\~{a}o: Use que $E(x'Sx) = tr(S E(xx'))$.
\item Num modelo linear de posto pleno, com $\hat p=(A'A)^{-1}A'y$ 
      estimando o par\^ametro $p$, mostre que
   \begin{enumerate}
   \item $Cov(\hat p)= {\sigma}^2 (A'A)^{-1}$.
   \item O erro quadr\'atico m\'edio, 
	 $MSE \equiv {\| y-P_Ay \|}^2 /(m-n)$,  
	 \'{e} um estimador n\~{a}o tendencioso de ${\sigma}^2$.
	 Sugest\~{a}o: Use $MSE= y'R_Ay/(m-n)$, onde $R_A=(I-P_A)$.  
   \end{enumerate}    
\end{enumerate}

%% file: apen5.tex
\chapter{\label{apen5}Programas}

\section{bigode.m} 
 
\renewcommand{\baselinestretch}{0.99}
\parskip 0.0cm
\begin{small} 
\begin{verbatim} 

%Mestrado Profissionalizante 
%Modelagem Matematica em Financas - Turma III
%Programacao dinamica: Bigode
%Daniel Granja

%Condicoes iniciais
infty = 1E10; eps = 1E-10;
h=2; q = 5; fmax = 3; kmax = 2;
s0 = 10; b0 = 20; sx = 5; c0 = 5;

%constantes independentes dos movimentos trinomiais
m1s = 1; m1b = 1; m2s = 1; m2b = 1; m3s = 1; m3b = 1;
fsu  = 1.3; psu = 0.3; psd = 0.7; 
fbu = 0.9; pbu = 0.6; pbd = 0.4; 
fbe = 0.0; fse = 0.0;

%simplificacao para binomial
psd=1-psu; pbd = 1 - pbu;

%constantes independentes
fbu=1.1; pbu=0.6; 
fsd=1/fsu; fbd=1/fbu;

%h=7;%Horizonte de tempo, isto eh, numero de periodos
fmax=3; %estoque fisico maximo de salsichas pereciveis
s0=10; %preco de partida
r0=20; %o fator de desconto,B eh definido como: 1/(1+r), 
%onde a taxa de juros eh ao ano

%-Opcoes

c0=3; %custo de uma opcao de salsicha para comprar 
%q salsichas a exercicio sx/salsicha. 
opt=1; %estoque maximo de opces. Dara a possibilidade 
%de formar q*opt salsichas longa-vida.
sx=5; %Preco de exercicio. 
%O custo de exercicio de q opcoes eh sx*q
q=3; %numero de salsichas no exercicio de uma opcao;
if (opt==0) q=0; end

%formacao dos estados possiveis da natureza
%passeio aleatorio binomial geometrico estacionario
fsu=1.3; psu=0.3; psd=1-psu;
fbu=1.1; pbu=0.6; pbd=1-pbu;
fsd=1/fsu;
fbd=1/fbu;
%constantes limitadoras para otimizacao:
infty=1E10;
eps=1E-10;

%arvore binomial para salsicha `a vista
S(1,1)=s0; PS(1,1)=1; dimst=1;
for t=2:(h+2)
   for m=1:t
      if (m<t) 
         S(t,m)=S(t-1,m)*fsd;
         if (m==1) PS(t,m)=PS(t-1,m)*psd;
         else
            PS(t,m)=PS(t-1,m)*psd+PS(t-1,m-1)*psu;
         end
      else 
         S(t,m)=S(t-1,m-1)*fsu;
         PS(t,m)=psu^(t-1);
      end
      dimst=dimst+1;
   end %m
end %t

%arvore binomial para taxa de juros de 6 meses, 
%mas expressa ao ano
R(1,1)=r0; PR(1,1)=1; dimRt=1;
for t=2:(h+2)
   for m=1:t
      if (m<t) 
         R(t,m)=R(t-1,m)*fbd;
         if (m==1) PR(t,m)=PR(t-1,m)*pbd;
         else
            PR(t,m)=PR(t-1,m)*pbd+PR(t-1,m-1)*pbu;
         end
         
      else 
         R(t,m)=R(t-1,m-1)*fbu; 
         PR(t,m)=pbu^(t-1);
      end
      dimRt=dimRt+1;
   end %m
end %t

%inicializa controle otimo
for t=(h+1):-1:1
   instante(t)=0;
   bestwl(t)=666;
   bestwk(t)=666;
   bestfoti(t)=666;
end%t    
instante(h+1)=0;

% prepara para condicao de contorno

%obs: como nao eh possivel no matlab declarar matriz 
%com indice zero, adiciono 1 nos limites de estoque
for k=1:(opt+1)
   for l=1:(k*q+fmax+1)
      for Rind=1:dimRt
         for sind=1:dimst
            
            fot(h+2,sind,Rind,l,k)=0;
            %a condicao de contorno eh para o 
            %instante seguinte ao final
            
         end%sind
      end%Rind
   end%l
end%k    

%bestfoti=infty;
estoquetp1(h+1)=0;
estoque_p_tpl(h+1)=0;

v=1;
%%%%%%%%%%%%%%%%%%%%%%%  
%Resolucao da equacao de Bellman para t>1 
%%%%%%%%%%%%%%%%%%%%%%%%
for k=1:(opt+1)
   %novamente, no matlab, nao posso partir de zero...
   %entao somo 1.Entenda-se que k=1 eh primeiro estado 
   %das quantidades das opcoes, onde tenho quantidade zero. 
   %Define-se qk como a quantidade, propriamente dita nesse estado
   qk=k-1;            
   
   % a equacao de Bellman eh resolvida do final para o inicio
   for t=(h+1):-1:2
      
      tp1=t+1; %instante `a frente do instante atual, t
      instante(t)=t
      %a cada loop em t, a funcao custo do instante t+1 
      %recebe o resultado da otimizacao do instante anterior:
      dimRtp1=dimRt;
      dimstp1=dimst;
      for j=1:(opt+1)
         for l=1:(j*q+fmax+1)
            for Rind=1:dimRtp1
               for sind=1:dimstp1
                  fotp1(t,sind,Rind,l,j)=fot(t+1,sind,Rind,l,j);
               end%sind
            end%Rind
         end%l
      end%k
      dimRt=t;%redimensionamento
      dimst=t;
      
      %inicia varredura dos estados       
      
      %estados definidos pelas quantidades possiveis de salsichas            
      for l=1:(fmax+1)
         %novamente, no matlab, nao posso partir de zero...entao adiciono 1.
         %Entenda-se que l=1 eh primeiro estado das quantidades de salsichas, 
         %onde tenho quantidade zero.
         %Define-se ql como a quantidade, propriamente dita nesse estado
         ql=l-1; 
         % if (ql==0) qlptp1=0; end
         qtot=ql+qk*q*0;
         %estados definidos pelas possibilidades de taxas de juros 
         %no instante presente              
         for mr=1:t 
            %como a arvore eh binomial, em cada instante de tempo 
            %ha t estados para r
            realr=R(t,mr);
            %estados definidos pelas possibilidades de precos 
            %das salsichas no mercado `a vista
            for ms=1:t
               reals=S(t,ms);
               bestfoti(t)=6666;
               %wk e wl sao as variaveis de controle
               %limites para numeros de opcoes executadas
               if (qk>0 & qtot<=(fmax+q*qk)) wkmax=1; else wkmax=0; end
               %estados de possibilidades de execucao de opcoes      
               for wk=0:wkmax

                  if (ql==0 & wk==0) wlmin=1; else wlmin=0; end
                  %o que pode ser comprado, se o estoque de salsichas 
                  %pereciveis estiver abaixo do maximo, eh o quanto falta 
                  %para chegar nesse maximo, mais um, pois esta eh a demanda
                  %para o periodo corrente:
                  if (ql<=fmax & wk==0) wlmax=(fmax-ql)+1; else wlmax=0; end
                  %estados definidos pelas posibilidades de quantidade de compra:
                  for wl=wlmin:wlmax
                     %equacao de evolucao dos estoques:
                     qtottp1=qtot+wk*q+wl-1;
                     estoqueauxtp1(t)=qtottp1;
                     %variavel auxiliar para ser usada na decisao do melhor
                     %qlptp1=wl-1;
                     ltottp1=qtottp1+1;
                     %variavel estado correspondente a quantidade qltp1
                     %equacao de evolucao dos estoques de opcoes:
                     qktp1=qk-wk;
                     ktp1=qktp1+1;
                     %variavel estado correspondente a quantidade qktp1
                     %custo do controle:
                     auxfoti=wk*q*sx+wl*reals;
                     %tratamento dos estados possiveis y=x(t+1)
                     %wr e ws sao as transicoes possiveis de t para t+1
                     for wr=mr:(mr+1)
                        % possibilidade seguinte eh para cima ou para baixo
                        %pwr=PR(t+1,wr)/PR(t,mr);
                        if (wr==mr) pwr=pbd; else pwr=pbu; end
                        %probabilidade de transicao para o proximo:
                        %para cima e para baixo
                        for ws=ms:(ms+1)%idem
                           if (ws==ms) pws=psd; else pws=psu; end
                           %pws=PS(t+1,ws)/PS(t,ms);
                           %adicionamos ao custo do controle o custo futuro 
                           %multiplicado pela probabilidade (tiramos a media, 
                           %portanto) de transicao do estado presente para os 
                           %possiveis estados futuros e descontamos pela taxa 
                           %de juros
                           auxfoti=auxfoti+(pwr*pws*(1/((1+realr/100)^(1/2))))...
                              *fotp1(t,ws,wr,ltottp1,ktp1);
                           teste(v)=auxfoti;
                           v=v+1;
                        end%ws
                     end%wr
                     
                     if((t<=(h+1)) & (t>=(h+1-fmax))& estoqueauxtp1(t)... 
                           <=(h+1-t)& estoqueauxtp1(t)>=0) 
                        if(auxfoti<bestfoti(t))
                           bestfoti(t)=auxfoti;
                           bestfotiescolhido(t)=auxfoti;
                           bestwk(t)=wk;
                           bestwl(t)=wl;
                           
                           estoque(t)=qtot;
                           
                           estoquetp1(t)=qtottp1;
                           best_num_op_estoque=qk;
                        end
                     else 
                        if(t<(h+1-fmax))
                           if(auxfoti<bestfoti(t))
                              bestfoti(t)=auxfoti;
                              bestfotiescolhido(t)=auxfoti; 
                              %variavel auxiliar para controlar a queda de bestfoti
                              bestwl(t)=wl;
                              bestwk(t)=wk;
                              estoque(t)=qtot;
                              
                              estoquetp1(t)=qtottp1;
                              best_num_op_estoque=qk;
                           end 
                        end%if
                     end%if   
                  end%wl
               end%wk
               fot(t,ms,mr,l,k)=bestfoti(t);
               wlot(t)=bestwl(t);
               wkot(t)=bestwl(t);
            end%ms
         end%mr
      end%l
   end%t
      
   %%%%%%%%%%%%%%%%%%%%%%%%%%%%%%%%%%%%%% 
   %Equacao de Bellman em t=1 (inicial) 
   %%%%%%%%%%%%%%%%%%%%%%%%%%%%%%%%%%
   
   t=1;
   tp1=t+1; 
   %instante `a frente do instante atual, t
   instante(t)=t
   %a cada loop em t, a funcao custo do instante t+1 
   %recebe o resultado da otimizacao do instante anterior:
   dimRtp1=dimRt;
   dimstp1=dimst;
   for j=1:(opt+1)
      for l=1:(j*q+fmax+1)
         for Rind=1:2
            for sind=1:2
               fotp1(t,sind,Rind,l,j)=fot(t+1,sind,Rind,l,j);
            end%sind
         end%Rind
      end%l
   end%k
   dimRt=t;%redimensionamento
   dimst=t;
   
   %inicia varredura dos estados

   %estados definidos pelas quantidades possiveis de salsichas:            
   for l=1:(fmax+1)
      %novamente, no matlab, nao posso partir de zero...entao somo 1.
      %Entenda-se que l=1 eh primeiro estado das quantidades de salsichas, 
      %onde tenho quantidade zero.
      %Define-se ql como a quantidade, propriamente dita nesse estado
      ql=l-1; 
      %qtot=q*qk+ql;
      %estados definidos pelas possibilidades de taxas de juros 
      %no instante presente
      mr=1;
      realr=R(t,mr);
      %estados definidos pelas possibilidades de precos das salsichas 
      %no mercado `a vista
      ms=1;
      reals=S(t,ms);
      bestfoti(t)=infty;
      %limites para o numero de opcoes compradas 
      if (qk<opt) wkmax=opt-qk; else wkmax=0; end
      for wk=0:wkmax
         %limites para compra no mercado fisico
         if(ql==0) wlmin=1; else wlmin=0; end
         if(ql<=(fmax)) wlmax=fmax-ql+1;else wlmax=0; end
         for wl=wlmin:wlmax
            %eq. evolucao dos estoques
            qltp1=ql+wl-1;
            estoqueauxtp1(t)=qtottp1;
            %variavel auxiliar para ser usada na decisao do melhor
            ltp1=qltp1+1;
            %variavel estado correspondente a quantidade qltp1
            qktp1=qk+wk;
            ktp1=qktp1+1;
            %variavel estado correspondente a quantidade qktp1
            %custo de controle
            auxfoti=wk*q*c0+wl*reals;
            %tratamento dos estados possiveis y=x(t+1)
            %wr e ws sao as transicoes possiveis de t para t+1
            for wr=mr:(mr+1)
               % possibilidade seguinte eh para cima ou para baixo
               %pwr=PR(t+1,wr)/PR(t,mr);
               if (wr==mr) pwr=pbd; else pwr=pbu; end
               for ws=ms:(ms+1)%idem
                  if (ws==ms) pws=psd; else pws=psu; end
                  %pws=PS(t+1,ws)/PS(t,ms);
                  %adicionamos ao custo do controle o custo futuro multiplicado  
                  %pela probabilidade (tiramos a media, portanto) 
                  %de transicao do estado presente para os possiveis
                  %estados futuros e descontamos pela taxa de juros
                  auxfoti=auxfoti+(pwr*pws*(1/((1+realr/100)^(1/2))))... 
                     *fotp1(t,ws,wr,ltp1,ktp1);
               end%ws
            end%wr
            if(auxfoti<bestfoti(t)) 
               bestfoti(t)=auxfoti;
               bestfotiescolhido(t)=auxfoti;
               bestwl(t)=ql;
               bestwk(t)=wk;
               estoque(t)=ql;
               estoquetp1(t)=qltp1;
               best_num_op_estoque=qk;
            end
         end%wl
      end%wk
      fot(t,ms,mr,l,k)=bestfoti(t);
      %wlot(t,ms,mr,l,k)=bestwl(t);
      %wkot(t,ms,mr,l,k)=bestwk(t);
      wlot(t)=bestwl(t);
      wkot(t)=bestwk(t);
   end%l
end%k 
               
\end{verbatim} 
\end{small}
\renewcommand{\baselinestretch}{1.1}
\parskip 0.2cm
\pagebreak

\section{simplex.m} 
 
\renewcommand{\baselinestretch}{0.99}
\parskip 0.0cm
\begin{small} 
\begin{verbatim}
  
function [Vo,b,xb,r] = driver(A,d,c)
% arquivo driver.m  driver e funcao simplex 

  clc;
  disp('Criterio para a escolha do indice residual:');
  disp('1 - menor j tal que z(j) > 0');
  disp('2 - j tal que z(j) eh maximo');
  criterio = input('1/2? ');
  disp('Deseja ver cada iteracao em um arquivo?');
  trace = input('Digite 1 para sim ou 0 para nao: ');
  if (trace)
    disp('Os resultados serao gravados no arquivo simplex.sai')
    disp('Tecle algo para continuar: ');
    pause;
    diary simplex.sai;
  end

  clc;
  disp('PPL: ');
  disp('Matriz A: '); disp (A);
  disp('Vetor  d: '); disp (d');
  disp('Vetor  c: '); disp (c);
  disp('Criterio para a escolha de j: ');
  if (criterio == 1)
    disp('menor j tal que z(j) > 0');
  else
    disp('j tal que z(j) eh maximo');
  end
  disp('');

  [m,n] = size(A);
  aux = diag(sign(d));
  Aaux = [A aux];
  czeros = zeros(1,n);
  uns = ones(1,m);
  caux = [czeros uns];
  r = 1:n;
  b = n+1:n+m;
  [Vo,b,xb,r] = simplex(Aaux,d,caux,b,r,criterio,trace);
  if (Vo == 0)
    raux = find(r <= n);
    r = r(raux);
    [Vo,b,xb,r] = simplex(A, d, c, b, r, criterio, trace);
  else
    disp('Problema sem solucao');
  end %if
  if (trace) 
    diary; 
  end
end

function [Vo,b,x,r] = simplex(A,d,c,b,r,criterio,trace)
% arquivo simplex.m ***********************************

  [m,n] = size(A);
  acabou = 0;
  while ~acabou
    B = A(:,b);
    R = A(:,r);
    if (det(B) == 0)
      disp('Erro: Det(B) = 0')
      return
    else 
      Rtil = B\R;
      dtil = B\d;
      z = c(b)*Rtil - c(r);
      if ( any(z > 0) )
        indices = find(z > 0);
        if criterio == 1 
          j = indices(1);    
        else 
          [aux,j] = max(z);   %maior declive
        end %if criterio
        k = find( Rtil(:,j) > 0 );
        epsilon = dtil(k) ./ Rtil(k,j);
        [aux,i] = min( epsilon );
        i = k(i);
        b = [b r(j)];
        r = [r b(i)];
        b(i) = [];
        r(j) = [];
      else % acabou 
        acabou = 1;
      end %if indices
      x = zeros(n,1);
      x(b) = A(:,b)\d;
      Vo = c * x;
      if (trace)
        disp('');
        disp('Vo: '); disp(Vo); 
        disp('indices da base: '); disp(b); 
        disp('indices residuais: '); disp(r);
        disp('Vertice: '); disp(x'), 
        disp(''); 
      end
    end % if det(B) 
  end %while
end

function C = conta(l,m,n) 
% todas as combinacoes de [l...m] n a n 

C=[]; 
if( l>m | m-l+1<n ) 
  C=[]; 
elseif( n==1 ) 
  C=[l:m]'; 
else 
  CA=[]; 
  CA= conta1(l,m-1,n-1); 
  [car,cac]= size(CA);  %rows,columns in CA 
  for j=1:car  
    CAJ= CA(j,:); 
    for k= (CAJ(cac)+1):m 
      C=[C;[CAJ,k]]; 
    end  
  end 
end 
\end{verbatim} 
\end{small}
\renewcommand{\baselinestretch}{1.1}
\parskip 0.2cm
\pagebreak

\section{mindist.m} 
 
\renewcommand{\baselinestretch}{0.88}
\parskip 0.0cm
\begin{small} 
\begin{verbatim} 
% Este eh o arquivo MATLAB "mindist.m".  
% Tudo q segue um "%" eh apenas comentario.
% Consulte o capitulo sobre a linguagem Matlab 

n = 5; 
% numero de vertices
C = [ 0  , 2  , 9E9, 1  , 2; 
      2  , 0  , 2  , 9E9, 0; 
      9E9, 2  , 0  , 3  , 4; 
      1  , 9E9, 3  , 0  , 0; 
      2  , 0  , 4  , 0  , 0;   ]; 
%matriz de custos
fo(1,1:n) = C(1,1:n); 
%custo do caminho de tamanho 1 
for t = 2:n-1; 
  fo(t,1:n) = 9E9*ones(1,n); 
  for j = 1:n 
    for i = 1:n 
      fo(t,j) = min( fo(t,j) , fo(t-1,i)+C(i,j)  );   
    end 
  end 
end   
fo,

% Este eh o arquivo MATLAB "dijk.m".  
n = 5; 
% numero de vertices
C = [ 0  , 2  , 9E9, 1  , 2; 
      2  , 0  , 2  , 9E9, 0; 
      9E9, 2  , 0  , 3  , 4; 
      1  , 9E9, 3  , 0  , 0; 
      2  , 0  , 4  , 0  , 0;   ]; 
%matriz de custos
inder = 2:n; 
%vertices cuja minima distancia nao eh a correta 
dcor = [1;0];  %distancias corretas   
fo = C(1,2:n); 
%custo do caminho de tamanho 1 
for t = 2:n-1; 
  [d,k] = min(fo);  %minimo da lista auxiliar 
  i = inder(k);     %vertice correspondente
  dcor = [dcor,[i;d]];  
  inder(k) = []; 
  fo(k) = [];          
  fo = min( fo , C(i,inder)+d*ones(1,n-t)  );   
end  
dcor = [dcor,[inder;fo]];  
dcor,  %lista de vertices e distancias, 
       %na ordem em que foram determinados  
\end{verbatim} 
\end{small}
\renewcommand{\baselinestretch}{1.1}
\parskip 0.2cm
\pagebreak

\section{Fatora\c{c}\~{a}o QR} 
 
\renewcommand{\baselinestretch}{0.9}
\parskip 0.0cm
\begin{small} 
\begin{verbatim} 

function rinv(dummy) 

global A U b r    
[m,n]=size(A);  

%basis reinversion  

for k=1:m
  U(:,k)= A(:,b(k)); 
end 

for j=1:m-1
  pivj = U(j,j);
  for i=j+1:m 
    
    %compute Givens rotation Q s.t. Q*[v1,v2]=[norm(2,v),0]
    %Q=[gcos, gsen; -gsen, gcos]
    %procedure is protective of overflow

    v1=pivj; v2=U(i,j);
    if( abs(v2)==0 ) 
      gcos=1; gsen=0;
    else 
      if( abs(v1)>=abs(v2) ) 
        tau= -v2/v1; 
        gcos= (1+tau*tau)^(-1/2); 
        gsen= gcos*tau;  
      else % abs(v2)>abs(v1) 
        tau= -v1/v2; 
        gsen= (1+tau*tau)^(-1/2); 
        gcos= gsen*tau; 
      end %else_abs
    end %else_v20

    % rotate rows U([j,i],j:m) by Q 
  
    pivj= gcos*v1 -gsen*v2;
    U(i,j)=0; 
   for k=j+1:m
      v1=U(j,k);
      v2=U(i,k); 
      U(j,k)= gcos*v1 -gsen*v2; 
      U(i,k)= gsen*v1 +gcos*v2; 
    end %fork 

  end %fori 
U(j,j)=pivj; 
end %forj 
 
%*********************************************************

function update(ip,jp)  

global A U b r 
[m,n]=size(A); 

% r(jp) enters the basis, 
% b(ip) exits the basis. 

a= A(:,r(jp));
y=qtm(a);
qrup(y,ip);
auxi=b(ip);
for k=ip:m-1
  b(k)=b(k+1);
end
b(m)=r(jp);   
r(jp)=auxi;

%******************************

function  qrup(x,k)

global A U b r 
[m,n]= size(A); 

%file qrup.m 
%Updates the U factor from the Hessemberg matrix obtained 
%by deleting column k of U, 
%and appending column x as the last column of U.  

EPSLN=1E-16;

for i=1:k-1 
  for j=k+1:m 
    U(i,j-1)=U(i,j); 
  end 
  U(i,m)=x(i,1);
end

for i=k:m-1 
  
  %compute Givens rotation Q s.t. Q*[v1,v2]=[norm(2,v),0]
  %Q=[gcos, gsen; -gsen, gcos]
  %procedure is protective of overflow

  v1=U(i,i+1); v2=U(i+1,i+1);
  if( abs(v2)<EPSLN ) 
    gcos=1; gsen=0;
  else 
    if( abs(v1)>=abs(v2) ) 
      tau= -v2/v1; 
      gcos= (1+tau*tau)^(-1/2); 
      gsen= gcos*tau; 
    else % abs(v2)>abs(v1) 
      tau= -v1/v2; 
      gsen= (1+tau*tau)^(-1/2); 
      gcos= gsen*tau; 
    end %else_abs
   end %else_v20

  % rotate rows U(i:i+1,:) by Q

  U(i,i)= gcos*v1 -gsen*v2;
  for j=i+2:m
    v1=U(i,j);
    v2=U(i+1,j); 
    U(i,j-1)= gcos*v1 -gsen*v2; 
    U(i+1,j)= gsen*v1 +gcos*v2; 
  end 

  %rotate elements x(i:i+1)
  v1=x(i,1);
  v2=x(i+1,1);
  U(i,m)= gcos*v1 -gsen*v2; 
  x(i+1,1)= gsen*v1 +gcos*v2; 
end 
U(m,m)=x(m); 

%********************************************************

function y = qtm(x)

global A U b r 
[m,n]=size(A); 
 
%file qtm.m y =Q'*x =inv(U')*B'*x  , B=QR

y=btm(x);
y=utim(y);

%****************************************

function y = uim(x) 

global A U b r 
[m,n]=size(A);   

%file uim.m    y=inv(U)*x    

y(m,1)=x(m,1)/U(m,m);
for i=m-1:-1:1 
  y(i,1)=( x(i,1) - U(i,i+1:m)*y(i+1:m,1) )/U(i,i);
end

%**************************************************

function y = utim(x) 

global A U b r 
[m,n]=size(A) 

%file utim.m  x=inv(U')*y

y=x;
for i=1:m-1 
  y(i)=y(i)/U(i,i); 
  y(i+1:m) = y(i+1:m,1) - y(i,1)*U(i,i+1:m)'; 
end 

y(m)=y(m)/U(m,m);

%********************************************

function y = btim(x) 

global A U b r 

%file btim.m  y = inv(B')*x = B*inv(U)*inv(U')*x 

y=utim(x);
y=uim(y);
y=bm(y); 

%***********************************************

function y = btm(x) 

global A U b r 
[m,n]= size(A); 

%file btm.m   y=B'*x 

for i=1:m 
  a=A(:,b(i)); 
  y(i,1)=a'*x;  
end 

%******************************************************

function  y = bim(x) 

global A U b r 

% file bim.m y= inv(B)*x =inv(U)*inv(U')*B'*x , B=QU  

y=btm(x);
y=utim(y);
y=uim(y);

%***************************************************

function y = bm(x) 

global A U b r 
[m,n]= size(A); 

%file bm.m   y=B*x  

y=zeros(m,1);
for j=1:m 
  a=A(:,b(j));
  y=y+x(j)*a;
end
 
%**************************************************

\end{verbatim} 
\end{small}
\renewcommand{\baselinestretch}{1.1}
\parskip 0.2cm
\pagebreak  

\section{depvital.m} 
 
\renewcommand{\baselinestretch}{0.84}
\parskip 0.0cm
\begin{small} 
\begin{verbatim}  

function fus= rank2(a1,a2) 

% F(t) is the component's cumulative 
% life probability distribution 
% F(t) = Pr(l<=t)
% Its complement is the survival probability distribution 
% Fc(t) = 1-F(t) = Pr(l>t) 
% The failure probability at the next 
% period x given the survival up to 
% current time t is 
% F(x|t) = (F(t+x)-F(t))/Fc(t) 
% = 1 -Fc(x|t)
% The failure rate, hazard rate or 
% force of mortality at age t is 
% h(t) = f(t)/Fc(t)  
% Integrating 
% I[0:x] h(t)dt = -log(Fc(x)) 
% Fc(x) = exp(-H(x))   
% H(x) = I[0:x] h(t)dt 

% A(:,1)= age 
% A(:,2)= h(t) 

nx=100; 
%maximum age at life table 

% generates test assuring h(nx)==1; 
% a= 1:nx; h= (1/nx)*a; h=h.^5;  %plot(h); 

% f= life density; h= haz.rate;
% a= age; c=complement; u=cumulative 
aux=0; 
for i=1:nx 
  aux= aux +h(i); 
  hu(i)= aux; 
  fuc(i)= exp(-hu(i)); 
  fu(i)= 1-fuc(i); 
end 

% 2 lifelong dependents    
% ak= current age of k-th depend  
% Xk= surviv. of k-th depend.  
% R2= sup{X1,X2} R1= inf{X1,X2}     
% Pr(R2<=t|a1,a2).
% = Pr(X1<=t|a1 and X2<=t|a2) 
% Pr(R1<=t|a1,a2)  
% = Pr(X1<=t|a1 or X2<=t|a2) 
% Pr(R1>t|a1,a2) 
% = Pr(X1>t|a1 and X2>t|a2) 

for t=1:100 
  if( (a1+t)>nx ) 
    fua1(t)=1; 
  else      %Pr(X1<=t|a1) 
    fua1(t) = ... 
    ((fu(a1+t)-fu(a1))/fuc(a1)); 
  end 
  if( (a2+t)>nx ) 
    fua2(t)=1; 
  else 
    fua2(t) = ... 
    ((fu(a2+t)-fu(a2))/fuc(a2)); 
  end 
  f2u(t)= fua1(t)*fua2(t);  
  f1u(t)= fua1(t) +fua2(t) -f1u(t); 
end 
fus=[fu;f2u;f1u;fua1;fua2]; 
plot(a',fu','--b',a',f2u,'-r',a',f1u,
..'-r',a',fua1','--k',a',fua2','--k'); 
title( ... 
..['Order statistics for survival', ... 
  int2str(a1),' and ',int2str(a2)]);  

%**************************************************

\end{verbatim} 
\end{small}
\renewcommand{\baselinestretch}{1.1}
\parskip 0.2cm

\pagebreak

\section{GRG} 
 
\renewcommand{\baselinestretch}{0.92}
\parskip 0.0cm
\begin{small} 
\begin{verbatim}

% Gradiente Reduzido Generalizado
% 
% Eduardo Oda   BMACC  2004 
% 
% Fazem parte desse pacote duas funcoes:
% 1. grg.m
%   Implementa o metodo em questao
% 
% 2. buscalinear.m
%   Implementa uma busca linear simples
% 
% Adicionalmente, seguem diretorios nomeados teste#, 
% onde '#' eh um numero, com exemplos para teste.
% 
% Para utilizar o exemplo, faca:
% 
%   $ cp teste1/* .
%   $ octave teste.m
% 
% Para definir um problema diferente, edite os arquivos de um dos 
% exemplos, copie-os para o mesmo diretorio dos arquivos principais 
% (grg.m e buscalinear.m) e execute:
%
%  $ octave teste.m
%%%%%%%%%%%%%%%%%%%%%%%%%%%%%%%%%%%%%%%%%%%%%%%%%%%%%%%%%%%%%%%%%%%%

%  Esta eh uma funcao em octave versao 2.0.16.92 que encontra um 
%  ponto x do R^n que seja uma solucao otima do problema
%  min  f(x)
%  s.a.  h(x)=0
%    l<=x<=u
%
%  com o metodo dos gradientes reduzidos generalizados (GRG),
%  recebendo como parametro:
%    - f(x) , f:R^n->R
%    - h(x), h:R^n->R^m
%    - l (n x 1)
%    - u (n x 1)
%
%  Adicionalmente a funcao recebe a variavel trace, que controla a 
%  exibicao das iteracoes, mostrando cada iteracao se valer 1 ou 
%  ocultando-as se valer 0.
%
%  As funcoes f e h, bem como seu gradiente (gradf) e jacobiano (jach),
%  respectivamente, devem ser definidas no octave. Preferencialmente devem
%  ser criados quatro arquivos no mesmo diretorio desta funcao:
%  a) f.m
%    define uma funcao que recebe um vetor x de dimensao [nx1] e
%    retorna o valor da funcao f calculada nesse ponto, um real.
%
%  b) gradf.m
%    define uma funcao que recebe um vetor x de dimensao [nx1] e
%    retorna o valor do gradiente da funcao f calculado nesse ponto,
%    um vetor de dimensao [nx1].
%
%  c) h.m
%    define uma funcao que recebe um vetor x de dimensao [nx1] e
%    retorna o valor da funcao h calculada nesse ponto, um vetor de
%    dimensao [mx1]. 
%
%  c) jach.m
%    define uma funcao que recebe um vetor x de dimensao [nx1] e
%    retorna o valor do jacobiano da funcao h calculado nesse ponto,
%    uma matriz de dimensao [mxn]. 
%
%  Junto com esta funcao seguem exemplos desses arquivos, use-os como base
%  para suas funcoes.
%
%  O retorno desta funcao eh um vetor (1 x 2) cujas componentes sao, na
%  ordem, o x otimo e o valor otimo.
%
%  Para utilizar essa funcao, inicie o octave, defina todos os parametros
%  descritos acima e execute o comando:
%
%  octave:> [x_otimo,v_otimo]=grg(x,l,u,trace)

function[x_otimo,v_otimo]=grg(x,l,u,trace)
  
  %%%%%%%%%%%%%%%%%%%%%%%%%%%%%%
  % Inicializando as variaveis %
  %%%%%%%%%%%%%%%%%%%%%%%%%%%%%%

  EPS=0.00000001;
  Jx=jach(x);
  [m,n]=size(Jx);
  count=1;

  %%%%%%%%%%%%%%%%%%%%%%%%%%%%%%%
  % Implementacao do metodo GRG %
  %%%%%%%%%%%%%%%%%%%%%%%%%%%%%%%

  while 1
    gradfx=gradf(x);

    % Procuramos uma matriz nao singular para ser base, dando
    % preferencia a indices cujas variaveis estejam mais distantes
    % dos bordos da caixa
    
    dif=abs(x-(u-l)/2);
    [odif i_dif]=sort(dif);
    i_base=[i_dif(1:m)];
    i_r=[i_dif(m+1:n)];
    i=m;  % indice do indice basico a ser trocado
    j=1;  % indice do indice residual a ser trocado
    while(abs(det(Jx(:,i_base)))<EPS)
      disp("");
      %desfaz
      if(i+1>m && j>1)
        aux=i_base(1);
        i_base(1)=i_r(j-1);
        i_r(j-1)=aux;
        % nao faz nada
      elseif(i+1<=m)
        aux=i_base(i+1);
        i_base(i+1)=i_r(j);
        i_r(j)=aux;
      end
      aux=i_base(i);
      i_base(i)=i_r(j);
      i_r(j)=aux;
      i=i-1;
      if(i==0)
        i=m;
        j=j+1;
        if(j>n-m)
          disp("Nao foi possivel encontrar uma base boa")
          break;
        end
      end
    end

    % Calculamos o vetor de gradientes reduzidos
%    z=gradfx(i_r)'-gradfx(i_base)'*inv(Jx(:,i_base))*Jx(:,i_r);
    z=gradfx(i_r)-gradfx(i_base)*inv(Jx(:,i_base))*Jx(:,i_r);
  
    % Definimos uma direcao para andar com a solucao viavel atual 
    v=zeros(n,1);
    i_neg=find(z<-EPS);
    if(size(i_neg)>0)
      i=find(x(i_r(i_neg))<u(i_r(i_neg)));
      v(i_r(i_neg(i)))=-z(i_neg(i))*diag(u(i_r(i_neg(i)))-x(i_r(i_neg(i))));
    end
    i_pos=find(z>EPS);
    if(size(i_pos)>0)
      i=find(x(i_r(i_pos))>l(i_r(i_pos)));
      v(i_r(i_pos(i)))=-z(i_pos(i))*diag(x(i_r(i_pos(i)))-l(i_r(i_pos(i))));
    end
    v(i_base)=-inv(Jx(:,i_base))*Jx(:,i_r)*v(i_r);
    v(find(abs(v)<EPS))=0;
    
    % Se a direcao for toda nula, o metodo termina, pois nao
    % podemos caminhar numa direcao que melhore a funcao objetivo
    if(abs(v)<EPS)
      break;
    end

    % Descobrimos o quanto devemos (e podemos) andar na direcao v
    eta(1)=0;
    e3n=inf;
    e3p=inf;
    i_neg=find(v<-EPS);
    if(size(i_neg)>0)
      e3n=min( (l(i_neg)-x(i_neg))./v(i_neg) );
    end
    i_pos=find(v>EPS);
    if(size(i_pos)>0)
      e3p=min( (u(i_pos)-x(i_pos))./v(i_pos) );
    end
    eta(3)=min(e3n,e3p);
    eta(2)=eta(3)/2;
    eta=buscalinear(eta,x,v,EPS);
    eta=eta(2);

    if(trace)
      printf("%%%%%%% Iteracao %d %%%%%%%%%%% \n",count);
      count++;
      disp("Ponto atual:")
      disp(x')
      disp("Indices da base:")
      disp(i_base')
      disp("Gradiente reduzido:")
      disp(v')
      disp("Eta:")
      disp(eta)
    end

    % Retornamos para a restricao h(x)=0 utilizando Newton-Raphson
    while eta>EPS
      x2=x+eta*v;
      Jx2=jach(x2);
  
      % Tentamos retornar para a restricao em, no maximo,
      % seis iteracoes do Newton-Raphson.
      % Segundo Lasdon, a experiencia mostra que chamadas
      % subsequentes ao Newton-Raphson geralmente nao
      % convergem.
      % Observe que isso nao interfere na generalidade do
      % metodo.
      for j=1:6
%        Dx=-inv(Jx2(:,i_base))*h(x2);
        Dx=-inv(Jx2(:,i_base))*h(x2)';
        if(abs(Dx)<EPS)
          break;
        end
        aux=x2(i_base)+Dx;
        % Verificamos se alguma restricao de caixa
        % nao seria honrada no retorno a restricao
        % h(x)=0
        if(any(aux<l(i_base)) || any(aux>u(i_base)))
          break;
        else
          x2(i_base)=aux;
          Jx2=jach(x2);
        end
      end
      % Se nao foi possivel retornar a restricao h(x)=0
      % diminuimos o tamanho do passo e fazemos uma nova
      % tentativa
      if(any(abs(Dx)>EPS))
        eta=eta/2;
      else
        x=x2;
        Jx=Jx2;
        break;
      end
    end

    % O tamanho do passo igual a zero indica que nao podemos 
    % melhorar a funcao objetivo caminhando nessa direcao,
    % entao o metodo termina
    if(eta<EPS)
      break;
    end
    x(find(abs(x)<EPS))=0;

  end
  x_otimo=x;
  v_otimo=f(x_otimo);
end
%%%%%%%%%%%%%%%%%%%%%%%%%%%%%%%%%%%%%%%%%%%%%%%%%%%%%%%%%%%%%%%%%%%%

%  Esta eh uma funcao em octave versao 2.0.16.92 que minimiza f(x+e*v),
%  com x e v fixos, fazendo uma busca linear.
%
%  Recebe como parametro os vetores x e v, ambos de dimensao [nx1], o
%  vetor eta, de dimensao [3x1], e um valor, abaixo do qual um numero 
%  em modulo eh considerado zero.
%
%  O vetor eta deve ser da forma eta(1)<=eta(2)<=eta(3).
%
%  Retorna o vetor eta de forma que min(f(x+e*v))=f(x+eta(2)*v)

function eta=buscalinear(eta,x,v,EPS)
  vf=[
    f(x+eta(1)*v);
    f(x+eta(2)*v);
    f(x+eta(3)*v);
  ];
  sf=vf(1)+vf(2)+vf(3);

  while 1 
    if(vf(2)-vf(1)<=EPS)
      if(vf(2)-vf(3)<=EPS)
        if(abs(vf(2)-vf(1))<EPS && abs(vf(2)-vf(3))<EPS)
          eta(2)=eta(3);
          break;
        end
        numerador=(
          vf(1)*(eta(3)^2-eta(2)^2)+
          vf(2)*(eta(1)^2-eta(3)^2)+
          vf(3)*(eta(2)^2-eta(1)^2)
          );
        denominador=(
          vf(1)*(eta(3)-eta(2))+
          vf(2)*(eta(1)-eta(3))+
          vf(3)*(eta(2)-eta(1))
          );
        eta4=numerador/(2*denominador);
        if(abs(eta4-eta(2))<EPS)
          break;
        end
        vf4=f(x+eta4*v);
        if(eta4<eta(2))
          eta(3)=eta(2);
          vf(3)=vf(2);
          eta(2)=eta4;
          vf(2)=vf4;
        else
          eta(1)=eta(2);
          vf(1)=vf(2);
          eta(2)=eta4;
          vf(2)=vf4;
        end
      else
        eta4=(eta(3)+eta(2))/2;
        if(abs(eta4-eta(2))<EPS)
          break;
        end
        vf4=f(x+eta4*v);
        eta(1)=eta(2);
        vf(1)=vf(2);
        eta(2)=eta4;
        vf(2)=vf4;
      end
    else
      eta4=(eta(1)+eta(2))/2;
      if(abs(eta4-eta(2))<EPS)
        break;
      end
      vf4=f(x+eta4*v);
      eta(3)=eta(2);
      vf(3)=vf(2);
      eta(2)=eta4;
      vf(2)=vf4;
    end
    sfn=vf(1)+vf(2)+vf(3);
    if((sf-sfn)<EPS)
      break;
    else
      sf=sfn;
    end
  end
end
%%%%%%%%%%%%%%%%%%%%%%%%%%%%%%%%%%%%%%%%%%%%%%%%%%%%%%%%%%%%%%%%%%%%

function y=f(x)
  y=-2*x(1)-4*x(2)+x(1)^2+x(2)^2+5;
end
        
function y=gradf(x)
  y(1)=-2+2*x(1);
  y(2)=-4+2*x(2);
  y(3)=0;
  y(4)=0;
end
                                
function y=h(x)
  y(1)=-x(1)+2*x(2)+x(3)-2;
  y(2)=x(1)+x(2)+x(4)-4;
end
                
function y=jach(x)
  y=[
  -1 2 1 0;
   1 1 0 1
  ];
end

l=[0;  0;  0;  0];
x=[0;  0;  2;  4];
u=[10;  10;  10;  10];

[x_otimo,v_otimo]=grg(x,l,u,0)

\end{verbatim} 
\end{small}
\renewcommand{\baselinestretch}{1.1}
\parskip 0.2cm
\pagebreak